\documentclass[aps,prd,twocolumn,preprintnumbers,showpacs,nofootinbib,amssymb]{
revtex4}
\usepackage{graphicx}
\usepackage{amsmath}
\usepackage{amssymb}
\usepackage{amsfonts}
\usepackage{bm}
\usepackage{color}

\def\be{\begin{equation}}
\def\ee{\end{equation}}
\def\bea{\begin{eqnarray}}
\def\eea{\end{eqnarray}}

\begin{document}

\title{Caloric curves of self-gravitating fermions in general
relativity}
\author{Giuseppe Alberti}
\affiliation{Laboratoire de Physique Th\'eorique,
Universit\'e Paul Sabatier, 118 route de Narbonne  31062 Toulouse, France}
\affiliation{
Living Systems Research, Roseggerstra\ss e 27/2, A-9020 Klagenfurt am
W\"{o}rthersee,
Austria}
\author{Pierre-Henri Chavanis}
\affiliation{Laboratoire de Physique Th\'eorique,
Universit\'e Paul Sabatier, 118 route de Narbonne  31062 Toulouse, France}

\begin{abstract}

We study the nature of phase transitions between gaseous and
condensed states
in the self-gravitating
Fermi gas at nonzero temperature in general relativity. The condensed states can
represent compact
objects such as white dwarfs, neutron stars, or dark matter fermion balls. The
caloric curves depend on two parameters:
the system size  $R$ and the particle number $N$. When $N<N_{\rm OV}$, where
$N_{\rm OV}$ is the Oppenheimer-Volkoff limit, there exists an equilibrium
state for any value of the temperature $T$ and of the energy $E$ as in the
nonrelativistic case [P.H. Chavanis, Int. J. Mod. Phys. B {\bf 
20}, 3113 (2006)]. Gravitational collapse is prevented
by quantum mechanics (Pauli's exclusion principle). When  $N>N_{\rm OV}$, there
is
no equilibrium state below a critical energy and below a critical
temperature. In that case, the system is expected to collapse
towards a
black hole. We plot the caloric curves of the general
relativistic Fermi gas, study the different types of phase
transitions that occur in the system, and determine the phase diagram
in the $(R,N)$ plane. The nonrelativistic results are recovered for
$N\ll N_{\rm OV}$ and $R\gg R_{\rm OV}$ with $NR^3$ fixed.  The classical
results are recovered for
$N\gg N_{\rm OV}$ and $R\gg R_{\rm OV}$ with $N/R$ fixed. We
discuss the commutation of the limits  $c\rightarrow
+\infty$ and $\hbar\rightarrow 0$. We study the relativistic
corrections to
the nonrelativistic caloric curves and the quantum corrections to the classical
caloric curves. We
highlight a situation of physical interest where a gaseous Fermi gas, by
cooling, first
undergoes a phase
transition towards a
compact object (white dwarf,
neutron star, dark matter fermion ball), then collapses into a black hole.
This situation occurs in the microcanonical ensemble when
$N_{\rm OV}<N<3.73\, N_{\rm OV}$.
We also relate the phase transitions from a gaseous state to a core-halo state 
in the microcanonical ensemble to the onset of red-giant structure  and to the 
supernova phenomenon.

\end{abstract}


\maketitle

\section{Introduction}

The study of phase transitions is an important problem in physics. Some examples
include solid-liquid-gas phase transitions,  superconducting and superfluid
transitions, Bose-Einstein
condensation, liquid-glass phase transition in
polymers, liquid crystal phases, Kosterlitz-Thouless transition etc.
Self-gravitating systems also undergo phase transitions but they are special due
to the unshielded long-range attractive
nature of the interaction \cite{paddy}. This leads to unusual
phenomena
such as negative specific heats, ensembles inequivalence, long-lived
metastable states, and gravitational collapse. A strict equilibrium
state can
exist only if the system is confined within a box, otherwise it has the tendency
to evaporate (this is already the case for an ordinary gas). On the other hand,
in order to define a
condensed phase we
need
to introduce a short-range
repulsion between the particles that opposes itself to the gravitational
attraction.\footnote{Without small-scale regularization, there is no equilibrium
state (global entropy maximum) in a strict sense \cite{antonov}. There can
exist, however, metastable gaseous states (local entropy maxima) that are
insensistive to the small-scale regularization
\cite{antonov,lbw}. These metastable
states  have a very long lifetime, scaling as $e^N$, where $N$ is the number of
particles in the system \cite{lifetime}. In practice this lifetime is much
larger
than the age of the Universe, making the metastable states fully relevant
in astrophysics \cite{ijmpb}.} In this paper, we
consider the case of self-gravitating fermions where an effective short-range
repulsion is due to quantum mechanics (Pauli's exclusion principle). The
object of this paper is to present a complete description of phase transitions
in the self-gravitating Fermi gas in general relativity. This study can have
applications in relation to the formation of compact objects such as white
dwarfs, neutron stars, dark matter stars, black holes etc. On the other
hand, the phase
transition from a gaseous state to a condensed state may be related to the
onset of red-giant structure  and to the  supernova phenomenon. We first start
by reviewing the literature on the subject. We focus our review on papers
that study phase transitions in the box-confined self-gravitating Fermi gas at
nonzero temperature.\footnote{The case of completely
degenerate self-gravitating fermions at $T=0$ and the case of  classical
(nondegenerate) self-gravitating systems are considered in our companion papers
\cite{paper1,paper2} where a detailed review of the literature is made.} We do
not review the immensely
vast literature related to
self-gravitating
fermions as models of white
dwarfs, neutron
stars, and dark matter halos. For a connection to this
literature, we refer to
\cite{chandrabook,shapiroteukolsky,btv,vss,rar,clm1,clm2, urbano,rsu} and
references therein. For a connection to the general 
literature on the statistical mechanics of
self-gravitating systems and systems with long-range interactions we refer to
the introduction of \cite{sd} and to the reviews
\cite{paddy,houches,katzrevue,ijmpb,cdr,campabook}.

The statistical mechanics of nonrelativistic self-gravitating
fermions at nonzero temperature enclosed within a box of radius $R$ was
first
studied by Hertel \&
Thirring (1971) \cite{htf}. They worked in the canonical ensemble and
rigorously proved that the mean field approximation (or effective field
approximation) and the Thomas-Fermi (TF)
approximation (which amounts to neglecting the quantum
potential) become exact in
a suitable thermodynamic limit $N\rightarrow +\infty$ where
$R\sim N^{-1/3}$, $T\sim N^{4/3}$,  $E\sim N^{7/3}$, $S\sim N$, and $F\sim
N^{7/3}$ (the scaling
$F\sim N^{7/3}$ was first obtained by
L\'evy-Leblond (1969) \cite{levyleblond} for the ground
state).\footnote{This is also equivalent to the usual
thermodynamic limit $N\rightarrow +\infty$ where $R\sim N^{1/3}$, $T\sim 1$,
$E\sim N$, $S\sim
N$ and $F\sim N$ with $G\sim N^{-2/3}$ (see Appendix \ref{sec_tl}).} This leads
to
the temperature-dependent TF equation.\footnote{It can be obtained
by combining the fundamental equation of hydrostatic equilibrium with the
Fermi-Dirac equation of state or, equivalently, by
substituting the Fermi-Dirac density into the Poisson equation;
see, e.g., 
Hertel (1977) \cite{hseul}. For that reason, the temperature-dependent TF
equation is sometimes called the Fermi-Dirac-Poisson
equation.}
The existence of the
TF limit for the thermodynamic functions of self-gravitating
fermions was proven by Hertel {\it et al.} (1972) \cite{hnt} for the
microcanonical and
canonical ensembles and by Messer (1979)
\cite{messer} for the grand canonical ensemble. The
convergence of the
quantum-statistically defined particle density towards the TF density was
proven by Baumgartner (1976) \cite{baumgartner}. He also showed that
there are no correlations in the thermodynamic limit. Narnhofer and Sewell
(1980) \cite{ns1} showed that
when
$N\rightarrow +\infty$ the equilibrium Gibbs distribution becomes a
tensor product of density functions of an ideal Fermi gas which minimize the TF
free energy functional. These density functions can be stable (global minima) or
metastable (local minima). Finally, Narnhofer and Sewell (1982) \cite{ns2}
showed
that  when
$N\rightarrow +\infty$ a quantum system of self-gravitating fermions is
described by the {\it classical} Vlasov equation \cite{bh}.

Hertel \& Thirring (1971) \cite{ht} studied numerically phase transitions in
the nonrelativistic self-gravitating Fermi gas in relation with the structure of
neutron stars.\footnote{The possibility of phase
transitions in
the self-gravitating Fermi gas at nonzero temperature was suggested in
the Appendix IV of Lynden-Bell and Wood (1969) \cite{lbw}.}
They assumed that the gas is enclosed within a box
and worked in the canonical ensemble. For a given number of particles $N$,
they
showed that a canonical first order phase transition arising from a  
multiplicity of solutions in the TF equation appears if the radius of
the
box is larger than a certain value
$R_{\rm CCP}(N)=12.8\, \hbar^2/(N^{1/3}Gm^3)$. This phase transition is
characterized by a
jump of energy (the energy $E=\partial(\beta F)/\partial\beta$, the
first derivative of $\beta F$
with respect to $\beta$, becomes
discontinuous) at a transition
temperature $T_t$ determined by a Maxwell construction
like in the theory of 
the van der Waals gas.\footnote{The phase transition
arises because the TF equation has two stable solutions  at the same
temperature that minimize the TF free energy. A rigorous
analytical
proof for the existence of this phase transition was given by Messer
(1981a,1981b) \cite{messerpt1,messerpt2} following numerical calculations by
Hertel (1977)
\cite{hseul}. When there are multiple
solutions
in the TF equation, they argue that one must choose the one
with the smallest value of free
energy.} This
corresponds to a transition
between a nearly homogeneous phase of
medium mass density (gaseous phase) and a phase with a high density core
surrounded by an
atmosphere of low density (condensed phase) when the system cools down below
$T_t$. Hertel \& Thirring (1971) \cite{ht} explained that this phase
transition replaces
the region of negative specific heats in the microcanonical ensemble (or
the piece of convex curvature in the entropy curve $S(E)$) which is associated
with unstable equilibrium states in the canonical ensemble. Therefore, the
microcanonical and canonical
ensembles are not equivalent \cite{hnt}. The region of negative specific heat in
the microcanonical ensemble is bridged by a phase transition in the canonical
ensemble.\footnote{Canonical phase transitions, associated with negative
specific heats, have also been found by Thirring (1970) \cite{thirring} in a
toy model of self-gravitating systems, by Aronson and
Hansen (1972) \cite{ah} for a self-gravitating hard spheres gas, by Carlitz
(1972) \cite{carlitz} for
hadronic matter, and by Hawking (1976) \cite{hawking} for black holes.}
Hertel \& Thirring (1971) \cite{ht}  applied their crude model of
neutron stars to a system of $N=10^{57}$ neutrons (the corresponding mass being
of the order of the solar mass) initially contained in a sphere of radius
$R=100\, {\rm km}$. The critical radius is $R_{\rm CCP}=43.1\, {\rm km}$. For
$R=100\, {\rm km}>R_{\rm CCP}$, the system undergoes a first order phase
transition below a
critical temperature $T_t=7.03\times 10^{10}\, {\rm K}$, collapses, and forms a
compact object (neutron star) containing almost all the mass. This compact
object has  approximately the same size, $R_C=4.51\,
\hbar^2/(N^{1/3}Gm^3)=15.1\, {\rm km}$, as a completely degenerate Fermi gas at
$T=0$  (equivalent, in their nonrelativistic
model, to a polytrope of index $n=3/2$) but it is surrounded by a
small isothermal atmosphere. This
gravitational phase transition could account for the implosion of the core in
the supernova phenomenon where the energy
is carried quickly by neutrinos.\footnote{Thirring (1970) \cite{thirring},
Hertel and Thirring (1971) \cite{ht} and Messer (1981) \cite{messerpt2} mention
the analogy between this phase transition and the formation of red giants and
supernovae. However, this analogy may not be fully correct because the phase
transition that they obtain just corresponds to an implosion. This is because
they work in the canonical ensemble and consider relatively small systems while
the phase transition leading to an implosion-explosion
phenomenon, associated with a core-halo structure, occurs in the microcanonical
ensemble for larger systems (see Ref. \cite{supernova} and Sec.
\ref{sec_ap}). Lynden-Bell and Wood (1968)
\cite{lbw}, considering a classical self-gravitating gas in the microcanonical
ensemble, find the emergence of a core-halo structure and relate it to the onset
of red giants.}

Gravitational phase transitions of fermionic matter were also studied
by Bilic \& Viollier (1997) \cite{bvn} in a cosmological
setting. They considered weakly interacting massive fermions of mass $17.2\,
{\rm keV/c^2}$ in the
presence of a large radiation-density background fixing the
temperature. They studied a halo of mass $M=10^9\, M_{\odot}$ and radius
$R=1.68\times 10^{-2}\, {\rm pc}>R_{\rm CCP}=6.00\times 10^{-3}\, {\rm pc}$.
When
the system cools down below a transition temperature $T_t=4.80\times 10^5\,
{\rm K}$, a condensed phase
emerges consisting of quasidegenerate supermassive fermion stars of mass $M\sim
10^9\, M_{\odot}$ and radius $R_C=2.10\times 10^{-3}\, {\rm pc}$.
They argued that these compact dark objects could play an important
role in structure formation in the early Universe. In particular,
these fermion stars  could explain, without
resorting to the
black hole hypothesis, some of the features observed around
supermassive compact dark objects which are reported to exist at the
centers of a number of galaxies including our own and quasistellar objects
(QSOs). On a technical point of view, their study is
analogous to the one
carried
out
by Hertel \& Thirring (1971) \cite{ht} for
neutron stars, i.e.,
they described the  canonical first order phase transtion between a ``gaseous''
phase and a
``condensed'' phase that appears below a transition temperature when the size of
the object is sufficiently large.

A detailed theoretical description of phase transitions in the
nonrelativistic self-gravitating Fermi gas
at nonzero temperature was given by Chavanis (2002) \cite{ijmpb}
(see also Refs. \cite{csmnras,pt,dark,ispolatov,rieutord,ptd}).\footnote{In
these papers, the statistical equilibrium state is
obtained by maximizing the Fermi-Dirac entropy $S$ at fixed mass $M$ and
energy $E$ in the microcanonical ensemble  and by minimizing  the Fermi-Dirac
free energy $F=E-TS$ at fixed mass $M$ in the canonical ensemble, where $S$ is
obtained from a combinatorial analysis taking into account the Pauli exclusion
principle. This leads to the TF (or Fermi-Dirac-Poisson) equation in a direct
manner. The study of the self-gravitating Fermi gas has also
applications in the statistical theory of violent relaxation developed by
Lynden-Bell \cite{lb} that also leads to a Fermi-Dirac-type
distribution \cite{csmnras}.} He showed that the caloric
curves $T(E)$ depend on a single control parameter
$\mu=\eta_0\sqrt{512\pi^4G^3MR^3}$
with $\eta_0=gm^4/h^3$ ($g$ is the spin multiplicity of the quantum states). For
a fixed particle number $N$, this paramerer can be seen as a measure of the size
of the system since $\mu\propto R^{3/2}$. Chavanis
\cite{ijmpb} studied in detail the nature of phase
transitions in the nonrelativistic self-gravitating Fermi gas in both
microcanonical
and canonical ensembles. He showed that there exist two
critical points (one in each ensemble) at which zeroth and first order phase
transitions
appear. The canonical critical point $\mu_{\rm CCP}=83$ at which canonical phase
transitions
appear is equivalent to the one previously found by Hertel and Thirring
(1971) \cite{ht}. The 
microcanonical critical point $\mu_{\rm MCP}=2670$  at which microcanonical
phase transitions
appear was not found previously. For $\mu\rightarrow +\infty$, one recovers the
caloric curve  of a nonrelativistic self-gravitating classical gas \cite{paddy}.
Chavanis
\cite{ijmpb,lifetime} argued  that first order phase transitions do not take
place in practice, contrary to previous claims \cite{ht,bvn}, because of the
very
long lifetime of metastable states for systems with long-range interactions.
Therefore, only zeroth order phase transitions take place at the spinodal points
where the metastable branches disappear. Recently, this study of phase
transitions was extended to the
nonrelativistic fermionic King model \cite{clm1,clm2}. This
model is more realistic as it avoids the need of an artificial box to confine
the system.

Gravitational phase transitions of fermionic matter in general relativity were
studied by Bilic and Viollier (1999) \cite{bvr}.\footnote{In
that case, the suitable thermodynamic limit corresponds to $N\rightarrow
+\infty$ where $R\sim N^{2/3}$, $T\sim N^{-1/3}$,  $E\sim N^{2/3}$, $S\sim
N$ and $F\sim N^{2/3}$ with $m\sim N^{-1/3}$ \cite{bvr}. This is also equivalent
to the usual
thermodynamic limit $N\rightarrow +\infty$ where $R\sim N^{1/3}$, $T\sim 1$,
$E\sim N$, $S\sim
N$ and $F\sim N$ with $G\sim N^{-2/3}$ (see Appendix \ref{sec_tl}).} They showed
that, at
some
critical temperature $T_t$, weakly
interacting massive fermionic matter with a total mass below the
Oppenheimer-Volkoff (OV) limit \cite{ov}
undergoes a first order gravitational phase transition from a diffuse to a
clustered state, i.e., a nearly completely degenerate fermion star. This is an
extension of
their previous paper \cite{bvr} in the Newtonian approximation. This
relativistic extension allowed them to consider situations where the mass of the
system is close to the OV limit so that the fermion star is strongly
relativistic. For fermions masses of $10$ to $25\, {\rm keV/c^2}$
they argued
that these fermions stars may well provide an alternative explanation for the
supermassive compact dark objects that are observed at galactic centers. Indeed,
a few Schwarzschild radii away from the object, there is little difference
between a supermassive black hole and a fermion star of the same mass near the
OV limit.\footnote{Some difficulties with the ``fermion ball''
scenario to provide an alternative to supermassive black holes at the centers of
the galaxies are pointed out in \cite{genzel}.} In their paper, they
considered fermionic particles of mass $m=17.2\,
{\rm keV/c^2}$ for which $N_{\rm
OV}=1.4254\times 10^{71}$, $N_{\rm
OV}m=2.1973\times 10^9\, M_{\odot}$, $M _{\rm OV}=2.1186\times 10^9\,
M_{\odot}$
and $R_{\rm OV}=8.88\times 10^{-4}\, {\rm pc}$. They studied a system of 
$N=0.95350\, N_{\rm OV}$ fermions, corresponding to a rest mass
$N m=2.0951\times 10^9\, M_{\odot}$ which is slightly
below the OV limit, in a sphere of size $R=29.789\, R_{\rm OV}=2.6391\times
10^{-2}\, {\rm pc}$.  The transition occurs at
$T_t=0.0043951\, mc^2=8.7725\times 10^5\, {\rm K}$. This leads to a fermion star
containing almost all the particles surrounded by a small atmosphere. If we
approximate the fermion star by a Fermi gas at $T=0$ containing all the rest
mass $\sim 2.0951\times 10^9\, M_{\odot}$, we find a radius $R_C=1.220\,
R_{\rm OV}=1.0809\times 10^{-3}\, {\rm pc}$ and a mass $M_C=0.9577\, M_{\rm
OV}=2.0290\times 10^9\, M_{\odot}$. 

The study of Bilic and Viollier \cite{bvr} is restricted to a unique value of
$R$ and $N$, with $N<N_{\rm OV}$, leading to a canonical phase transition. The
object of this paper is to perform a more general study of  phase transitions
in the self-gravitating Fermi gas in general relativity for arbitrary values of
$R$ and $N$. In particular, we would like to determine what
happens when $N>N_{\rm OV}$, or what happens for larger values of $R$ 
where a
microcanonical phase transition is expected.

The paper is organized as follows. In Sec. \ref{sec_basic}, we present the basic
equations describing a general relativistic Fermi gas at statistical equilibrium
in a box. In Sec. \ref{sec_cc}, we expose general notions concerning the
construction of the caloric curves and the description of phase transitions. In
Sec. \ref{sec_pl}, we recall the results previously obtained in the
nonrelativistic and classical limits. In Sec.
\ref{sec_first}, we consider the case $R_{\rm CCP}<R<R_{\rm MCP}$ where the
system undergoes a canonical phase transition from a gaseous phase to a
condensed phase when $N_{\rm CCP}<N<N_{e}$. In Sec. \ref{sec_second}, we
consider the case $R>R_{\rm MCP}$ where the system undergoes
a canonical phase transition   when
$N_{\rm CCP}<N< N_{e}$ and a microcanonical phase transition when 
$N_{\rm MCP}<N<N_{f}$ (we find that $N_{e}\sim N_{\rm OV}$ and  $N_{f}\sim
3.73\, N_{\rm OV}$). In Sec. \ref{sec_vlr}, we consider the case
of very large radii  $R\gg R_{\rm MCP}$ where extreme core-halo configurations
with a high central density appear. They correspond to the solutions computed in
\cite{btv,rar,clm2} in connection to the ``fermion ball'' scenario.
However, following \cite{clm2}, we point out that these solutions are
thermodynamically unstable (hence very unlikely). In
Secs. \ref{sec_dix} and \ref{sec_un},  we consider the
cases $R_{\rm OV}<R<R_{\rm CCP}$ and $R<R_{\rm OV}$ 
where there is no phase transition. In Sec. \ref{sec_rn}, we present the
complete phase diagram of the  general relativistic Fermi gas in the $(R,N)$
plane. In Sec. \ref{sec_ncl}, we
recover the nonrelativistic  and classical results as particular limits of our
general study and we discuss the commutation of the limits $\hbar\rightarrow 0$
and $c\rightarrow +\infty$. In Sec. \ref{sec_rqc}, we study the relativistic
corrections to
the nonrelativistic caloric curves and the quantum corrections to the classical
caloric curves. In Sec. \ref{sec_ap}, we consider astrophysical applications of
our results
in relation to the formation of white dwarfs, neutron stars, dark matter
fermion stars, and black holes. We also connect the phase transitions found in
our study with the onset of the red-giant structure and with the  supernova
phenomenon.

\section{Basic equations of a general relativistic Fermi gas}
\label{sec_basic}

In this section, we give the basic equations describing the structure of a
general relativistic Fermi gas at nonzero temperature (see 
\cite{bvcqg,bvr,papiertheorique} for their derivation). Using the
normalized
variables introduced in Appendix \ref{sec_dq}, the local number
density $n(r)$, the energy density $\epsilon(r)$, the pressure $P(r)$ and
the temperature
$T(r)$ are related to the gravitational potential $\Phi(r)$ by 
\begin{equation}
\label{b1}
n(r)=\frac{1}{\pi^2}\int_0^{+\infty}
\frac{y^2\,
{\rm d}y}{1+e^{-\alpha}e^{|\alpha|\sqrt{(y^2+1)/(\Phi(r)+1)}}},
\end{equation}
\begin{equation}
\label{b2}
\epsilon(r)=\frac{1}{\pi^2}\int_0^{+\infty}
\frac{y^2\sqrt{1+y^2}\,
{\rm d}y}{1+e^{-\alpha}e^{|\alpha|\sqrt{(y^2+1)/(\Phi(r)+1)}}},
\end{equation}
\begin{equation}
\label{b3}
P(r)=\frac{1}{3\pi^2}\int_0^{+\infty}
\frac{y^4\,
{\rm d}y}{\sqrt{1+y^2}\left
\lbrack 1+e^{-\alpha}e^{|\alpha|\sqrt{(y^2+1)/(\Phi(r)+1)}}\right \rbrack},
\end{equation}
\begin{equation}
\label{b4}
T(r)=\frac{1}{|\alpha|}\sqrt{\Phi(r)+1},
\end{equation}
where
\begin{equation}
\label{b5}
\alpha=\frac{\mu(r)}{T(r)}=\frac{\mu_{\infty}}{T_{\infty}}
\end{equation}
is a quantity that is uniform throughout the system. These equations define the
equation of state of a relativistic Fermi gas in parametric form.

The Tolman-Oppenheimer-Volkoff (TOV) equations, which correspond to the 
equations of hydrostatic equilibrium
in general relativity, can be written as
\begin{equation}
\label{b6}
\frac{{\rm d}\Phi}{{\rm d}r}=-2\left\lbrack \Phi(r)+1\right \rbrack
\frac{M(r)+4\pi P(r)r^3}{r^2
\left \lbrack 1-\frac{2M(r)}{r}\right \rbrack},
\end{equation}
\begin{equation}
\label{b7}
\frac{{\rm d}M}{{\rm d}r}=4\pi\epsilon(r)r^2,
\end{equation}
where $M(r)$ is the mass-energy within the sphere of radius $r$.
They have to be
solved with the  boundary conditions
\begin{equation}
\label{b8}
M(0)=0,\qquad \Phi(0)=\Phi_0>-1.
\end{equation}
We assume that the system is confined within a box of radius $R$. The total
mass of the gas and the  total particle number are given by
\begin{equation}
\label{b9}
M=M(R)=\int_0^R \epsilon(r)4\pi r^2\, {\rm d}r,
\end{equation}
\begin{equation}
\label{b10}
N=\int_0^R n(r)\left \lbrack 1-\frac{2M(r)}{r}\right \rbrack^{-1/2}4\pi r^2\,
{\rm d}r.
\end{equation}
The temperature at infinity is given
by 
\begin{equation}
\label{b11}
T_{\infty}=T(R)\left
(1-\frac{2M}{R}\right )^{1/2},
\end{equation}
where $T(R)$ is the temperature of the system on the edge of the box. Using Eq.
(\ref{b4}), we
obtain
\begin{equation}
\label{b12}
T_{\infty}=\frac{1}{|\alpha|}\sqrt{\Phi(R)+1}\left
(1-\frac{2M}{R}\right )^{1/2}.
\end{equation}
The entropy is given by
\begin{equation}
\label{b13}
S=\int_0^R \frac{P+\epsilon}{T}\left \lbrack 1-\frac{2M(r)}{r}\right
\rbrack^{-1/2}4\pi
r^2\, {\rm d}r-\alpha N.
\end{equation}
Finally, the free energy is given by
\begin{equation}
\label{b14}
F=E-T_{\infty} S,
\end{equation}
where  $E=M-N$ is the binding energy.\footnote{The binding energy is usually
defined as $E_{\rm b}=Nmc^2-Mc^2$. Here, for convenience, we define it with the
opposite sign, i.e., $E=Mc^2-Nmc^2$. In the Newtonian limit, $M\simeq Nm$ and
$E$ reduces to the usual energy $E=K+W$ which is the sum of the kinetic and
potential (gravitational) energies.}

\section{Caloric curves and phase transitions}
\label{sec_cc}

In order to study the phase transitions in the general relativistic
Fermi gas we have to determine the caloric curves
$T_{\infty}(E)$ relating the temperature at infinity $T_{\infty}$ to the energy
$E$. These caloric curves depend on two parameters $R$ and $N$. The manner to
obtain these caloric curves is detailed in Appendix \ref{sec_conn}. In order to
make the connection
with the nonrelativistic results \cite{ijmpb}, we shall plot the caloric curves
in
terms of the dimensionless parameters $\eta$ (inverse temperature) and $\Lambda$
(minus energy) defined by
\begin{equation}
\label{b15}
\eta=\frac{\beta GNm^2}{R} \qquad {\rm and}\qquad  \Lambda=\frac{-ER}{GN^2m^2},
\end{equation}
where $\beta=1/(k_{\rm B} T_{\infty})$ and $E=Mc^2-Nmc^2$. In terms of
our normalized  variables, they reduce to
\begin{equation}
\label{b16}
\eta=\frac{\beta N}{R} \qquad {\rm and}\qquad \Lambda=\frac{-ER}{N^2},
\end{equation} 
where $\beta=1/T_{\infty}$ and $E=M-N$.  We shall therefore plot
the caloric curves $\eta(\Lambda)$ as a function of $R$
and $N$.

We recall that for systems with long-range interactions, such as
self-gravitating systems, the statistical ensembles are not equivalent. In this
paper, we shall consider the microcanonical and canonical ensembles
separately.

In the microcanonical ensemble, the system is isolated so that its energy $E$
is conserved. It serves as a control parameter. A stable equilibrium state is a
(local) maximum of entropy $S$ at fixed energy $E$ and particle number $N$.
A minimum, or a saddle point, of entropy is unstable. The global maximum of
entropy corresponds to the most probable state (the one
that is the most represented at the microscopic level). The
microcanonical caloric curve gives
the temperature at infinity $1/T_{\infty}=\partial S/\partial E$ as a function
of the energy $E$.

In the
canonical ensemble, the system is in contact with a heat bath so that its
temperature at infinity $T_{\infty}$ is fixed. It serves as a control
parameter. A stable equilibrium state is a
(local) minimum of free energy $F$ at fixed temperature $T_{\infty}$ and
particle number $N$. A maximum, or a saddle point, of free energy is unstable.
The global minimum  of free energy corresponds to the most probable state.
The caloric curve gives
the average energy $E=\partial(\beta F)/\partial\beta$ as a function
of the temperature at infinity $T_{\infty}$.

The equilibrium states are the same in the microcanonical and
canonical ensembles. This is because an extremum (first variations) of entropy
at fixed energy and particle number coincides with an extremum of free energy at
fixed particle number. However, their stability  (second variations) may differ
in the microcanonical and canonical ensembles. A configuration that is stable in
the canonical ensemble is necessarily stable in the microcanoniocal ensemble
but the converse is wrong. As a corollary we recall that the specific heat
$C=dE/dT_{\infty}=Nk_B \eta^2
d\Lambda/d\eta$ of stable equilibrium states is always positive in the canonical
ensemble while it can be positive or negative in the microcanonical ensemble
(for systems with long-range interactions).

The stability of the solutions can be determined by using the
Poincar\'e turning point criterion \cite{poincare}. We refer to the papers of
Katz \cite{katzpoincare1,katzpoincare2} for a presentation and a generalization
of this criterion, and for its application to the nonrelativistic classical
self-gravitating gas. This method was applied  to the
nonrelativistic self-gravitating Fermi gas in \cite{ijmpb}. We use the
same method in the
present paper.

In the discussion of the caloric curves, we shall only consider
stable states. An equilibrium state that is a local, but
not a global, extremum  of the relevant thermodynamical potential (entropy in
the microcanonial ensemble and free energy in the canonical ensemble) is said to
be metastable. A global extremum  of the thermodynamical potential is said
to be fully stable. For systems with short-range interactions, metastable
states have a short lifetime so that the caloric curve should contain
only fully stable states. However, for systems with long-range interactions, the
metastable states have a very long lifetime scaling as $e^N$ which is usually
much
longer than the age of the Universe. As a result, metastable states
can be as much, or even more, relevant than fully
stable states \cite{lifetime}. The selection between a fully stable state or
a metastable
state depends on the initial condition and on a notion of basin of attraction.
In this paper, we shall not distinguish between metastable and fully stable
states. The physical caloric curve should contain all types of stable
equilibrium states.\footnote{The existence, or nonexistence, of fully stable
states for self-gravitating fermions in general relativity is an
interesting problem by itself but it will not be considered in the present
paper (see the Remark at the end of Sec. \ref{sec_trq} showing that this
problem is not trivial).}

For real systems, that are not in a box, the natural evolution proceeds along
the series of equilibria towards larger and larger density
contrasts.\footnote{The reason is that, for
real systems (globular clusters, dark matter halos...) such
as those described by the King model, the Boltzmann or Fermi-Dirac entropy
(resp. the Boltzmann or
Fermi-Dirac free energy)  increases (resp. decreases) with the concentration
parameter; see Fig. 5 of \cite{cohn} and  Fig. 46 of \cite{clm2}. Note that,
surprisingly, for box-confined systems
this is the opposite; see Fig 3 of \cite{pt}.}
In general, this corresponds to lower and lower
temperatures and energies.\footnote{This is explicitly shown in
Figs. \ref{LambdaR_R50_N0p399PH} and
\ref{etaR_R50_N0p399PH} below. Note that this result is valid only for
mid and low energies and temperatures. At very high energies and temperatures,
where the system behaves as a self-gravitating radiation, the density
contrast increases with the energy and the temperature (see
Figs. 2 and 3 of \cite{paper2}) implying that the natural evolution of the
system is
towards higher and
higher energies and temperatures. This situation has been discussed in
\cite{paper2} and will not be considered here.} Therefore, in
the
discussion of the caloric curves, we shall describe the evolution of the system
starting from high energies and high temperatures, and reducing the temperature
and
the energy until an instability takes place.

\section{Particular limits}
\label{sec_pl}

In this section, we briefly recall well-known results that
correspond to particular limits of the general relativistic Fermi gas.

\subsection{The nonrelativistic $+$ classical limit}

\begin{figure}
\begin{center}
\includegraphics[clip,scale=0.3]{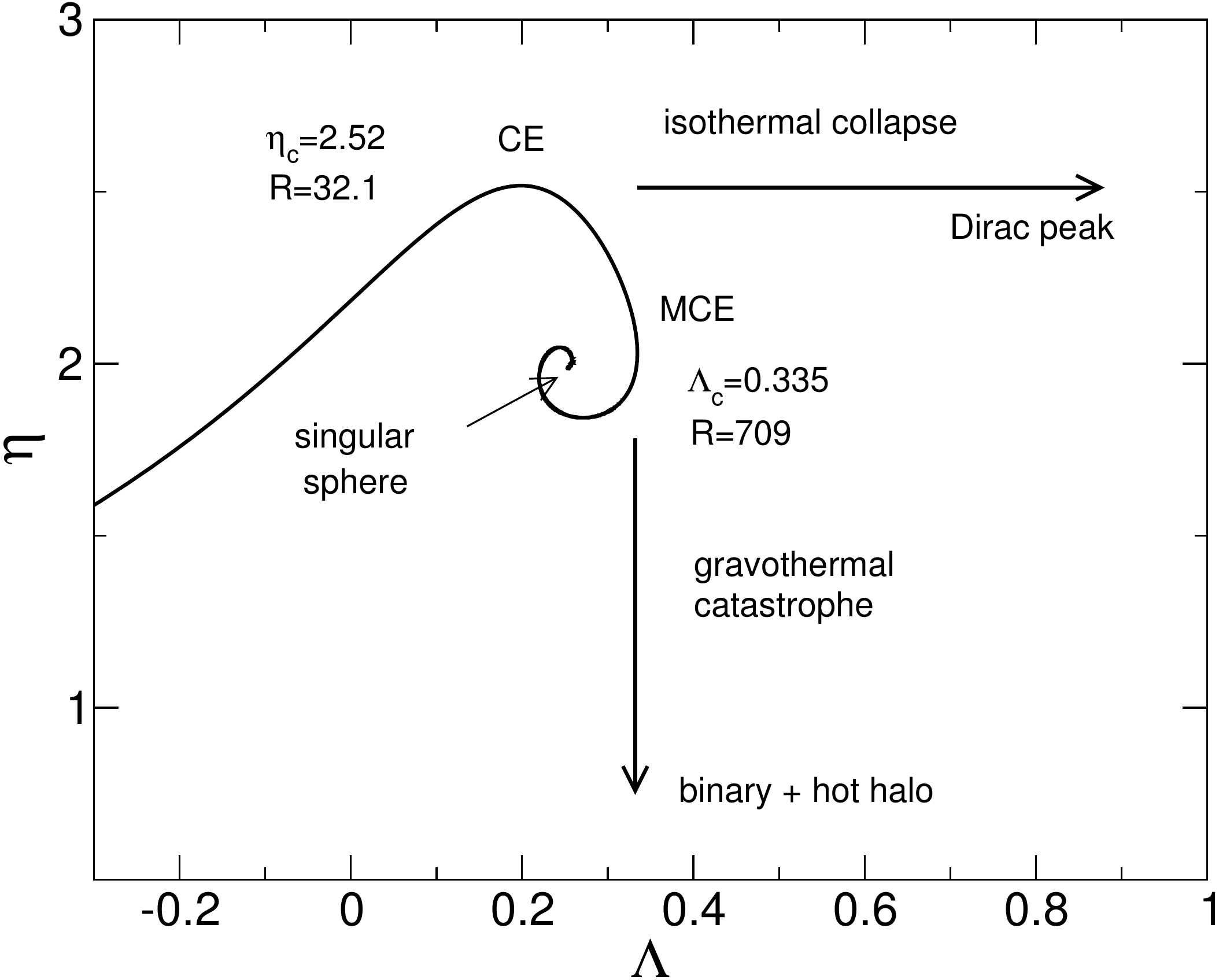}
\caption{Caloric curve of the nonrelativistic classical self-gravitating
gas.}
\label{etalambda}
\end{center}
\end{figure}

The thermodynamics of a nonrelativistic classical self-gravitating gas has been
studied in detail in \cite{antonov,lbw,katzpoincare1,paddyapj,aa}. The
caloric
curve $\eta(\Lambda)$ forms a spiral (see Fig. \ref{etalambda}). In the
microcanonical ensemble, there is no equilibrium state below a critical energy
$E_c$ corresponding to $\Lambda_c=0.335$. In that case, the system undergoes a
gravothermal catastrophe (core collapse) leading to a binary
star surrounded by a hot halo \cite{lbe,inagaki,cohn}. In the canonical
ensemble, there is
no equilibrium state below a
critical temperature $T_c$, corresponding to $\eta_c=2.52$. In
that case, the system undergoes an isothermal collapse leading to a Dirac peak
containing all the mass \cite{post}.

\subsection{The nonrelativistic limit}
\label{sec_sunl}

The thermodynamics of the nonrelativistic self-gravitating Fermi gas has been
studied in detail in \cite{ijmpb}. It is shown that the caloric curves
$\eta(\Lambda)$ depend on a single control parameter (it should not be
confused with the chemical potential):
\begin{equation}
\label{nlq1}
\mu=\eta_0\sqrt{512\pi^4G^3NmR^3},\qquad \eta_0=\frac{gm^4}{h^3}.
\end{equation}
It can be written as \cite{ijmpb}:
\begin{equation}
\label{nlq2}
\mu=17.3\left (\frac{R}{R_0}\right )^{3/2},\quad
R_0=0.181\frac{h^2}{Gm^{8/3}g^{2/3}M^{1/3}},
\end{equation}
or as
\begin{equation}
\label{nlq3}
\mu=17.3\left (\frac{M}{M_0}\right )^{1/2},\quad
M_0=5.97\times 10^{-3} \frac{h^6}{G^3m^8g^2R^3},
\end{equation}
where $R_0$ (resp. $M_0$) is the radius (resp. mass) of a fermion star of mass
$M$ (resp. radius $R$) at $T=0$ (see Appendix \ref{sec_bf}).
Introducing
the normalized
variables of Appendix \ref{sec_dq}, this parameter becomes
\begin{equation}
\label{nl1}
\mu=\frac{4\sqrt{2}}{\pi}(NR^3)^{1/2}.
\end{equation}

\begin{figure}
\begin{center}
\includegraphics[clip,scale=0.3]{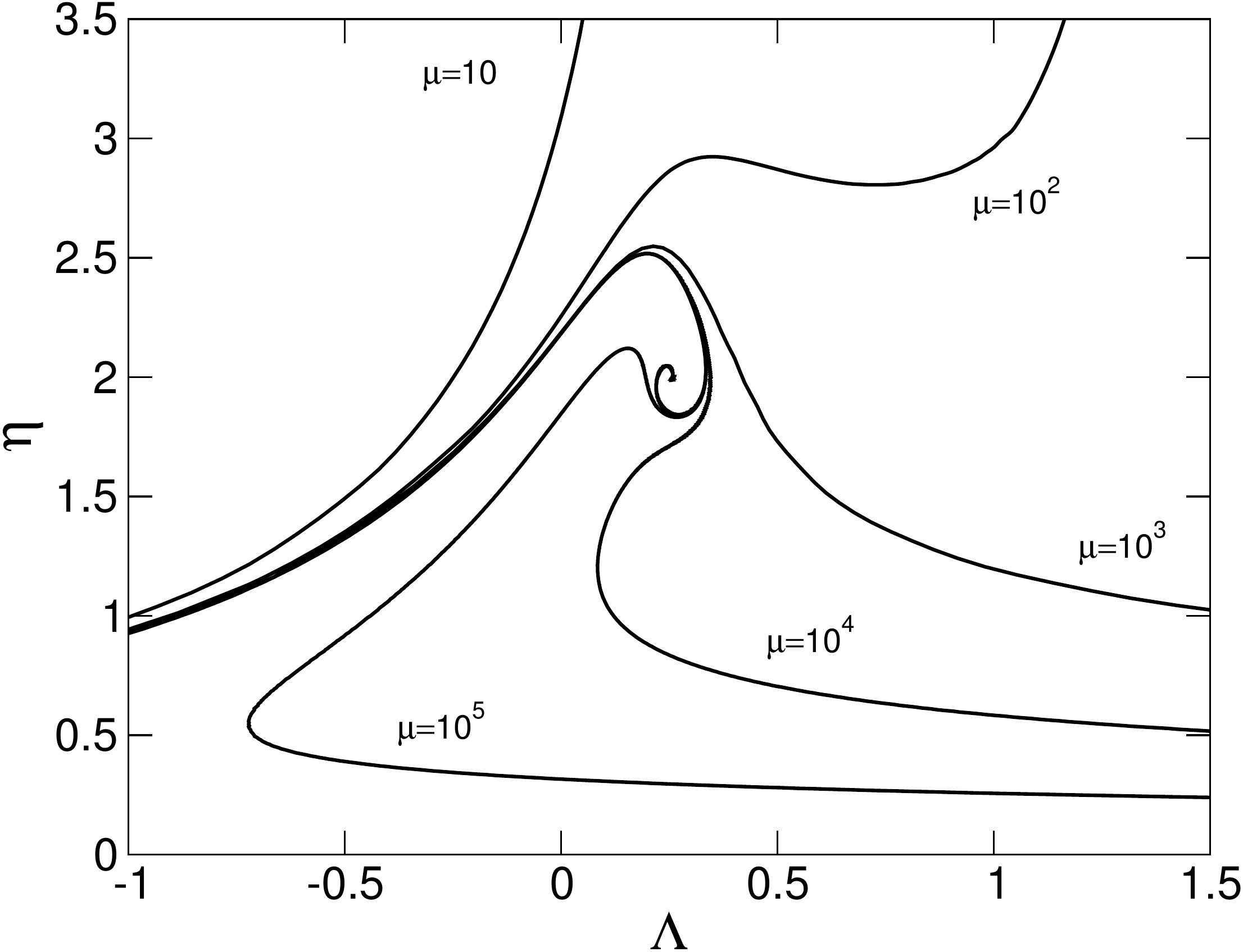}
\caption{Caloric curves of the nonrelativistic self-gravitating Fermi gas for
different values of
$\mu=(4\sqrt{2}/{\pi})(NR^3)^{1/2}$.}
\label{multimu2}
\end{center}
\end{figure}

Some caloric curves are represented in Fig. \ref{multimu2}. They display a
canonical critical point at $\mu_{\rm CCP}=83$  and a microcanonical
critical point at $\mu_{\rm MCP}=2670$. When $\mu<\mu_{\rm CCP}=83$ there is no
phase
transition. When
$\mu_{\rm CCP}=83<\mu<\mu_{\rm MCP}=2670$ the system displays zeroth and first
order
canonical phase transitions. When $\mu>\mu_{\rm MCP}=2670$ the system displays
zeroth
and first order canonical and
microcanonical phase transitions. When
$\mu\rightarrow +\infty$ we
recover the caloric curve  of the nonrelativistic classical 
self-gravitating gas (spiral) represented in Fig. \ref{etalambda}. When
$\mu<+\infty$ there is a
statistical
equilibrium state for any accessible value of energy and
temperature. The
gravitational collapse of the  nonrelativistic classical 
self-gravitating gas (gravothermal catastrophe in the microcanonical ensemble
and isothermal collapse in the canonical ensemble) is prevented by quantum
mechanics (Pauli's exclusion principle). 

For a given box radius, the nonrelativistic canonical phase transition appears
when 
\begin{equation}
\label{nl2}
N>N_{\rm CCP}(R)\equiv \left (\frac{\pi\mu_{\rm CCP}}{4\sqrt{2}}\right
)^2\frac{1}{R^3}=\frac{2125}{R^3}.
\end{equation}
If we consider the general relativistic problem, we must require $N\ll
N_{\rm OV}$, where $N_{\rm OV}=0.39853$ is the OV limit, for the validity of
the nonrelativistic treatment. Therefore, we will see
the nonrelativistic canonical phase transition for $N_{\rm CCP}(R)<N\ll N_{\rm
OV}$
provided that
\begin{equation}
\label{nl3}
R\gg R^{\rm approx.}_{\rm CCP}\equiv \left (\frac{\pi\mu_{\rm
CCP}}{4\sqrt{2}}\right
)^{2/3}\frac{1}{N_{\rm OV}^{1/3}}=17.5.
\end{equation}
In comparison $R_{\rm OV}=3.3569$. This argument just provides
an order of
magnitude of the radius $R_{\rm CCP}$ above which a
canonical phase transition appears for $N>N_{\rm CCP}(R)$. By solving the
general relativistic equations, we find that the exact value is
$R_{\rm CCP}^{\rm exact}=12.0$ (see Sec. \ref{sec_rn}). 

For a given box radius, the nonrelativistic microcanonical phase transition
appears when
\begin{equation}
\label{nl4}
N>N_{\rm MCP}(R)\equiv\left (\frac{\pi\mu_{\rm MCP}}{4\sqrt{2}}\right
)^2\frac{1}{R^3}=\frac{2.20\times 10^6}{R^3}.
\end{equation}
If we consider the  general relativistic problem, using the same argument as
before, we
will see
the nonrelativistic microcanonical phase transition for $N_{\rm MCP}(R)<N\ll
N_{\rm OV}$
provided that
\begin{equation}
\label{nl5}
R\gg R^{\rm approx.}_{\rm MCP}\equiv \left (\frac{\pi\mu_{\rm
MCP}}{4\sqrt{2}}\right
)^{2/3}\frac{1}{N_{\rm OV}^{1/3}}=177.
\end{equation}
This argument just provides an order of
magnitude of the radius $R_{\rm MCP}$ above which a
microcanonical phase transition appear for $N>N_{\rm MCP}(R)$. By solving
the
general relativistic equations, we find that the exact value is $R_{\rm
MCP}^{\rm exact}=92.0$ (see Sec. \ref{sec_rn}).

\subsection{The classical
limit}
\label{sec_suncl}

The thermodynamics of a classical self-gravitating gas in
general relativity has been
studied in detail in Refs. \cite{roupas} and \cite{paper2}. This corresponds to
the nondegenerate
limit of the general relativistic  Fermi gas. It is shown that the caloric
curves $\eta(\Lambda)$  depend on a single control parameter
\begin{equation}
\label{clq1}
\nu=\frac{GNm}{Rc^2}.
\end{equation}
It can be written as
\begin{equation}
\label{clq2}
\nu=\frac{R^*_S}{2R},\qquad R^*_S=\frac{2GNm}{c^2},
\end{equation}
or as
\begin{equation}
\label{clqb}
\nu=\frac{N}{2N^*_S},\qquad N^*_S=\frac{Rc^2}{2Gm},
\end{equation}
where $R^*_S$ can be interpreted as a sort of
Schwarzschild radius defined with the rest mass $Nm$ instead of the mass $M$
(reciprocally, $N^*_S m$ is a sort of Schwarzschild rest mass of an object of
radius $R$).
Introducing the normalized variables of Appendix \ref{sec_dq} this parameter
becomes
\begin{equation}
\label{cl1}
\nu=\frac{N}{R}.
\end{equation}

\begin{figure}
\begin{center}
\includegraphics[clip,scale=0.3]{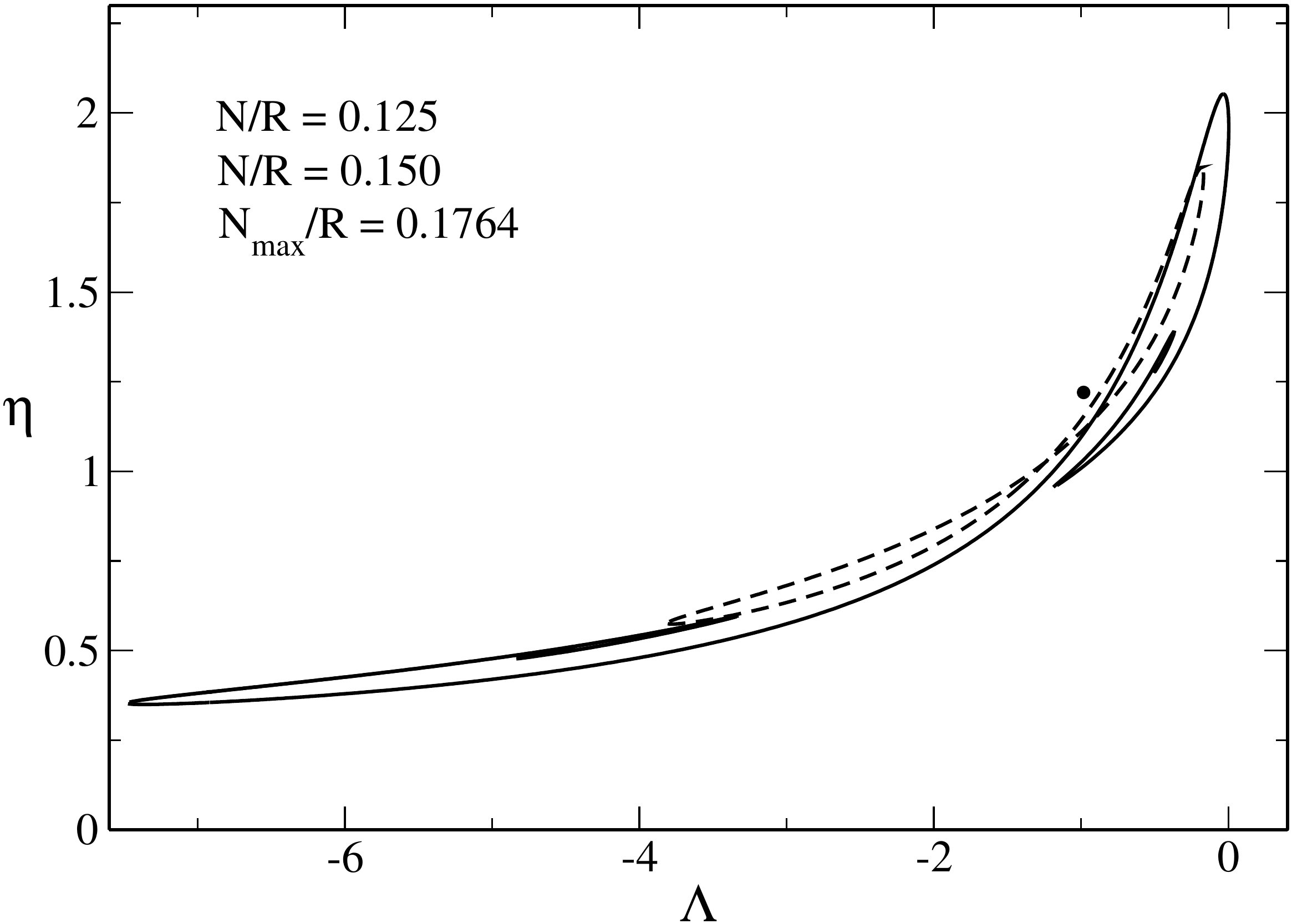}
\caption{Caloric curves of the classical self-gravitating gas in general
relativity for different values of $\nu=N/R$.}
\label{total_droitePH}
\end{center}
\end{figure}

Some caloric curves are represented in Fig. \ref{total_droitePH}. 
When $\nu\rightarrow 0$ ($N\ll N_S^*$ or $R\gg R_S^*$), we
recover the caloric curve of the
nonrelativistic classical
self-gravitating gas (spiral) represented in Fig. \ref{etalambda}. When
$0<\nu<\nu'_S=0.128$ the caloric curve has the
form
of a
double spiral exhibiting a collapse at low energies and low temperatures
(cold spiral) and at high energies and high temperatures (hot
spiral).\footnote{The hot spiral corresponds to an
ultrarelativistic classical
gas \cite{roupas} which is similar to a form of radiation described by an
equation of state
$P=\epsilon/3$ \cite{sorkin,aarelat1,aarelat2} (see \cite{paper2} for a
detailed discussion).} When
$\nu'_S=0.128<\nu<\nu_S=0.1415$ the two spirals are amputed (truncated) and
touch
each other.
When
$\nu_S=0.1415<\nu<\nu_{\rm max}=0.1764$ the two spirals disappear and the
caloric
curve
makes a loop resembling to the symbol ``$\infty$''. As $\nu$ increases, the loop
shrinks more and more and, when $\nu=\nu_{\rm max}=0.1764$, it reduces to a
point
located at
$(\Lambda_*,\eta_*)=(-0.9829,1.2203)$. When
$\nu>\nu_{\rm max}=0.1764$, no equilibrium state is possible.

For a given box radius, the spirals touch each other when
\begin{equation}
\label{cl2}
N>N'_S(R)=\nu'_S R=0.128\, R
\end{equation}
and they form a loop when
\begin{equation}
\label{cl3}
N>N_S(R)=\nu_S R=0.1415\, R.
\end{equation}
The caloric curve reduces to a point when
\begin{equation}
\label{cl4}
N=N_{\rm max}(R)=\nu_{\rm max} R=0.1764\, R.
\end{equation}

If we consider the truly quantum problem, we must require $N\gg
N_{\rm OV}=0.39853$ for the validity of the classical (nondegenerate) treatment.
Therefore, we will
see the double spiral and its evolution described previously for  $N_{\rm
OV}\ll N<N_{\rm max}$ provided that 
\begin{equation}
\label{cl5}
R\gg \frac{N_{\rm OV}}{\nu_{\rm max}}=2.259.
\end{equation}
We note that  $N_{\rm OV}/\nu_{\rm max}=2.259$ is of the order of $R_{\rm
OV}=3.3569$.

{\it Remark:} For a given box radius $R$, coming back to
dimensional variables,
equilibrium states exist only when $N\le N_{\rm max}(R)=0.1764\,
Rc^2/Gm=0.3528\, N_S^*$. Inversely, for a given number of particles $N$, 
equilibrium states exist only when $R\ge R_{\rm min}(N)=5.669\,
GNm/c^2=2.834\, R^*_S$. The
nonrelativistic limit
corresponds to $N\ll N_{\rm max}(R)\sim Rc^2/2Gm\sim N_S^*$ or $R\gg R_{\rm
min}(N)\sim 2GNm/c^2\sim R^*_S$.
These results are valid in the classical limit. For
small systems, quantum effects will come into play. If we argue that $N_{\rm
max}=\nu_{\rm max} Rc^2/Gm \sim N_{\rm OV}$ when
$R\sim R_{\rm OV}$, or equivalently  $R_{\rm min}=GNm/\nu_{\rm max}c^2\sim
R_{\rm OV}$ when $N\sim N_{\rm OV}$, we find that $\nu_{\rm max}
\sim G N_{\rm OV}m/R_{\rm OV}c^2 \sim 0.1187$.
This may justify the order of magnitude of this constant. Alternatively, we may
just remark that $\nu_{\rm max}=GN_{\rm max}m/Rc^2=0.1764$
is of the same order as $GN_{\rm OV}m/R_{\rm OV}c^2=0.1187$.

\subsection{Summary}

Before treating the general case, let us summarize the previous results.

{\it Nonrelativistic $+$ classical limit.} For a given box radius $R$ and
particle number $N$ the system undergoes a
catastrophic collapse towards a singularity at low temperatures in the canonical
ensemble and at low energies in the microcanonical ensemble.

{\it Nonrelativistic limit.} For a given box radius $R$ there
is no
phase transition when $N<N_{\rm CCP}(R)$, the system can undergo a
canonical phase
transition when $N_{\rm CCP}(R)<N<N_{\rm MCP}(R)$, and the system can
undergo a canonical and a microcanonical phase transition when $N>N_{\rm
MCP}(R)$. For a given particle number $N$, there is no phase
transition when $R<R_{\rm CCP}(N)$, the system can undergo a canonical phase
transition when $R_{\rm
CCP}(N)<R<R_{\rm MCP}(N)$, and the system can undergo a canonical and a
microcanonical phase
transition when $R>R_{\rm MCP}(N)$. Here, $R_{\rm CCP}(N)=12.9/N^{1/3}$ and
$R_{\rm MCP}(N)=130/N^{1/3}$ are the reciprocal of $N_{\rm CCP}(R)=2125/R^3$ and
$N_{\rm MCP}(R)=2.20\times 10^6/R^3$. There is an equilibrium
state at all temperatures $T\ge 0$ in the canonical
ensemble and at all accessible energies $E\ge E_{\rm min}$ (where $ E_{\rm
min}$ is the energy of the ground state) in the microcanonical ensemble. 

{\it Classical  limit.} For a given box radius $R$, the
caloric curve has the form of a double spiral when
$N<N_{\rm S}'(R)$, the spirals touch each
other when $N_{\rm S}'(R)<N<N_{\rm S}(R)$, the caloric curve makes a 
loop when $N_{\rm S}(R)<N<N_{\rm max}(R)$, and there is no equilibrium state
when   $N>N_{\rm max}(R)$. For a given particle number $N$, the
caloric curve has the form of a double spiral when
$R>R_{\rm S}'(N)$, the spirals touch each
other when $R_{\rm S}(N)<R<R'_{\rm S}(N)$, the caloric curve makes a 
loop when $R_{\rm min}(N)<R<R_{\rm S}(N)$, and there is no equilibrium state
when   $R<R_{\rm min}(N)$. Here, $R_{\rm S}'(N)=7.81\, N$, $R_{\rm
S}(N)=7.07\, N$ and $R_{\rm min}(N)=5.67\, N$  are the reciprocal of $N_{\rm
S}'(R)=0.128\, R$, $N_{\rm S}(R)=0.1415\, R$ and $N_{\rm max}(R)=0.1764\, R$.
The system undergoes a
catastrophic collapse towards a singularity at both low and high temperatures in
the canonical
ensemble and at both  low and high energies in the microcanonical ensemble.

\section{The case $R_{\rm CCP}<R<R_{\rm MCP}$}
\label{sec_first}

In this section, we study the general relativistic Fermi gas in the case
$R_{\rm CCP}=12.0<R<R_{\rm MCP}=92.0$  where only a canonical phase transition
may occur (see Fig. \ref{phase_phase_Newton_corrPH} below). For
illustration, we select $R=50$. For this value of $R$, the
canonical phase transition occurs above $N_{\rm
CCP}=0.0170$.

\subsection{The case $N<N_{\rm CCP}$}
\label{sec_tr}

\begin{figure}
\begin{center}
\includegraphics[clip,scale=0.3]{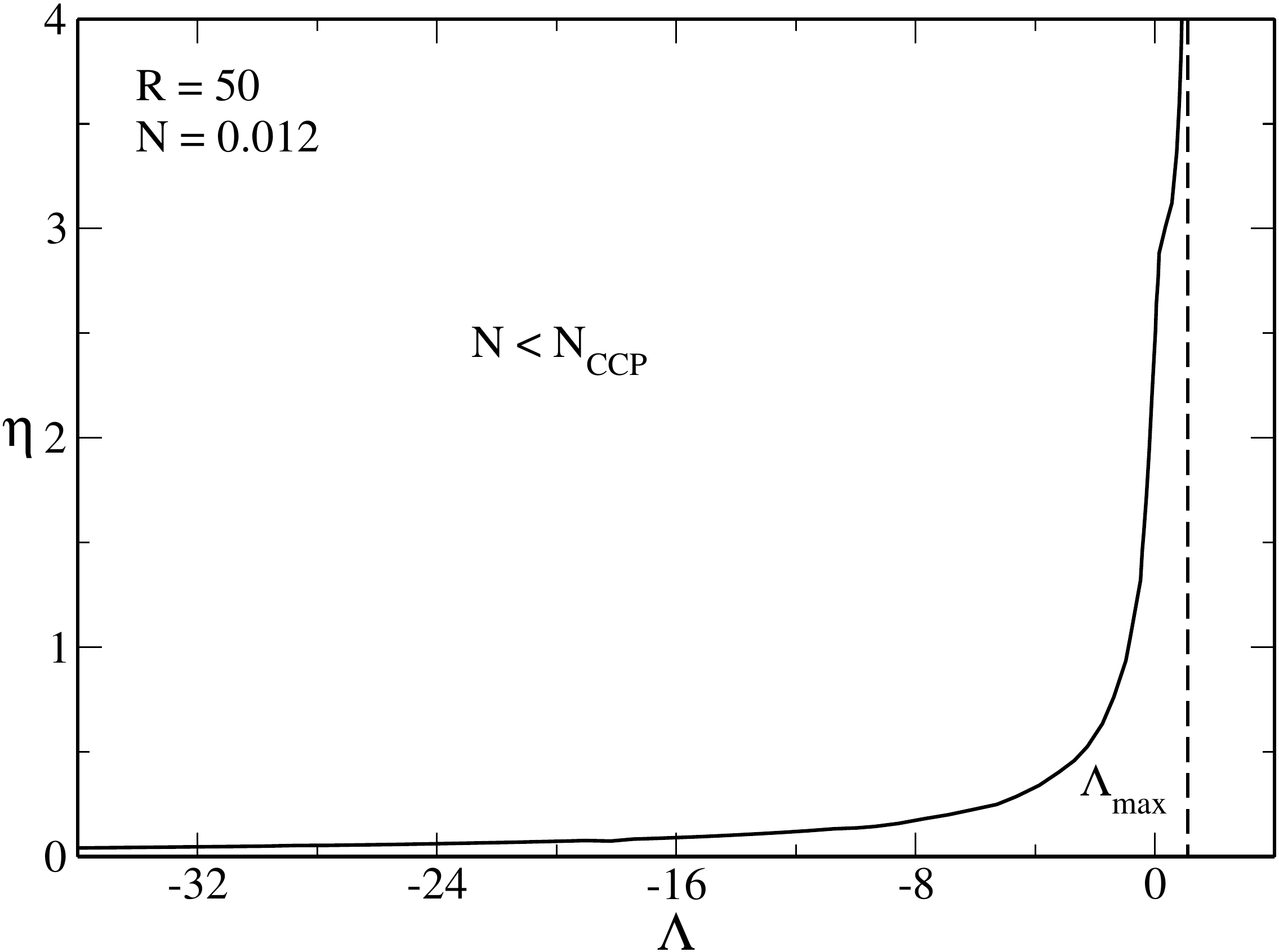}
\caption{Caloric curve for $N<N_{\rm CCP}=0.0170$ (specifically $R = 50$ and $N
= 0.012$).}
\label{kcal_R50_N0p012_unified2PH}
\end{center}
\end{figure}

In Fig. \ref{kcal_R50_N0p012_unified2PH} we have plotted the caloric curve for
$N<N_{\rm
CCP}=0.0170$.
Since $N\ll N_{\rm OV}=0.39853$, this caloric curve coincides with the one
obtained in the nonrelativistic limit \cite{ijmpb} except at very high
energies and very high temperatures (see the Remark at the end of this
section).\footnote{As
discussed in Sec. \ref{sec_ncl} the nonrelativistic limit corresponds to
$N\rightarrow 0$ and $R\rightarrow +\infty$ in such a way that $NR^3$ is fixed
(in more
physical terms $N\ll N_{\rm OV}$ and $R\gg R_{\rm OV}$ with $NR^3$
fixed).}

The series of equilibria $\eta(\Lambda)$ is monotonic.  According to the
Poincar\'e theory of linear series
of equilibria, all
the equilibrium states are stable. The
statistical ensembles (microcanonical and canonical) are
equivalent. The
caloric curve
presents the following features:

(i) There is no phase
transition and no gravitational collapse.

(ii) The specific heat is always positive. The entropy
versus energy curve (not represented) is concave.

The evolution of the system is the following.
At high energies and high temperatures, the system is
nondegenerate (Boltzmannian). As the energy and the temperature are reduced, the
system becomes more and more centrally condensed. At intermediate energies and
intermediate temperatures, the Fermi gas is partially degenerate (see Appendix
\ref{sec_av}). At $T=0$, the Fermi gas
is completely degenerate. This cold nonrelativistic fermion
ball, equivalent to a polytrope of index $n=3/2$, is similar to a 
nonrelativistic white dwarf.
This is the state of minimum energy $E_{\rm min}$ (ground
state). Since there
is a stable equilibrium state at $T=0$ (i.e. $\eta\rightarrow +\infty$) with a
finite energy $E_{\rm min}$, the caloric curve $\eta(\Lambda)$
presents a vertical asymptote at $\Lambda=\Lambda_{\rm max}$.\footnote{In the
nonrelativistic limit $\Lambda_{\rm
max}=0.0950 N^{1/3}R$ (see Appendix \ref{sec_bf}). More
generally, a
complete characterization of the ground state of the self-gravitating Fermi gas,
in the nonrelativistic and relativistic regimes, taking into account the
presence of the box is given in \cite{paper1}.}

{\it Remark:} At very high energies and very high temperatures, the
system is relativistic even though $N\ll N_{\rm OV}$. In that case, we
recover the hot spiral studied in \cite{roupas,paper2}. 
As a result,  the complete caloric
curve of
the general relativistic Fermi gas presents a region of negative specific
heat and a region
of ensemble inequivalence at very high energies and very high temperatures.
The system undergoes a gravitational collapse above $E_{\rm
max}$ in the microcanonocal ensemble and above
$T_{\rm max}$ in the canonical ensemble. We note that quantum mechanics cannot
prevent
such a gravitational collapse since it takes place at very high energies and
very
high temperatures where the system is nondegenerate. As a result, the system is
expected to collapse towards a black hole. For $N\rightarrow 0$,
it is shown in \cite{paper2} that $\Lambda_{\rm
min}\sim -0.246/N^2 \rightarrow -\infty$ and $\eta_{\min}\sim 18.3
N^2\rightarrow 0$ so that the hot spiral is rejected at
infinity.\footnote{In terms of dimensional variables this correponds
to $E_{\rm max}\rightarrow 0.24631 Rc^4/G$ and $k_BT_{\rm max}\sim 0.0547
Rc^4/NG$.} For small values of $N$ ($N\ll
N_{\rm max}$) the hot spiral occurs at very negative values of $\Lambda$ and
at very small values of $\eta$. This is why we do not see it in Fig. 
\ref{kcal_R50_N0p012_unified2PH} (it is
outside of the frame since $\Lambda_{\rm
min}\simeq -1708$ and $\eta_{\min}\simeq 2.63\times 10^{-3}$). The hot spiral
becomes visible only for larger values of
$N$ ($N\lesssim N_{\rm max}$) as in Fig. \ref{kcal_R50_N4_unifiedPH} below. In
this
paper, we shall not discuss the hot spiral specifically since it has been
described
in detail in \cite{roupas,paper2}.

\subsection{The case $N_{\rm CCP}<N<N_{1}$}
\label{sec_ccp1}

\begin{figure}
\begin{center}
\includegraphics[clip,scale=0.3]{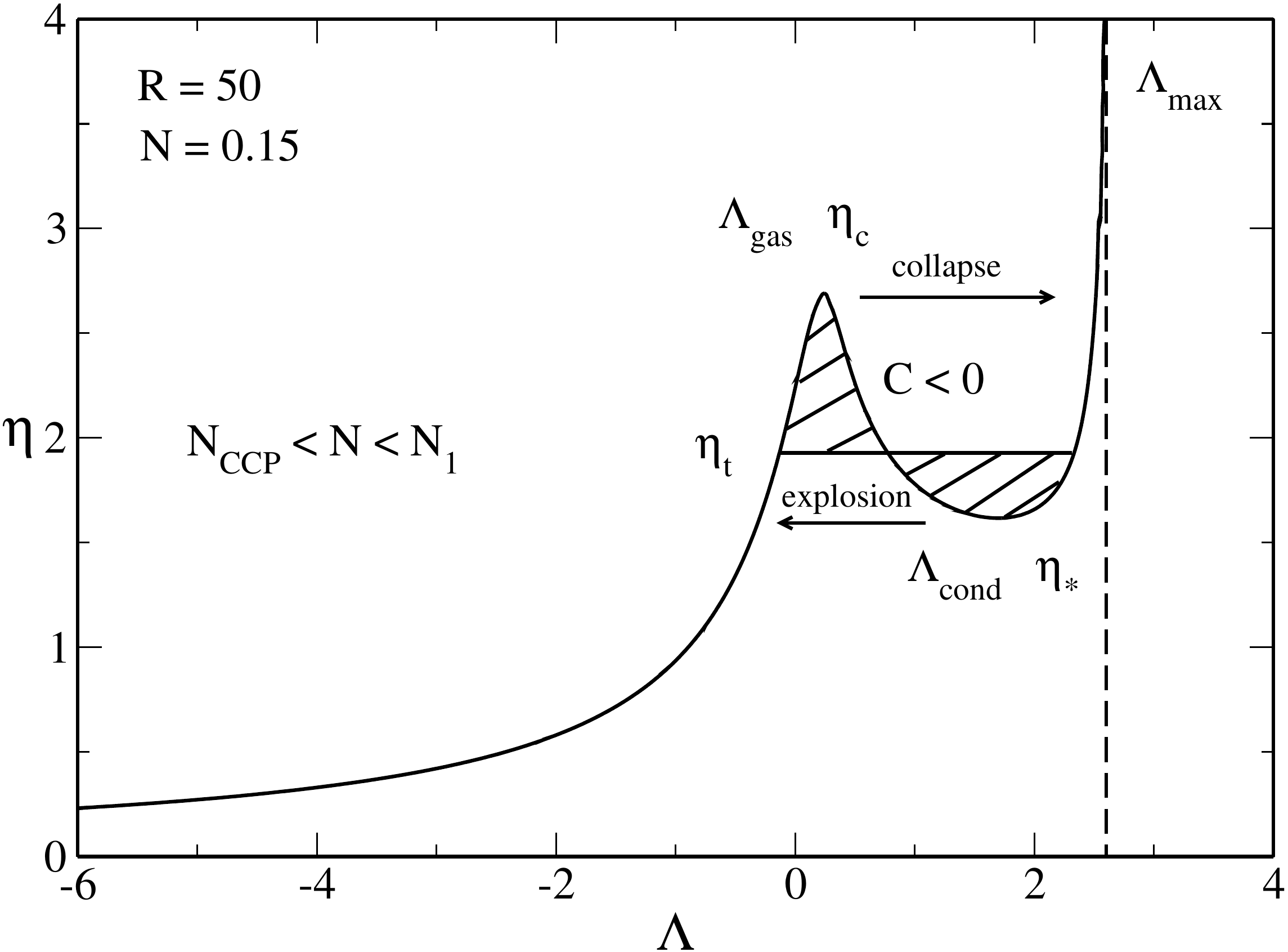}
\caption{Caloric curve for $N_{\rm
CCP}=0.0170<N<N_{1}=0.18131$ (specifically $R = 50$ and $N =
0.15$).}
\label{Xkcal_R50_N0p15_unified2blackPH}
\end{center}
\end{figure}

In Fig. \ref{Xkcal_R50_N0p15_unified2blackPH} we have plotted the caloric
curve for  $N_{\rm CCP}=0.0170<N<N_{1}=0.18131$. Since $N\ll N_{\rm
OV}=0.39853$,
the
caloric curve coincides with the one obtained in the nonrelativistic limit
\cite{ijmpb}. The novely with respect to the previous case is that the
caloric curve has a $N$-shape structure leading to canonical
phase transitions
and ensembles inequivalence. This $N$-shape structure appears at $N=N_{\rm
CCP}=0.0170$ where the caloric curve presents a horizontal inflexion point.
Let us
consider
the microcanonical and canonical
ensembles successively (see \cite{ijmpb} for a more detailed discussion).

\subsubsection{Microcanonical ensemble}

The curve
$\eta(\Lambda)$ is univalued.  According
to the Poincar\'e theory, the whole series of equilibria is stable. The
caloric curve
presents the following features:

(i) There is no phase transition and no gravitational collapse.

(ii) There is a region of negative specific heats  between $\Lambda_{\rm
gas}$ and $\Lambda_{\rm cond}$. In this
range of intermediate energies the system is purely self-gravitating,
i.e., it almost
does not feel the quantum pressure  (Pauli exclusion principle) nor the pressure
of the box. The negative specific heat leads to a convex intruder (dip) in the
entropy versus energy curve (see Fig. 25 of \cite{ijmpb}).

The evolution of the system in the microcanonical
ensemble is the following.  Let us start from high energy states and
decrease the energy. At high energies, the system is
almost homogeneous. As energy
decreases, and especially when we enter in the region of negative specific
heats, the system becomes more and more concentrated and partially degenerate.
At the minimum energy
$E_{\rm min}$  (ground state) the system is completely degenerate. There is no
phase transition,
just a progressive clustering of the system until the ground state is reached.

\subsubsection{Canonical ensemble}

The curve $\Lambda(\eta)$ is multivalued leading to the possibility of phase
transitions in
the canonical ensemble. The left branch up to
$\eta_c$ corresponds to the gaseous phase and the right branch after $\eta_*$
corresponds to the condensed phase. According to the Poincar\'e turning point
criterion, these equilibrium states are stable while the
equilibrium states on the
intermediate branch between  $\eta_c$ and  $\eta_*$ are unstable. These
equilibrium states have a core-halo
structure (see below) and a negative specific heat.
This is a sufficient (but not necessary) condition of instability in the
canonical ensemble. The caloric curve presents the following features:

(i) When $\eta<\eta_*$ there are only gaseous states. When $\eta>\eta_c$
there are only condensed states. When $\eta_*<\eta<\eta_c$ there exist gaseous
and
condensed states at the same temperature. A first order
phase transition is expected at a transition temperature
$\eta_t$ determined by the
Maxwell construction (see Fig. \ref{Xkcal_R50_N0p15_unified2blackPH}) or by the
equality of the free energy of the gaseous and condensed phases (see Fig. 28 of
\cite{ijmpb}). When 
$\eta_*<\eta<\eta_t$ the gaseous states have a lower free energy than the
condensed states. When  $\eta_t<\eta<\eta_c$ the condensed states have a
lower free energy than the gaseous states. However, the first order phase
transition does not take place in practice because of the very long lifetime of
the metastable states. 

(ii) There is a zeroth order phase transition at $\eta_c$ from the gaseous
phase to the condensed phase. It corresponds to a gravitational collapse
(isothermal collapse)
 ultimately halted by quantum degeneracy.

(iii) There is a zeroth order phase transition at $\eta_*$ 
from the condensed phase to the gaseous phase. It corresponds to an 
explosion ultimately halted by the boundary of the box.

The evolution of the system in the canonical
ensemble is the following.  Let us start from high temperature states and
decrease the temperature. At high temperatures
the system is in the gaseous phase. At $\eta=\eta_t$, the system is expected to
undergo a first order phase transition from the gaseous phase to the condensed
phase. However, in practice, this phase
transition does not take place because the metastable gaseous states
have a very long lifetime. At $\eta=\eta_c$ the system
collapses towards the condensed phase. Complete gravitational collapse is
prevented by quantum mechanics. The system reaches an equilibrium state
similar
to a nonrelativistic white dwarf (fermion ball). If we now  increase the
temperature the system
remains in the condensed phase until the point $\eta_*$ (again, the
first order phase transition expected at $\eta_t$ does not take place
because  the metastable condensed states have a very long lifetime) at which it
explodes and
returns to the gaseous phase. We have thus decribed an hysteretic cycle in the
canonical ensemble \cite{ijmpb}.

\subsubsection{Density profiles}
\label{sec_dp}

\begin{figure}
\begin{center}
\includegraphics[clip,scale=0.3]{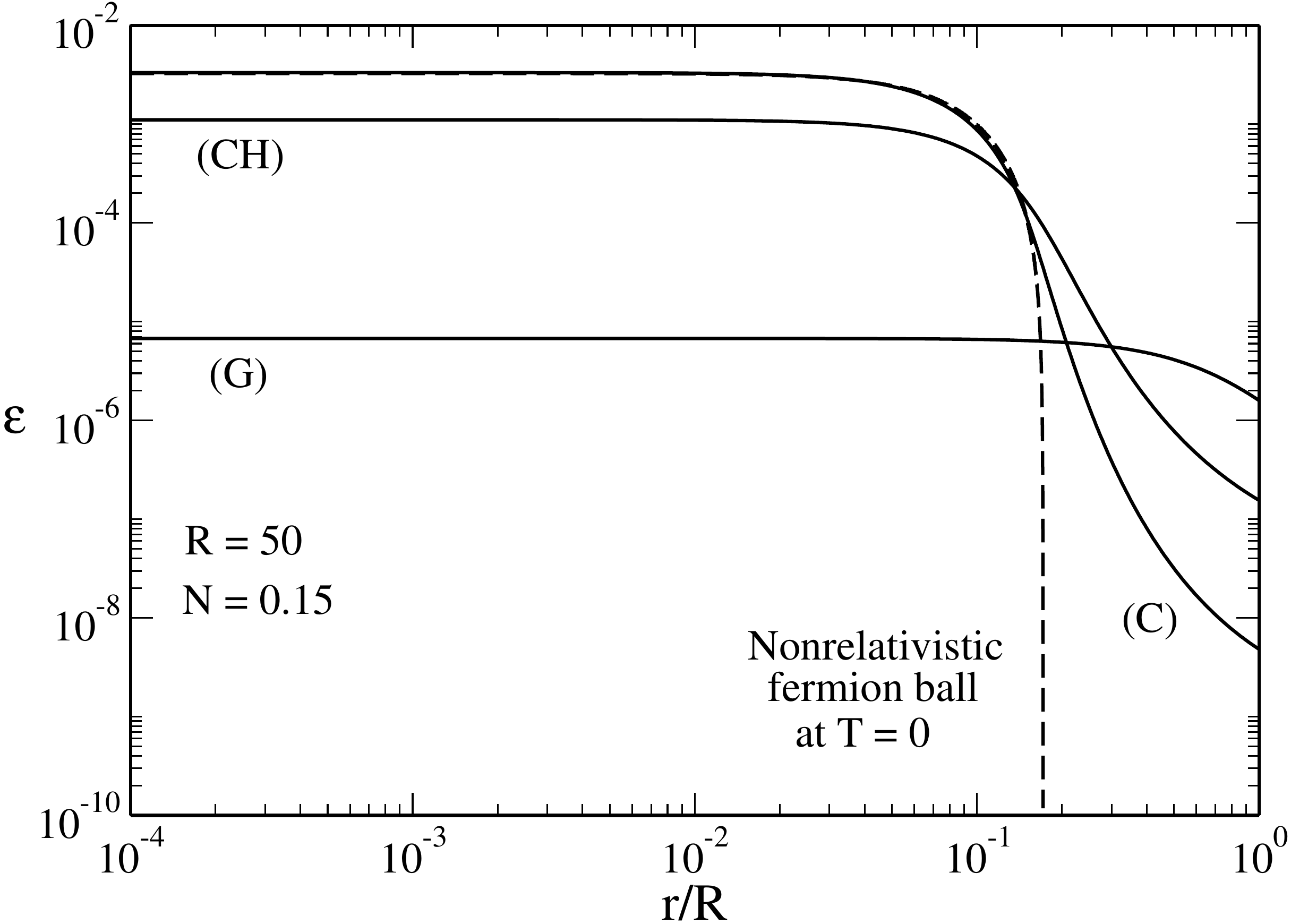}
\caption{Energy density profiles of gaseous, core-halo and condensed 
states at
$\eta=\eta_t$ (specifically $\eta_t=1.9285$). The dashed line represents the
density profile of a nonrelativistic fermion ball at
$T=0$ (similar to a nonrelativistic white dwarf).}
\label{profile_R50_N0p15_newPH}
\end{center}
\end{figure}

In Fig. \ref{profile_R50_N0p15_newPH} we have plotted the density
profiles of the
gaseous (G), core-halo (CH) and condensed (C) states at the transition
point $\eta_t$. We note
that the energy density
is very low confirming that we are in the nonrelativistic regime.

(i) In the gaseous phase (high energies and high temperatures), quantum
mechanics is negligible and the density profile is dilute. The
equilibrium state results from the competition between the
gravitational attraction and the thermal pressure. The gaseous
equilibrium state (G) is almost uniform because the temperature is high so that
the thermal pressure overcomes the gravitational attraction. In that case, the
gas is held by the walls of the box. 

(ii) In the condensed phase (low energies and low temperatures), thermal
effects are negligible and the density profile is very compact. The
equilibrium state results from the competition between the
gravitational attraction and the quantum pressure arising from the Pauli
exclusion principle. The condensed equilibrium state (C) almost coincides with 
a nonrelativistic fermion ball at $T=0$ containing all the mass
(see \cite{ijmpb} and Appendix \ref{sec_thermoqce}). It is
similar
to a nonrelativistic white dwarf
corresponding to a polytrope $n=3/2$. In that case,
gravitational collapse is prevented by quantum mechanics and the confining box
is not
necessary. At small but finite temperatures, we see in Fig.
\ref{profile_R50_N0p15_newPH} 
that the dashed
line corresponding to a polytrope $n=3/2$ provides a good fit to the core of the
distribution. There is a small isothermal atmosphere that becomes thiner and
thiner  as the temperature is reduced.

(iii) The intermediate state (CH) has a sort of core-halo
structure with a degenerate core and an isothermal atmosphere. The
equilibrium state results from the competition between the
gravitational attraction, the thermal pressure, and the
quantum pressure. The pressure of the box and
the quantum pressure  have a weak effect on the equilibrium of the system so it
essentially behaves as a 
self-gravitating isothermal gas. This is why it presents a negative specific
heat.

Let us recall that the these three equilibrium states have the
same temperature but different energies. The core-halo state (CH)  
is unstable in the canonical ensemble while it is stable in the microcanonical
ensemble. It lies in a region of negative specific heats. The gaseous and
condensed states (G) and (C) are stable in both ensembles.

\subsection{The case $N_{1}<N<N_{\rm OV}$}
\label{sec_trq}

\begin{figure}
\begin{center}
\includegraphics[clip,scale=0.3]{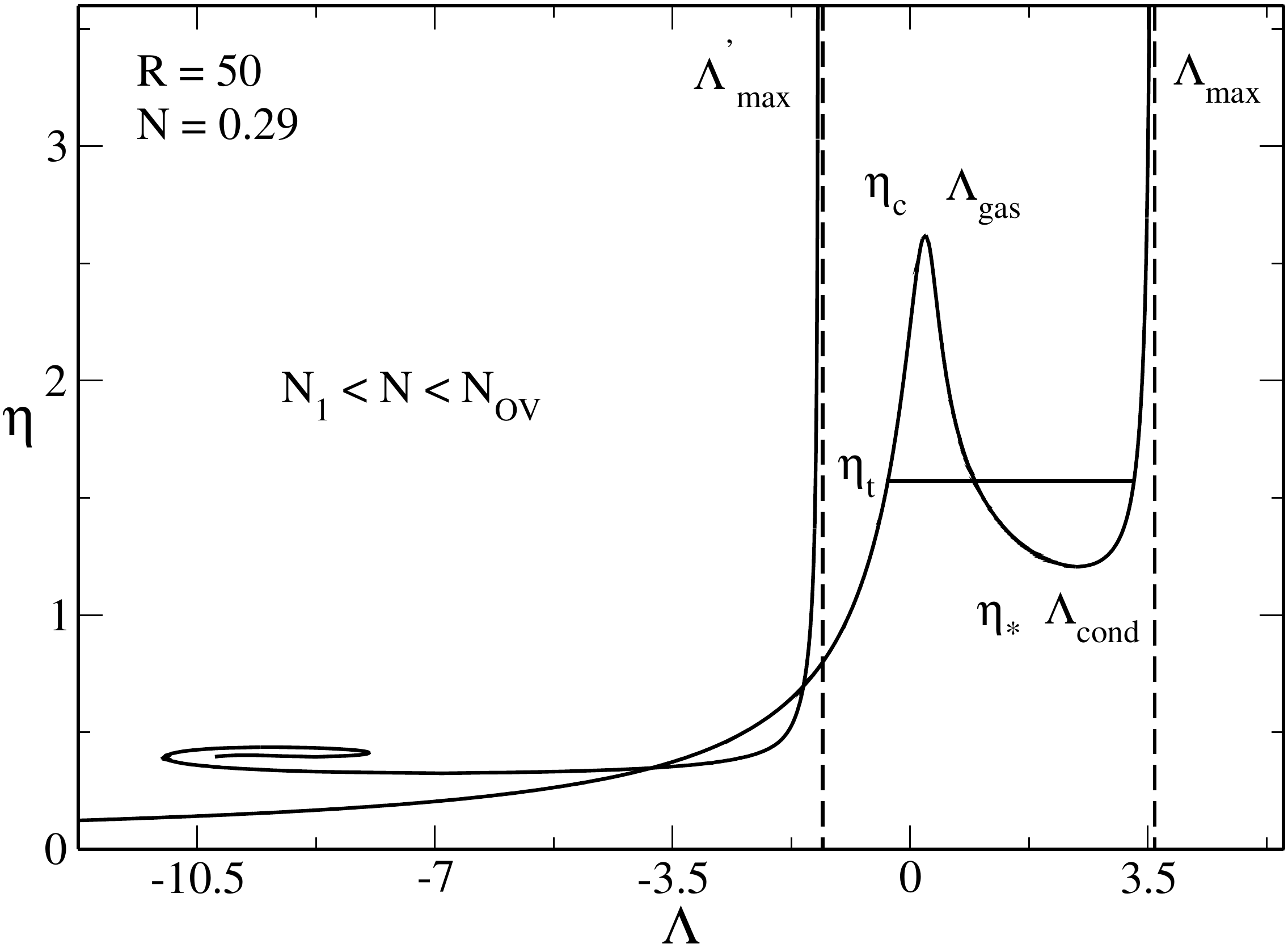}
\caption{Caloric curve for $N_{1}=0.18131<N<N_{\rm
OV}=0.39853$ (specifically $R = 50$ and $N = 0.29$).}
\label{kcal_R50_N0p29_unifiedPH}
\end{center}
\end{figure}

In Fig. \ref{kcal_R50_N0p29_unifiedPH} we have plotted the caloric
curve for $N_{1}=0.18131<N<N_{\rm OV}=0.39853$. The novelty with respect to the
previous case is the existence of a secondary branch presenting an asymptote at 
$\Lambda'_{\rm max}$. This secondary branch appears suddently at $N=N_1=0.18131$
(at that point $\Lambda'_{\rm max}=-0.536\, R$ and 
$\Lambda_{\rm max}=0.0570\,
R$). As detailed in \cite{paper1}, for $N_1<N<N_{\rm OV}$, there exists another
equilibrium state at $T=0$ (i.e. $\eta\rightarrow +\infty$) corresponding to a
completely degenerate fermion ball distinct from the ground state. This
secondary equilibrium state is unstable.\footnote{Actually,
for $N>N_1$, there can exist several  unstable equilibrium
states at $T=0$ (up to an infinity) that have more and more modes of
instability. They are related to the spiral structure of the mass-radius
relation of the general relativistic Fermi gas at $T=0$
\cite{shapiroteukolsky,paper1}.
They give rise to additional branches (with vertical asymptotes) in the caloric
curve. We shall not consider these unstable solutions here, except for the
less unstable  one already mentioned.} Its
mass is larger than the mass of the stable ground state so that $\Lambda'_{\rm
max}\le \Lambda_{\rm max}$. According to the Poincar\'e theory, all the
configurations of the secondary
branch are
unstable.\footnote{The spiral present on the left of this secondary branch
will
ultimately
become the cold spiral of Refs. \cite{roupas,paper2} when $N$ will be
sufficiently large
(see below).} Therefore, the presence of this secondary branch does
not qualitatively change the description of the caloric curve made  in Sec.
\ref{sec_ccp1}.

\begin{figure}
\begin{center}
\includegraphics[clip,scale=0.3]{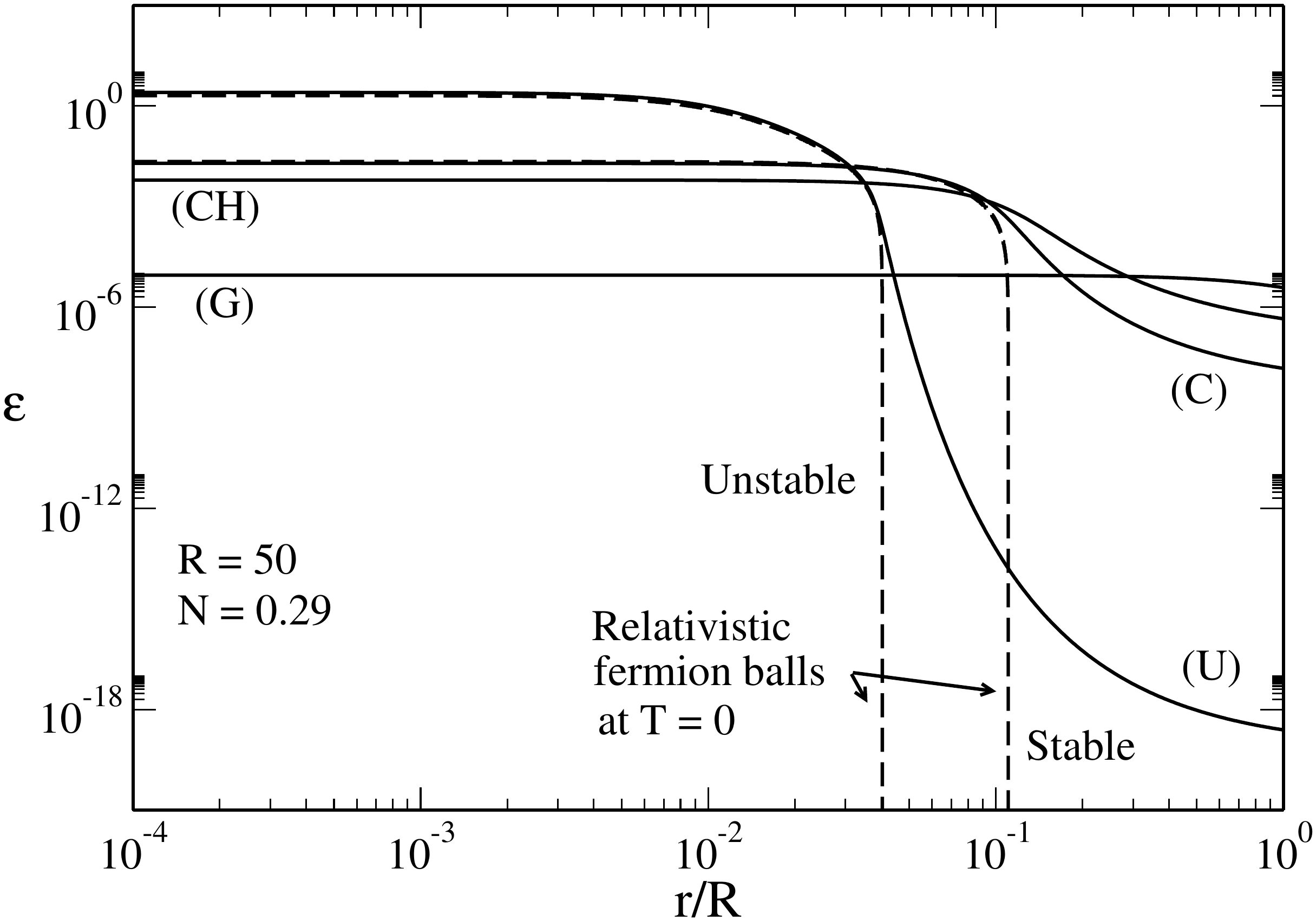}
\caption{Density profiles of gaseous, core-halo and (stable and
unstable) condensed states at
$\eta=\eta_t$ (specifically $\eta_t = 1.5722$). The dashed lines represent the
density profiles of stable and unstable general relativistic fermion balls at
$T=0$ (similar to neutron stars).}
\label{profile_R50_N0p29_newPH}
\end{center}
\end{figure}

However, for $N>N_1$, relativistic effects start to become important. This has
some consequences on the interpretation of the density profiles. In Fig.
\ref{profile_R50_N0p29_newPH} we have plotted the different density profiles at
$\eta_t$. We see that the energy density is low for the gaseous state (G) and
for the core-halo state (CH) indicating that we are in the nonrelativistic
regime. By contrast, the energy density is relatively high for the
stable condensed state (C) and for the unstable condensed state (U) indicating
that we are in the relativistic regime. The
condensed states almost coincide with a general relativistic fermion
ball at $T=0$ containing  all the mass (see Appendix
\ref{sec_thermoqce}). They
are similar to stable and
unstable neutron stars \cite{ov}. At small but finite temperatures, we see in
Fig.
\ref{profile_R50_N0p29_newPH} that the dashed
line obtained from the OV theory provides a good fit to the
core of the distribution. There is a small atmosphere
(containing a little mass) that becomes thinner
and thinner as
the temperature is reduced.

{\it Remark:} In Fig. \ref{kcal_R50_N0p29_unifiedPH}, when the
temperature is low enough, we find four solutions. The solutions (G) and (C)
are
stable (local minima of free energy) while the solutions (CH) and (U) are
unstable
(saddle points of free energy). Since we have an even number of extrema, this
suggests that there is no global minimum of free energy (naively, this results
from
simple topological arguments if we plot a curve $f(x)$ with two minima and two
maxima). The stable equilibrium state with the lowest value of free energy
may be only metastable, not fully stable. This is consistent
with the result of
Zel'dovich \cite{zel446} who showed that, at $T=0$, the OV equilibrium states
are
only metastable. In Fig. \ref{Xkcal_R50_N0p15_unified2blackPH}, when
$\eta_*<\eta<\eta_c$,  we find three solutions. The solutions (G) and (C)
are
stable (local minima of free energy) while the solution (CH) is unstable (saddle
point of free energy). Since we
have an odd number of extrema, this suggests that the solution with the lowest
value of free energy is a global minimum. This is the case in
Newtonian
gravity \cite{ijmpb}. However, this is not quite clear in
general relativity since the result of
Zel'dovich \cite{zel446} still applies for $N<N_1$.  Therefore, the
existence of a
global minimum of free energy (fully stable state) in general relativity is not
trivial and would require a more careful study. Anyway, for practical purposes,
metastable states are very relevant (possibly more relevant than fully stable
states) so we shall determine all types of stable equilibrium states,
disregarding whether they are fully stable or just metastable.

\subsection{The case $N_{\rm OV}<N<N_e$}
\label{sec_deb}

\begin{figure}
\begin{center}
\includegraphics[clip,scale=0.3]{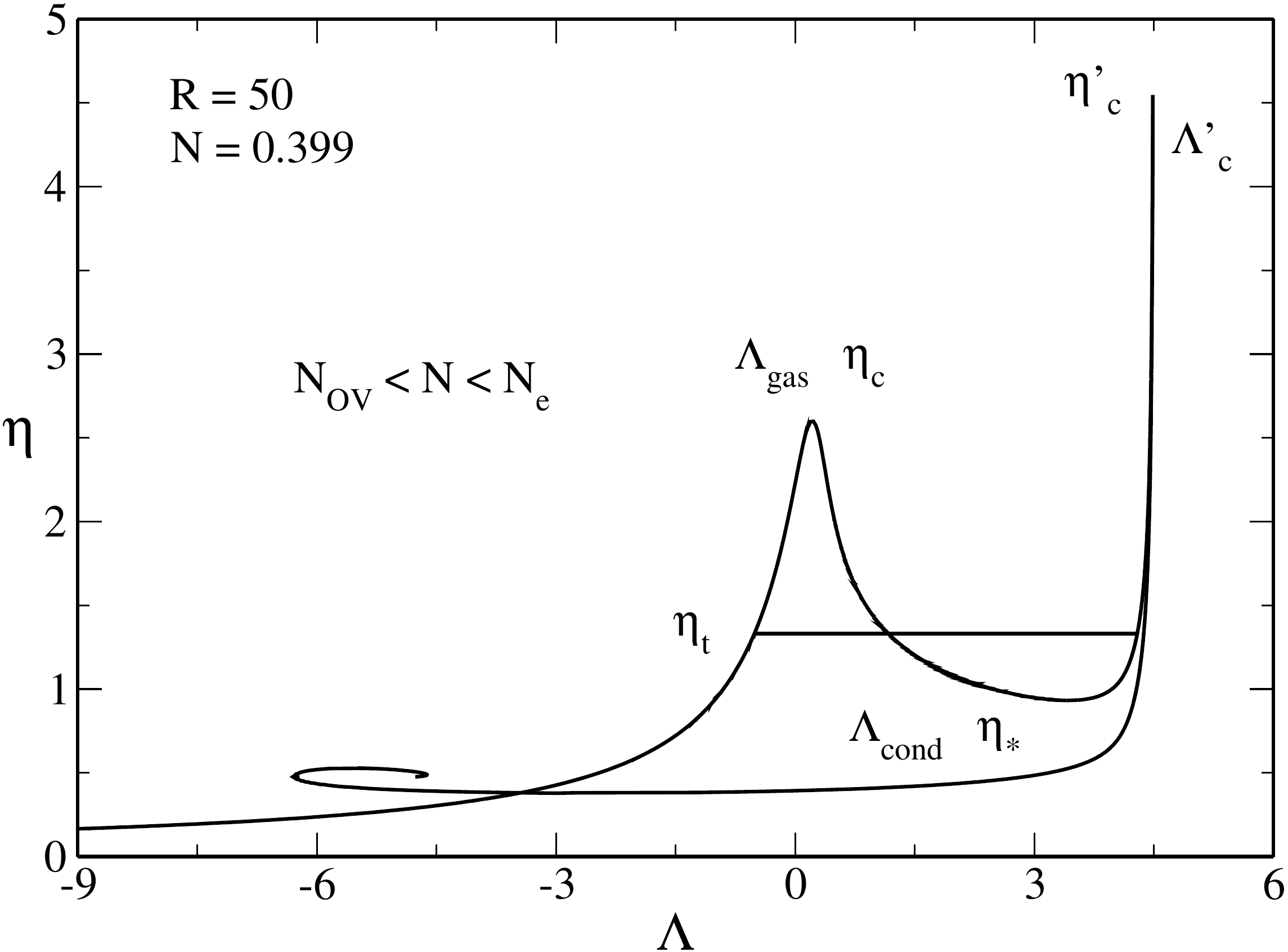}
\caption{Caloric curve for  $N_{\rm OV}=0.39853<N<N_e=0.40002$
(specifically $R = 50$ and $N = 0.399$).}
\label{kcal_R50_N0p399_unifiedPH}
\end{center}
\end{figure}

In Fig. \ref{kcal_R50_N0p399_unifiedPH} we have plotted the caloric
curve for $N_{\rm OV}=0.39853<N<N_e=0.40002$. The novelty with respect to the
previous case is
that the two branches have merged. The merging occurs at $N=N_{\rm OV}$ at which
the two asymptotes $\Lambda'_{\rm max}$ and $\Lambda_{\rm max}$ coincide (at
that point  $\Lambda_{\rm max}=\Lambda'_{\rm max}=0.08985\, R$). This
is the highest value of $N$ at which there exist an equilibrium state at
$T=0$ (ground state). When
$N>N_{\rm OV}$ there is no equilibrium state
at $T=0$ (no ground state) anymore \cite{ov}. In that case, the caloric curve
presents a turning
point of
temperature at $\eta'_c$ and a turning point of energy at  $\Lambda'_c$. As a
result, there is no equilibrium state at $\eta>\eta'_c$ in the canonical
ensemble, i.e., below a critical temperature. Similarly,  there is no
equilibrium state at $\Lambda>\Lambda'_c$ in the microcanonical ensemble, i.e.,
below a critical energy. This means that when
the system becomes strongly relativistic (i.e. when $N>N_{\rm OV}$) quantum
mechanics is not able to prevent gravitational collapse at low temperatures and
low
energies. This is a generalization of the result first obtained at $T=0$ by
Oppenheimer and Volkoff \cite{ov} in the context of neutron stars.

\subsubsection{Microcanonical ensemble}

Let us first consider the microcanonical ensemble. The curve
$\eta(\Lambda)$ is multivalued.  According to
the Poincar\'e
turning point criterion, the series of equilibria is stable up to
$\Lambda_c'$ and then becomes unstable.  The caloric curve
presents the following features:

(i) There is no phase transition  (there is only one
stable equilibrium state
for each $\Lambda<\Lambda'_c$).

\begin{figure}
\begin{center}
\includegraphics[clip,scale=0.3]{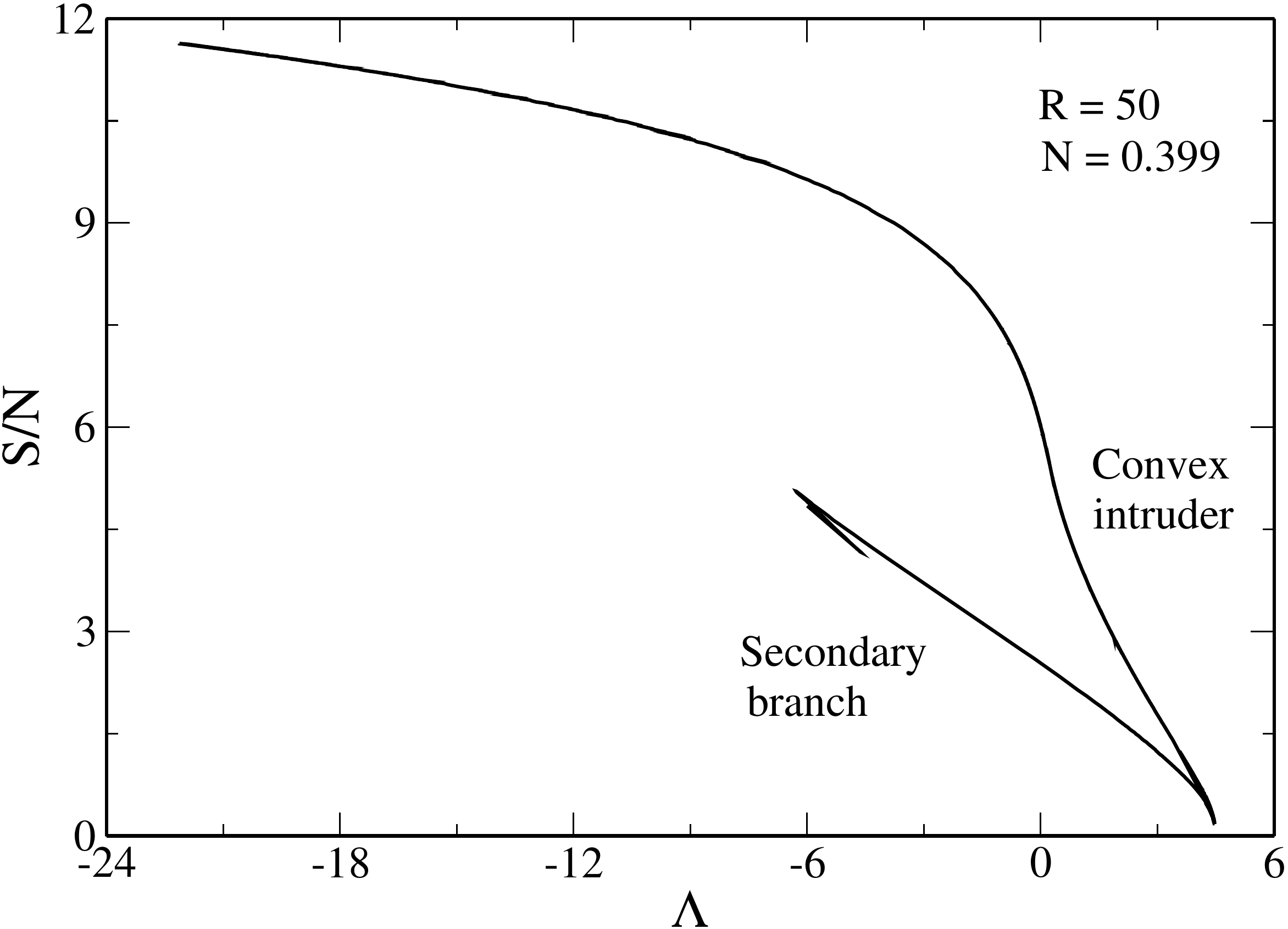}
\caption{Entropy per fermion as a function of the normalized energy for
$N_{\rm OV}<N<N_e$ (specifically $R = 50$
and $N = 0.399$). We can check that the unstable equilibrium 
states (saddle points of entropy) have an entropy lower than
the stable equilibrium states (entropy maxima).}
\label{SLambda_R50_N0p399_colorPH}
\end{center}
\end{figure}

(ii) There are
two regions of negative specific heats, one  between $\Lambda_{\rm
gas}$ and $\Lambda_{\rm cond}$ (as before) and another one 
between  $\Lambda'_{\rm
gas}$ (the energy corresponding to $\eta'_c$) and
$\Lambda'_{c}$. We note that this second region  of negative specific heats is
extremely tiny. In  Fig.
\ref{SLambda_R50_N0p399_colorPH} we clearly
see the convex intruder (dip) associated with the first region of specific heat.
The convex intruder associated with the second region of specific heat is
imperceptible.

(iii) There is a catastrophic collapse
at $\Lambda_c'$ towards a black hole.\footnote{For simplicity, when there is
no equilibrium state, we shall say that the system forms a black hole.
Actually, as discussed in Paper II, it is not completely clear that the system
will always form a black hole in that case. We leave this interesting problem
open to future works.}

\begin{figure}
\begin{center}
\includegraphics[clip,scale=0.3]{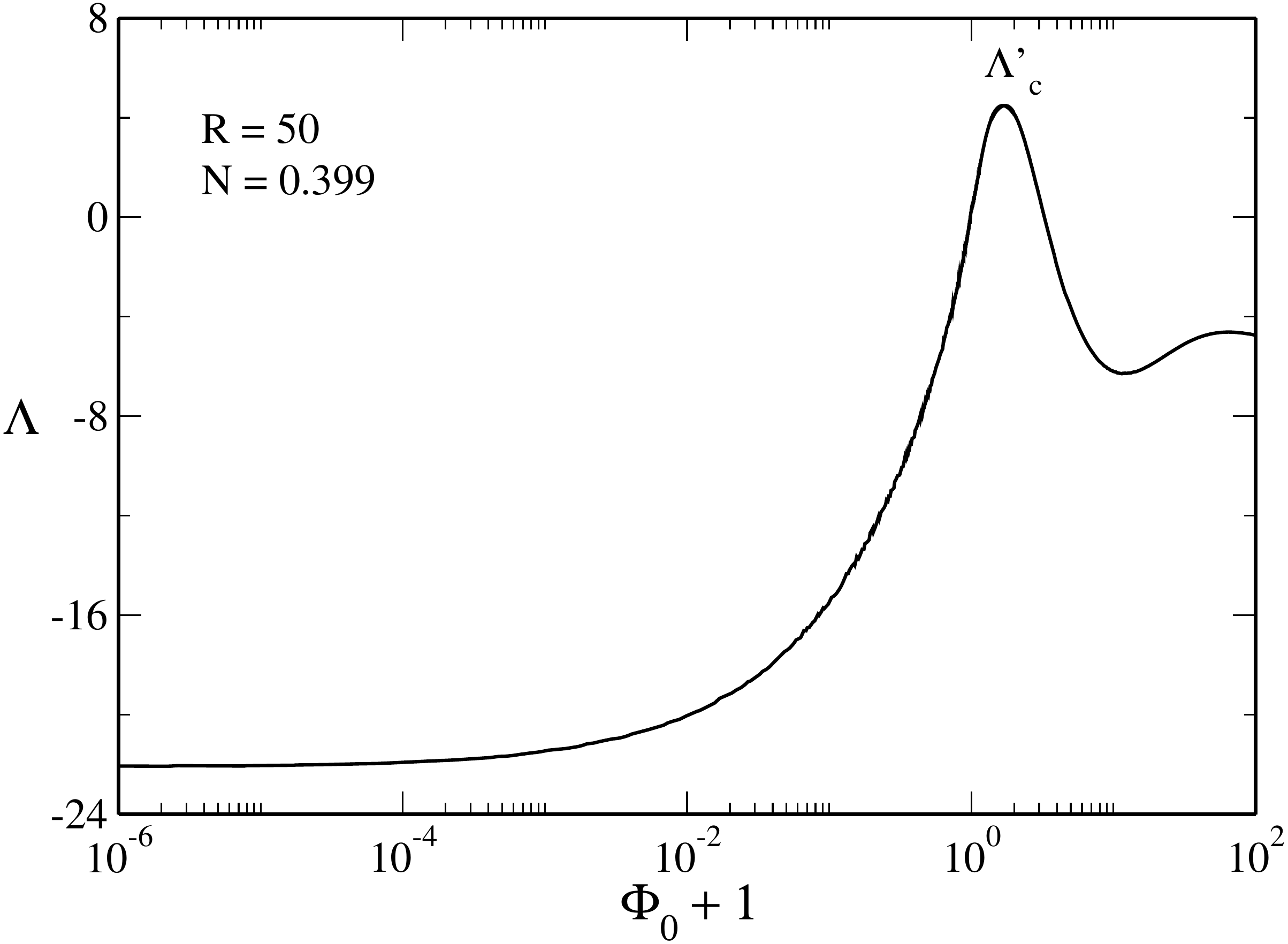}
\caption{Normalized energy as a function of the central potential $\Phi_0$
for $N_{\rm OV}<N<N_e$ (specifically $R = 50$
and $N = 0.399$).}
\label{LambdaPhi0_R50_N0p399PH}
\end{center}
\end{figure}

In Fig. \ref{LambdaPhi0_R50_N0p399PH} we have plotted the relation 
$\Lambda(\Phi_0)$
between the normalized energy
and the central potential. We can
see that $\Phi_0$ increases
monotonically along the series of equilibria. The curve $\Lambda(\Phi_0)$
presents a peak at $\Lambda'_c$ then displays damped oscillations. These
oscillations
correspond to the
unstable equilibrium states forming the spiral of 
the
caloric curve.

\begin{figure}
\begin{center}
\includegraphics[clip,scale=0.3]{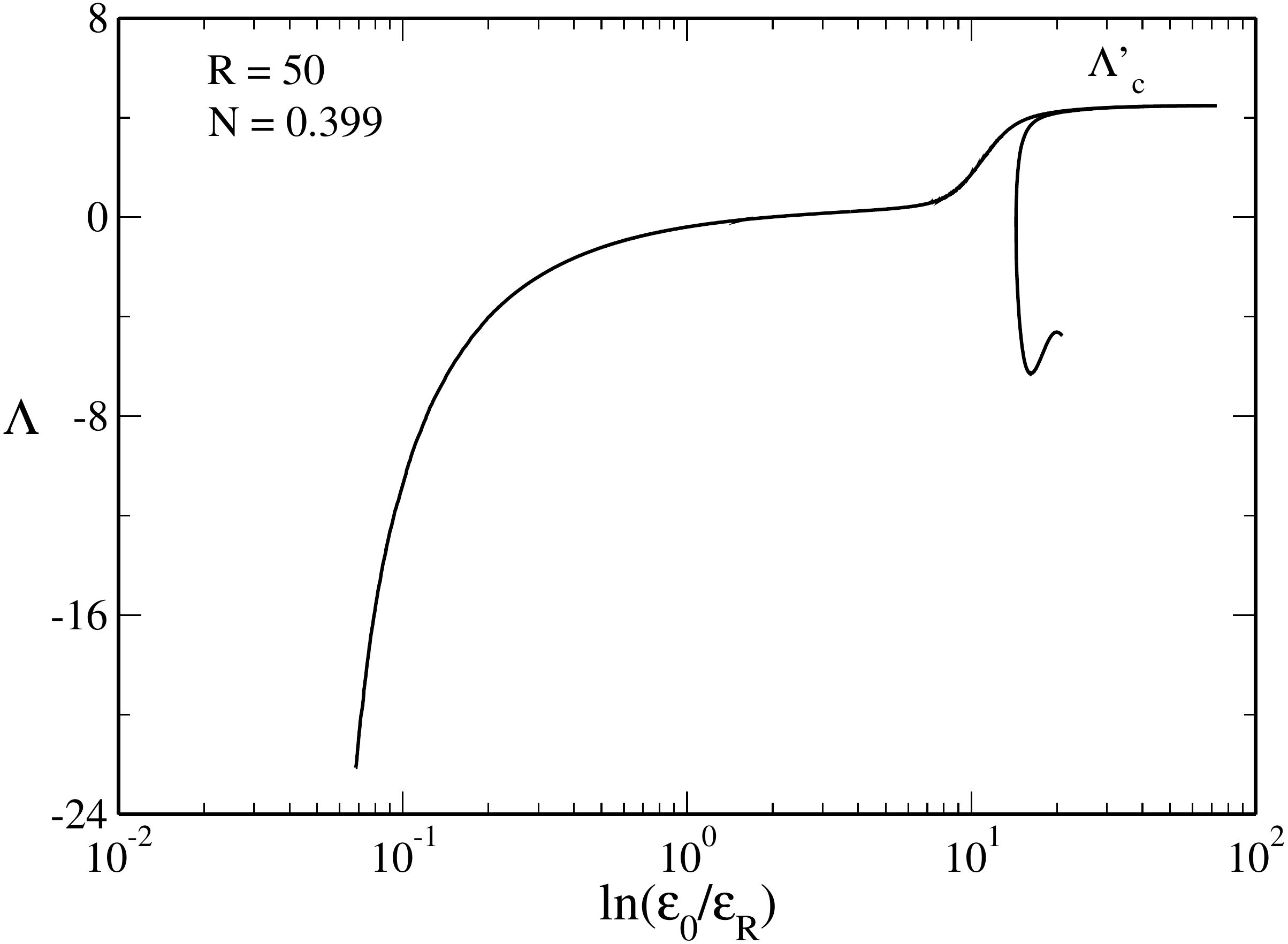}
\caption{Normalized energy as a function of the energy density contrast
$\mathcal{R} =
\epsilon_0/\epsilon_R$ for $N_{\rm OV}<N<N_e$ (specifically $R = 50$
and $N = 0.399$).}
\label{LambdaR_R50_N0p399PH}
\end{center}
\end{figure}

In Fig. \ref{LambdaR_R50_N0p399PH} we have plotted the relation
$\Lambda({\cal R})$ between
the normalized energy
and the energy density contrast ${\cal
R}=\epsilon_0/\epsilon_R$. We can see that ${\cal
R}$ increases monotonically along the series of equilibria up
to $\Lambda'_c$. Then, on the unstable branch, it displays a more complicated
behavior.

The evolution of the system in the microcanonical
ensemble is the following. Let us start from high energy states and decrease the
energy. As energy decreases, the system becomes more and more concentrated. The
central potential and the density contrast increase. If we keep decreasing the
energy
there comes a point $E'_c$ at which the system undergoes a gravitational
collapse towards a black hole. This is an instability of general relativistic
origin which has no counterpart in the Newtonian theory.

\subsubsection{Canonical ensemble}

\begin{figure}
\begin{center}
\includegraphics[clip,scale=0.3]{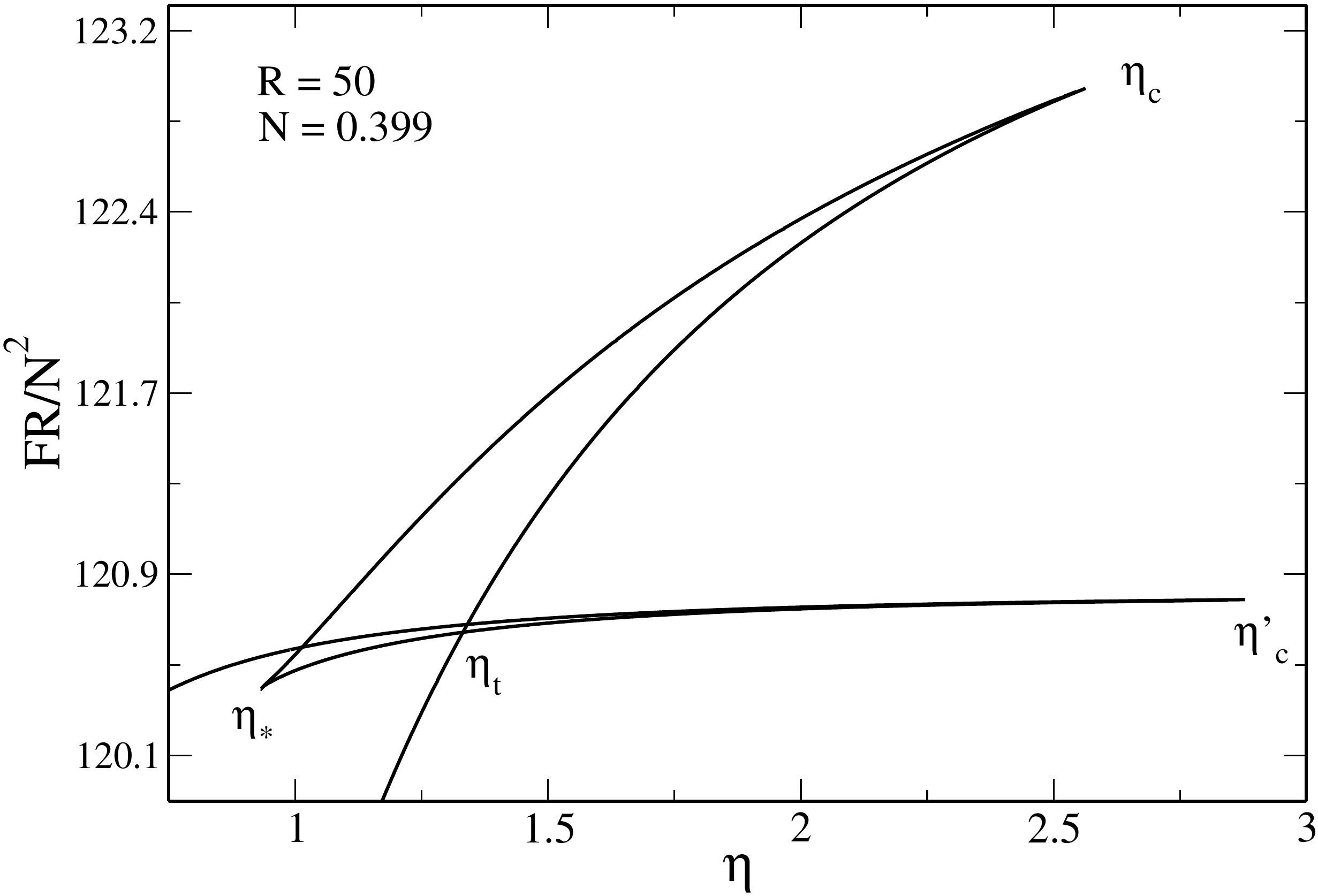}
\caption{Normalized free energy as a function of the normalized inverse
temperature for
$N_{\rm OV}<N<N_e$ (specifically $R = 50$
and $N = 0.399$). The first derivative of $\beta F$ with respect
to $\beta$ is discontinuous at the transition temperature $\beta_t$. This
corresponds to a first order phase transition, connecting
the gaseous phase to the condensed phase, which is associated with a jump of
energy $E=\partial(\beta F)/\partial\beta$ in the caloric curve. On the other
hand, $\beta F$ is discontinuous at the spinodal points $\eta_c$ and
$\eta_*$. This corresponds to zeroth order phase transitions
which are associated with a jump of free energy.
We can check that the unstable equilibrium 
states (saddle points of free energy) between $\eta_c$ and $\eta_*$ have a free
energy higher than
the stable equilibrium states (minima of free energy).
 However, the unstable
equilibrium 
states after $\eta'_c$ can have a free energy lower than
the stable equilibrium states before $\eta_c$.}
\label{Feta_R50_N0p399PH}
\end{center}
\end{figure}

We now consider the canonical ensemble. The
function $\Lambda(\eta)$
is multivalued. According to the
Poincar\'e
turning point criterion, the series of equilibria is stable  up
to $\eta_c$,
becomes unstable between $\eta_c$ and $\eta_*$,
is stable again between
$\eta_*$ and $\eta'_c$ and becomes unstable again after $\eta'_c$.
The caloric curve
presents the following features:

(i) When $\eta<\eta_*$ there are only gaseous states. When $\eta_c<\eta<\eta'_c$
there are only condensed states. When $\eta_*<\eta<\eta_c$ there exist gaseous
and condensed states at the same temperature. A first order phase transition is
expected at a transition temperature $\eta_t$ determined by the Maxwell
construction (see Fig. \ref{kcal_R50_N0p399_unifiedPH}) or by the equality of
the 
free energy of the gaseous and condensed phases (see
Fig. \ref{Feta_R50_N0p399PH}). When
$\eta_*<\eta<\eta_t$ the gaseous states have a lower free energy than the
condensed states. When  $\eta_t<\eta<\eta_c$ the condensed states have a
lower free energy than the gaseous states.  However, the
first order phase
transition does not take place in practice because of the very long lifetime of
the metastable states.

(ii) There is a zeroth order phase transition at $\eta_c$ from the gaseous
phase to the condensed phase. It corresponds to a gravitational
collapse (isothermal collapse) ultimately halted by quantum
degeneracy.

(iii) There is a zeroth order phase transition at $\eta_*$ 
from the condensed phase to the gaseous phase. It corresponds to an
explosion ultimately halted by the boundary of the box.

(iv) There is a catastrophic collapse at $\eta'_c$ from the condensed
phase to a black hole.

\begin{figure}
\begin{center}
\includegraphics[clip,scale=0.3]{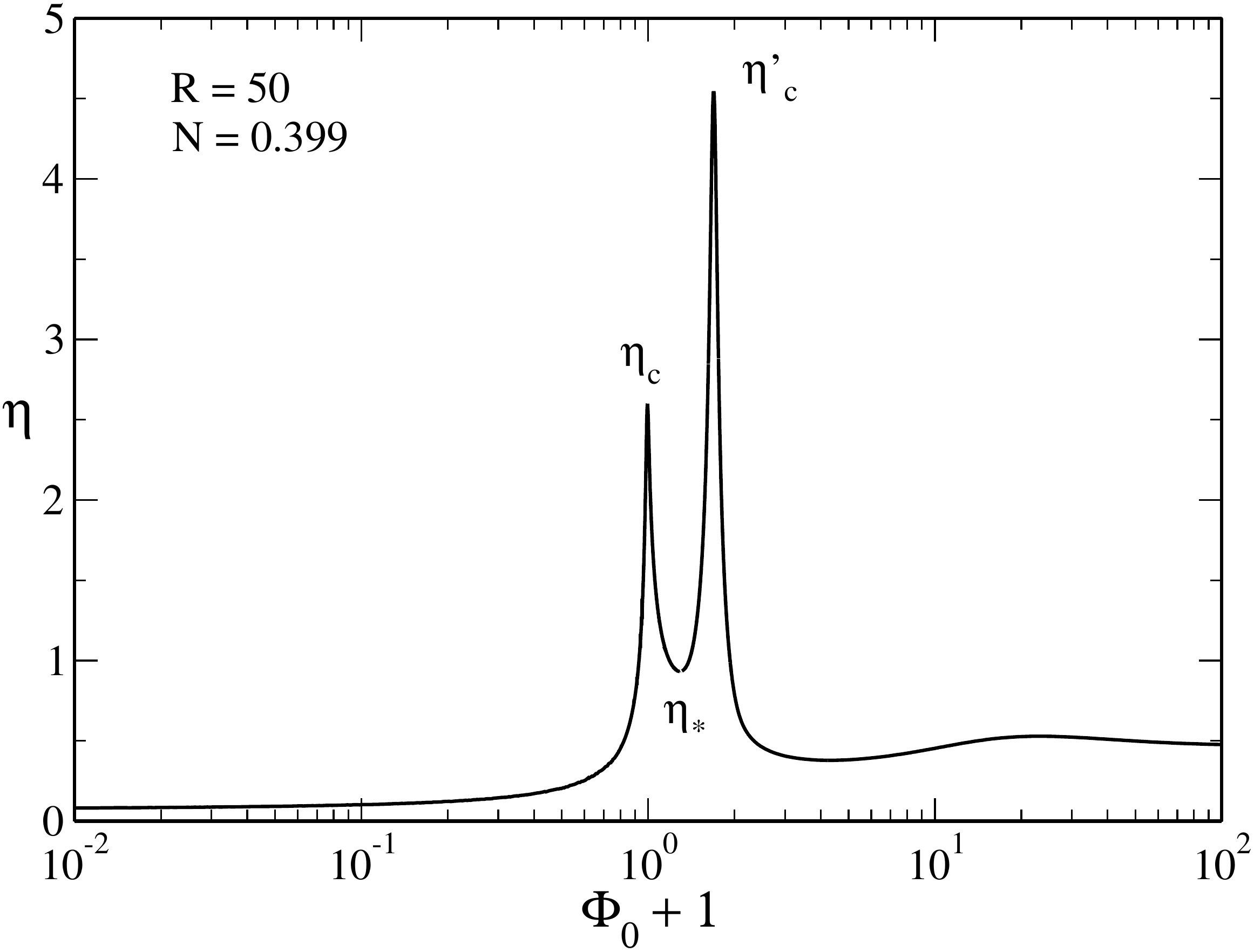}
\caption{Normalized inverse temperature as a function of the central potential
$\Phi_0$
for $N_{\rm OV}<N<N_e$ (specifically $R = 50$
and $N = 0.399$).}
\label{etaPhi0_R50_N0p399PH}
\end{center}
\end{figure}

In Fig. \ref{etaPhi0_R50_N0p399PH} we have plotted the relation 
$\eta(\Phi_0)$
between the inverse temperature
and the central potential. We
see that $\Phi_0$ increases
monotonically along the series of equilibria. The curve $\eta(\Phi_0)$
presents a first peak at $\eta_c$ and a second peak at $\eta'_c$. Then, it
displays damped oscillations.  They correspond to
unstable equilibrium states associated with the spiral of the
caloric curve.

\begin{figure}
\begin{center}
\includegraphics[clip,scale=0.3]{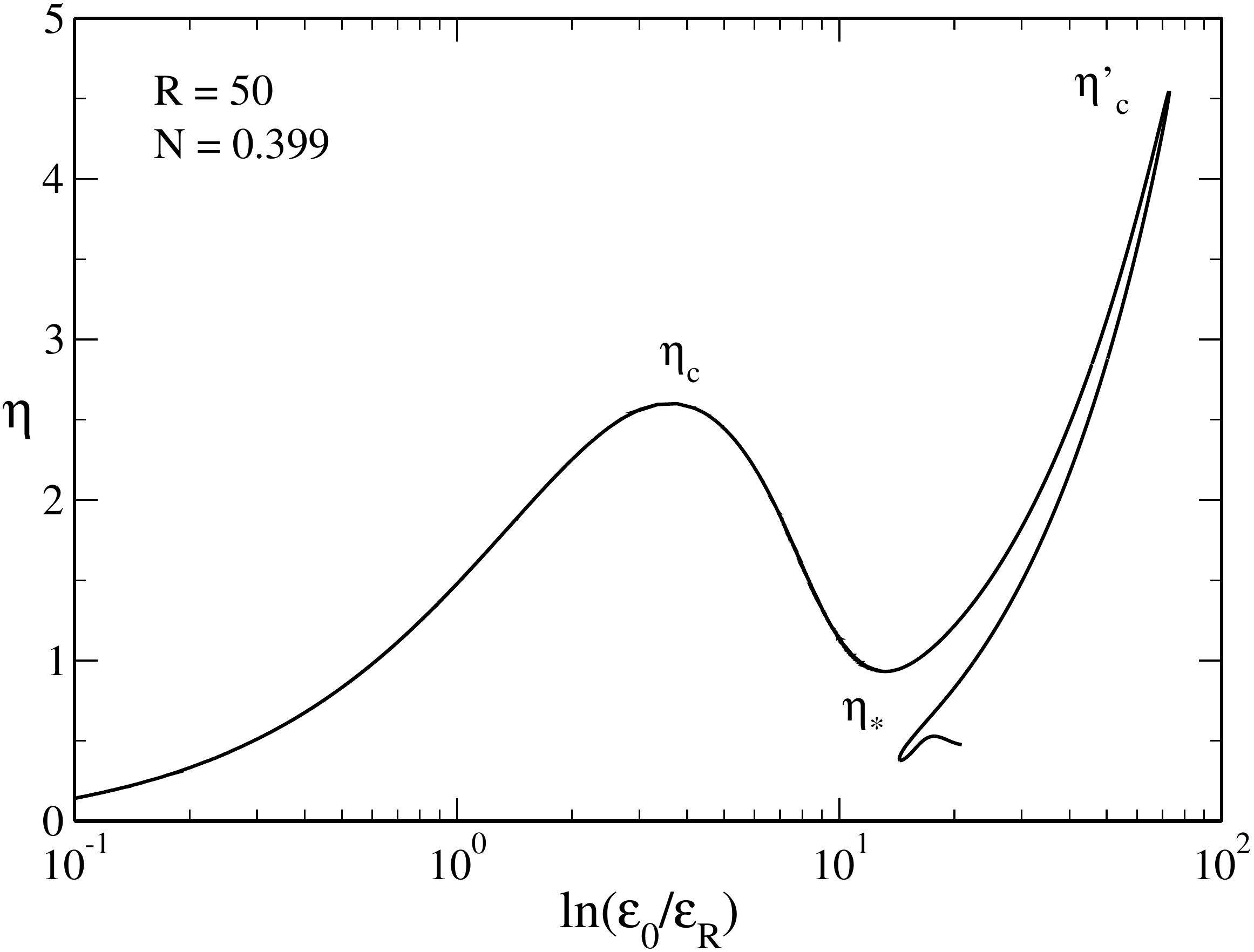}
\caption{Normalized inverse temperature as a function of the energy density
contrast
$\mathcal{R} =
\epsilon_0/\epsilon_R$ 
for $N_{\rm OV}<N<N_e$ (specifically $R = 50$
and $N = 0.399$). }
\label{etaR_R50_N0p399PH}
\end{center}
\end{figure}

In Fig. \ref{etaR_R50_N0p399PH} we have plotted the relation
$\eta({\cal R})$ between
the normalized inverse temperature
and the energy density contrast ${\cal
R}=\epsilon_0/\epsilon_R$. We can see that ${\cal
R}$ increases monotonically along the series of equilibria up
to $\eta'_c$. Then, on the second unstable branch, it displays a more
complicated
behavior.

The evolution of the system in the canonical
ensemble in the following. Let us start from high temperature states and
decrease the
temperature. At high temperatures, the system is in the gaseous phase. At
$\eta=\eta_t$, we
expect
the system to undergo a  first order phase transition from the gaseous phase to
the condensed phase. However, in practice, this phase transition does not take
place because the metastable gaseous states have a very long lifetime. The
physical
transition occurs at the critical temperature $\eta_c$ (spinodal point) at which
the gaseous phase
disappears. At that point the system undergoes a zeroth order phase
transition (collapse) from the gaseous
phase to the condensed phase.  If we keep
decreasing the temperature there comes another critical point $\eta'_c$ at
which the system undergoes a catastrophic
collapse from the condensed phase to a black hole. This is an instability
of general relativistic
origin which has no counterpart in the Newtonian theory. Inversely, if we
increase the temperature, the system
displays a zeroth order phase
transition (explosion) at $\eta_*$ from the
condensed phase to the gaseous phase.

\subsection{The case $N_e<N<N'_e$}

\begin{figure}
\begin{center}
\includegraphics[clip,scale=0.3]{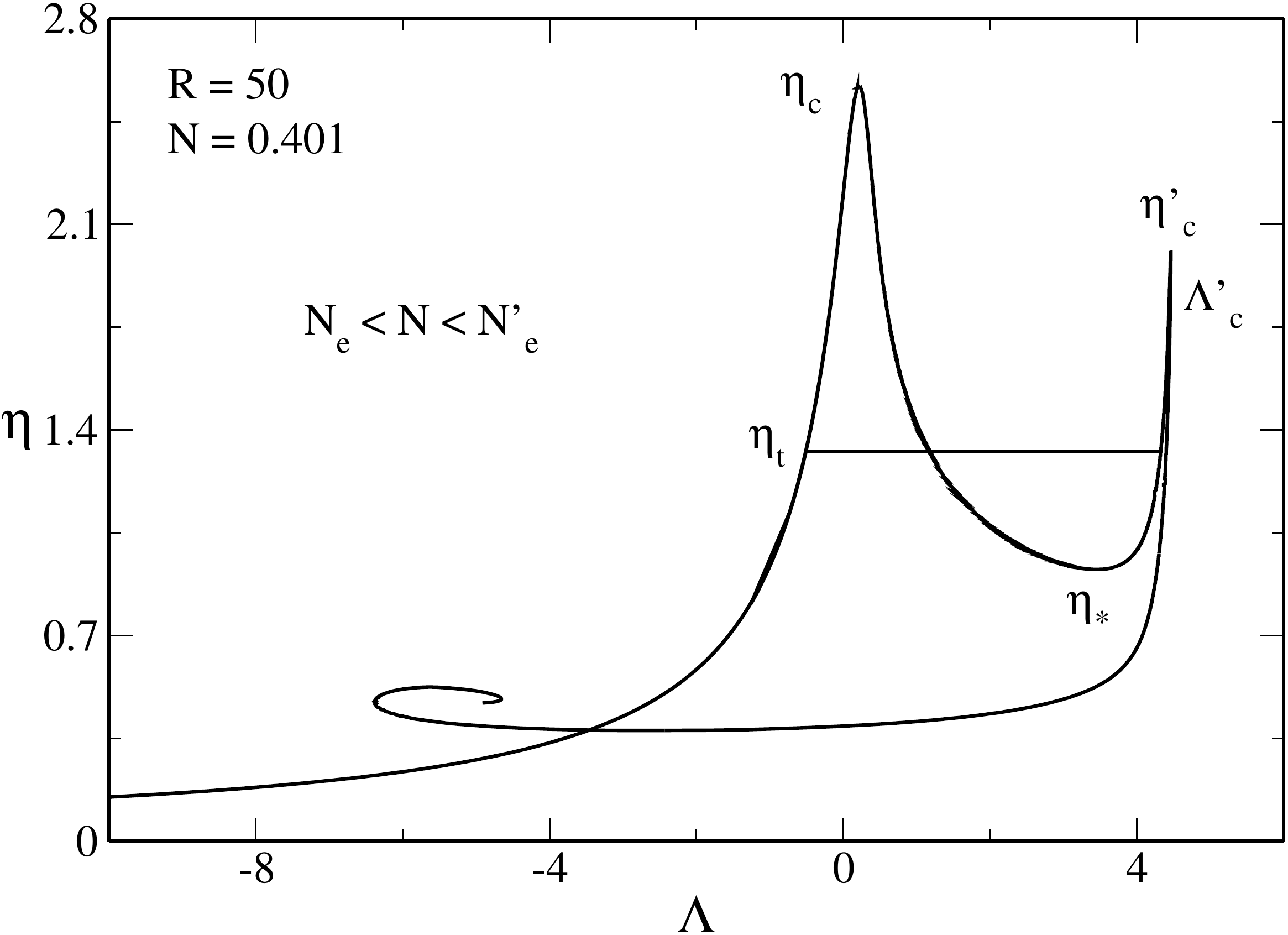}
\caption{Caloric curve for $N_e=0.40002<N<N'_e=0.40469$  (specifically $R = 50$
and $N = 0.401$).}
\label{Xkcal_R50_N0p401_unifiedPH}
\end{center}
\end{figure}

In Fig. \ref{Xkcal_R50_N0p401_unifiedPH} we have plotted the caloric
curve for $N_e=0.40002<N<N'_e=0.40469$. The novelty with
respect to the
previous case is that now $\eta'_c$ is smaller than $\eta_c$ (they become equal
when  $N=N_e=0.40002$).

\begin{figure}
\begin{center}
\includegraphics[clip,scale=0.3]{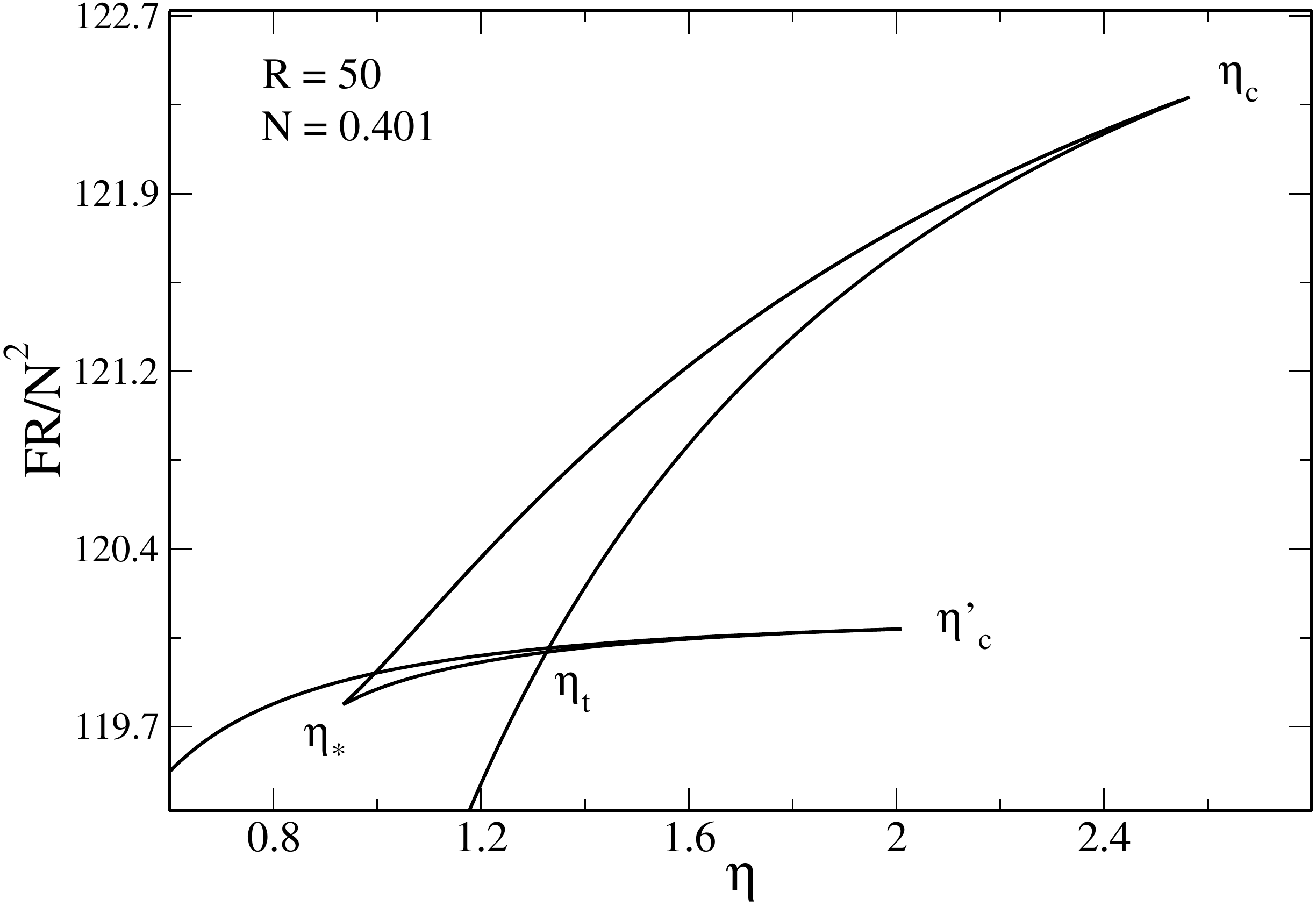}
\caption{Normalized free energy as a function of the inverse temperature for
$N_e<N<N'_e$ 
(specifically $R = 50$ and $N = 0.401$).}
\label{Feta_R50_N0p401_colorPH}
\end{center}
\end{figure}

The description in the microcanonical ensemble
is the same as
before.

In the canonical ensemble, the caloric curve presents the following
features:

(i) When $\eta<\eta_*$
and when $\eta'_c<\eta<\eta_c$ there are only gaseous states. When
$\eta_*<\eta<\eta'_c$ there exist gaseous and condensed states at the same
temperature. A first
order phase transition is expected  at a transition temperature  $\eta_t$
determined by the Maxwell construction (see Fig.
\ref{Xkcal_R50_N0p401_unifiedPH}) or by the equality of the free energy of the
two phases (see
Fig. \ref{Feta_R50_N0p401_colorPH}). When
$\eta_*<\eta<\eta_t$ the gaseous states have a lower free energy than the
condensed states. When $\eta_t<\eta<\eta'_c$ the condensed states
have a lower free energy than the gaseous states. However,
the first order phase
transition does not take place in practice because of the very long lifetime of
the
metastable states.

(ii) There is a catastrophic collapse at $\eta_c$ from the gaseous phase
to a black hole.

(iii) There is a  catastrophic  collapse at $\eta'_c$ from the condensed phase
to a black hole.

(iv) There is a zeroth order phase transition at $\eta_*$ from
the condensed
phase to the gaseous phase. It correspond to an explosion
ultimately halted by the boundary of the box.

The evolution of the system in the canonical
ensemble is the following. Let us start from high temperature states and
decrease the
temperature.  At high temperatures, the system is in the gaseous phase. At
$\eta=\eta_t$ the system is expected to undergo a  first order
phase transition from the gaseous phase to
the condensed phase. However, this phase transition does not take place in
practice. At $\eta=\eta_c$ the system undergoes a catastrophic collapse towards
a black hole.  A condensed phase exists for
$\eta_*<\eta<\eta'_c$ but it is not clear how it can be reached in practice.

\subsection{The case $N'_e<N<N_*$}

\begin{figure}
\begin{center}
\includegraphics[clip,scale=0.3]{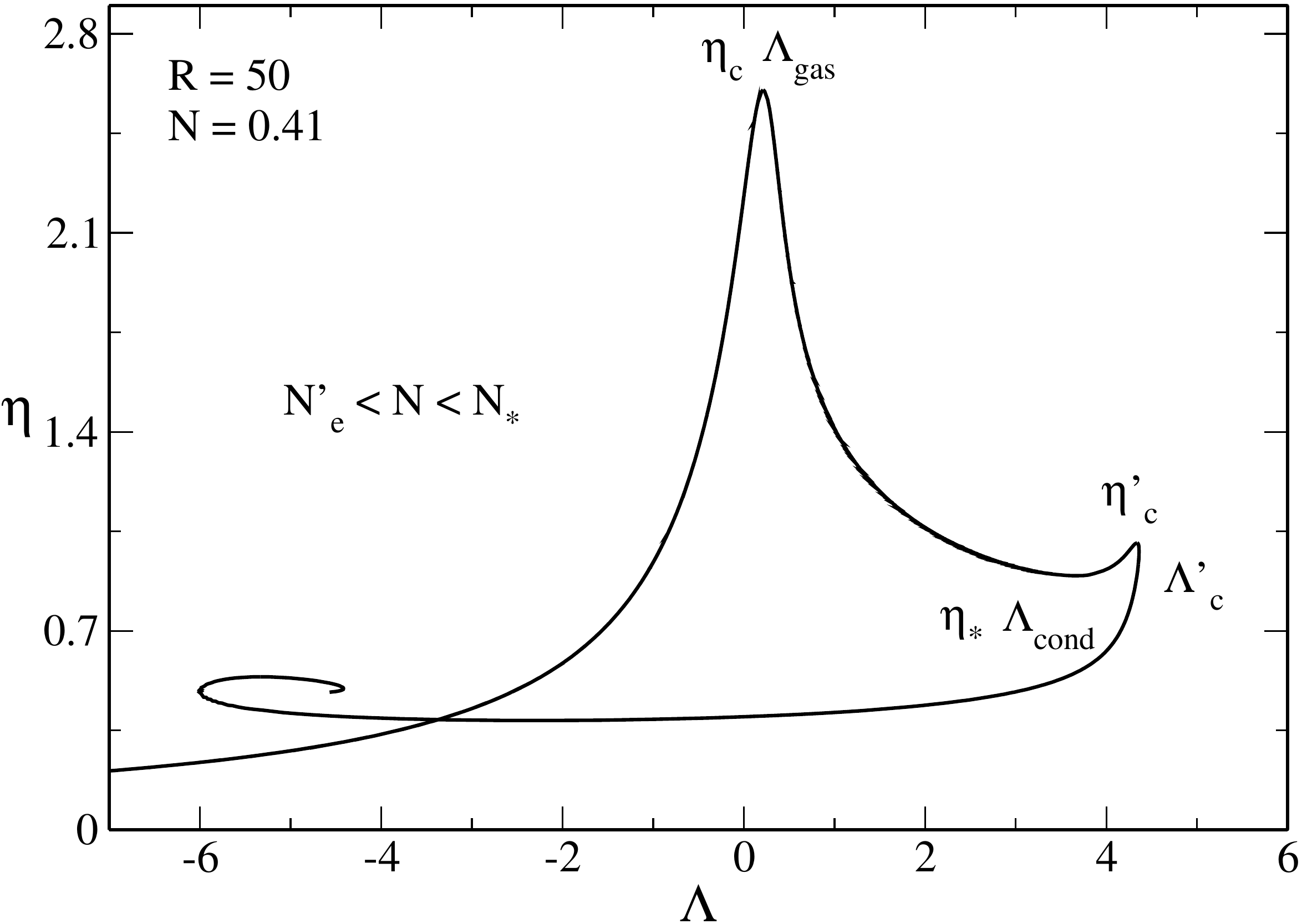}
\caption{Caloric curve for $N'_e=0.40469<N<N_*=0.41637$ (specifically $R = 50$
and
$N =
0.41$).}
\label{kcal_R50_N0p41_unifiedPH}
\end{center}
\end{figure}

In Fig. \ref{kcal_R50_N0p41_unifiedPH} we have plotted the caloric
curve for $N'_e=0.40469<N<N_*=0.41637$, where $N'_e$ is defined such that 
$\eta'_c=\eta_t$.

\begin{figure}
\begin{center}
\includegraphics[clip,scale=0.3]{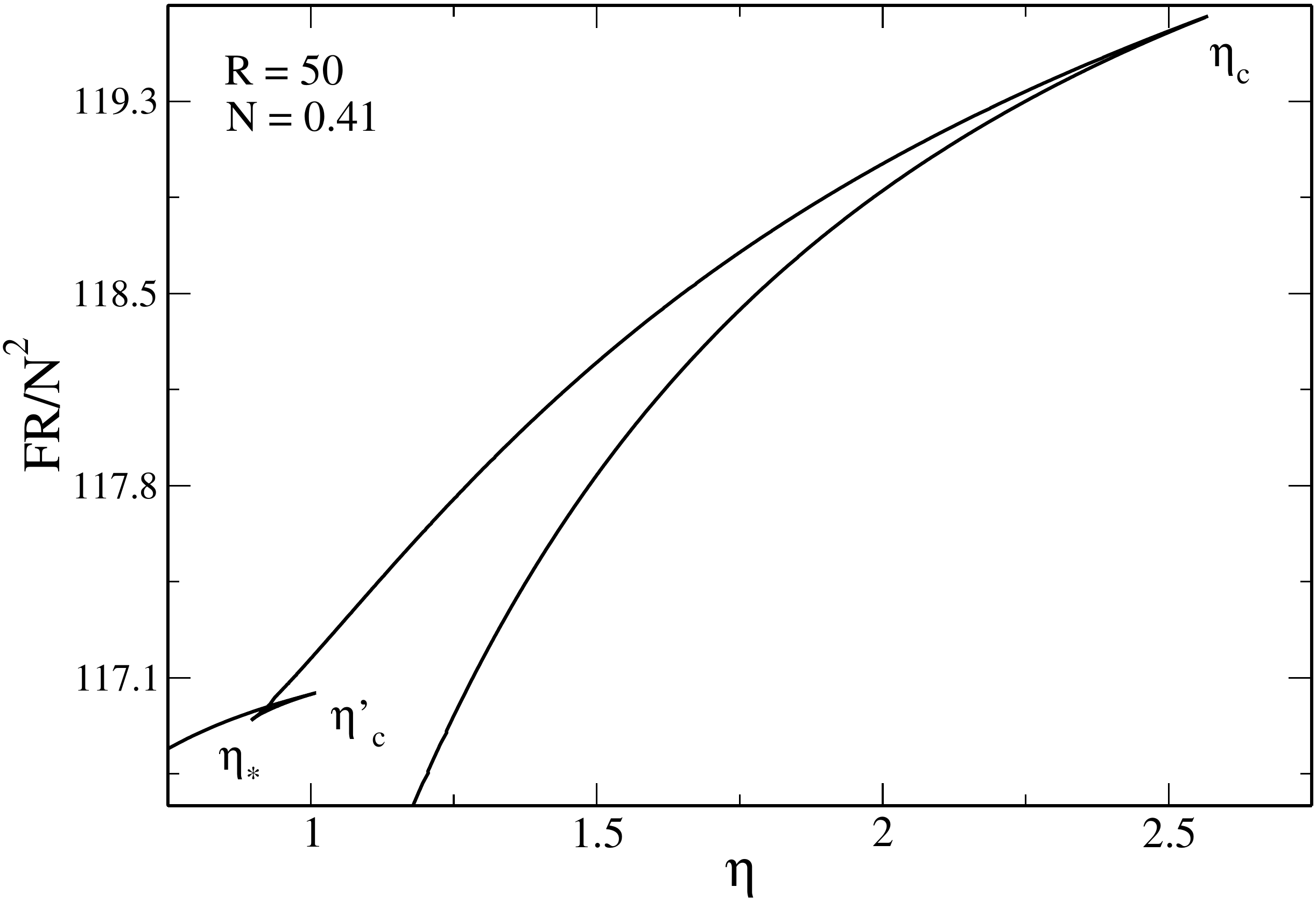}
\caption{Normalized free energy as a function of the inverse temperature for
$N'_e<N<N_*$
(specifically $R = 50$ and $N =
0.41$).}
\label{Feta_R50_N0p41PH}
\end{center}
\end{figure}

The description in the microcanonical
ensemble is the same as
before.

In the canonical ensemble, the caloric curve presents the following features:

(i) When $\eta<\eta_*$
and when $\eta'_c<\eta<\eta_c$ there are only gaseous states. When
$\eta_*<\eta<\eta'_c$ there exist gaseous and condensed states at the same
temperature. However,
there is no first order phase transition, even in theory, because we cannot
satisfy the  Maxwell
construction (see Fig. \ref{kcal_R50_N0p41_unifiedPH}) or the equality of the
free energy of the gaseous and condensed
phases (see Fig. \ref{Feta_R50_N0p41PH}). When
$\eta_*<\eta<\eta'_c$ the
gaseous states always have a lower free energy than the condensed states (see
Fig. \ref{Feta_R50_N0p41PH}). Therefore, although there are several
stable equilibrium states when $\eta_*<\eta<\eta'_c$ there is no phase
transition
from one phase to the other. This is a particularity of the relativistic
situation.

(ii) There is a  catastrophic collapse at $\eta_c$ from the gaseous phase
to a
black hole.

(iii) There is a catastrophic collapse at $\eta'_c$  from the condensed
phase to a
black hole.

(iv) There is a zeroth order phase transition at $\eta_*$ from
the condensed
phase to the gaseous phase. It corresponds to an explosion ultimately
halted by the boundary of the box.

The evolution of the system is the same as described previously.

\subsection{The case $N>N_*$}
\label{sec_fromto}

\begin{figure}
\begin{center}
\includegraphics[clip,scale=0.3]{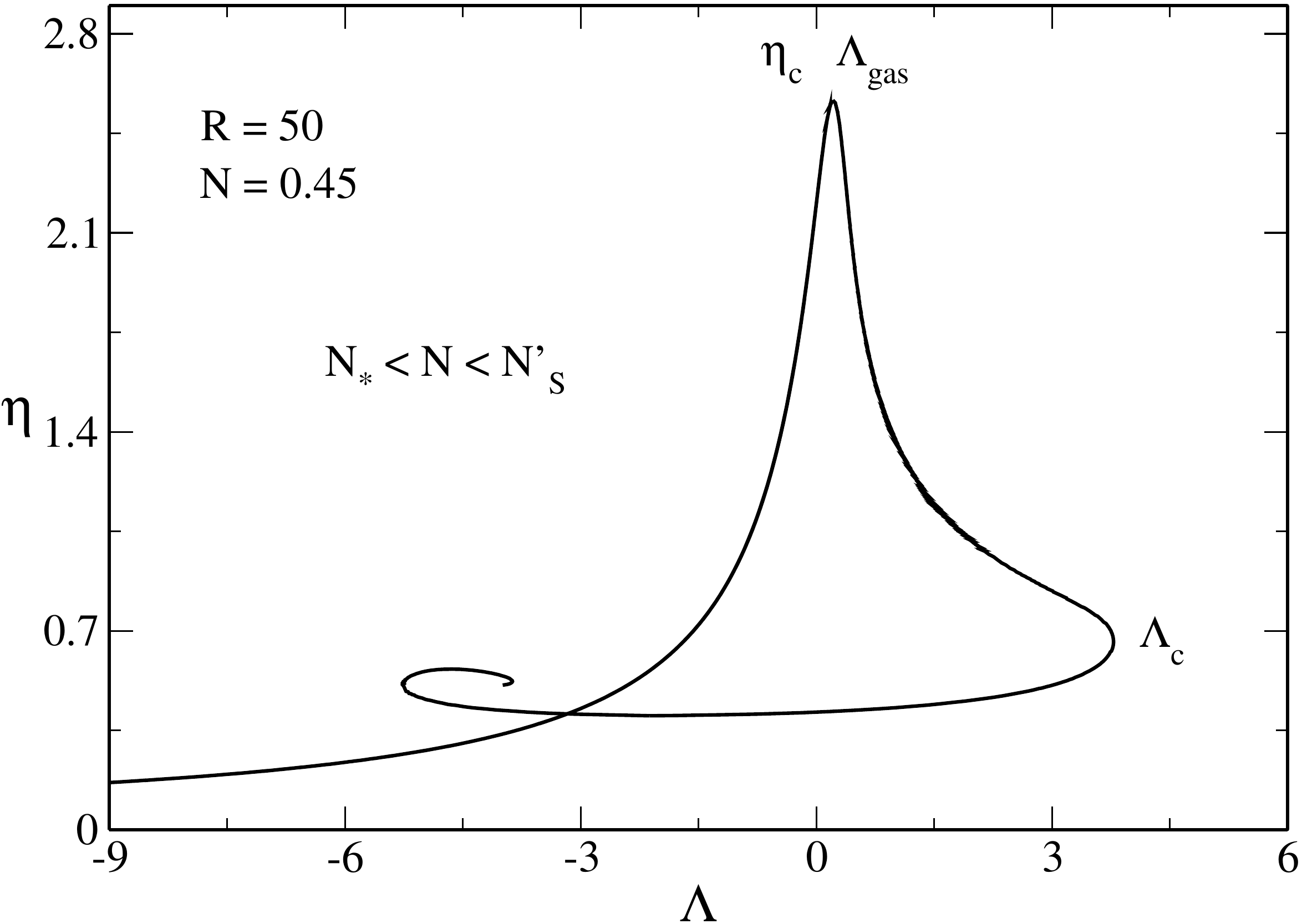}
\caption{Caloric
curve for  $N_*=0.41637<N<N'_S=6.40$ (specifically 
$R = 50$ and $N = 0.45$).}
\label{Xkcal_R50_N0p45_unified2PH}
\end{center}
\end{figure}

In Fig. \ref{Xkcal_R50_N0p45_unified2PH} we have plotted the caloric
curve for  $N>N_*=0.41637$, where $N_*$ is defined such that $\eta'_c=\eta_*$.
From
that moment, we denote the minimum energy by $\Lambda_c$ instead of
$\Lambda'_c$.

\subsubsection{Microcanonical ensemble}

Let us first consider the microcanonical ensemble. The curve
$\eta(\Lambda)$ is multivalued. According to
the Poincar\'e
turning point criterion, the series of equilibria is stable up to
$\Lambda_c$ and then becomes unstable. The caloric curve
presents the following features:

(i) There is no phase transition (there is only one stable equilibrium state
for each $\Lambda<\Lambda_c$). 

(ii) There is a region of negative specific heats between $\Lambda_{\rm
gas}$ and $\Lambda_{c}$.

(iii) There is a catastrophic collapse at $\Lambda_c$ towards  a black
hole.

The evolution of the system is the same as described previously.

\subsubsection{Canonical ensemble}

We now consider the canonical ensemble. The function
$\Lambda(\eta)$ is multivalued. According to
the Poincar\'e
turning point criterion, the series of equilibria is stable  up to
$\eta_c$ and then becomes unstable.  The caloric curve
presents the following features:

(i) There is no phase transition (there is only one stable equilibrium state
for each $\eta<\eta_c$). 

(ii) There is a catastrophic collapse at $\eta_c$
towards a black hole.

The evolution of the system is the same as described previously. The only
difference  is
that the condensed phase has
disappeared.

\subsection{Larger values of $N$}
\label{sec_fin}

\begin{figure}
\begin{center}
\includegraphics[clip,scale=0.3]{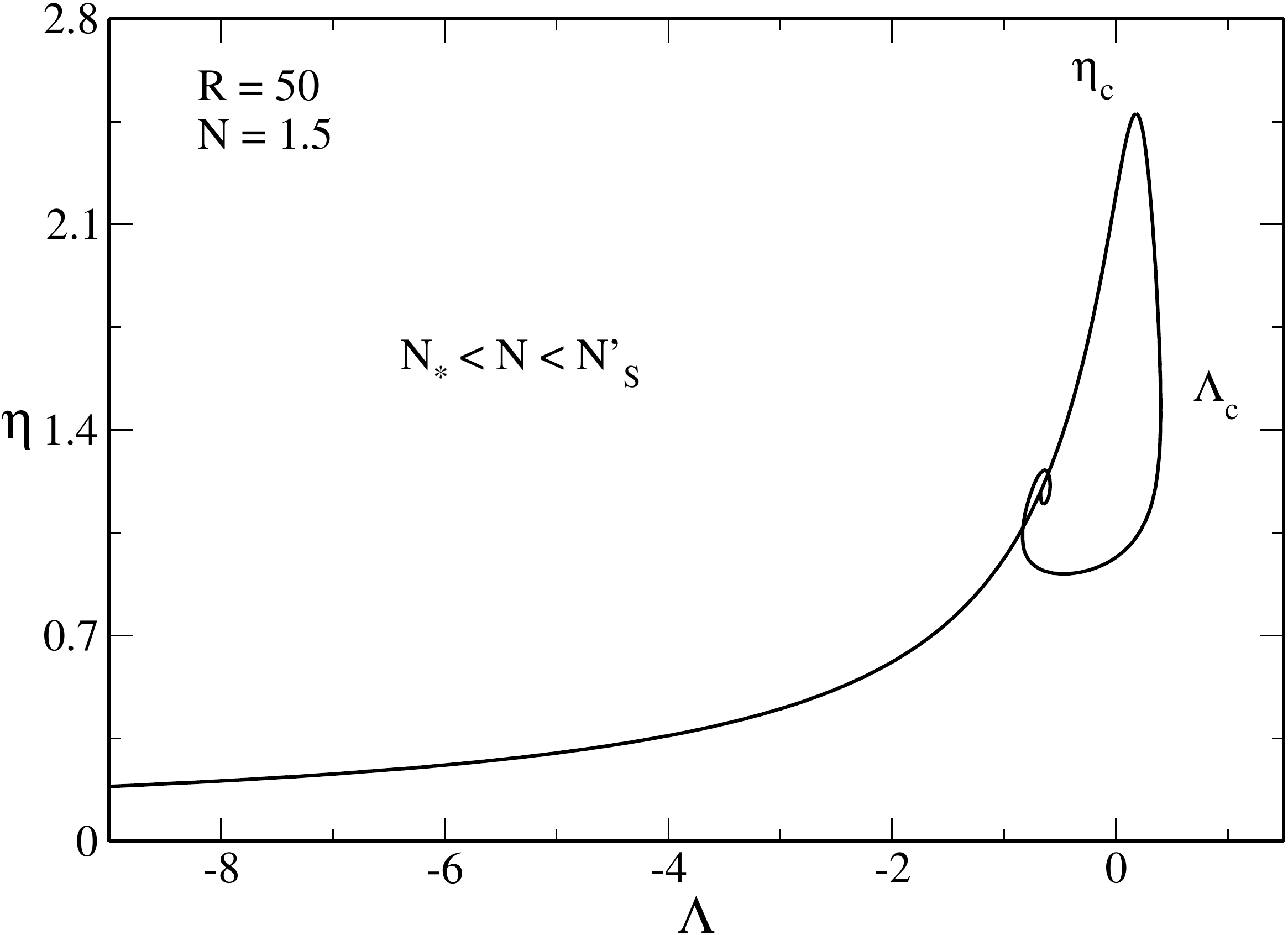}
\caption{Caloric
curve for $N_*<N<N'_S$ (specifically $R =
50$ and
$N = 1.5$).}
\label{kcal_R50_N1p5_unifiedPH}
\end{center}
\end{figure}

\begin{figure}
\begin{center}
\includegraphics[clip,scale=0.3]{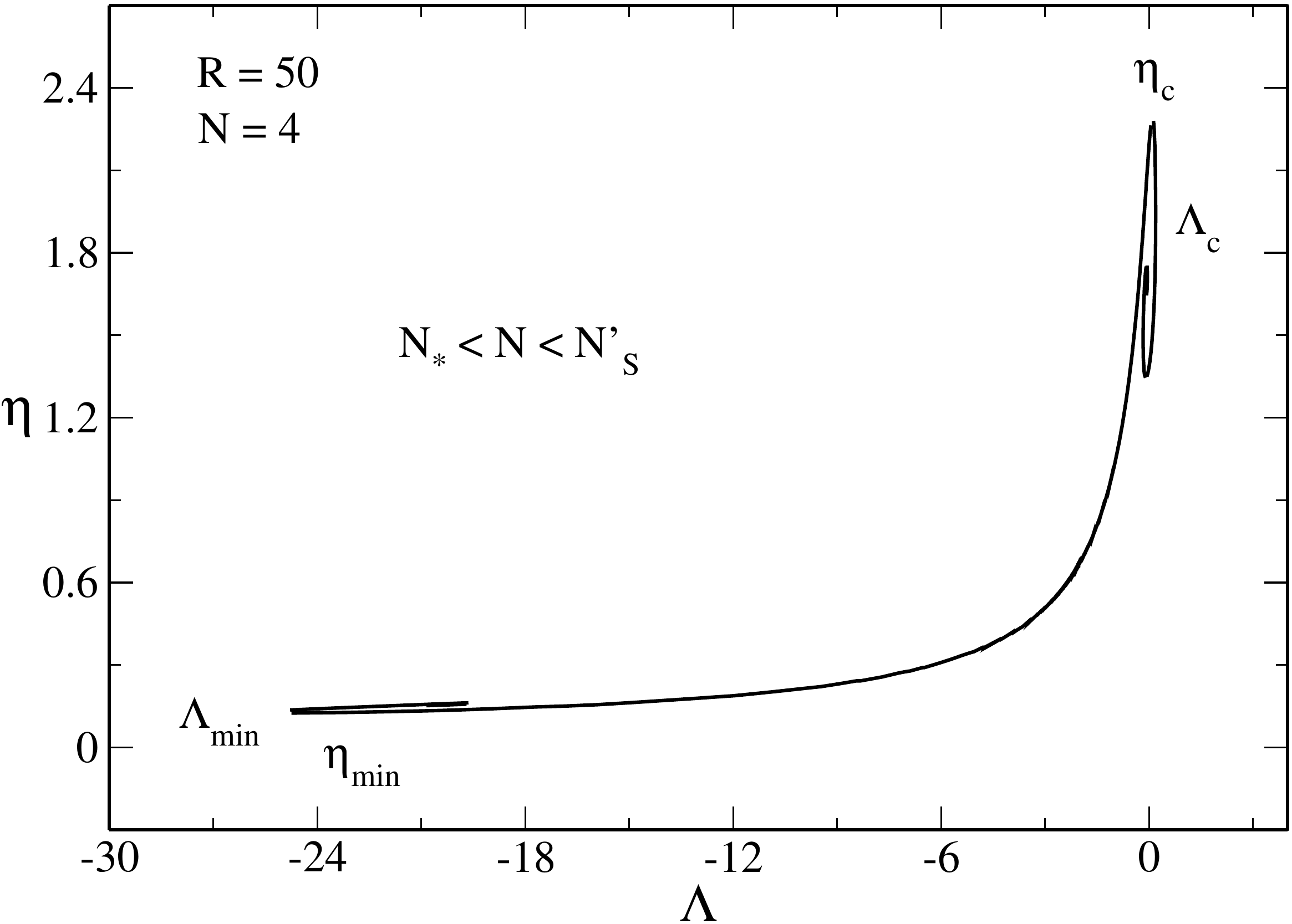}
\caption{Caloric curve for $N_*<N<N'_S$ (specifically  $R = 50$
and $N = 4$).}
\label{kcal_R50_N4_unifiedPH}
\end{center}
\end{figure}

In Figs. \ref{kcal_R50_N1p5_unifiedPH} and \ref{kcal_R50_N4_unifiedPH} we have
plotted the caloric curves for larger values of $N$. When
$N\gg N_{\rm OV}=0.39853$,  the system is
nondegenerate and we recover the results of \cite{roupas,paper2} for a
classical general relativistic gas described by the Boltzmann
distribution.\footnote{As
discussed in Sec. \ref{sec_ncl} the classical limit corresponds to $N\rightarrow
+\infty$ and $R\rightarrow +\infty$ in such a way that $N/R$ is fixed (in more
physical terms $N\gg N_{\rm OV}$  and $R\gg R_{\rm OV}$ with $N/R$ fixed).} 
The caloric curve exhibits a double spiral. When
$N<N'_S=6.40$ (see Fig. 7 of \cite{paper2}) the two
spirals are separated. When  $N'_S=6.40<N<N_S=7.08$ (see Fig. 8
of \cite{paper2}) the two
spirals are amputed (truncated) and touch each other. When $N_S=7.08<N<N_{\rm
max}=8.821$ (see Fig. 9 of \cite{paper2}) the spirals
disappear
and the caloric curve makes a 
``loop''. When $N\rightarrow N_{\rm max}$, the caloric curve reduces to a 
``point'' located at
$(\Lambda_*,\eta_*)=(-0.9829,1.2203)$.

\subsection{The canonical phase diagram}

\begin{figure}
\begin{center}
\includegraphics[clip,scale=0.3]{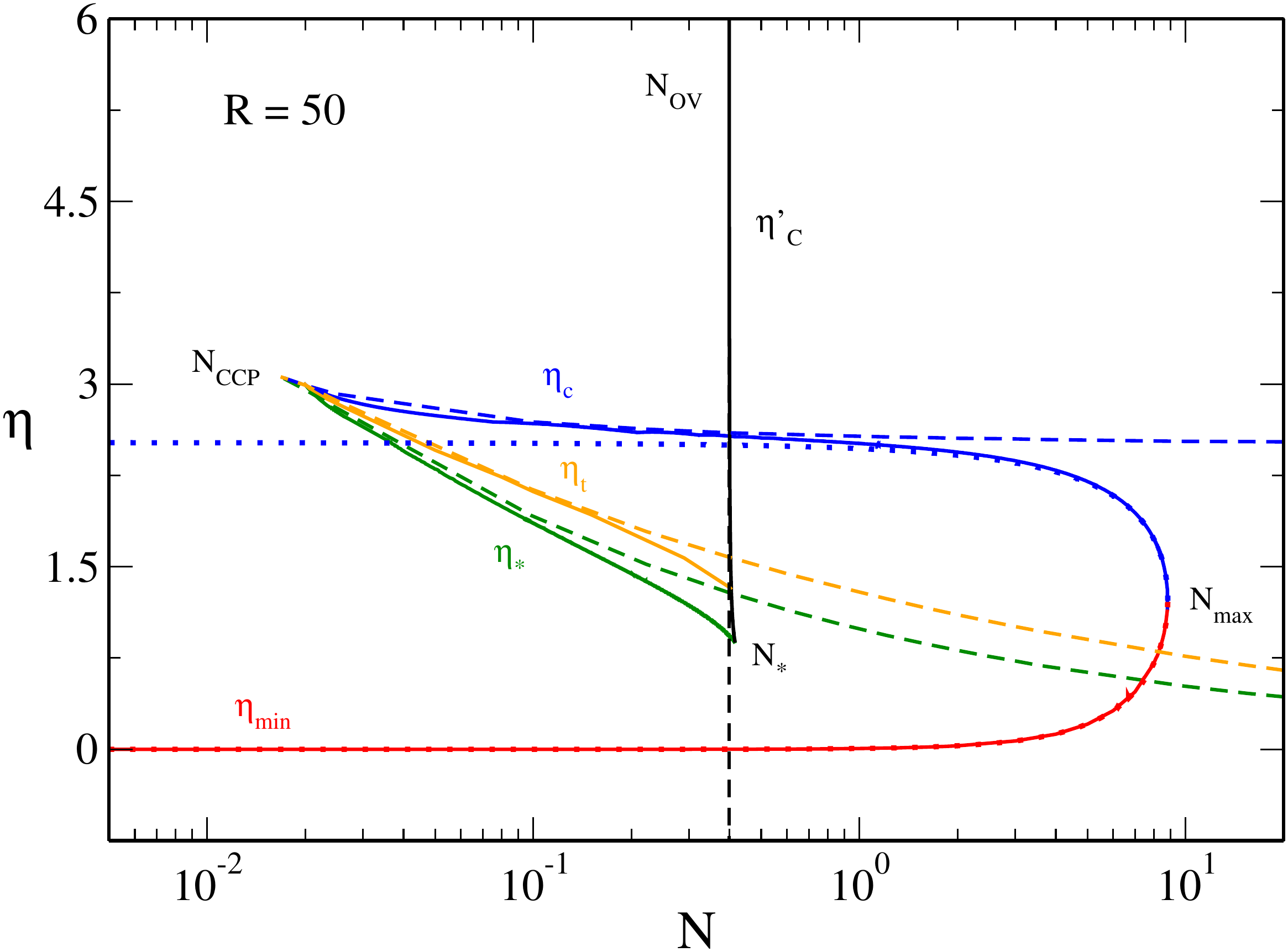}
\caption{Canonical phase diagram  for  $R_{\rm CCP}=12.0<R<R_{\rm MCP}=92.0$
(specifically $R = 50$).}
\label{Xphase_eta_R50_def_new2PH}
\end{center}
\end{figure}

\begin{figure}
\begin{center}
\includegraphics[clip,scale=0.3]{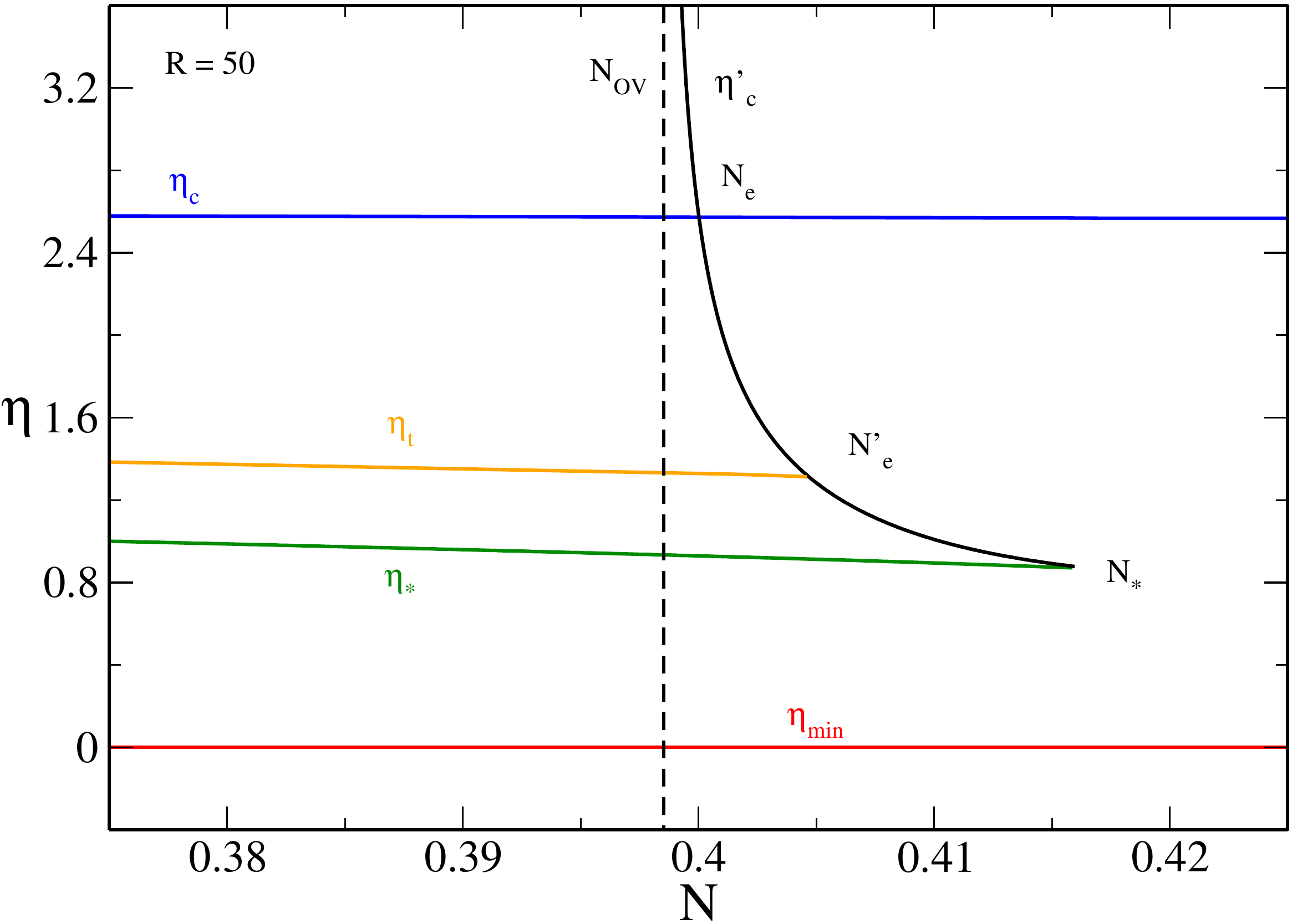}
\caption{Zoom of the canonical phase diagram  for  $R_{\rm CCP}<R<R_{\rm MCP}$
(specifically $R = 50$) in the
region near $N_{\rm OV}$. For $N\rightarrow N_{\rm OV}^+$, we
find that
$\eta'_c\sim 0.104 \, (N-N_{\rm OV})^{-1/2}$.}
\label{Xphase_eta_R50_defZOOM_new2PH}
\end{center}
\end{figure}

In Figs. \ref{Xphase_eta_R50_def_new2PH} and \ref{Xphase_eta_R50_defZOOM_new2PH}
we
have represented the canonical phase diagram corresponding to $R_{\rm
CCP}=12.0<R<R_{\rm MCP}=92.0$ (specifically $R=50$). It shows the evolution  of
the critical
temperatures
$\eta_{\rm min}$, $\eta_c$, $\eta_*$, $\eta_t$, $\eta'_c$ with
$N$. We can
clearly see the canonical critical point at $N_{\rm CCP}=0.0170$ at which the
canonical
phase transition appears. We also see the point $N_{\rm OV}=0.39853$ above which
quantum mechanics
is not able to prevent gravitational collapse above $\eta'_c(N)$. Finally, we
see
the point
$N_{\rm max}=8.821$ above which there is no equilibrium state anymore.

The nonrelativistic limit \cite{ijmpb} corresponds to the dashed lines. It
provides a very good approximation of
$\eta_c$, $\eta_*$ and $\eta_t$ for $N\ll N_{\rm OV}$. As
we approach $N_{\rm OV}$ general relativity must be taken into account.

The classical  limit \cite{roupas,paper2} corresponds to the dotted
lines.
It provides a very good
approximation of $\eta_{\rm min}$ (hot
spiral) for any $N$. It also provides a very good
approximation of $\eta_c$ (cold spiral) for $N\gg N_{\rm OV}$. As we
approach $N_{\rm
OV}$ quantum mechanics must be taken into account.

\subsection{The microcanonical phase diagram}

\begin{figure}
\begin{center}
\includegraphics[clip,scale=0.3]{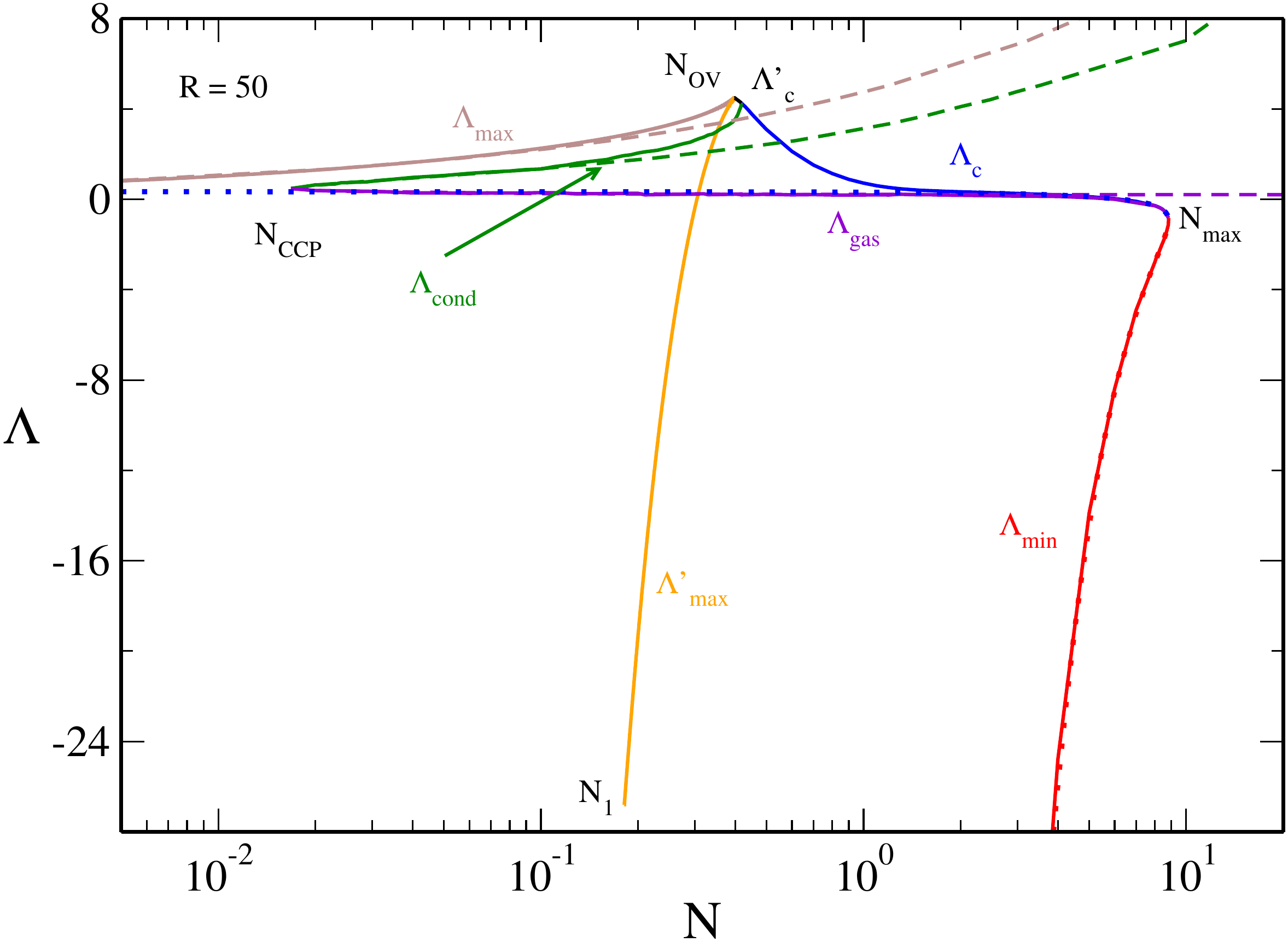}
\caption{Microcanonical phase diagram   for  $R_{\rm
CCP}=12.0<R<R_{\rm MCP}=92.0$
(specifically $R = 50$).}
\label{Xphase_Lambda_R50_def4PH}
\end{center}
\end{figure}

\begin{figure}
\begin{center}
\includegraphics[clip,scale=0.3]{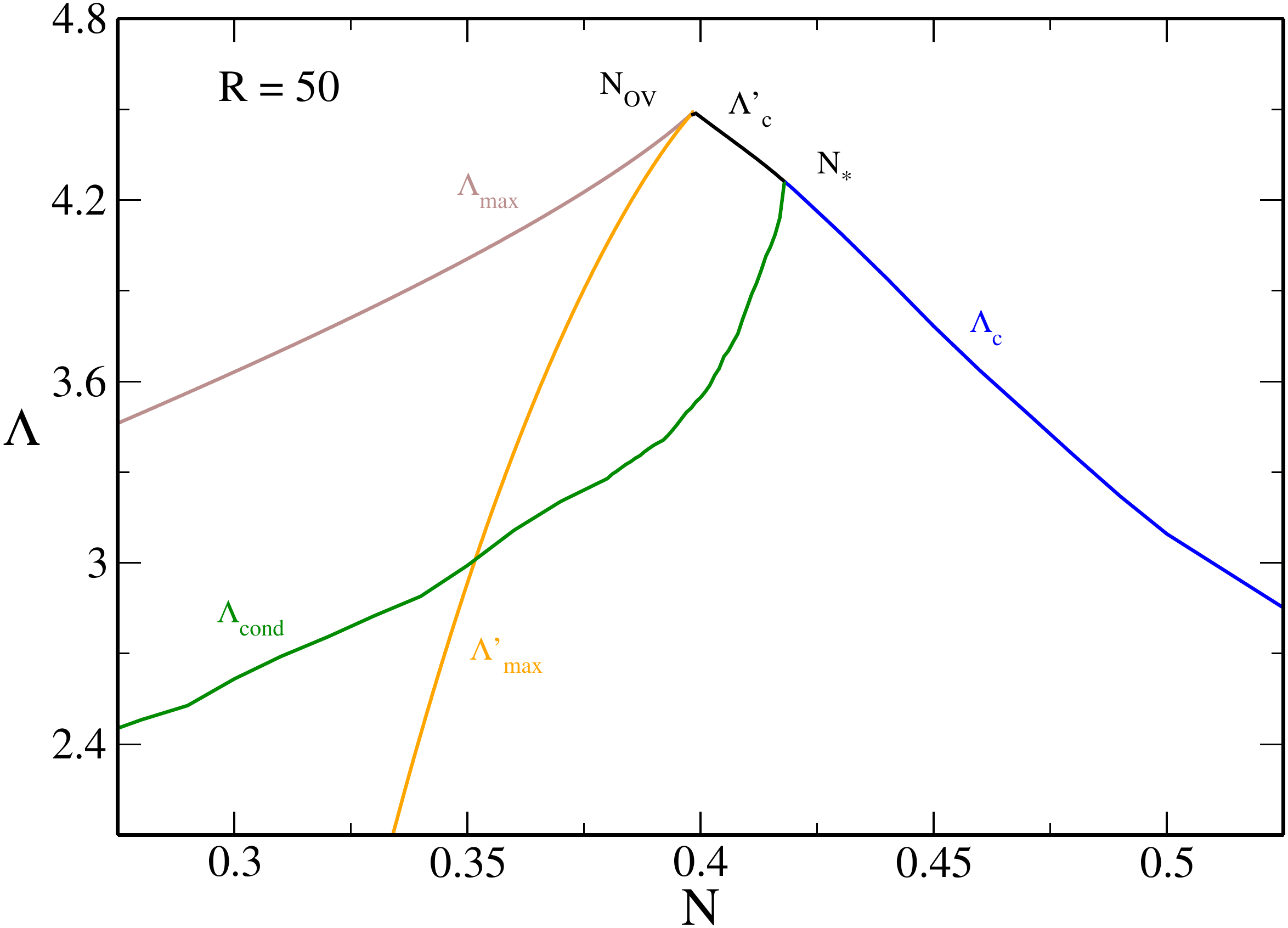}
\caption{Zoom of the microcanonical phase diagram for  $R_{\rm CCP}<R<R_{\rm
MCP}$ (specifically $R = 50$) in the
region near $N_{\rm OV}$.}
\label{phase_Lambda_R50_def2PH}
\end{center}
\end{figure}

In Figs. \ref{Xphase_Lambda_R50_def4PH} and \ref{phase_Lambda_R50_def2PH} we
have represented the microcanonical phase diagram corresponding to 
$R_{\rm
CCP}=12.0<R<R_{\rm MCP}=92.0$ (specifically $R=50$). It shows the evolution of
the critical energies
$\Lambda_{\rm min}$, $\Lambda_{\rm max}$, $\Lambda'_{\rm max}$, $\Lambda_{\rm
gas}$, $\Lambda_{\rm cond}$, $\Lambda'_c$, $\Lambda_c$ with $N$. We can
clearly see the canonical critical point at $N_{\rm CCP}=0.0170$ at
which the
region of negative specific heat (associated with the canonical phase
transition) appears.  We also see the
point
$N_{\rm OV}=0.39853$ above which quantum mechanics
is not able to prevent gravitational collapse above $\Lambda'_c(N)$, and the
point $N_{\rm max}=8.821$ above which there is no equilibrium state anymore.

The nonrelativistic limit \cite{ijmpb} corresponds to the dashed lines. It
provides
a very good approximation of $\Lambda_{\rm max}$, $\Lambda_{\rm
gas}$ and $\Lambda_{\rm cond}$ for $N\ll N_{\rm OV}$. As
we approach  $N_{\rm OV}$ general
relativity must be taken into account.

The classical limit \cite{roupas,paper2} corresponds to the dotted lines.
It provides a
very good
approximation of $\Lambda_{\rm min}$ (hot spiral) for any $N$. It also provides
a very good
approximation of  $\Lambda_c$ (cold spiral) for $N\gg N_{\rm OV}$. As we
approach $N_{\rm
OV}$ quantum mechanics must be taken into account.

{\it Remark:} we recall that the minimum energy above which
equilibrium states
exist is $\Lambda_{\rm max}$ (ground state) when $N<N_{\rm OV}$ and $\Lambda'_c$
or  $\Lambda_c$  when $N>N_{\rm OV}$. From Fig. \ref{Xphase_Lambda_R50_def4PH}
we note that $\Lambda_{\rm max}(N)$
increases with $N$ while $\Lambda'_c(N)$ and $\Lambda_c(N)$ decrease with
$N$. We also note that the system would be a black hole if
$M>Rc^2/2G$, i.e., $M>R/2$ in terms of dimensionless variables. Using Eq.
(\ref{b15}), this leads to the condition
\begin{equation}
\Lambda<-\frac{\left (\frac{R}{2}-N\right )R}{N^2}\equiv\Lambda_{\rm BH}(N,R).
\end{equation}
One can locate the black hole energy curve $\Lambda_{\rm BH}(N,R)$ in Fig.
\ref{Xphase_Lambda_R50_def4PH}. It behaves as $\Lambda_{\rm BH}\sim
-R^2/2N^2\rightarrow -\infty$ when $N\rightarrow 0$ and as $\Lambda_{\rm BH}\sim
R/N\rightarrow 0^+$ when $N\rightarrow +\infty$. It vanishes at $N=R/2$ and has
a maximum $(\Lambda_{\rm BH})_{\rm max}=1/2$ at $N=R$. One can show that the
black hole energy curve never intersects the curves of Fig.
\ref{Xphase_Lambda_R50_def4PH} so that the system is never a black hole (see
\cite{paper2} for a detailed discussion).

\section{The case $R>R_{\rm MCP}$}
\label{sec_second}

We now study the case $R>R_{\rm MCP}=92.0$ where the system can display a
canonical phase transition (as before) and a microcanonical phase
transition (see Fig. \ref{phase_phase_Newton_corrPH} below). For illustration we
take $R=600$. In that case, the canonical
phase transition appears above $N_{\rm CCP}=9.84\times 10^{-6}$ and the
microcanonical phase transition appears above $N_{\rm MCP}=1.02\times 10^{-2}$.

The description of the caloric curves in the  canonical ensemble is the
same as before. Therefore,  in the following, we only consider the
microcanonical ensemble. In addition, we focus on what is new and do not
treat in detail the situations that are similar to those described previously.

\subsection{The case $N<N_{\rm MCP}$}

When $N<N_{\rm MCP}=1.02\times 10^{-2}$, the discussion is the same as in
Sec. \ref{sec_first} ($R=50$). The canonical phase transition appears at 
$N_{\rm CCP}=9.84\times
10^{-6}$. Since $N_{\rm MCP}\ll N_{\rm OV}$, we
are in the nonrelativistic regime \cite{ijmpb}.

\subsection{The case $N_{\rm MCP}<N<N_{\rm 1}$}

\begin{figure}
\begin{center}
\includegraphics[clip,scale=0.3]{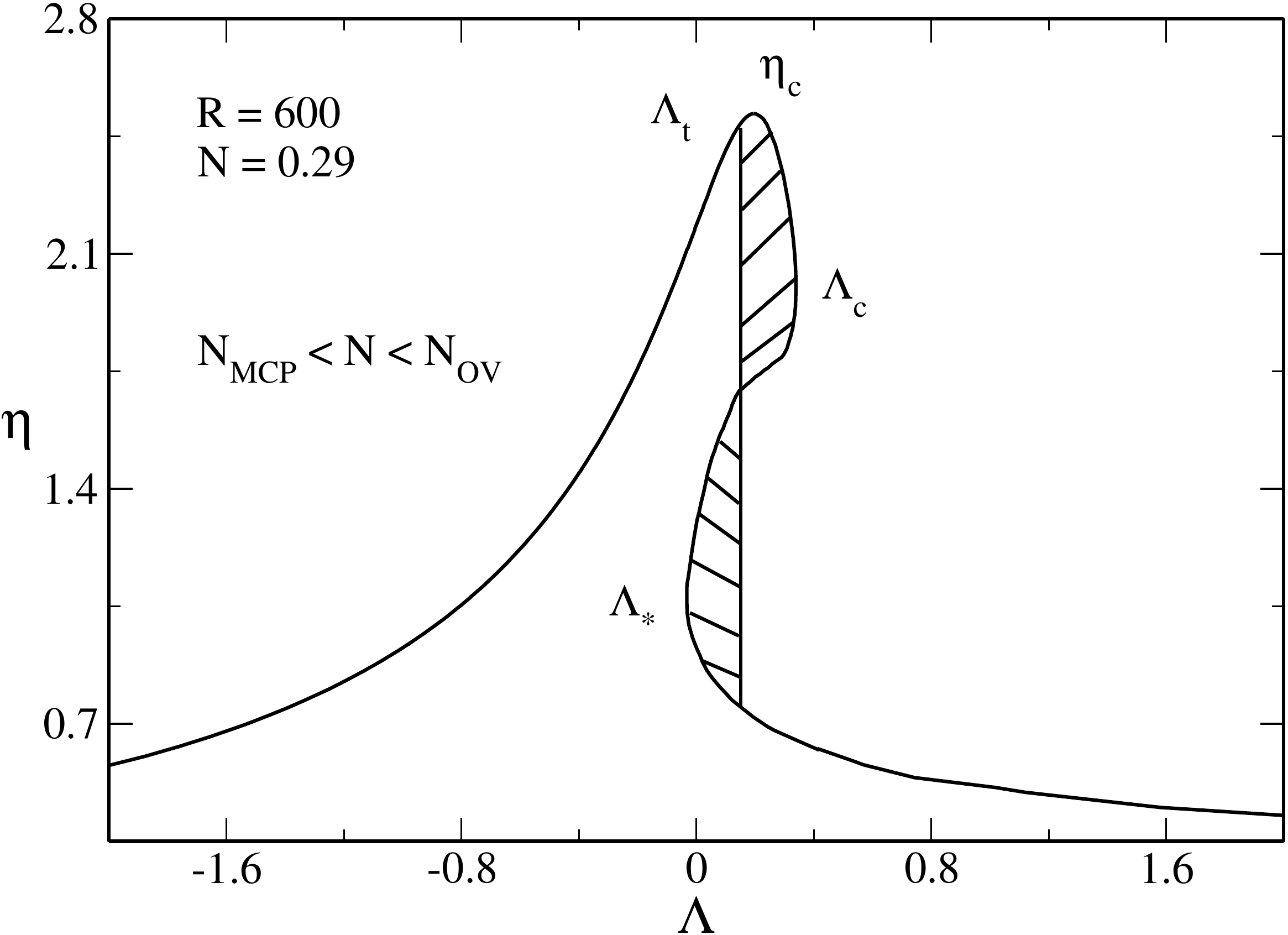}
\caption{Caloric curve for $N_{\rm MCP}=1.02\times 10^{-2}<N<N_{\rm
OV}=0.39853$ (specifically $R = 600$ and $N =
0.29$).}
\label{Xkcal_R600_N0p29_unified_bnewPH}
\end{center}
\end{figure}

In Fig. \ref{Xkcal_R600_N0p29_unified_bnewPH} we have plotted the caloric
curve for $N_{\rm MCP}=1.02\times 10^{-2}<N<N_{\rm OV}=0.39853$. Since
$N\ll N_{\rm
OV}$, the caloric curve coincides with the one obtained in the nonrelativistic
limit \cite{ijmpb}. It has a
$Z$-shape structure leading to a microcanonical phase
transition.\footnote{The caloric curve resembles a
dinosaur's neck \cite{ijmpb}. However, in Fig.
\ref{Xkcal_R600_N0p29_unified_bnewPH} the dinosaur has
no ``chin''. The ``chin'' appears at  $N_{\rm chin}=0.5062$  as explained in
Appendix \ref{sec_chin}. The presence, or not, of the
``chin'' has no physical consequence since it concerns a region of the caloric
curve where the equilibrium states are unstable.} This
$Z$-shape
structure
appears at $N=N_{\rm MCP}=1.02\times 10^{-2}$ at which the caloric curve
presents a
vertical inflexion point. The caloric curve continues up to $\Lambda_{\rm max}$
(outside the frame of the figure) at which it
presents an asymptote.

The curve
$\eta(\Lambda)$ is multivalued leading to the possibility of phase transitions
in the
microcanonical ensemble.  The upper
branch up to $\Lambda_c$
corresponds to the gaseous phase
and the lower branch after  $\Lambda_*$ corresponds to the condensed phase.
According to the Poincar\'e turning point criterion, these equilibrium
states are stable while
the equilibrium states on the
intermediate branch between  $\Lambda_c$ and  $\Lambda_*$ are unstable.
The caloric curve presents
the following features:

(i) When $\Lambda<\Lambda_*$ there are only gaseous states. When
$\Lambda>\Lambda_c$
there are only condensed states. When $\Lambda_*<\Lambda<\Lambda_c$ there exist
gaseous and
condensed states with the same energy. A
first order microcanonical
phase transition is expected at a transition energy $\Lambda_t$ determined by
the
Maxwell construction (see Fig. \ref{Xkcal_R600_N0p29_unified_bnewPH}) or by the
equality of the entropy of the gaseous and condensed phases (see Fig. 18 of
\cite{ijmpb}). When 
$\Lambda_*<\Lambda<\Lambda_t$ the gaseous states have a higher entropy than
the
condensed states. When  $\Lambda_t<\Lambda<\Lambda_c$ the condensed states have
a
higher entropy than the gaseous states. However, the first order phase
transition does not take place in practice because of the very long lifetime of
the metastable states. 

(ii) There is a zeroth order phase transition at $\Lambda_c$ from the gaseous
phase to the condensed phase. It corresponds to a gravitational
collapse (gravothermal catastrophe) ultimately halted by
quantum degeneracy.

(iii) There is a zeroth order phase transition at $\Lambda_*$ 
from the condensed phase to the gaseous phase. It corresponds to an
explosion ultimately halted by the boundary of the box.

(iv) There are two regions of negative specific heats, one between $\Lambda_{\rm
gas}$ and $\Lambda_c$ and another one between $\Lambda_*$ and $\Lambda_{\rm
cond}$. 

The evolution of the system in the microcanonical
ensemble is the following. Let us start from high energies and decrease the
energy. At high energies,
the system is in the gaseous phase. At
$\Lambda=\Lambda_t$
we expect
the system to undergo a  first order phase transition from the gaseous phase to
the condensed phase. However, in practice, this phase
transition does not take place because the metastable gaseous states
have a very long lifetime.  At
$\Lambda=\Lambda_c$ the system collapses towards the condensed phase. Complete
gravitational collapse is prevented by quantum
mechanics. The system reaches an equilibrium state similar to a
nonrelativistic white dwarf (fermion ball) surrounded by an isothermal
atmosphere. If we now increase the energy the system remains in the
condensed phase (again, the first order phase transition expected at $\Lambda_t$
does not take place
because  the metastable condensed states have a very long lifetime)
until the point $\Lambda_*$ at which it explodes and returns to the gaseous
phase. We have thus described an hysteretic cycle in the microcanonical
ensemble \cite{ijmpb}.

\begin{figure}
\begin{center}
\includegraphics[clip,scale=0.3]{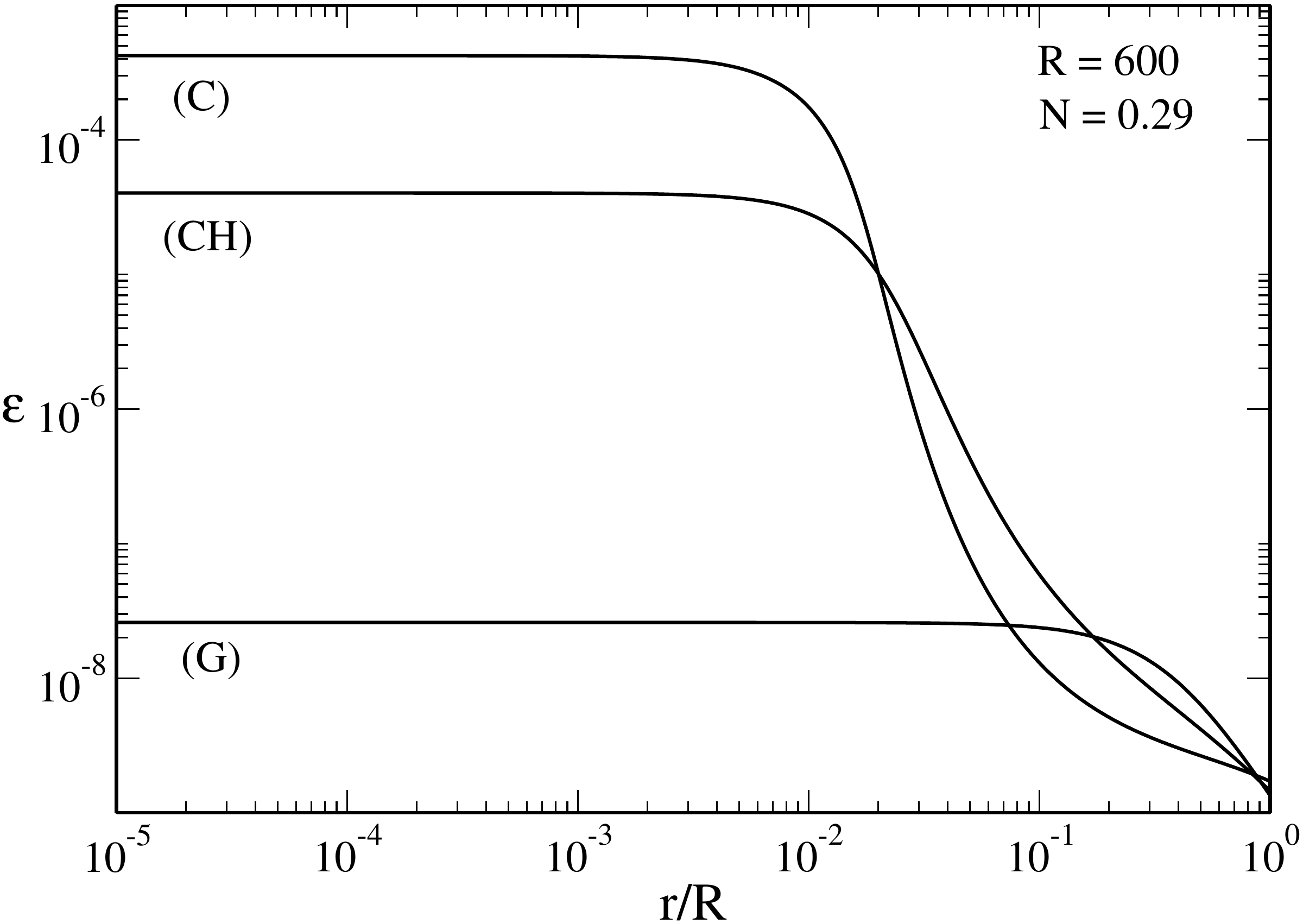}
\caption{Energy density profiles of gaseous, core-halo and condensed states at
$\Lambda=\Lambda_t$ (specifically $\Lambda_t=0.151$).}
\label{density_profiles_R600_N0p29PH}
\end{center}
\end{figure}

In Fig. \ref{density_profiles_R600_N0p29PH} we have
plotted the density profiles of the
gaseous (G), core-halo (CH) and condensed (C) states at the transition
point $\Lambda_t$. We note
that the energy density
is very low confirming that we are in the nonrelativistic regime. The
discussion is essentially the same as in Sec. \ref{sec_dp} with the difference
that the
fermion ball (similar to a nonrelativistic cold white dwarf) that forms in
the condensed phase
contains only a fraction ($\sim 1/4$) of the mass (see
\cite{ijmpb}, Sec. \ref{sec_ap}
and
Appendix \ref{sec_thermoqmce}). The rest of the
mass is diluted in a hot halo. This core-halo
structure is reminiscent of a red-giant (see Sec.
\ref{sec_ap}).

\subsection{The case $N_{1}<N<N_{\rm OV}$}

The second branch with an asymptote at $\Lambda'_{\rm max}$ appears at
$N_1=0.18131$ but
this does
not change the discussion since this branch is made of unstable equilibrium
states. From that moment, the system starts to be strongly
relativistic.

\subsection{The case $N_{\rm OV}<N<N_{f}$}

\begin{figure}
\begin{center}
\includegraphics[clip,scale=0.3]{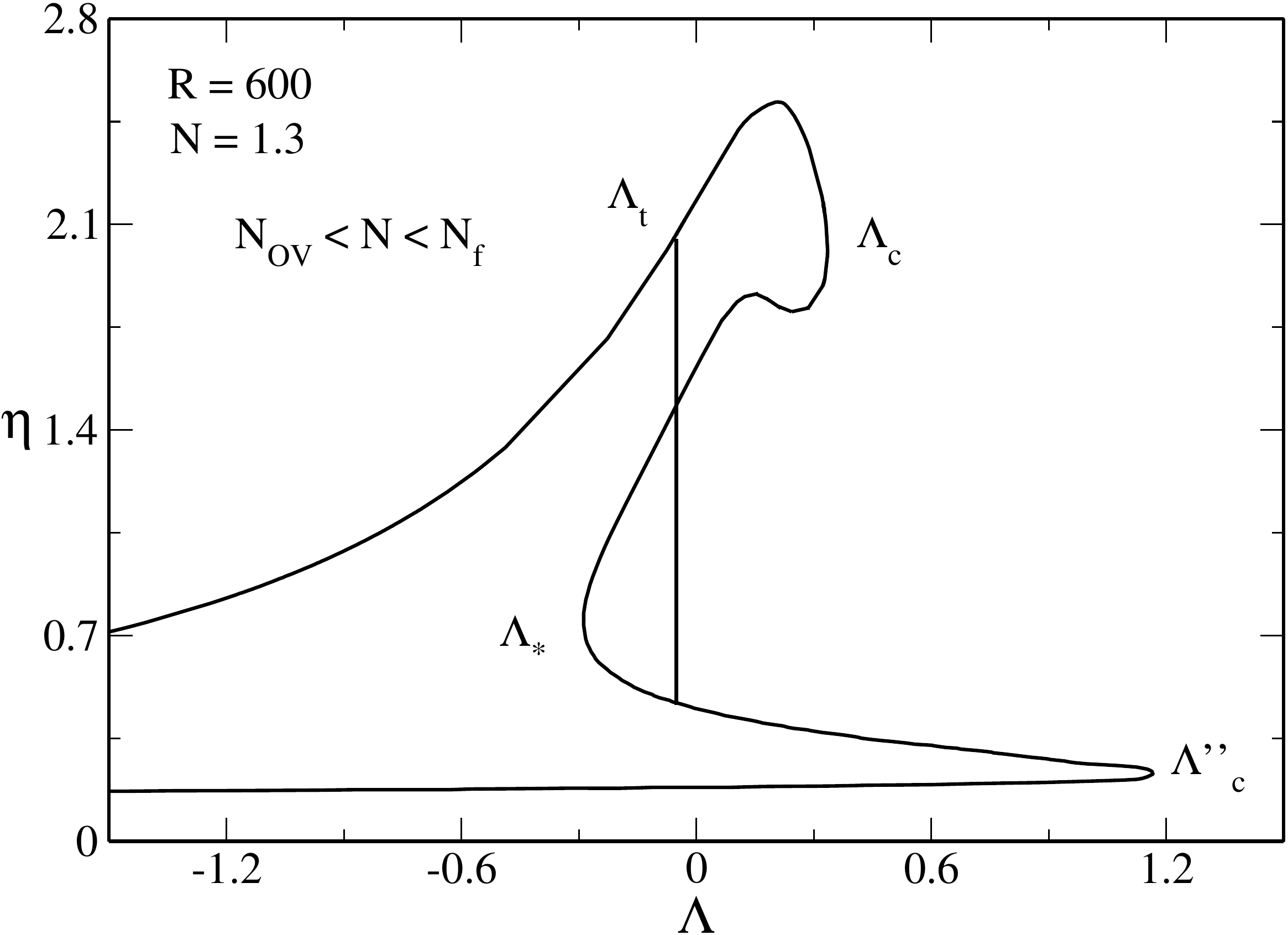}
\caption{Caloric curve for $N_{\rm OV}=0.39853<N<N_f=1.4854$ (specifically
$R=600$ and $N
= 1.3$). }
\label{kcal_R600_N1p3_unifiedPH}
\end{center}
\end{figure}

In Fig. \ref{kcal_R600_N1p3_unifiedPH} we have plotted the caloric
curve for $N_{\rm OV}=0.39853<N<N_f=1.4854$.\footnote{We note that the ``chin''
of the dinosaur has appeared since $N=1.3>N_{\rm chin}=0.5062$.} The novelty
with respect to the
previous case is that the two branches have merged. As a result there is no
ground state anymore (see Sec. \ref{sec_deb}). The caloric curve
presents a
turning point of energy which
corresponds to the minimum energy. When $N<N_*=0.405$ we call it
$\Lambda'_c$ and when $N>N_*$ we call it $\Lambda''_c$ (see Sec.
\ref{sec_fromto} for the
definition of $N_*$). In the following, to be specific, we assume that
$N>N_*$  but the discussion is essentially the same for $N<N_*$.

According to
the Poincar\'e
turning point criterion, the series of equilibria is stable
up to $\Lambda_c$,
becomes unstable between $\Lambda_c$ and $\Lambda_*$,
becomes stable again between
$\Lambda_*$ and $\Lambda''_c$ and becomes unstable again after $\Lambda''_c$.
 The caloric curve presents the following features:

\begin{figure}
\begin{center}
\includegraphics[clip,scale=0.3]{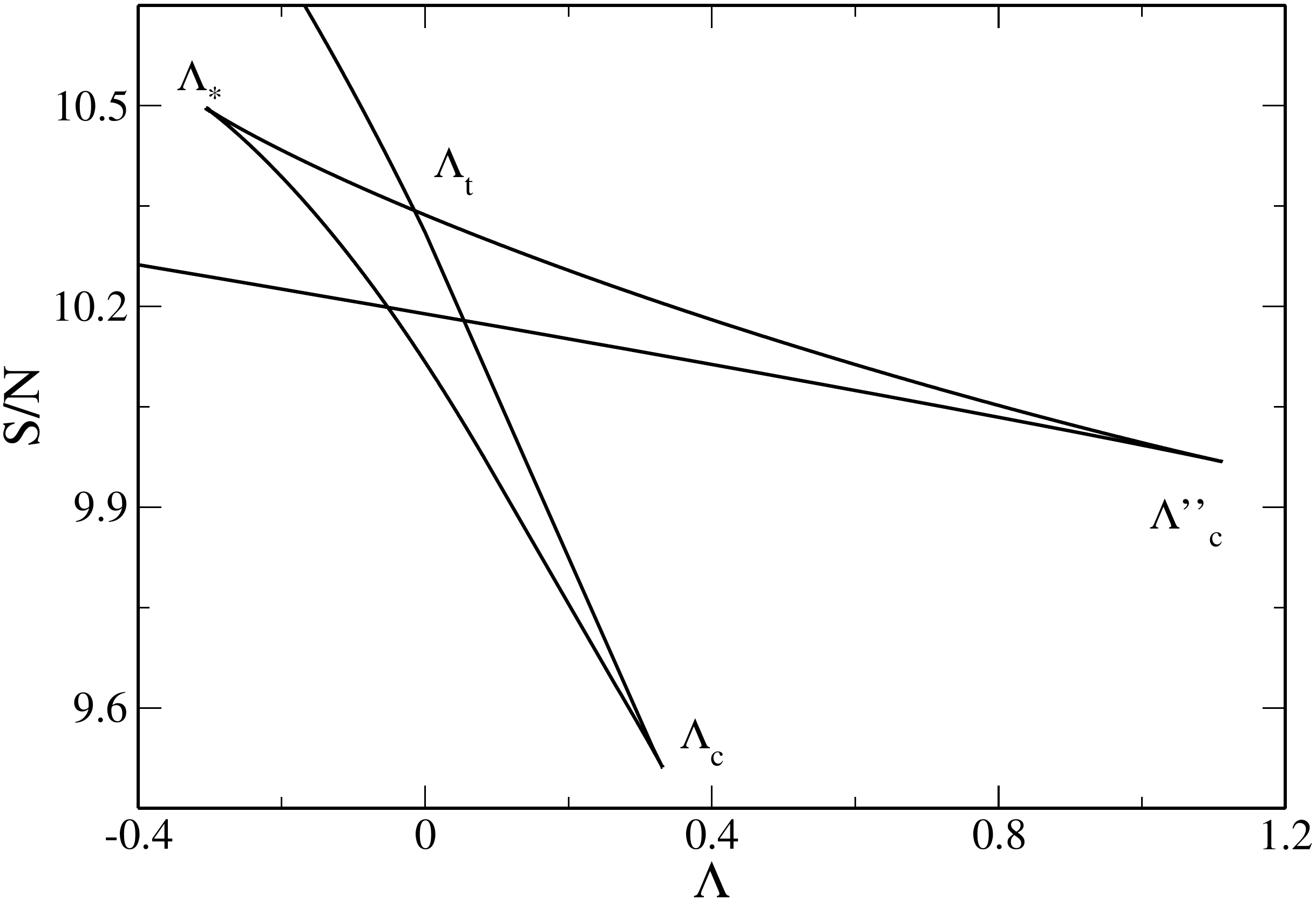}
\caption{Entropy per fermion as a function of the normalized energy for $N_{\rm
OV}<N<N_{f}$
(specifically $R = 600$ and 
$N = 1.3$). The first derivative of $S$ with respect
to $E$ is discontinuous at the transition energy $E_t$. This
corresponds to a first order phase transition, connecting the
gaseous phase to the condensed phase, which is associated with a jump of
temperature $\beta=\partial S/\partial E$ in the caloric curve. On the
other
hand, $S$ is discontinuous at the spinodal points $E_c$ and
$E_*$. This corresponds to zeroth order phase
transitions which are associated with a jump of entropy. We can
check that
the unstable equilibrium 
states (saddle points of entropy) between $\Lambda_c$ and $\Lambda_*$ have an
entropy lower than
the stable equilibrium states (maxima of entropy). However, the  unstable
equilibrium 
states after $\Lambda''_c$ can have an entropy higher than
the stable equilibrium states before $\Lambda_c$.}
\label{XSLambda_R600_N1p3PH}
\end{center}
\end{figure}

(i) When $\Lambda<\Lambda_*$ there are only gaseous states. When
$\Lambda_c<\Lambda<\Lambda''_c$
there are only condensed states. When $\Lambda_*<\Lambda<\Lambda_c$ there exist
gaseous and
condensed states with the same energy. A first order phase transition is
expected at a transition energy $\Lambda_t$ determined by the  Maxwell
construction (see Fig. \ref{kcal_R600_N1p3_unifiedPH}) of by the equality of
the entropy of the gaseous and condensed phases (see
Fig. \ref{XSLambda_R600_N1p3PH}).  When
$\Lambda_*<\Lambda<\Lambda_t$ the gaseous states have a higher entropy than
the
condensed states. When  $\Lambda_t<\Lambda<\Lambda_c$ the condensed states have
a
higher entropy than the gaseous states.  However, the first order
phase
transition does not take place in practice because of the very long lifetime of
metastable states.

(ii) There is a zeroth order phase transition at $\Lambda_c$ from the gaseous
phase to the condensed phase. It corresponds to a gravitational
collapse (gravothermal catastrophe)  ultimately halted by
quantum degeneracy.

(iii) There is a zeroth order phase transition at
$\Lambda_*$ 
from the condensed phase to the gaseous phase. It corresponds to a
explosion ultimately halted by the boundary of the box.

(iv) There is a catastrophic collapse at $\Lambda''_c$ from the condensed
phase to a black hole.

(v) There are two regions of negative specific heats, one between $\Lambda_{\rm
gas}$ and $\Lambda_c$ and another one between $\Lambda_*$ and $\Lambda''_c$.

The evolution of the system in the microcanonical
ensemble in the following. Let us start from high energies and decrease the
energy.  At high energies, the
system is in the gaseous phase. At
$\Lambda=\Lambda_t$, we
expect
the system to undergo a  first order phase transition from the gaseous phase to
the condensed phase. However, in practice, this phase transition does not take
place because the metastable gaseous states have a very long lifetime. The
physical
transition occurs at the critical energy $\Lambda_c$ (spinodal point) at which
the gaseous phase
disappears. At that point the system undergoes a zeroth order phase
transition (collapse)  from the
gaseous phase to the condensed phase. 
If we keep
decreasing the energy there comes another critical point $\Lambda''_c$ at
which the system undergoes a catastrophic
collapse from the condensed phase to a black hole. This is an instability
of general relativistic
origin which has no counterpart in the Newtonian theory. Inversely, if we
increase the energy, the system
displays a zeroth order phase
transition (explosion) at $\Lambda_*$ from the
condensed phase to the gaseous phase.

\begin{figure}
\begin{center}
\includegraphics[clip,scale=0.3]{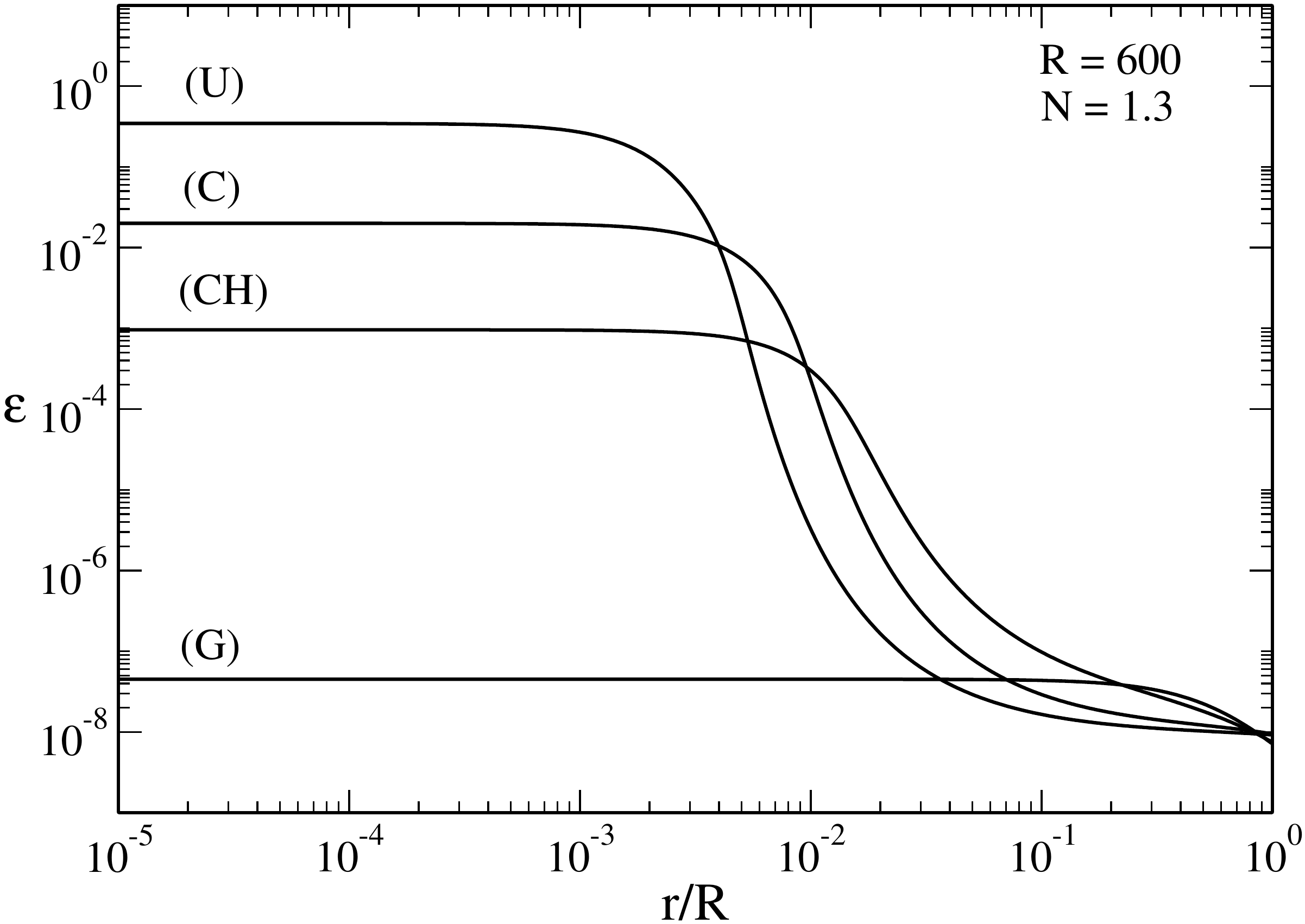}
\caption{Density profiles of gaseous, core-halo and (stable and unstable)
condensed states at $\Lambda=\Lambda_t$ (specifically $\Lambda_t=-0.0510$).}
\label{density_profiles_R600_N1p3PH}
\end{center}
\end{figure}

In Fig.
\ref{density_profiles_R600_N1p3PH} we have plotted the different density
profiles at
$\Lambda_t$. We see that the energy density is low for the gaseous state (G) and
for the core-halo state (CH) indicating that we are in the nonrelativistic
regime. By contrast, the energy density is relatively high for the
stable condensed state (C) and for the unstable condensed state (U) indicating
that we are in the relativistic regime. The
discussion is essentially the same as in Sec. \ref{sec_trq} with the difference
that the
fermion ball (similar to a general relativistic cold neutron star) that
forms in
the condensed phase
contains only a fraction ($\sim 1/4$) of the mass (see  Sec.
\ref{sec_ap} and Appendix
\ref{sec_thermoqmce}). The rest of the
mass is diluted in a hot halo. This core-halo
structure is reminiscent of a supernova (see Sec.
\ref{sec_ap}).

\subsection{The case $N_{f}<N<N'_{f}$}

\begin{figure}
\begin{center}
\includegraphics[clip,scale=0.3]{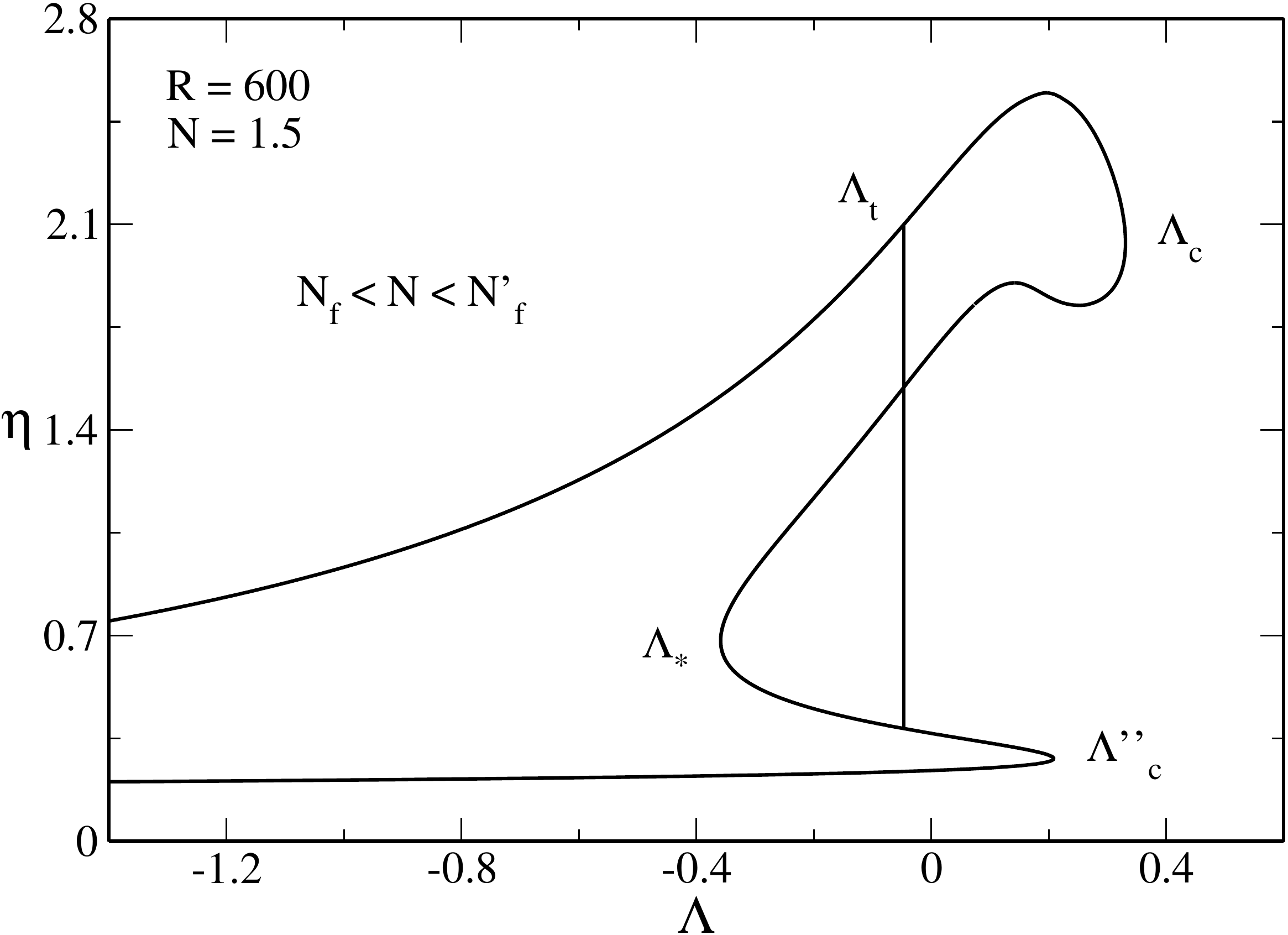}
\caption{Caloric curve for $N_f=1.4854<N<N'_f=1.619$ (specifically
$R = 600$
and $N =1.5$).}
\label{kcal_R600_N1p5_unifiedPH}
\end{center}
\end{figure}

In Fig. \ref{kcal_R600_N1p5_unifiedPH} we have plotted the caloric
curve for $N_f=1.4854<N<N'_f=1.619$. The novelty with respect to the
previous case is that now $\Lambda''_c$ is smaller than $\Lambda_c$ (they become
equal when $N=N_f$).

\begin{figure}
\begin{center}
\includegraphics[clip,scale=0.3]{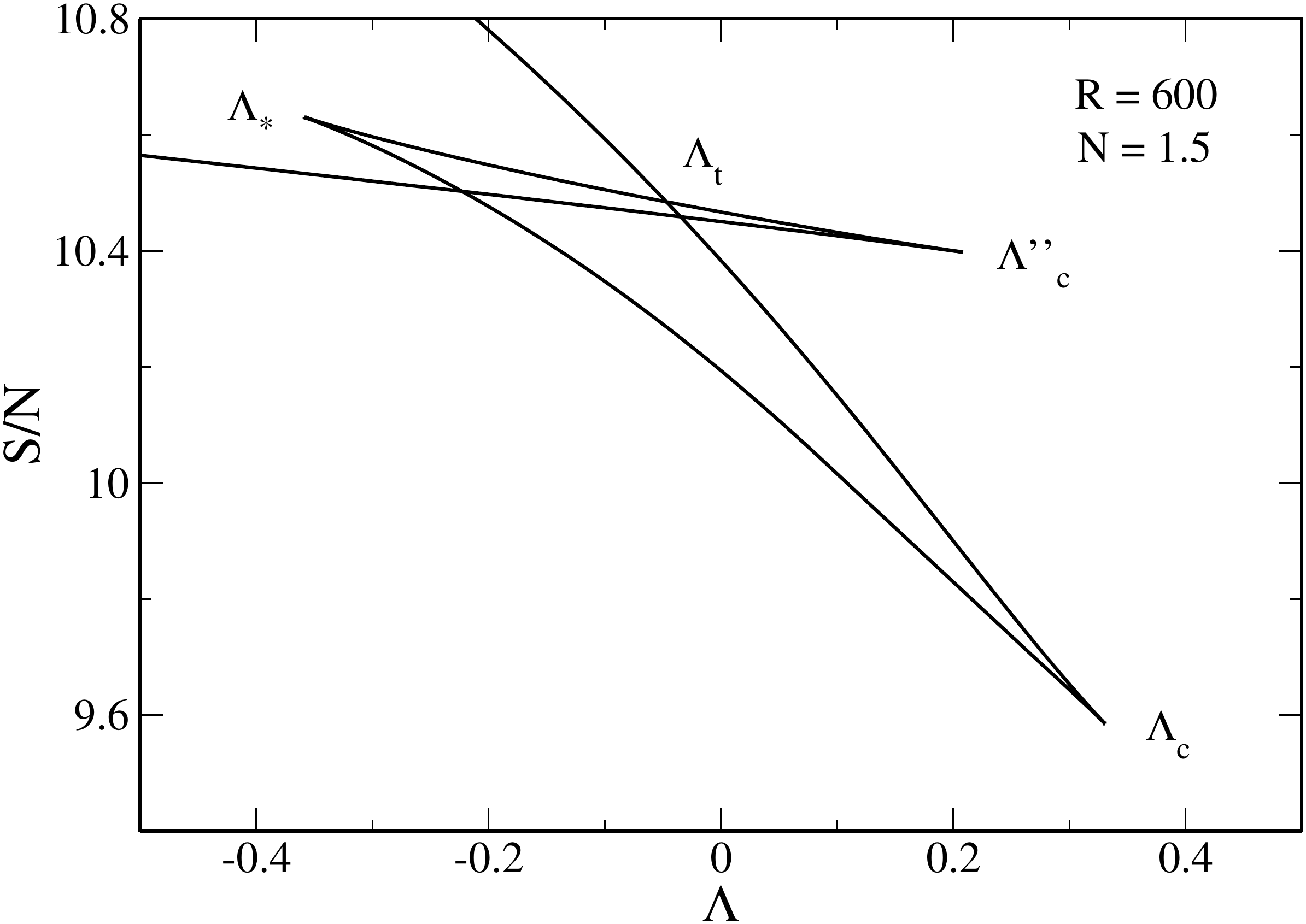}
\caption{Entropy per fermion as a function of the normalized energy  for
$N_f=1.4854<N<N'_f=1.619$ (specifically
$R = 600$
and $N =1.5$).}
\label{SLambda_R600_N1p5PH}
\end{center}
\end{figure}

The caloric curve presents the following features:

(i) When $\Lambda<\Lambda_*$ and $\Lambda''_c<\Lambda<\Lambda_c$ there are only
gaseous states. When
$\Lambda_*<\Lambda<\Lambda''_c$ there exist gaseous and condensed states with
the same energy. A
first order phase transition is expected  at a transition energy $\Lambda_t$
determined by the Maxwell construction (see Fig. \ref{kcal_R600_N1p5_unifiedPH})
or by the equality of the entropy of the
two phases (see Fig.  \ref{SLambda_R600_N1p5PH}). When
$\Lambda_*<\Lambda<\Lambda_t$ the
gaseous states have a higher entropy than the condensed states. When
$\Lambda_t<\Lambda<\Lambda''_c$ the condensed states
have a higher entropy than the gaseous states. However,
the first order phase
transition does not take place
in practice because of the very long lifetime of the metastable states.

(ii) There is a catastrophic collapse at $\Lambda_c$ from the gaseous
phase to a
black hole.

(iii) There is a  catastrophic collapse at $\Lambda''_c$ from the
condensed phase to a
black hole.

(iv) There is a zeroth order phase transition at $\Lambda_*$ from the condensed
phase to the gaseous phase. It corresponds to an explosion ultimately halted by
the boundary of the box. 

(v) There are two regions of negative specific heats, one between $\Lambda_{\rm
gas}$ and $\Lambda_c$ and another one between $\Lambda_*$ and $\Lambda''_c$.

The evolution of the system in the microcanonical ensemble is
the following. Let us start from high energies and decrease the
energy.   At high energies, the system is in the gaseous phase. At
$\Lambda=\Lambda_t$ the system is expected to undergo a  first order
phase transition from the gaseous phase to
the condensed phase. However, this phase transition does not take place in
practice. At 
$\Lambda=\Lambda_c$ the system undergoes a catastrophic
collapse towards a black hole.  A condensed phase exists for
$\Lambda_*<\Lambda<\Lambda''_c$ but it is not clear
how it can be reached in practice.

\subsection{The case $N'_f<N<N'_*$}

\begin{figure}
\begin{center}
\includegraphics[clip,scale=0.3]{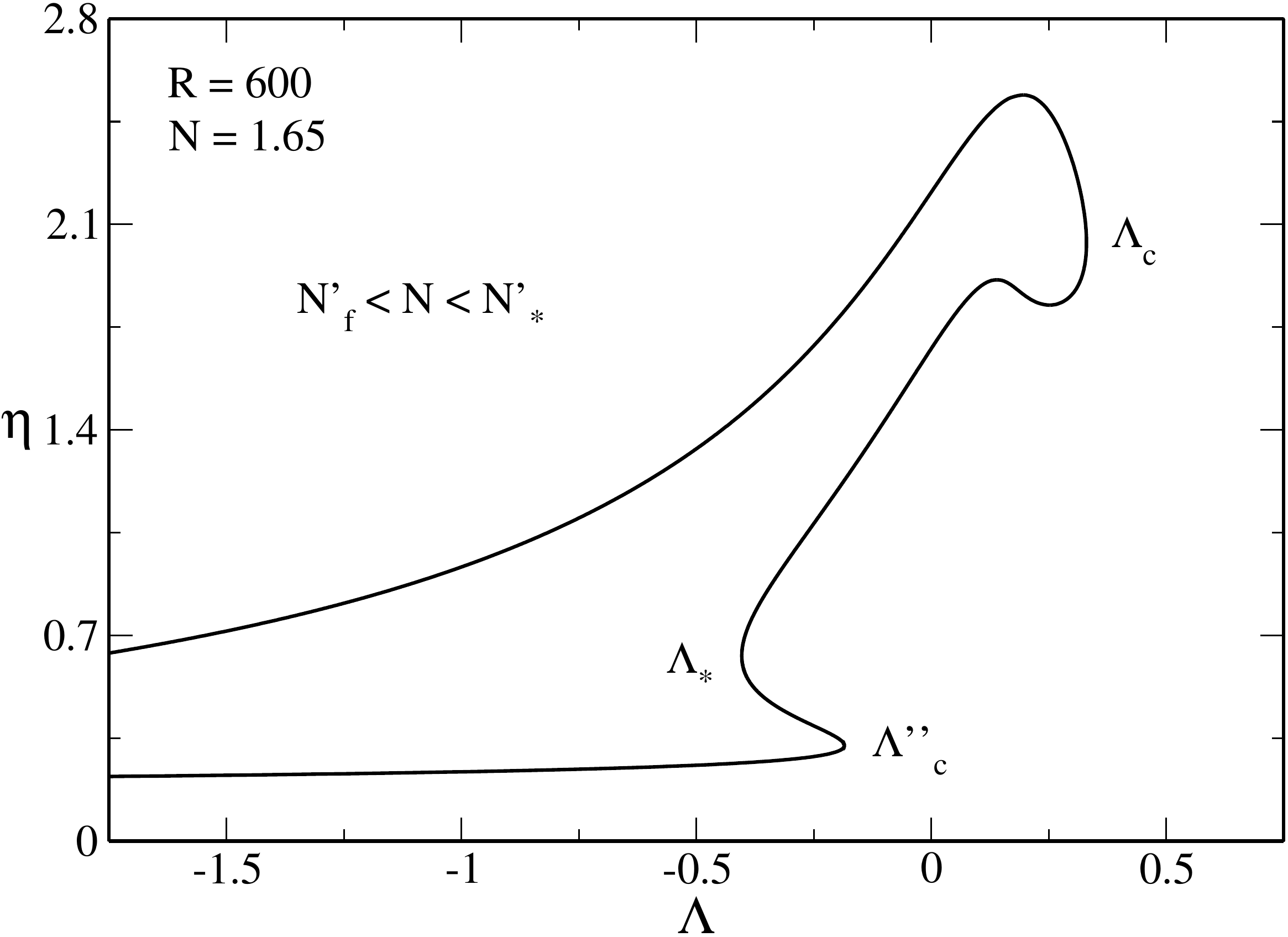}
\caption{Caloric curve for $N'_f=1.619<N<N'_*=1.9000$
(specifically $R = 600$ and $N = 1.65$).}
\label{kcal_R600_N1p65_unifiedPH}
\end{center}
\end{figure}

In Fig. \ref{kcal_R600_N1p65_unifiedPH} we have plotted the caloric
curve for $N'_f=1.619<N<N'_*=1.9000$, where $N'_f$ is defined such
that $\Lambda''_c=\Lambda_t$.

\begin{figure}
\begin{center}
\includegraphics[clip,scale=0.3]{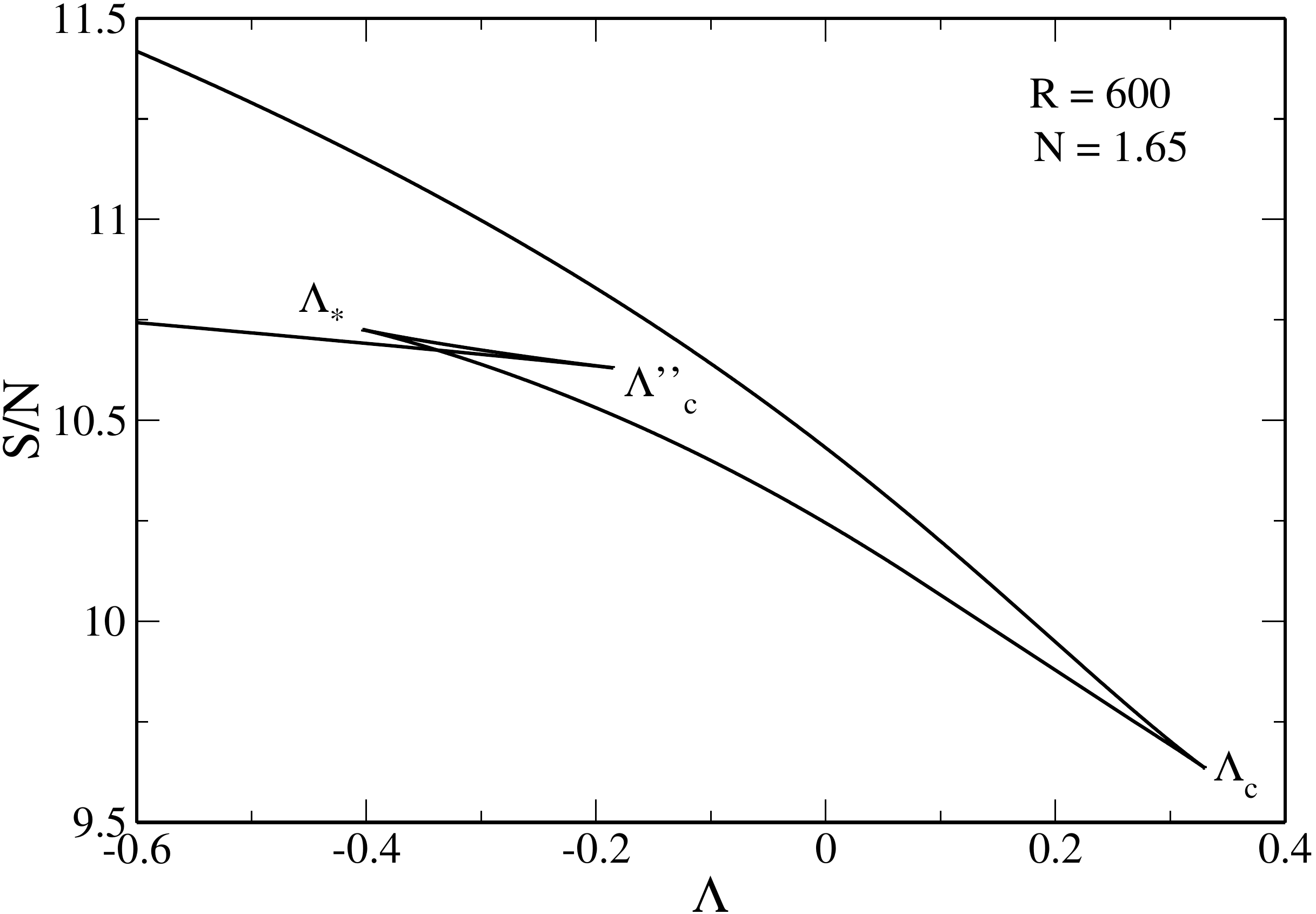}
\caption{Entropy per fermion as a function of the normalized energy for
$N'_f=1.619<N<N'_*=1.9000$
(specifically $R = 600$ and $N = 1.65$).}
\label{SLambda_R600_N1p65PH}
\end{center}
\end{figure}

The caloric curve presents the following features:

(i) When $\Lambda<\Lambda_*$ and when $\Lambda''_c<\Lambda<\Lambda_c$ there are
only
gaseous states. When 
$\Lambda_*<\Lambda<\Lambda''_c$ there exist gaseous and condensed states with
the same energy.
However, there is no first order phase
transition, even in theory, because we cannot satisfy the  Maxwell
construction (see Fig. \ref{kcal_R600_N1p65_unifiedPH}) or the equality of the
entropy of the gaseous and condensed
phases (see Fig. \ref{SLambda_R600_N1p65PH}). When
$\Lambda_*<\Lambda<\Lambda''_c$
the gaseous states always have an entropy higher than the condensed states.
Therefore, although there are
several
stable equilibrium states when $\Lambda_*<\Lambda<\Lambda''_c$ there is no phase
transition from one phase to the other. This is a particularity of the
relativistic situation.

(ii) There is a catastrophic collapse at $\Lambda_c$ from the gaseous phase
to a
black hole.

(iii) There is a  catastrophic  collapse at $\Lambda''_c$  from the condensed
phase to a
black hole. 

(iv)  There is a zeroth order phase transition at $\Lambda_*$ from the condensed
phase to the gaseous phase. It corresponds to an explosion ultimately halted by
the boundary of the box. 

(v) There are two regions of negative specific heats, one between $\Lambda_{\rm
gas}$ and $\Lambda_c$ and another one between $\Lambda_*$ and $\Lambda''_c$.

The evolution of the system is the same as described previously.

\subsection{The case $N'_*<N<N_{\rm max}$}
\label{sec_tmr}

\begin{figure}
\begin{center}
\includegraphics[clip,scale=0.3]{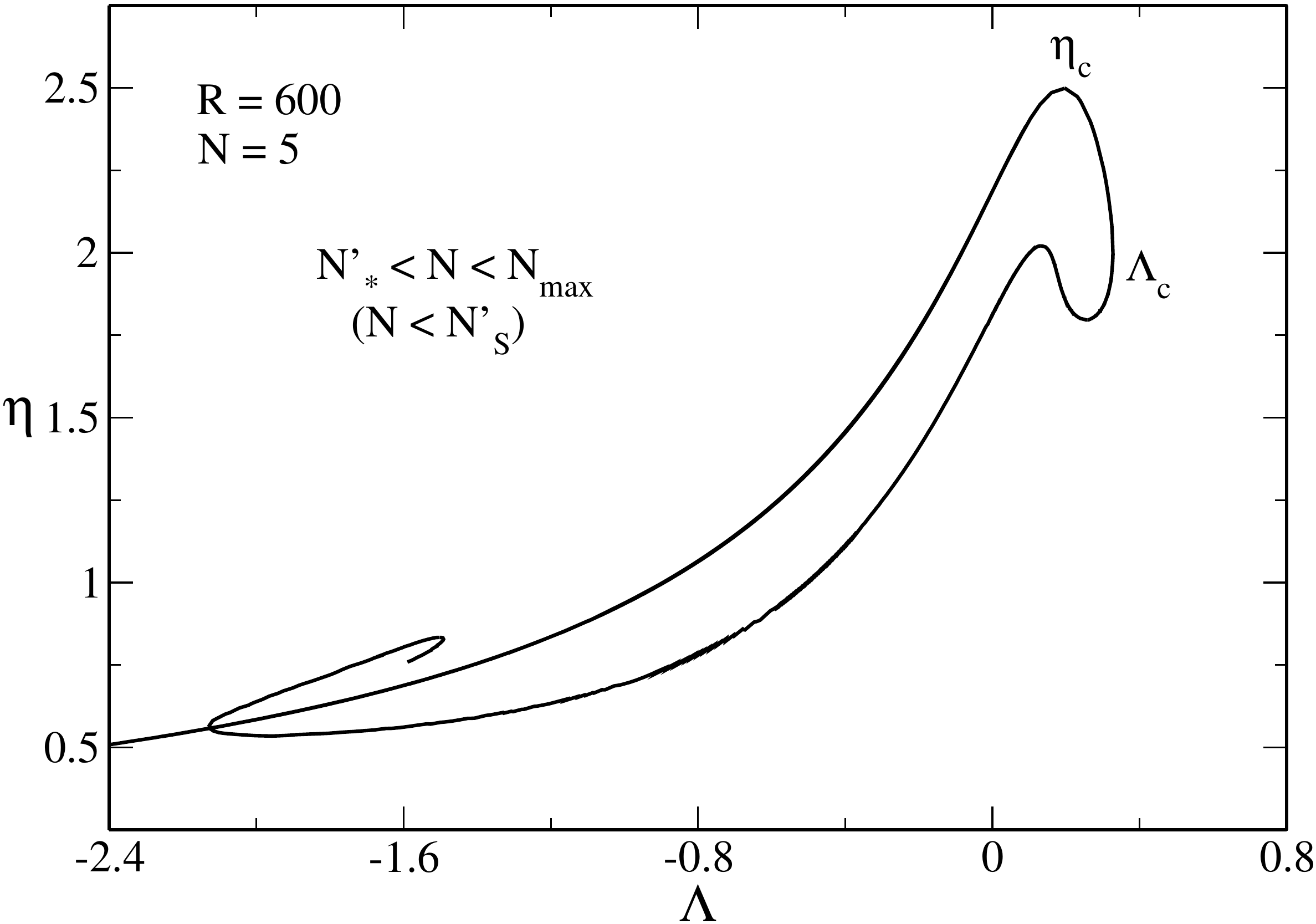}
\caption{Caloric curve for  $N'_*=1.9000<N<N_{\rm
max}=105.9$ (specifically $R = 600$ and
$N = 5$).}
\label{kcal_R600_N5_unified2PH}
\end{center}
\end{figure}

In Fig. \ref{kcal_R600_N5_unified2PH} we have plotted the caloric
curve for  $N'_*=1.9000<N<N_{\rm max}=105.9$, where $N'_*$ is
defined such that $\Lambda''_c=\Lambda_*$.  From that moment, we denote the
minimum energy by $\Lambda_c$ instead of $\Lambda''_c$. The novelty with respect
to the
previous case is that there is
no condensed phase anymore. The discussion is the same as in Secs.
\ref{sec_fromto} and \ref{sec_fin}.

\subsection{The microcanonical phase diagram}

\begin{figure}
\begin{center}
\includegraphics[clip,scale=0.3]{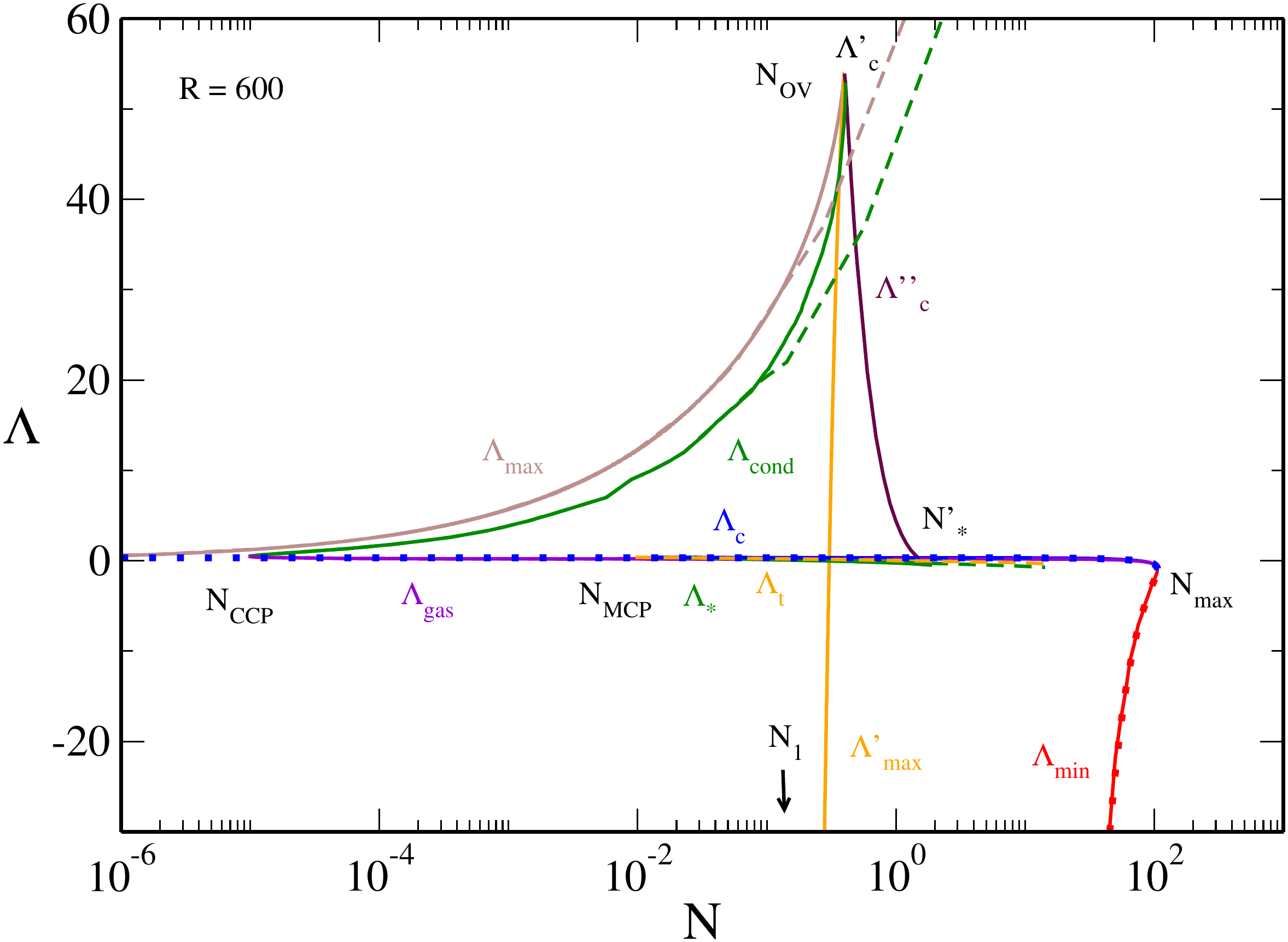}
\caption{Microcanonical phase diagram for $R>R_{\rm MCP}=92.0$ (specifically $R
=
600$).}
\label{Xphase_Lambda_R600_def4aPH}
\end{center}
\end{figure}

\begin{figure}
\begin{center}
\includegraphics[clip,scale=0.3]{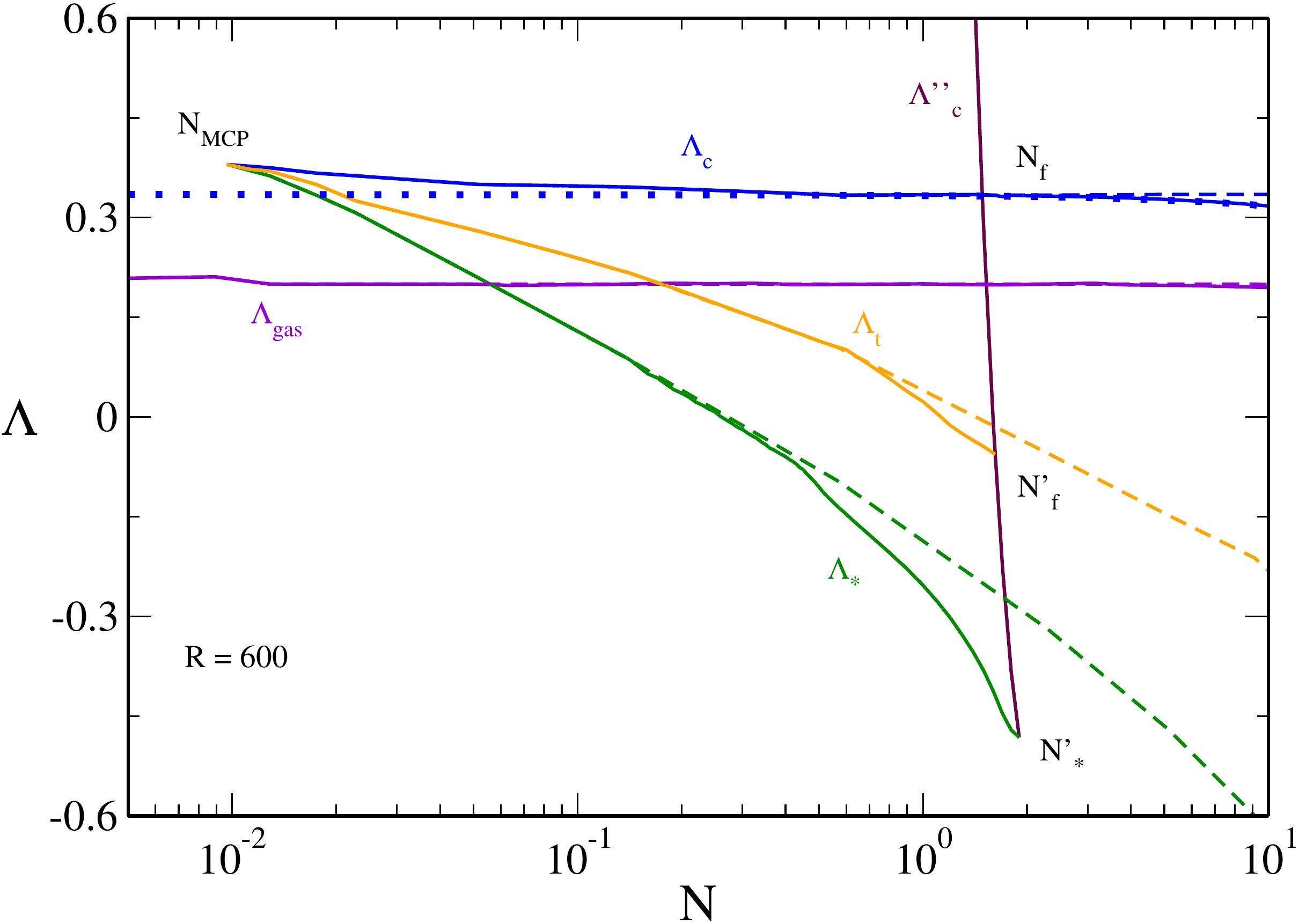}
\caption{Zoom of the microcanonical phase diagram  for $R>R_{\rm MCP}$
(specifically $R =
600$) in the region near
$N_{\rm OV}$.}
\label{Xphase_Lambda_R600_def4bPH}
\end{center}
\end{figure}

In Figs. \ref{Xphase_Lambda_R600_def4aPH} and \ref{Xphase_Lambda_R600_def4bPH}
we
have represented the microcanonical phase diagram corresponding to
$R>R_{\rm MCP}$. It shows the evolution of the critical energies
$\Lambda_{\rm min}$, $\Lambda_{\rm max}$, $\Lambda_{\rm gas}$, $\Lambda_{\rm
cond}$, $\Lambda_c$, $\Lambda_*$, $\Lambda_t$, $\Lambda'_{\rm max}$, 
$\Lambda'_c$ and $\Lambda''_c$ with $N$. We can
clearly see the canonical critical point at $N_{\rm CCP}=9.84\times 10^{-6}$ at
which the
region of negative specific heat (associated with the canonical phase
transition) appears and the microcanonical critical point at $N_{\rm
MCP}=1.02\times 10^{-2}$
at which the microcanonical
phase transition appears. We also see the point  $N_{\rm OV}=0.39853$ above
which
quantum
mechanics
is not able to prevent gravitational collapse above $\Lambda'_c(N)$ or
$\Lambda''_c(N)$. Finally, we
see the point $N_{\rm max}=105.9$ above which there is no equilibrium state
anymore.

The nonrelativistic limit \cite{ijmpb} corresponds to the dashed lines. It
provides a very good approximation of
$\Lambda_{\rm max}$, $\Lambda_{\rm gas}$, $\Lambda_{\rm
cond}$, $\Lambda_c$, $\Lambda_*$, and $\Lambda_t$ for $N\ll
N_{\rm OV}$. As
we approach $N_{\rm OV}$ general relativity must be taken into account.

The classical limit \cite{paper2} corresponds to the dotted
lines.
It provides a very good
approximation of $\Lambda_{\rm min}$ (hot spiral) for any $N$. It also provides
a very good
approximation of  $\Lambda_c$ (cold spiral) for $N\gg N_{\rm OV}$. As we
approach $N_{\rm
OV}$ quantum mechanics must be taken into account.

{\it Remark:} From Fig. \ref{Xphase_Lambda_R600_def4aPH} we note
that $\Lambda_{\rm max}(N)$
increases with $N$ while $\Lambda_c(N)$, $\Lambda'_c(N)$ and $\Lambda''_c(N)$
decrease with
$N$.

\section{The case $R\gg R_{\rm MCP}$}
\label{sec_vlr}

For very large radii ($R\gg R_{\rm MCP}$), a spiral, winding
then unwinding, appears in the caloric curve at the location of the ``head'' of
the dinosaur  (this is similar to Fig. 22 of \cite{ijmpb} and Fig. 44 of
\cite{clm2}). However, this spiral is made of unstable states. Therefore, if we
restrict ourselves to stable equilibrium states, this mathematical complication
(the
proliferation of unstable states associated with the spiral) does not change the
previous discussion.

{\it Remark:} The equilibrium states that are deep in the
spiral have a pronounced ``core-halo'' structure with a large central
density (see Fig. 45 of \cite{clm2}). These core-halo
states correspond to the configurations computed by Bilic {\it et al.}
\cite{btv} and, more recently, by Ruffini {\it et al.} \cite{rar} and Chavanis
{\it et al.} \cite{clm2}. They consist in a large nondegenerate isothermal
atmosphere harboring a small ``fermion ball''. These solutions look
very attractive at first sight because they could provide a self-consistent
model of DM halos in which the fermion ball would mimic the presence of a
supermassive black hole at the centers of the galaxies (an idea originally
proposed in \cite{btv}). However, as argued in \cite{clm2}, these extreme
core-halo structures are thermodynamically unstable (see Secs. VI-VIII of
\cite{clm2} for a detailed discussion).\footnote{By contrast, less extreme
core-halo configurations, such as the solution (CH) computed in Fig.
\ref{profile_R50_N0p15_newPH}, can be stable in the microcanonical ensemble.
They have  a negative specific heat.} These core-halo states are dynamically
(Vlasov) stable meaning that if we artificially prepare the system in such a
state, it will remain in this state for a long time. However, since these
extreme  core-halo states are thermodynamically unstable, they are
very unlikely (from a thermodynamical point of view) to appear spontaneously.
The fermion ball is like a  ``critical droplet'' in nucleation processes.
This
may be a problem for the fermion ball scenario to mimic the effect of a  black
hole, as
mentioned in \cite{clm2}. Other problems with the fermion ball scenario are
pointed out in \cite{genzel}.

\section{The case $R_{\rm OV}<R<R_{\rm CCP}$}
\label{sec_dix}

We now study the case $R<R_{\rm CCP}=12.0$ where there is no phase
transition (see Fig. \ref{phase_phase_Newton_corrPH} below). In this section, we
assume $R>R_{\rm OV}=3.3569$ so that $N_{\rm
OV}$ and $N_{\rm max}$ are
relatively well separated.
For illustration, we take $R=10$.

\subsection{The case $N<N_1$}

When $N<N_1=0.18131$ the caloric curve is similar to that shown in
Fig. \ref{kcal_R50_N0p012_unified2PH}. It is monotonic and presents
an asymptote at
$\Lambda_{\rm max}$. The discussion is similar to that given in Sec.
\ref{sec_tr}.

\subsection{The case $N_1<N<N_{\rm OV}$}

\begin{figure}
\begin{center}
\includegraphics[clip,scale=0.3]{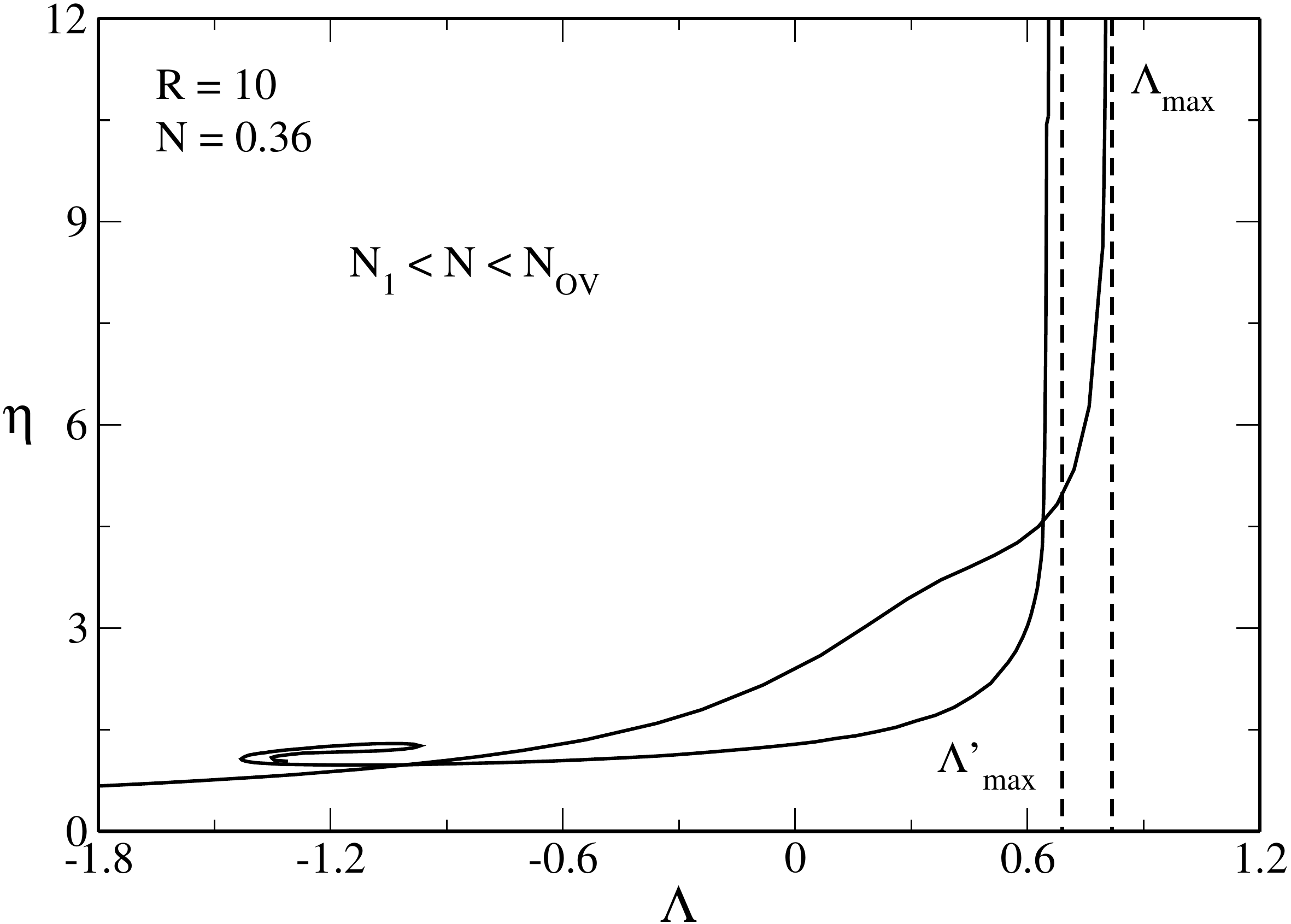}
\caption{Caloric curve for $N_1=0.18131<N<N_{\rm OV}=0.39853$ 
(specifically $R =
10$ and $N = 0.36$).}
\label{kcal_R10_N0p36_unifiedPH}
\end{center}
\end{figure}

In Fig. \ref{kcal_R10_N0p36_unifiedPH} we have plotted the caloric curve for
$N_1=0.18131<N<N_{\rm OV}=0.39853$. The difference with the case treated in Sec.
\ref{sec_trq} is that
there is no canonical phase transition.  The caloric
curve is monotonic\footnote{We
see a sort of inflexion
of the curve which signals  the  canonical first
order phase transition that
appears at larger radii $R>R_{\rm CCP}$.} and presents an asymptote at
$\Lambda_{\rm max}$. There is
another branch  presenting an asymptote at $\Lambda'_{\rm max}$ but it is made
of
unstable states. The series of equilibria of the main branch is stable in both
ensembles. The specific heat is always positive. There is no phase transition
and no gravitational collapse. The
ensembles are equivalent.

\subsection{The case $N_{\rm OV}<N<N_{\rm max}$}

\begin{figure}
\begin{center}
\includegraphics[clip,scale=0.3]{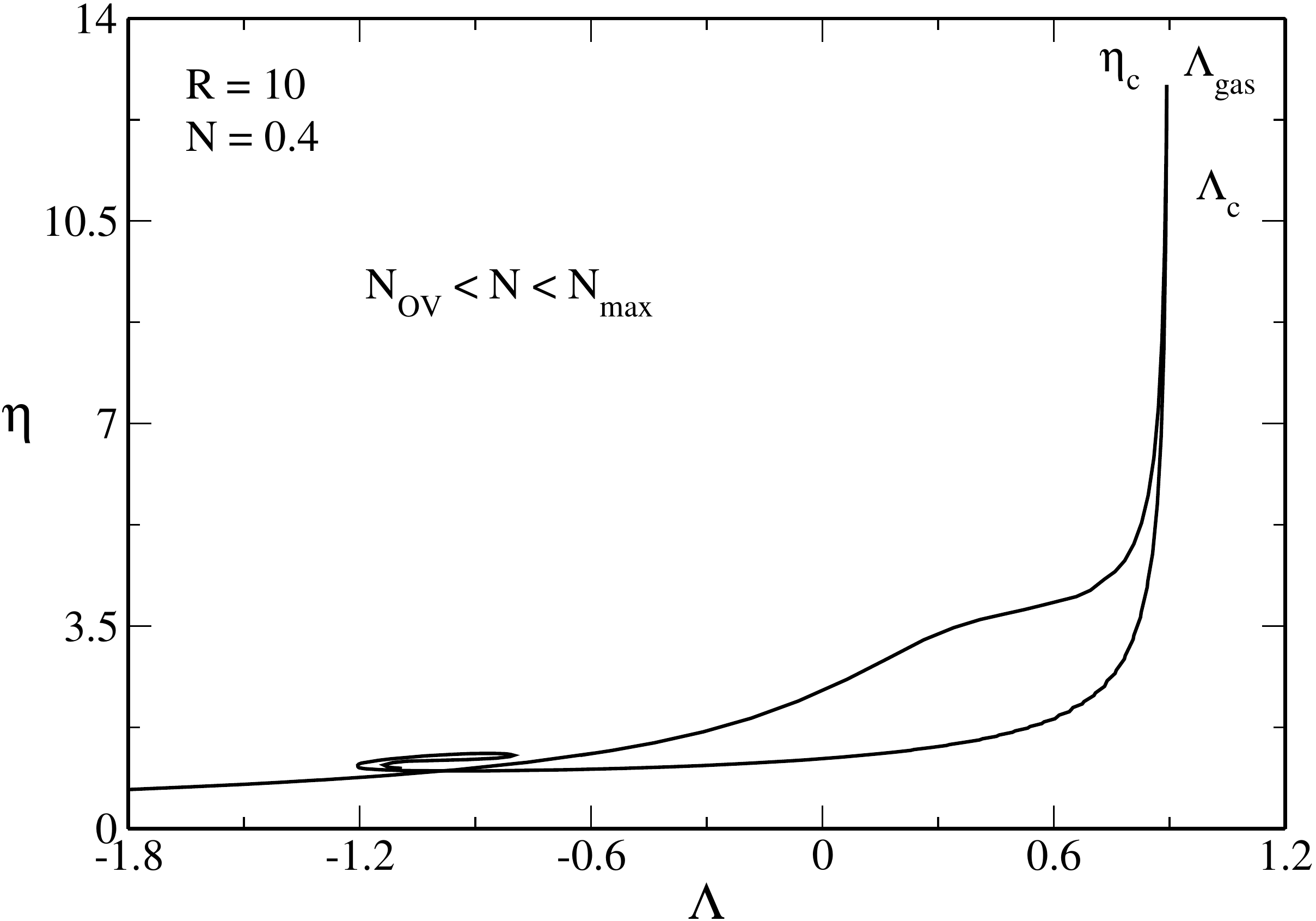}
\caption{Caloric curve for $N_{\rm OV}=0.39853<N<N_{\rm
max}=1.764$ (specifically  $R = 10$ and $N = 0.4$).  }
\label{kcal_R10_N0p4_unifiedPH}
\end{center}
\end{figure}

In Fig. \ref{kcal_R10_N0p4_unifiedPH}  we have plotted the caloric curve for
$N_{\rm OV}=0.39853<N<N_{\rm max}=1.764$. The difference with the cases treated
in Secs. \ref{sec_deb}-\ref{sec_fin} is that
there is no phase transition. When $N>N_{\rm OV}$ the two
asymptotes
have merged leading to a turning point of temperature at $\eta_c$ and a turning
point of energy at $\Lambda_c$. According to the
Poincar\'e turning point
criterion, the series of equilibria is stable up to $\eta_c$ in the canonical
ensemble and up to $\Lambda_c$ in the microcanonical ensemble.

In the microcanonical ensemble, the caloric curve presents the following
features:  

(i) There is no phase transition.

(ii) There is a region of negative specific heats between $\Lambda_{\rm
gas}$ and $\Lambda_c$.

(iii) There is  a catastrophic collapse at $\Lambda_c$ towards a
black hole.

In the canonical ensemble, the caloric curve presents the following
features: 
 
(i) There is no phase transition.

(ii) There is a catastrophic collapse at $\eta_c$ towards a
black hole.

\subsection{The phase diagrams}

\begin{figure}
\begin{center}
\includegraphics[clip,scale=0.3]{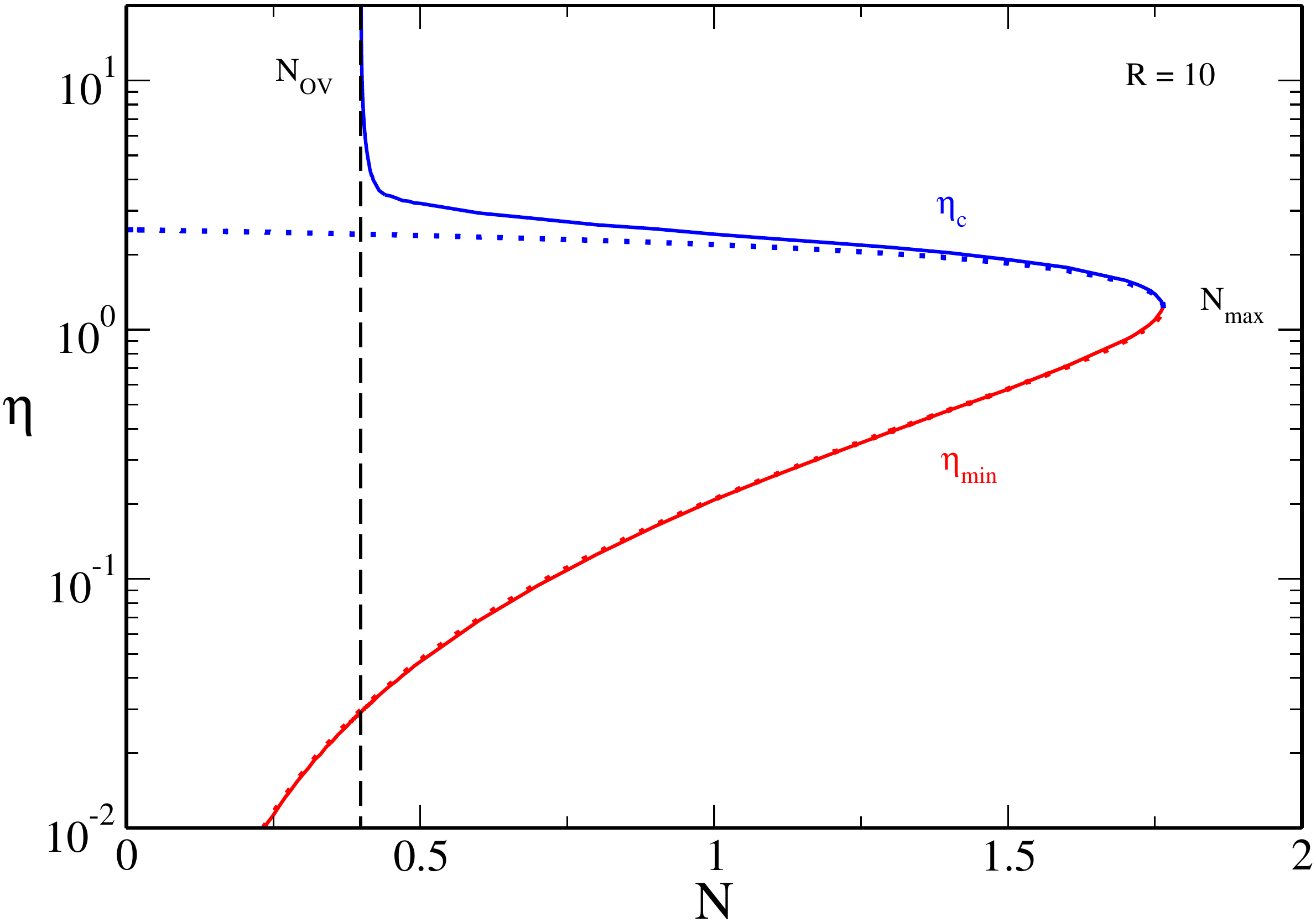}
\caption{Canonical phase diagram for $R_{\rm OV}<R<R_{\rm CCP}$ (specifically $R
=
10$). For $N\rightarrow
N_{\rm OV}^+$, we
find that
$\eta_c\sim 0.516 \, (N-N_{\rm OV})^{-1/2}$.}
\label{Xphase2_R10_newPH}
\end{center}
\end{figure}

In Fig. \ref{Xphase2_R10_newPH} we
have represented the canonical phase diagram corresponding to $R<R_{\rm CCP}$.
It shows the evolution of the critical temperatures
$\eta_{\rm max}$ and $\eta_c$ with $N$. We see the point $N_{\rm OV}$ above
which
quantum
mechanics
is not able to prevent gravitational collapse above $\eta_c$. We also 
see the point $N_{\rm max}$ above which there is no equilibrium state anymore.

The classical limit \cite{paper2} corresponds to the dotted
lines.
It provides a very good
approximation of $\eta_{\rm max}$ (hot spiral) for any $N$. It also provides a
very good
approximation of $\eta_c$ (cold spiral) for $N\gg N_{\rm OV}$. As we
approach $N_{\rm
OV}$ quantum mechanics must be taken into account.

\begin{figure}
\begin{center}
\includegraphics[clip,scale=0.3]{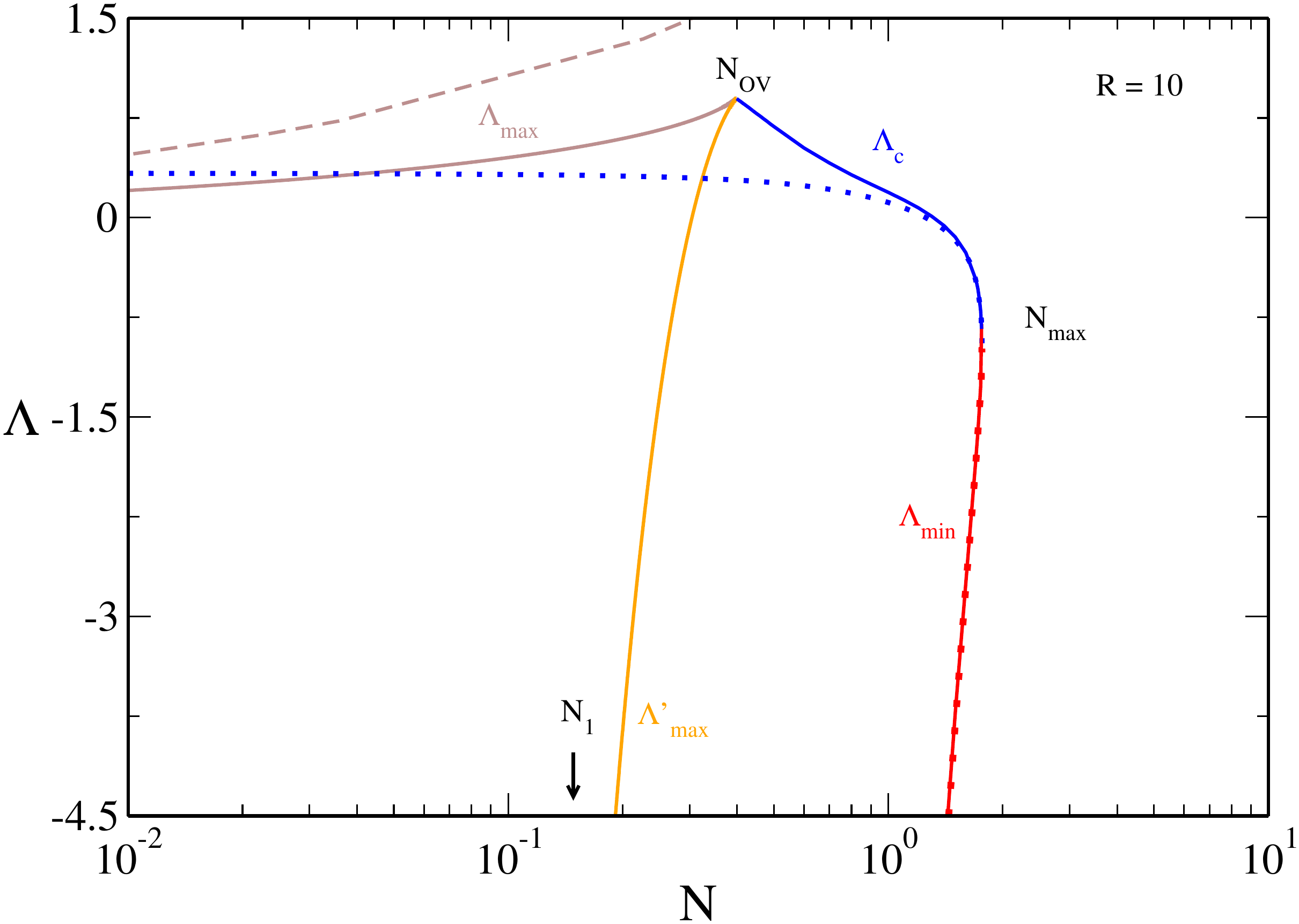}
\caption{Microcanonical phase diagram for $R_{\rm OV}<R<R_{\rm CCP}$
(specifically $R =
10$).}
\label{Xphase1_R10_newPH}
\end{center}
\end{figure}

In Fig. \ref{Xphase1_R10_newPH} we
have represented the microcanonical phase diagram corresponding to $R<R_{\rm
CCP}$. It shows the evolution of the critical energies
$\Lambda_{\rm min}$, $\Lambda_{\rm max}$, $\Lambda_c$ and
$\Lambda'_{\rm max}$  with $N$.  We see the point $N_{\rm OV}$ above which
quantum
mechanics
is not able to prevent gravitational collapse above $\Lambda_c$. We also
see the point $N_{\rm max}$ above which there is no equilibrium state anymore.

The nonrelativistic limit \cite{ijmpb} corresponds to the dashed lines. It
provides a very good approximation of $\Lambda_{\rm max}$ for
$N\ll N_{\rm OV}$ (this is not apparent in the figure but the curves coincide
for smaller values of $N$). As
we approach $N_{\rm OV}$ general relativity must be taken into account.

The classical limit  \cite{paper2} corresponds to
the dotted
lines.
It provides a very good
approximation of $\Lambda_{\rm min}$ (hot spiral) for any $N$. It also provides
a
very good
approximation of $\Lambda_c$  (cold spiral) for $N\gg N_{\rm OV}$. As we
approach $N_{\rm
OV}$ quantum mechanics must be taken into account.

\section{The case $R<R_{\rm OV}$}
\label{sec_un}

We now study the case $R<R_{\rm OV}=3.3569$. In that case, $N_{\rm OV}^b\simeq
N_{\rm max}$ and $N_{1}^b\simeq N'_{S}$ (see Fig.
\ref{phase_phase_Newton_corrPH} below).\footnote{We note that for small values
of $R$, the values of  $N_{1}^b$ and $N_{\rm OV}^b$ (as well as $\Lambda_{\rm
max}^b$, $(\Lambda'_{\rm max})^b$...) are affected by the presence of the box.
This is because the fermion
ball at $T=0$ and $E=E_{\rm min}$  (ground state) is confined by the walls of
the box instead of being self-confined (see
\cite{paper1} for a detailed study). This is why we have added the letter $b$
on these quantities.} For illustration, we take
$R=1$.

\subsection{The case $N<N_1^b\simeq N'_{S}$}

\begin{figure}
\begin{center}
\includegraphics[clip,scale=0.3]{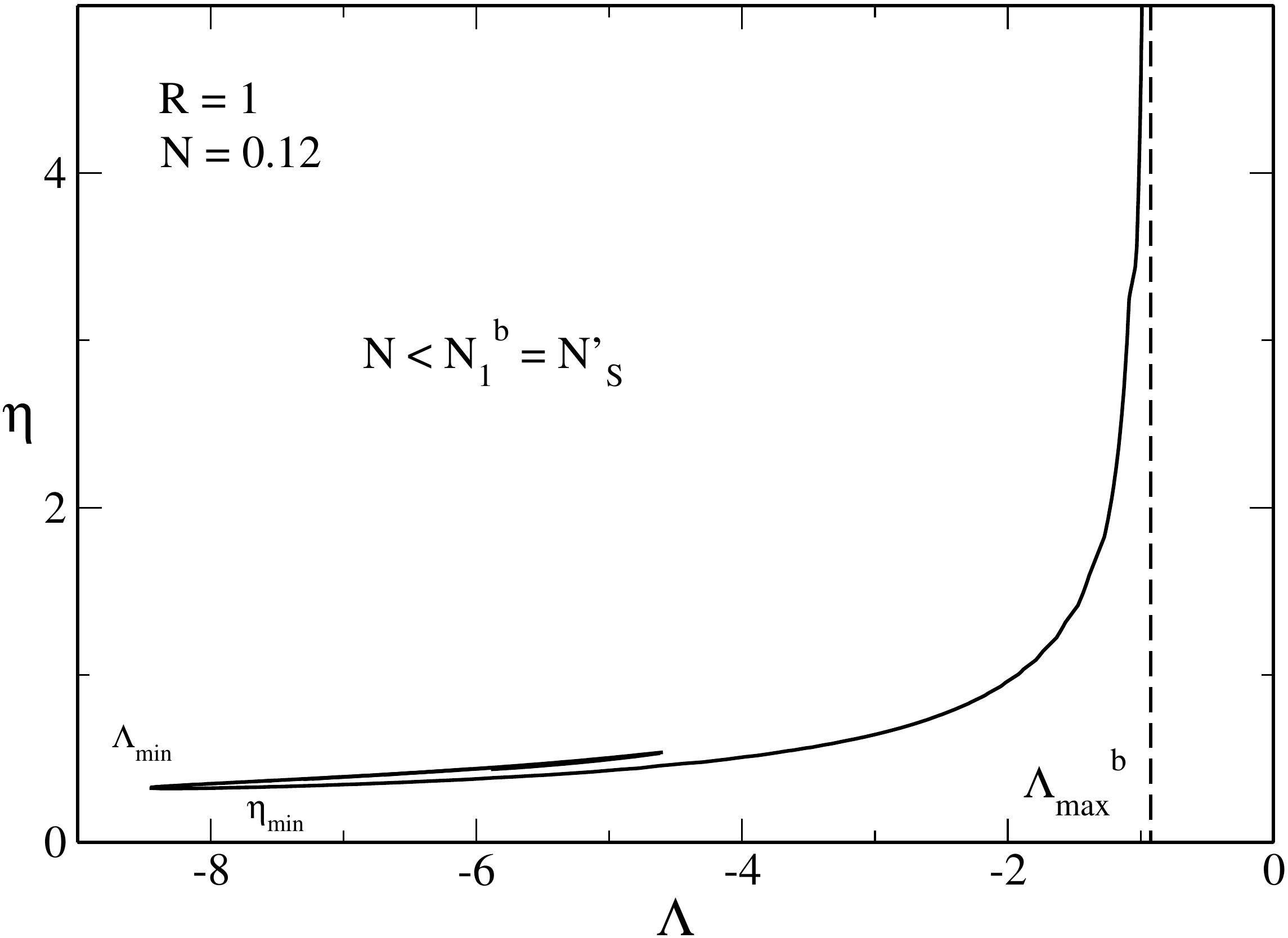}
\caption{Caloric curve for  $N<N_1^b=0.13627\simeq N'_{S}$ (specifically  $R =
1$ and $N
= 0.12$).}
\label{kcal_R1_N0p12_intersectionsPH}
\end{center}
\end{figure}

In Fig. \ref{kcal_R1_N0p12_intersectionsPH} we have plotted the caloric curve
for
$N<N_{1}^b=0.13627\simeq N'_{S}$. It is similar to that shown in
Fig. \ref{kcal_R50_N0p012_unified2PH}. Since $N$ is close to $N_{\rm max}$ (see
below), we
clearly see the
hot spiral that was outside the frame of Fig. \ref{kcal_R50_N0p012_unified2PH}.
According to the Poincar\'e turning point criterion, the series of
equilibria is
stable up to the maximum temperature $T_{\rm max}$ (corresponding to
$\eta_{\rm min}$) in the 
the canonical
ensemble and up to the maximum energy $E_{\rm max}$ (corresponding to
$\Lambda_{\rm min}$) in the
microcanonical ensemble. Above $T_{\rm max}$ and  $E_{\rm max}$ the system
collapses into a black hole as discussed in \cite{roupas,paper2}. If we
restrict ourselves to small and mid temperatures and energies 
(as in the preceding sections), there is no phase transition
and no
gravitational collapse. The specific heat is always positive and the ensembles
are equivalent.

\subsection{The case $N_1^{\rm b}\simeq N'_{S}<N<N_{\rm OV}^{\rm b}\simeq N_{\rm
max}$}
\label{sec_mardi}

\begin{figure}
\begin{center}
\includegraphics[clip,scale=0.3]{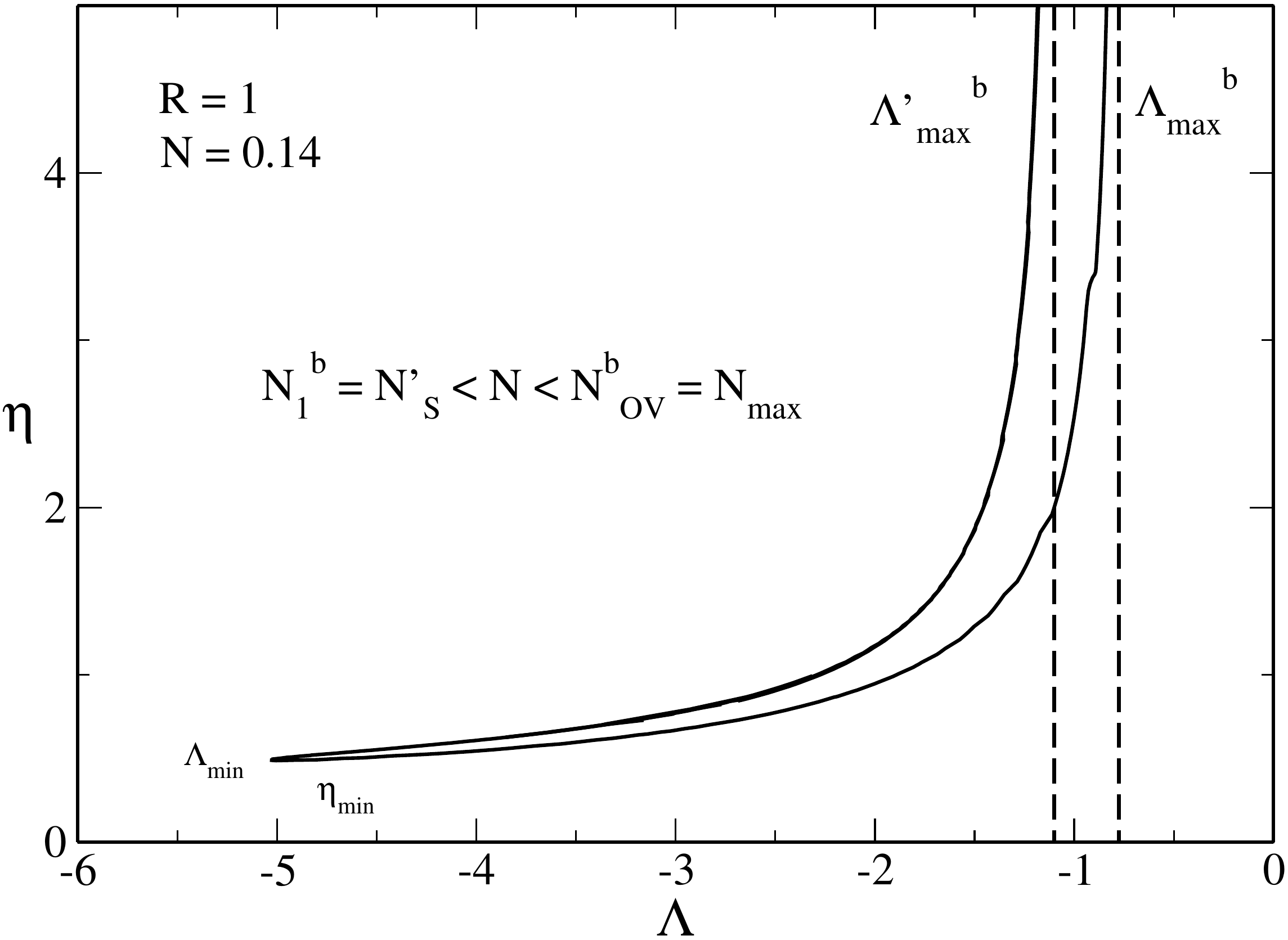}
\caption{Caloric curve for $N_1^b=0.13627\simeq
N'_{S}<N<N_{\rm OV}^b=0.2015 \simeq N_{\rm max}$  (specifically  $R=1$
and $N = 0.14$).}
\label{kcal_R1_N0p14_intersectionsPH}
\end{center}
\end{figure}

In Fig. \ref{kcal_R1_N0p14_intersectionsPH} we have plotted the caloric curve
for
$N_1^b=0.13627\simeq N'_{S}<N<N_{\rm OV}^b=0.2015\simeq N_{\rm max}$. In that
case,
we have two asymptotes at $\Lambda_{\rm max}^{\rm b}$ and $(\Lambda_{\rm
max}^{\rm b})'$ and a
turning point of energy and temperature at $\Lambda_{\rm min}$ and $\eta_{\rm
min}$. According to the Poincar\'e turning point criterion, the series of
equilibria is
stable along the main branch up to the maximum temperature $T_{\rm max}$
(corresponding to
$\eta_{\rm min}$) in the 
the canonical
ensemble and up to the maximum energy $E_{\rm max}$ (corresponding to
$\Lambda_{\rm min}$) in the
microcanonical ensemble. As before, if we
restrict ourselves to small and mid temperatures and energies, we conclude
that there is no phase transition and no
gravitational
collapse. The specific heat is always
positive and the ensembles
are equivalent.

\subsection{The phase diagrams}

\begin{figure}
\begin{center}
\includegraphics[clip,scale=0.3]{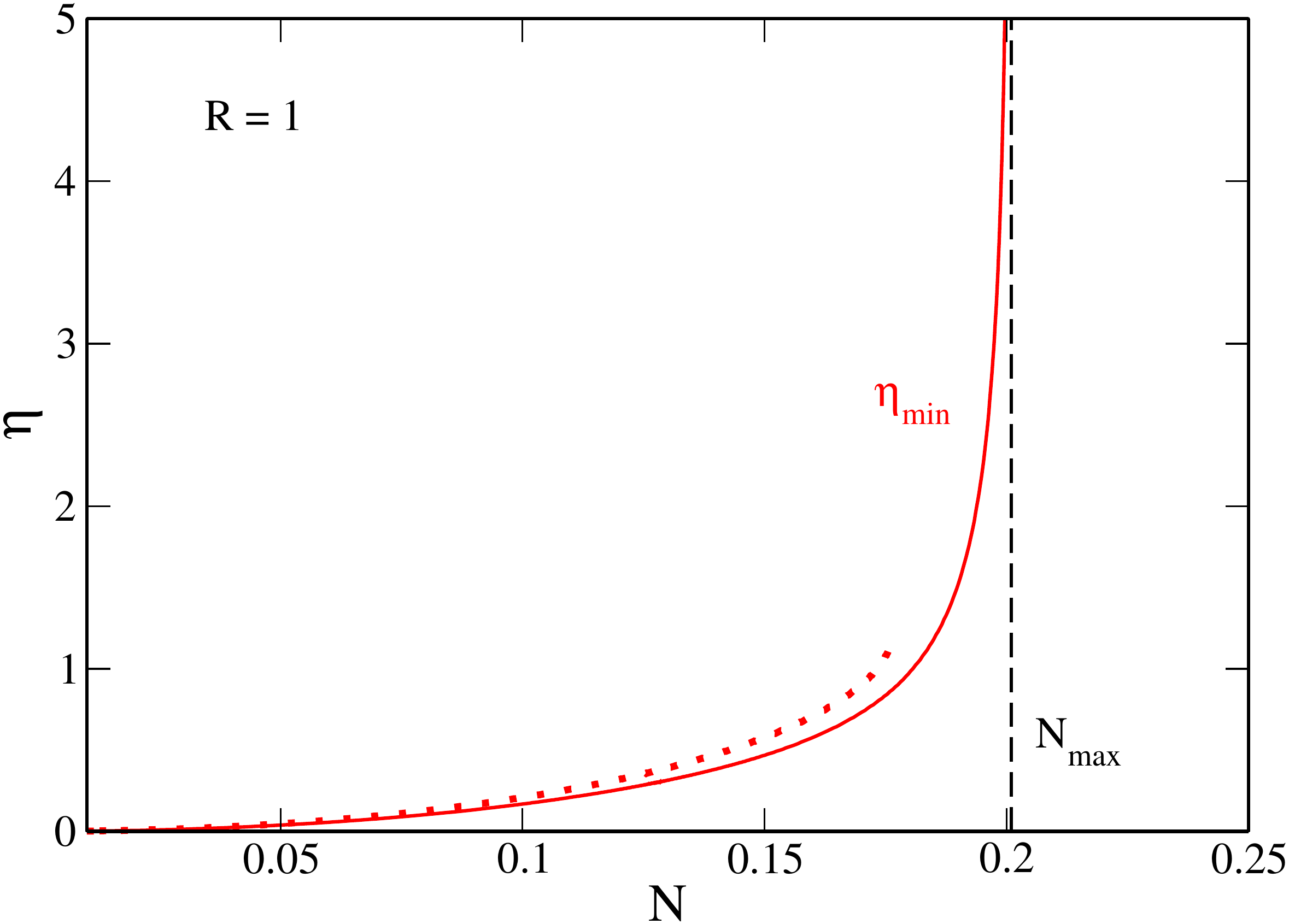}
\caption{Canonical
phase
diagram for $R<R_{\rm OV}$ (specifically $R = 1$). For $N\rightarrow
N_{\rm max}^-$, we
find that
$\eta_{\rm min}\sim 0.01 \, (N_{\rm max}-N)^{-1}$.}
\label{comparison_etamin_boltzmannPH}
\end{center}
\end{figure}

\begin{figure}
\begin{center}
\includegraphics[clip,scale=0.3]{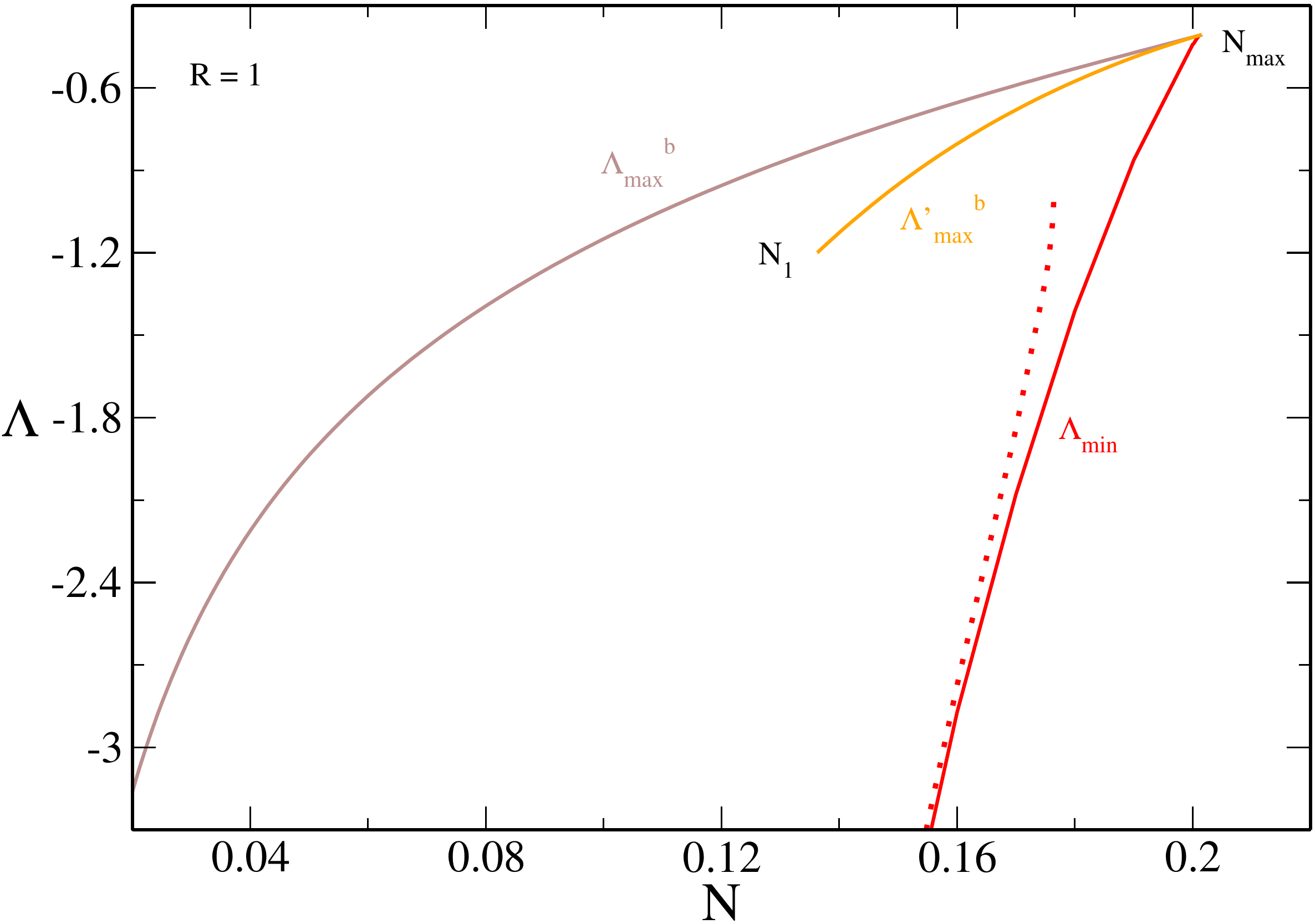}
\caption{Microcanonical phase
diagram for  $R<R_{\rm OV}$ (specifically $R = 1$).}
\label{Xphase_lambda_R1bisPH}
\end{center}
\end{figure}

In Fig. \ref{comparison_etamin_boltzmannPH} we
have represented the canonical phase diagram corresponding to $R<R_{\rm OV}$.
It shows the evolution of the critical temperature
$\eta_{\rm min}$ with
$N$. We note that  $\eta_{\rm min}$ diverges to $+\infty$ when $N\rightarrow 
N_{\rm
OV}^b=0.2015\simeq N_{\rm
max}$ meaning that the caloric curve goes up to infinity and
disappears. The classical limit
\cite{paper2}
corresponds to the dotted line. It provides a good approximation for $N\ll
N_{\rm max}$. It is not valid for $N\sim N_{\rm max}$. This shows that when
$R<R_{\rm OV}$  quantum
effects are still important close to 
$N_{\rm
max}$. This is because, in the present case, $N_{\rm
max}\simeq N_{\rm
OV}^b$ while in the previous examples  $N_{\rm
max}\gg N_{\rm
OV}$.

In Fig. \ref{Xphase_lambda_R1bisPH} we
have represented the microcanonical phase diagram corresponding to
$R<R_{\rm OV}$.
It shows the evolution of the critical energies
$\Lambda_{\rm min}$, $\Lambda^b_{\rm max}$ and  $(\Lambda^{'}_{\rm max})^b$ 
with
$N$.  The classical limit \cite{paper2}
corresponds to the dotted line. It provides a good approximation for $N\ll
N_{\rm max}$. It is not valid for $N\sim N_{\rm max}$. When
$R<R_{\rm OV}$  quantum
effects are still important close to  $N_{\rm
max}$.

\section{The $(R,N)$ phase diagram}
\label{sec_rn}

\begin{figure}
\begin{center}
\includegraphics[clip,scale=0.3]{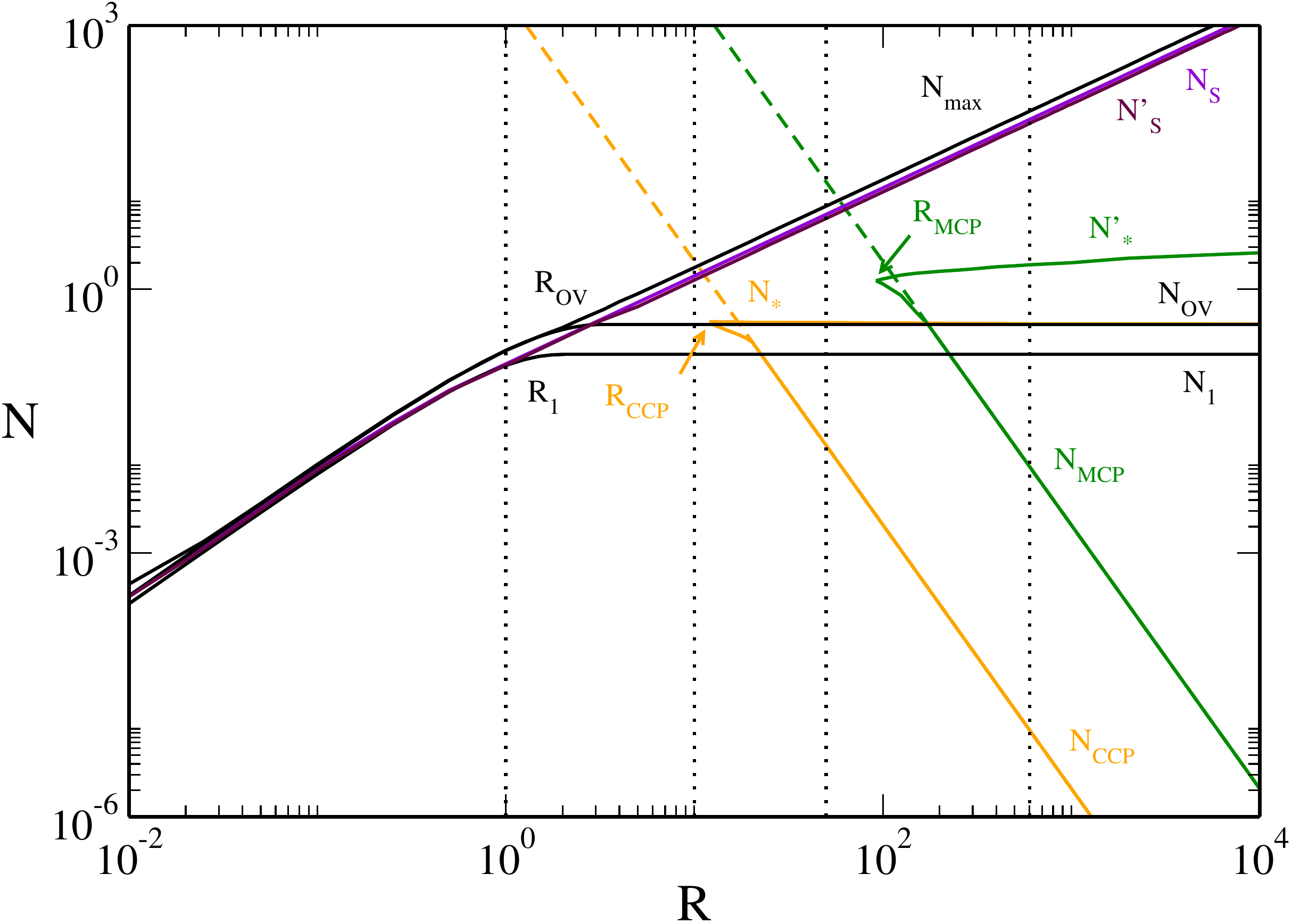}
\caption{
The $(R,N)$ phase diagram of the general relativistic Fermi gas. The
characteristic
particle numbers and radii are
defined in the text. The dashed lines correspond to the nonrelativistic
self-gravitating Fermi gas \cite{ijmpb}. The dotted lines correspond to the
radii $R=1$, $R=10$, $R=50$ and $R=600$ considered in the paper. 
 Note that the quantities $N_{e}(R)$ and $N'_{e}(R)$
have not
been represented because they are extremely close to
$N_{*}(R)$. Similarly, the quantities $N_{f}(R)$ and $N'_{f}(R)$
have not
been represented because they are extremely close to
$N'_{*}(R)$.}
\label{phase_phase_Newton_corrPH}
\end{center}
\end{figure}

We can summarize our results by plotting the characteristic particle
numbers $N_{\rm X}(R)$ encountered in our study  as a
function of the box radius $R$. This leads to the $(R,N)$ phase diagram
represented in
Fig. \ref{phase_phase_Newton_corrPH}.

\subsection{Characteristic particle numbers and characteristic radii}
\label{sec_chara}

Let us briefly recall the meaning of the characteristic particle numbers
appearing on this
diagram.\footnote{We present the 
characteristic particle numbers by order of appearance in the caloric curves 
as we increase $N$ for a given value of $R$. To fix the ideas, we take a large
radius so that all kinds of phase transitions are present. We start from
a small value of $N$. In that case, the caloric curve is monotonic with an
asymptote at $\Lambda_{\rm max}(N,R)$ corresponding to the stable ground state
of the self-gravitating Fermi gas at $T=0$ ($\eta=\infty$). We then increase
$N$ until we meet the different characteristic  particle numbers signaling a
topological change of the caloric curve.}

(i) When $R>R_{\rm CCP}=12.0$,  $N_{\rm CCP}(R)$ is the particle number at
which the canonical
phase transition appears, i.e., the particle number at which $\eta_c$ and
$\eta_*$ appear in the
caloric curve. When  $R\gg R_{\rm CCP}$, the
function  $N_{\rm CCP}(R)$ is given by the relation $N_{\rm CCP}(R)\sim
2.12\times 10^3 R^{-3}$ obtained in the
nonrelativistic study of \cite{ijmpb}. 

(ii) When $R>R_{\rm MCP}=92.0$, $N_{\rm MCP}(R)$ is the particle number at
which the
microcanonical
phase transition appears, i.e., the particle number at which $\Lambda_c$
and $\Lambda_*$ appear in
the caloric curve. When $R\gg R_{\rm MCP}$,
the function
$N_{\rm MCP}(R)$ is given by the relation $N_{\rm MCP}(R)\sim 2.20\times 10^6
R^{-3}$ obtained in the nonrelativistic study of \cite{ijmpb}. 

(iii) $N_{1}(R)$ is the particle number at which the unstable equilibrium
states at $T=0$ appear, i.e., the particle number at which the second
branch with an asymptote at
$\Lambda'_{\rm max}(N,R)$ appears in the caloric curve. The function $N_{1}(R)$
is studied in \cite{paper1}. When $R>R_1=2.0556$, the fermion star is
self-confined and we have
the standard value
$N_1=0.18131$ of the OV theory. When $R<R_{1}$, the fermion star is
box-confined and we find that
$N_{1}^{\rm b}(R)$ decreases as $R$ decreases. When $R\rightarrow 0$, we find
that $N_1^{\rm b}(R)\sim 0.2492\, R^{3/2}$ \cite{paper1}.

(iv) $N_{\rm OV}(R)$ is the particle number above
which there 
is no equilibrium
state at $T=0$ (no ground state) anymore. At $N=N_{\rm OV}(R)$ the asymptotes 
$\Lambda_{\rm
max}(N,R)$ and $\Lambda'_{\rm
max}(N,R)$ merge. When  $N>N_{\rm OV}(R)$ they are  replaced by a turning
point $\eta'_c$ in temperature and by a turning point $\Lambda'_{c}$ in energy.
The function $N_{\rm OV}(R)$ is studied in \cite{paper1}.
When $R>R_{\rm OV}=3.3569$, the fermion star is self-confined and we have the
standard value $N_{\rm
OV}=0.39853$ of the OV theory.
When $R<R_{\rm OV}$, the fermion star is box-confined, and we find that $N_{\rm
OV}^{\rm b}(R)$ decreases as $R$ decreases.  When
$R\rightarrow 0$, we find
that $N_{\rm OV}^{\rm b}(R)\sim 0.3104\, R^{3/2}$ \cite{paper1}.

(v) When $R>R_{\rm CCP}=12.0$, $N_{e}(R)$ is the particle number at
which
the zeroth order phase
transition in the canonical ensemble disappears, i.e., the
particle number at which $\eta'_c=\eta_c$. 

(vi) When $R>R_{\rm CCP}=12.0$,  $N'_{e}(R)$  is the particle number at
which the first order phase transition in the canonical ensemble disappears, 
i.e., the particle number at
which $\eta'_c=\eta_t$.

(vii) When $R>R_{\rm CCP}=12.0$, $N_{*}(R)$ is the  particle number at
which
the condensed
phase disappears in the canonical ensemble, i.e., the  particle number at which
$\eta'_c=\eta_*$. 

(viii) When $R>R_{\rm MCP}=92.0$,  $N_{f}(R)$ is the  particle number at
which the zeroth order phase
transition in the microcanonical ensemble disappears,
i.e., the  particle number at
which $\Lambda''_c=\Lambda_c$.

(ix) When $R>R_{\rm MCP}=92.0$, $N'_{f}(R)$  is the  particle number at
which the first order
microcanonical phase transition disappears, i.e., the  particle number at
which
$\Lambda''_c=\Lambda_t$.

(x)  When $R>R_{\rm MCP}=92.0$, $N_{*}'(R)$ is the particle number at  which
the condensed
phase  disappears in the microcanonical ensemble, i.e., the  particle number at
which
$\Lambda''_c=\Lambda_*$. 

(xi)  When  $R>R_{1}=2.0556$,  $N_{\rm S}'(R)$ is the value of the particle
number above which the two
spirals of the caloric curve are amputed (truncated) and touch
each other. When 
$R\gg
R_{1}$, the function  $N_{\rm S}'(R)$ is given by the relation $N_{\rm
S}'(R) \sim 0.128\, R$ obtained in the classical study of \cite{paper2}. When
$R<R_{1}=2.0556$, we find that $N_{\rm S}'(R)\simeq
N_1^b(R)$ \cite{paper1}.  Above that value, the caloric curve looks like Fig.
\ref{kcal_R1_N0p14_intersectionsPH} instead of looking like a double spiral.

(xii) When  $R>R_{1}=2.0556$, $N_{\rm S}(R)$ is the value of the
particle number above which there is no spiral anymore and the caloric curve
makes a ``loop''. When $R\gg
R_{\rm OV}$, the function  $N_{\rm S}(R)$ is given by the relation  $N_{\rm
S}(R) \sim 0.1415\, R$ obtained in the classical study of \cite{paper2}.
 When   $R<R_{1}=2.0556$, we find that $N_{\rm S}(R)\simeq N_{\rm
S}'(R)\simeq N_1^b(R)$ \cite{paper1}.

(xiii) $N_{\rm max}(R)$ is the maximum particle
number below which an
equilibrium state may exist for certain values of energy and temperature.
For $N>N_{\rm max}(R)$ there is no
equilibrium state, whatever the energy and the temperature. When $R\gg
R_{\rm OV}$, the function  $N_{\rm max}(R)$ is given by the relation 
$N_{\rm max}(R)\sim 0.1764\, R$ obtained
in the classical study of \cite{roupas,paper2}. When  $R<R_{\rm
OV}=3.3569$, we find that 
 $N_{\rm max}(R)\simeq N_{\rm OV}^b(R)$ \cite{paper1}. We note
that $N_{\rm max}\gg N_{\rm OV}=0.39853$ when $R\gg R_{\rm OV}=3.3569$ while
$N_{\rm max}\ll N_{\rm OV}$ when $R\ll R_{\rm OV}$. The change of regime
takes place at $R\sim R_{\rm OV}$ where $N_{\rm max}\sim N_{\rm OV}$ (see the
Remark at the end of Sec. \ref{sec_suncl}).

\subsection{Summary of the main results when $R$ is fixed and $N$ is varied}

The $(R,N)$ phase diagram exhibits two critical points
at $(R_{\rm CCP},N_{\rm CCP})=(12.0,0.424)$  and 
 $(R_{\rm MCP},N_{\rm MCP})=(92.0,1.25)$.  $R_{\rm CCP}=12.0$ is
the radius above which the system experiences
a canonical phase transition when $N_{\rm
CCP}(R)<N<N_e(R)$. $R_{\rm MCP}=92.0$ is the radius above which the system 
experiences a microcanonical phase transition when
$N_{\rm MCP}(R)<N<N_f(R)$. Below, we summarize the essential
features of the microcanonical and canonical phase transitions found for the
self-gravitating Fermi gas in general relativity. In this section, we consider
the situation where $R$ is fixed and $N$ is varied. We recall that there
is a possible equilibrium state only for $N<N_{\rm max}(R)$.
We have
\begin{equation}
\label{summ1}
N_{\rm max}(R)\sim 0.3104\, R^{3/2} \qquad (R\ll R_{\rm
OV}),
\end{equation}
\begin{equation}
\label{summ2}
N_{\rm max}(R)\sim 0.1764\, R \qquad (R\gg R_{\rm
OV}).
\end{equation}
As in the previous sections, we do not consider the case of
very high energies and very high
temperatures which has been treated in \cite{roupas,paper2}.

\subsubsection{$R<R_{\rm CCP}$}

When $N<N_{\rm OV}(R)$, there is no phase transition and no catastrophic
collapse (see Fig. \ref{kcal_R10_N0p36_unifiedPH}). When
$N_{\rm OV}(R)<N<N_{\rm max}(R)$, there
is no phase transition but there is a catastrophic collapse towards a
black hole at $\eta_c(N,R)$ in the canonical
ensemble and at $\Lambda_c(N,R)$ in the microcanonical ensemble (see Fig.
\ref{kcal_R10_N0p4_unifiedPH}).

\subsubsection{$R_{\rm CCP}<R<R'_{\rm CCP}$}

\begin{figure}
\begin{center}
\includegraphics[clip,scale=0.3]{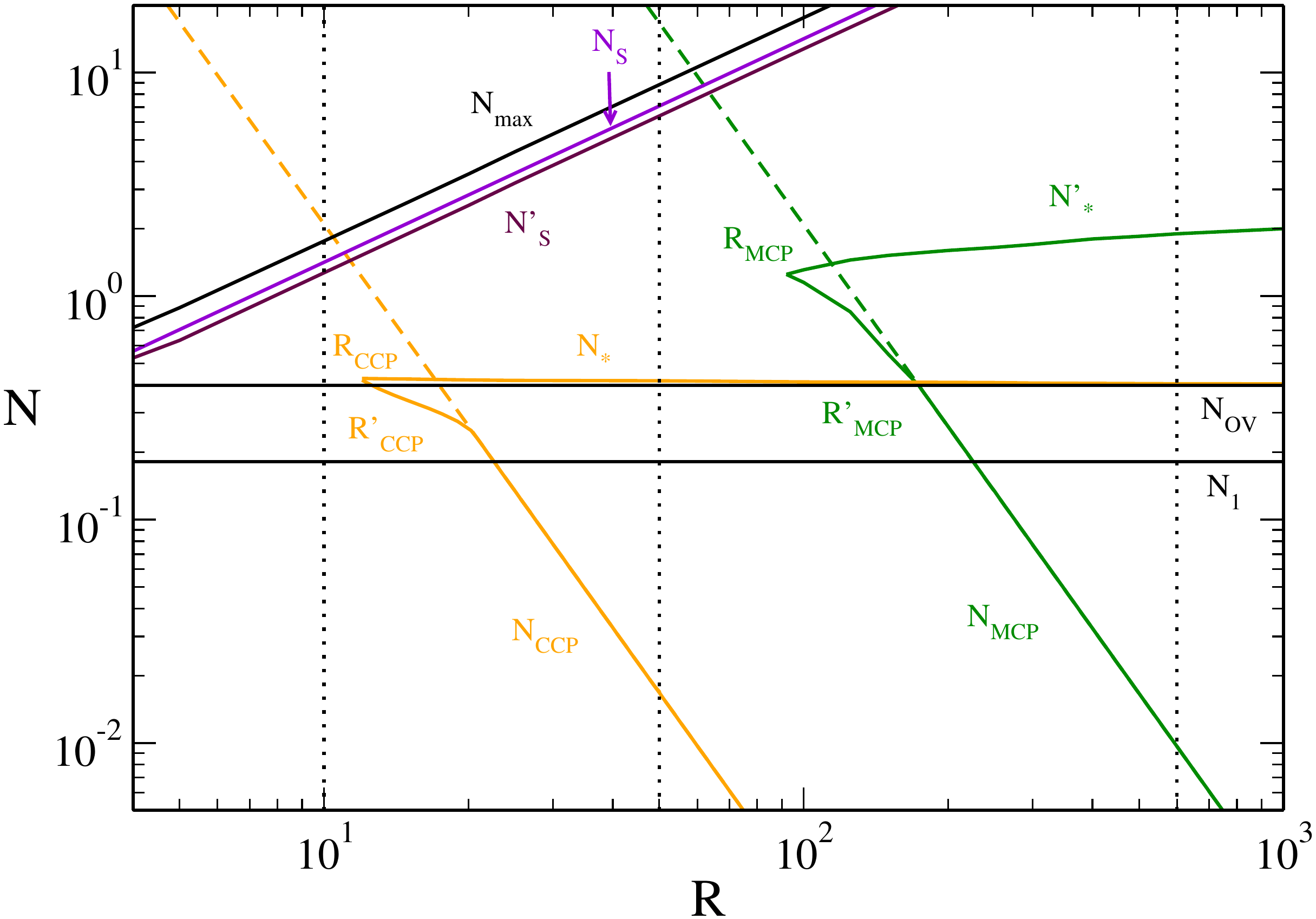}
\caption{Zoom of the $(R,N)$ phase diagram of the general relativistic Fermi
gas. When $R_{\rm CCP}=12.0<R<R'_{\rm CCP}=12.6$ (resp. $R>R'_{\rm
CCP}=12.6$) the
canonical phase transition appears after (resp. before) $N_{\rm OV}$. 
When $R_{\rm MCP}=92.0<R<R'_{\rm MCP}=173$ (resp. $R>R'_{\rm MCP}=173$) 
the
microcanonical phase transition appears after (resp. before) $N_{\rm OV}$.}
\label{phase_phase_Newton_corrPHzoom}
\end{center}
\end{figure}

This case, in which $N_{\rm CCP}>N_{\rm OV}$ (see Fig.
\ref{phase_phase_Newton_corrPHzoom}), was not treated explicitly in Secs.
\ref{sec_first}-\ref{sec_un}.

In the canonical ensemble when $N<N_{\rm OV}$, there is
no phase transition and no catastrophic
collapse. When
$N_{\rm OV}<N<N_{\rm CCP}(R)$, there is no phase transition but there is
a catastrophic collapse toward a black hole at
$\eta_c(N,R)$. When $N_{\rm CCP}(R)<N<N_{e}$, there is a zeroth order phase
transition from
the gaseous phase
to the condensed phase at $\eta_c(N,R)$ and a catastrophic collapse from the
condensed phase to a black
hole at $\eta'_c(N,R)$. When
$N>N_{e}(R)$, there is no phase transition but there is
a catastrophic collapse from the gaseous phase to a black hole at
$\eta_c(N,R)$. 

In the microcanonical ensemble, the situation is the same as before.

\subsubsection{$R_{\rm CCP}<R<R_{\rm MCP}$}

In the canonical ensemble when $N<N_{\rm CCP}(R)$
there is no phase
transition and no catastrophic collapse (see Fig.
\ref{kcal_R50_N0p012_unified2PH}). When $N_{\rm CCP}(R)<N<N_{\rm OV}$ 
there is a zeroth order phase
transition from the gaseous phase to the condensed phase at $\eta_c(N,R)$ and
no catastrophic collapse (see Figs. \ref{Xkcal_R50_N0p15_unified2blackPH}
and \ref{kcal_R50_N0p29_unifiedPH}). When
$N_{\rm OV}<N<N_e(R)$ there is a zeroth order phase transition from
the gaseous phase
to the condensed phase at $\eta_c(N,R)$ and a catastrophic collapse from the
condensed phase to a black
hole at $\eta'_c(N,R)$ (see Fig. \ref{kcal_R50_N0p399_unifiedPH}). When
$N>N_e(R)$ there is no phase transition but there
is a catastrophic collapse from
the gaseous phase to a
black hole at $\eta_c(N,R)$ (see Figs.
\ref{Xkcal_R50_N0p401_unifiedPH}, \ref{kcal_R50_N0p41_unifiedPH},
\ref{Xkcal_R50_N0p45_unified2PH}, \ref{kcal_R50_N1p5_unifiedPH}
and \ref{kcal_R50_N4_unifiedPH}).

In the microcanonical ensemble, the situation is the same as before.

\subsubsection{$R_{\rm MCP}<R<R_{\rm MCP}'$}

This case, in which $N_{\rm MCP}>N_{\rm OV}$ (see Fig.
\ref{phase_phase_Newton_corrPHzoom}), was not treated explicitly in Secs.
\ref{sec_first}-\ref{sec_un}.

In the canonical ensemble, the situation is the same as before so we focus on
the microcanonical ensemble. When $N<N_{\rm OV}$ there is no phase
transition and no catastrophic collapse. When $N_{\rm OV}<N<N_{\rm MCP}(R)$ 
there is no phase transition but there is a catastrophic  collapse towards a
black hole at $\Lambda_c(N,R)$.
When
$N_{\rm MCP}(R)<N<N_f(R)$ there is a zeroth order phase transition
from the gaseous
phase
to the condensed phase at $\Lambda_c(N,R)$ and a catastrophic collapse from
the condensed phase to a
black
hole at $\Lambda'_c(N,R)$ or $\Lambda''_c(N,R)$. When $N>N_f(R)$ there is no
phase transition but there is  a catastrophic collapse
from the gaseous phase to a
black hole at $\Lambda_c(N,R)$.

\subsubsection{$R>R_{\rm MCP}'$}

In the canonical ensemble, the situation is the same as before so we focus on
the microcanonical ensemble. When $N<N_{\rm MCP}(R)$ there is no phase
transition and no catastrophic collapse. When $N_{\rm
MCP}(R)<N<N_{\rm OV}$ 
there is a zeroth
order phase
transition from the gaseous phase to the condensed phase at $\Lambda_c(N,R)$
and no catastrophic collapse (see Fig. \ref{Xkcal_R600_N0p29_unified_bnewPH}).
When
$N_{\rm OV}<N<N_f(R)$ there is a zeroth order phase transition
from the gaseous
phase
to the condensed phase at $\Lambda_c(N,R)$ and a catastrophic collapse from
the condensed phase to a
black
hole at $\Lambda''_c(N,R)$ (see Fig.
\ref{kcal_R600_N1p3_unifiedPH}). When $N>N_f(R)$ there is no
phase transition but there is a catastrophic collapse
from the 
gaseous phase to a
black hole at $\Lambda_c(N,R)$ (see Figs.
\ref{kcal_R600_N1p5_unifiedPH}, \ref{kcal_R600_N1p65_unifiedPH} and
\ref{kcal_R600_N5_unified2PH}).

\subsection{Summary of the main results when $N$ is fixed and $R$ is varied}

We now consider the situation where  $N$ is fixed and $R$ is
varied.\footnote{This is, for example, the viewpoint adopted by Hertel and
Thirring \cite{ht} as recalled in the introduction (see also Sec.
\ref{sec_ap}).} These
results can be deduced from the $(R,N)$ phase diagram of Fig.
\ref{phase_phase_Newton_corrPH} by taking the inverse of the
functions $N_{\rm X}(R)$. This leads to the $(N,R)$ phase
diagram. We note that there is
a possible equilibrium state only for $R>R_{\rm min}(N)$ where $R_{\rm min}(N)$
is the inverse function of $N_{\rm max}(R)$. We have
\begin{equation}
\label{summ1b}
R_{\rm min}(N)\sim 2.181\, N^{2/3} \qquad (N\ll N_{\rm OV}),
\end{equation}
\begin{equation}
\label{summ2b}
R_{\rm min}(N)\sim 5.669\, N\qquad (N\gg N_{\rm OV}).
\end{equation}

\subsubsection{$N<N_{\rm OV}$}

In the canonical ensemble when $R_{\rm min}(N)<R<R_{\rm
CCP}(N)$
there is no phase transition and no catastrophic collapse.  When $R>R_{\rm
CCP}(N)$ (with $R_{\rm
CCP}(N)\sim 12.8\, N^{-1/3}$  in the nonrelativistic limit $N\ll N_{\rm OV}$)
there
is a zeroth order phase transition from
the gaseous phase
to the condensed phase at $\eta_c(N,R)$ and no catastrophic collapse.

In the microcanonical ensemble when $R_{\rm min}(N)<R<R_{\rm
MCP}(N)$ (with $R_{\rm
MCP}(N)\sim 130\, N^{-1/3}$  in the nonrelativistic limit $N\ll N_{\rm OV}$)
there is
no phase transition and no catastrophic collapse.  When $R>R_{\rm MCP}(N)$
there
is a zeroth order phase transition from
the gaseous phase
to the condensed phase at $\Lambda_c(N,R)$ and  no catastrophic
collapse.

This is essentially like in the nonrelativistic limit \cite{ijmpb}. Relativistic
corrections occur only close to $N_{\rm OV}$ and/or close to $R_{\rm min}(N)$.

\subsubsection{$N_{\rm OV}<N<N_{\rm CCP}\simeq N_e$}

In the canonical ensemble when $R_{\rm min}(N)<R<R_{\rm
CCP}(N)$
there is no phase transition but there is  a catastrophic collapse towards a
black hole at $\eta_c(N,R)$. When $R>R_{\rm CCP}(N)$
there
is a zeroth order phase transition from
the gaseous phase
to the condensed phase at $\eta_c(N,R)$ and a catastrophic collapse from the
condensed phase to a black hole at $\eta'_c(N,R)$. 

In the microcanonical ensemble when
$R_{\rm min}(N)<R<R_{\rm MCP}(N)$
there is no phase transition but there is a catastrophic collapse towards a
black hole at $\Lambda_c(N,R)$.
When $R>R_{\rm
MCP}(N)$ there is a zeroth order phase transition from
the gaseous phase
to the condensed phase at $\Lambda_c(N,R)$ and  a catastrophic collapse
from the condensed phase to a black hole at $\Lambda''_c(N,R)$.

\subsubsection{$N_{\rm CCP}\simeq N_e<N<N_{\rm MCP}\simeq N_f$}

In the canonical ensemble, when $R>R_{\rm min}(N)$ there is
no phase transition but
there is a catastrophic collapse to a black hole  at $\eta_c(N,R)$.

In the microcanonical ensemble, the situation is the same as before.

\subsubsection{$N>N_{\rm MCP}$}

In the canonical ensemble, the situation is the same as before. In the
microcanonical ensemble when $R>R_{\rm min}(N)$
there is no phase transition but there is  a catastrophic collapse towards a
black hole at $\Lambda_c(N,R)$. 

This is essentially like in the classical limit
\cite{roupas,paper2}.

\section{The nonrelativistic and classical limits}
\label{sec_ncl}

In this section, we consider the nonrelativistic ($c\rightarrow
+\infty$) and classical ($\hbar\rightarrow 0$) limits and study
the commutation of these limits.

\subsection{An apparent paradox related to the commutation of the limits
$\hbar\rightarrow 0$ and $c\rightarrow
+\infty$}

The commutation of the limits $\hbar\rightarrow 0$ and $c\rightarrow
+\infty$ leads to an apparent paradox. This can be seen from the 
expression of the maximum OV particle number given by
\begin{equation}
\label{dq2qq}
N_{\rm OV}=0.39853\, \sqrt{\frac{2}{g}}\left (\frac{\hbar c}{G}\right
)^{3/2}\frac{1}{m^3}.
\end{equation}

(i) If we take the nonrelativistic limit $c\rightarrow +\infty$
first \cite{ijmpb}, we find
that $N_{\rm
OV}\rightarrow +\infty$. Therefore, we always have  $N<N_{\rm
OV}$. As a result,
there is always
an equilibrium state at low temperatures and low energies,
whatever the value of $\hbar$, i.e., even if we consider the classical limit
$\hbar\rightarrow 0$.\footnote{In the canonical
ensemble, when $\hbar\ll 1$ and $T<T_c$, the equilibrium state
corresponds to a fermion ball containing most of the mass. When
$\hbar\rightarrow 0$ the fermion ball contains all the mass and its radius
goes to zero. In that case, we get a Dirac peak of mass $M$
(see Appendix \ref{sec_thermoqce}). In the
microcanonical ensemble, when
$\hbar\ll 1$ and $E<E_c$, the equilibrium state corresponds to a
fermion ball containing a fraction of the total mass surrounded by a hot halo.
When $\hbar\rightarrow 0$ the mass of the fermion ball and its radius go
to zero while its potential energy goes to $-\infty$. As a result, the
temperature of
the halo goes to $+\infty$. In that case, we
get a Dirac peak of zero mass
 and infinite potential energy (binary) surrounded by an
infinitely hot
halo (see Appendix \ref{sec_thermoqmce}). Note that 
for $\hbar>0$, as small as one pleases, there is always
a
regular equilibrium state.}

(ii) If we take the classical limit $\hbar\rightarrow 0$ first
\cite{roupas,paper2}, we
find that $N_{\rm OV}\rightarrow 0$. Therefore, we always have $N>N_{\rm
OV}$. As a result, the system undergoes  a catastrophic collapse at low
temperatures and low energies, whatever the value of $c$,
i.e., even if we consider the nonrelativistic limit
$c\rightarrow +\infty$.

Therefore, if we consider a nonrelativistic classical gas ($c\rightarrow
+\infty$ and $\hbar\rightarrow 0$), the first argument tells us
that there is an  equilibrium state at low temperatures and
low energies while the second argument tells us that there is no equilibrium
state at low temperatures and 
low energies. How can we reconcile these two apparent contradictory situations?
In the next
two subsections, we re-express these results in terms of
dimensionless variables, and in the third subsection we provide a solution of
this apparent paradox.

\subsection{When the nonrelativistic limit $c\rightarrow +\infty$ is taken
before the classical limit $\hbar\rightarrow 0$}
\label{sec_nonrel}

\begin{figure}
\begin{center}
\includegraphics[clip,scale=0.3]{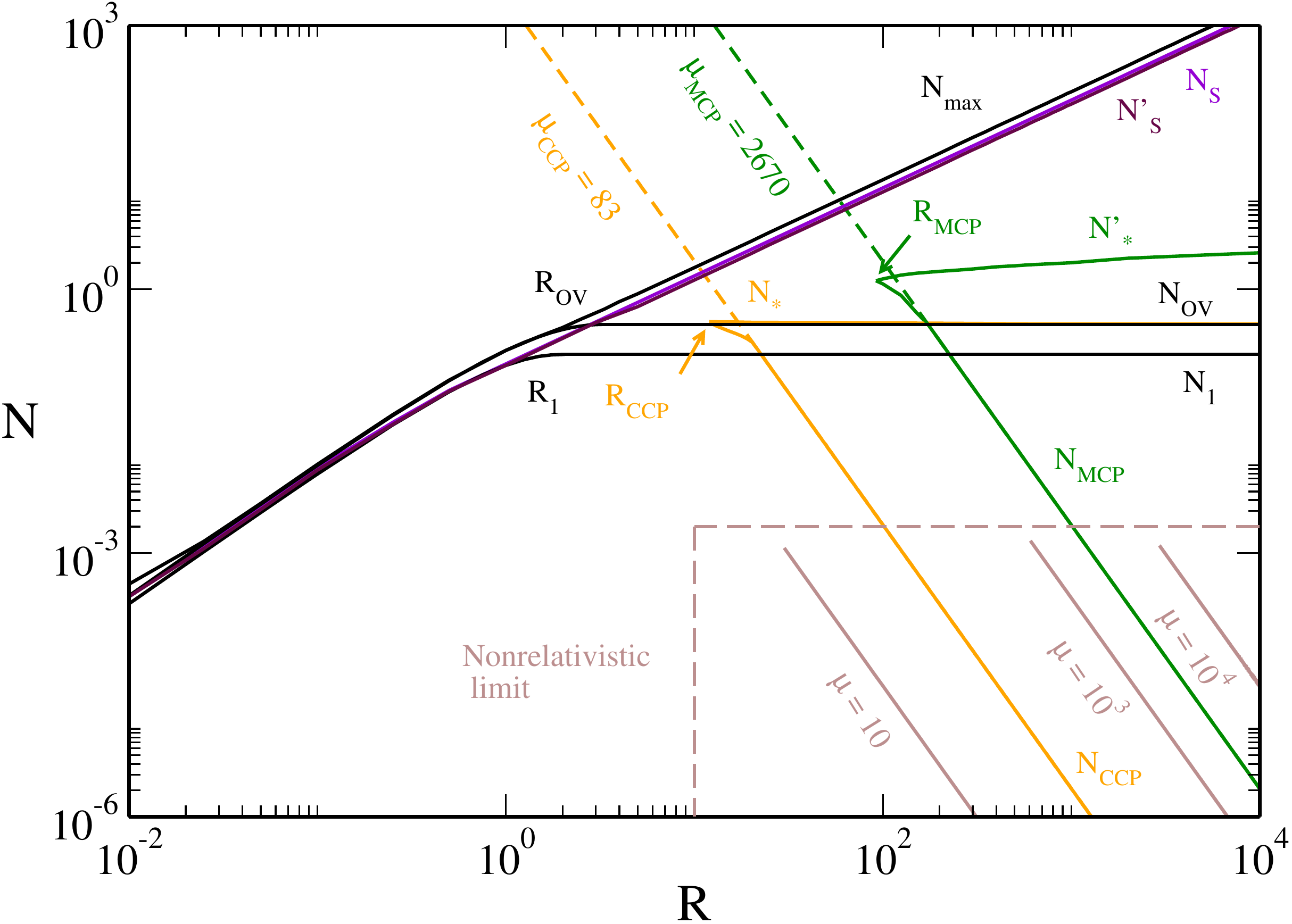}
\caption{Lines of constant $\mu$ in the $(R,N)$ phase diagram of Fig.
\ref{phase_phase_Newton_corrPH} characterizing the
nonrelativistic limit $N\rightarrow 0$ and $R\rightarrow +\infty$ with fixed
$\mu=(4\sqrt{2}/\pi)(NR^3)^{1/2}$. We have chosen $\mu=10$,
$\mu=10^3$ and $\mu=10^4$ (brown lines) corresponding to the values appearing in
Fig. 14 of \cite{ijmpb}.}
\label{phase_phaseMUPH}
\end{center}
\end{figure}

In this subsection, we consider the situation where the nonrelativistic limit
($c\rightarrow +\infty$) is taken before the classical limit ($\hbar\rightarrow
0$). Using the dimensionless variables of Appendix \ref{sec_dq}, the
nonrelativistic limit corresponds to
$N\rightarrow
0$ and
$R\rightarrow +\infty$ in such a way that $NR^3$ is fixed.\footnote{This scaling
is obtained in order to keep the parameter $\mu$ defined in \cite{ijmpb} fixed
(see also Sec. \ref{sec_sunl}). Coming back to dimensional variables, the
nonrelativistic limit corresponds to $N\ll N_{\rm OV}\sim (\hbar c/G)^{3/2}/m^3$
and $R\gg R_{\rm OV}\sim (\hbar^3/Gc)^{1/2}/m^2$ with
$\mu^2\sim NR^3m^9G^3/\hbar^6$ fixed. This is consistent with the fact that
$N_{\rm OV}\rightarrow +\infty$ and $R_{\rm OV}\rightarrow 0$ when $c\rightarrow
+\infty$.} This scaling defines an ensemble of
parallel
lines of constant
$\mu=(4\sqrt{2}/\pi)(NR^3)^{1/2}$ in the bottom right panel of
Fig. \ref{phase_phaseMUPH}. In the nonrelativistic limit, the caloric
curves
are the same for any couple of points $(R,N)$ belonging to a given $\mu$-line.
As $\mu$ increases,  the
$\mu$-lines move to the right and the system becomes more and more
classical \cite{ijmpb}.

The phase transitions occuring in a nonrelativistic self-gravitating 
Fermi gas have been studied in \cite{ijmpb}. When $\mu<\mu_{\rm CCP}=83$
there is no phase transition (see Fig. 14 of \cite{ijmpb}). When
$\mu_{\rm CCP}=83<\mu<\mu_{\rm MCP}=2670$ there is  a
canonical phase transition (see Fig. 31 of \cite{ijmpb}). When $\mu>\mu_{\rm
MCP}=2670$
there are canonical and microcanonical phase transitions (see Fig. 21 of
\cite{ijmpb}). 
The classical limit corresponds to  $\mu\rightarrow +\infty$ (see Fig. 22 of
\cite{ijmpb})

\begin{figure}
\begin{center}
\includegraphics[clip,scale=0.3]{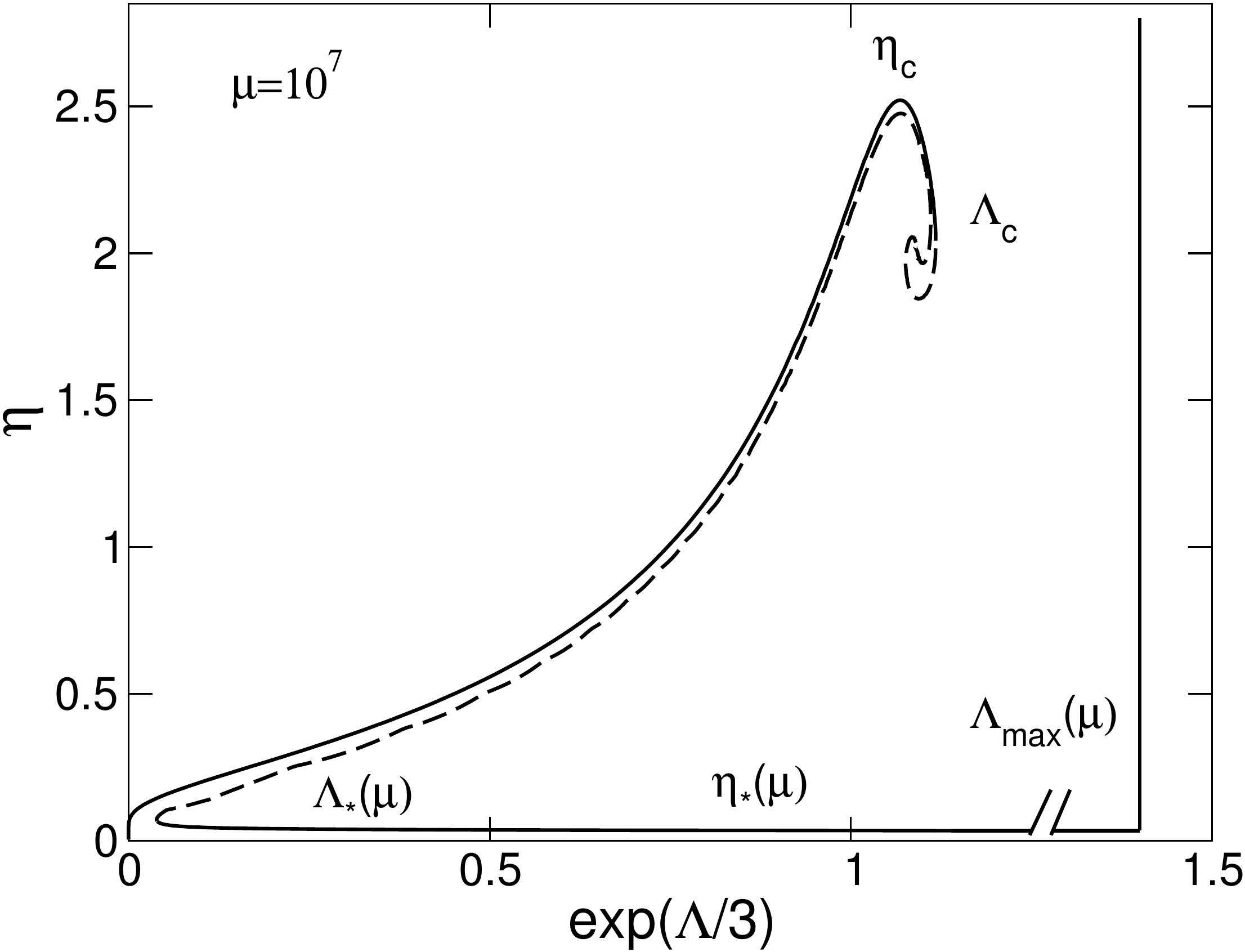}
\caption{Caloric curve of nonrelativistic self-gravitating fermions in
the classical limit $\mu\rightarrow +\infty$ (here $\mu=10^7$).}
\label{limite}
\end{center}
\end{figure}

For large but finite values of $\mu$ (see Fig. \ref{limite}), the series of
equilibria forms a spiral
which finally unwinds, progresses backward along an inverse spiral
until
$\Lambda_*(\mu)$, turns right, forms a lower branch, and finally tends
towards an asymptote at $\Lambda_{\rm max}(\mu)=0.0642\, \mu^{2/3}$ where
$\eta\rightarrow +\infty$
(ground state). When $\mu\rightarrow +\infty$, the direct spiral tends to a
limit curve ($\eta_c(\mu)\rightarrow 2.52$, $\Lambda_c(\mu)\rightarrow 0.335$),
the inverse spiral coincides with
the direct
spiral, the turning point $\Lambda_*(\mu)$ is
pushed
towards $-\infty$, the turning point $\eta_*(\mu)$ is
pushed
towards $0$, the lower branch coincides with the $x$-axis ($\eta=0$)
and the asymptote at $\Lambda_{\rm max}(\mu)$ is pushed
towards $+\infty$. In this limit, we recover the standard
nonrelativistic
classical
caloric curve of Fig. \ref{etalambda} plus a singular branch at $\eta=0$
corresponding to a Dirac peak of zero mass but infinite potential
energy surrounded by a halo of infinite
temperature, and a singular branch at $\Lambda=\Lambda_{\rm max}\rightarrow
+\infty$ corresponding to a Dirac peak containing all the mass (ground state).

\subsection{When the classical limit $\hbar\rightarrow 0$ is taken before
the nonrelativistic limit $c\rightarrow +\infty$}
\label{sec_class}

\begin{figure}
\begin{center}
\includegraphics[clip,scale=0.3]{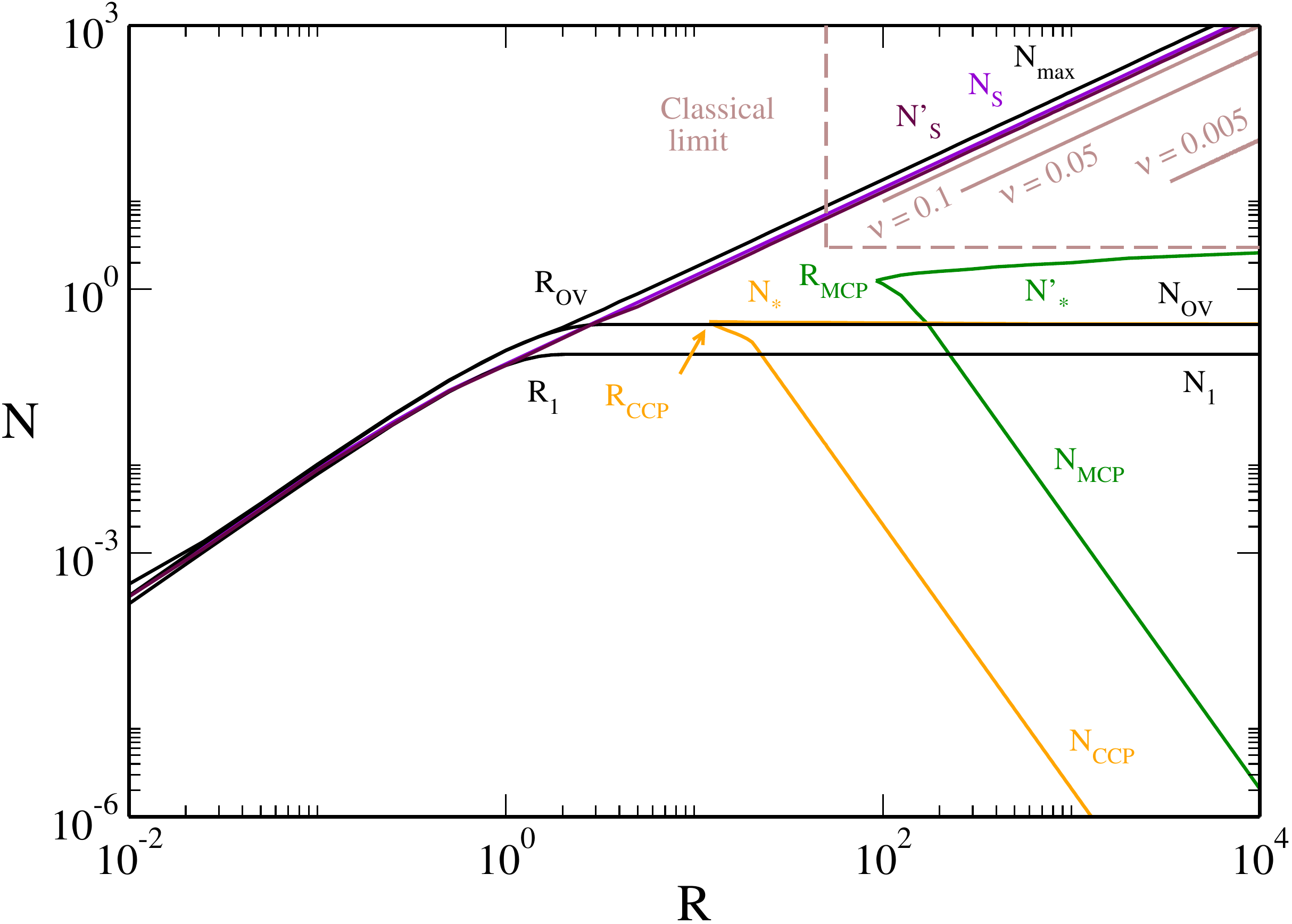}
\caption{Lines of constant $\nu$ in the $(R,N)$ phase diagram of Fig.
\ref{phase_phase_Newton_corrPH} characterizing the classical limit $N\rightarrow
+\infty$ and $R\rightarrow +\infty$ with fixed
$\nu=N/R$.  We have chosen $\nu=0.005$, 
$\nu=0.05$ and $\nu=0.1$ (brown lines) corresponding to the values appearing in
Fig. 12 of \cite{paper2}.}
\label{phase_phaseNUPH}
\end{center}
\end{figure}

In this section, we consider the situation where the classical
limit ($\hbar\rightarrow 0$) is taken before the nonrelativistic limit
($c\rightarrow +\infty$). Using the dimensionless variables of Appendix
\ref{sec_dq}, the classical limit
corresponds to $N\rightarrow
+\infty$ and $R\rightarrow +\infty$ in such a way that $N/R$ is
fixed.\footnote{This scaling
is obtained in order to keep the parameter $\nu$ defined in \cite{roupas,paper2}
fixed
(see also Sec. \ref{sec_suncl}). Coming back to dimensional variables, the
classical limit corresponds to $N\gg N_{\rm OV}\sim (\hbar c/G)^{3/2}/m^3$
and $R\gg R_{\rm OV}\sim (\hbar^3/Gc)^{1/2}/m^2$ with
$\nu\sim GNm/Rc^2$ fixed. This is consistent with the fact that
$N_{\rm OV}\rightarrow 0$ and $R_{\rm OV}\rightarrow 0$ when $\hbar\rightarrow
0$.} This scaling defines an ensemble of parallel
lines of constant
$\nu=N/R$ in the upper right panel of Fig. \ref{phase_phaseNUPH}. In the
classical limit,
the caloric curves
are the same for any couple of points $(R,N)$ belonging to a given $\nu$-line.
As $\nu$ decreases, the $\nu$-lines move to the right and the system becomes
less and less relativistic \cite{roupas,paper2}.

The classical  general relativistic
self-gravitating gas has been
studied in \cite{roupas,paper2}. When
$\nu<\nu'_S=0.128$ the caloric curve displays a  double spiral
(see Fig. 7 of \cite{paper2}). When
$\nu'_S=0.128<\nu<\nu_S=0.1415$
the two spirals are truncated (see Fig. 8 of \cite{paper2}).
When $\nu_S=0.1415<\nu<\nu_{\rm max}=0.1764$
the caloric curve makes
a loop (see Fig. 9 of \cite{paper2}). When  $\nu=\nu_{\rm
max}=0.1764$ the caloric curve reduces to a point and
disappears (see Fig. 15 of \cite{paper2}).
There is a gravitational collapse at low energies and low 
temperatures (cold spiral) and at high energies and high
temperatures (hot spiral).  The
nonrelativistic limit corresponds to
$\nu\rightarrow
0$ (see Fig. 12 of \cite{paper2}).

\begin{figure}
\begin{center}
\includegraphics[clip,scale=0.3]{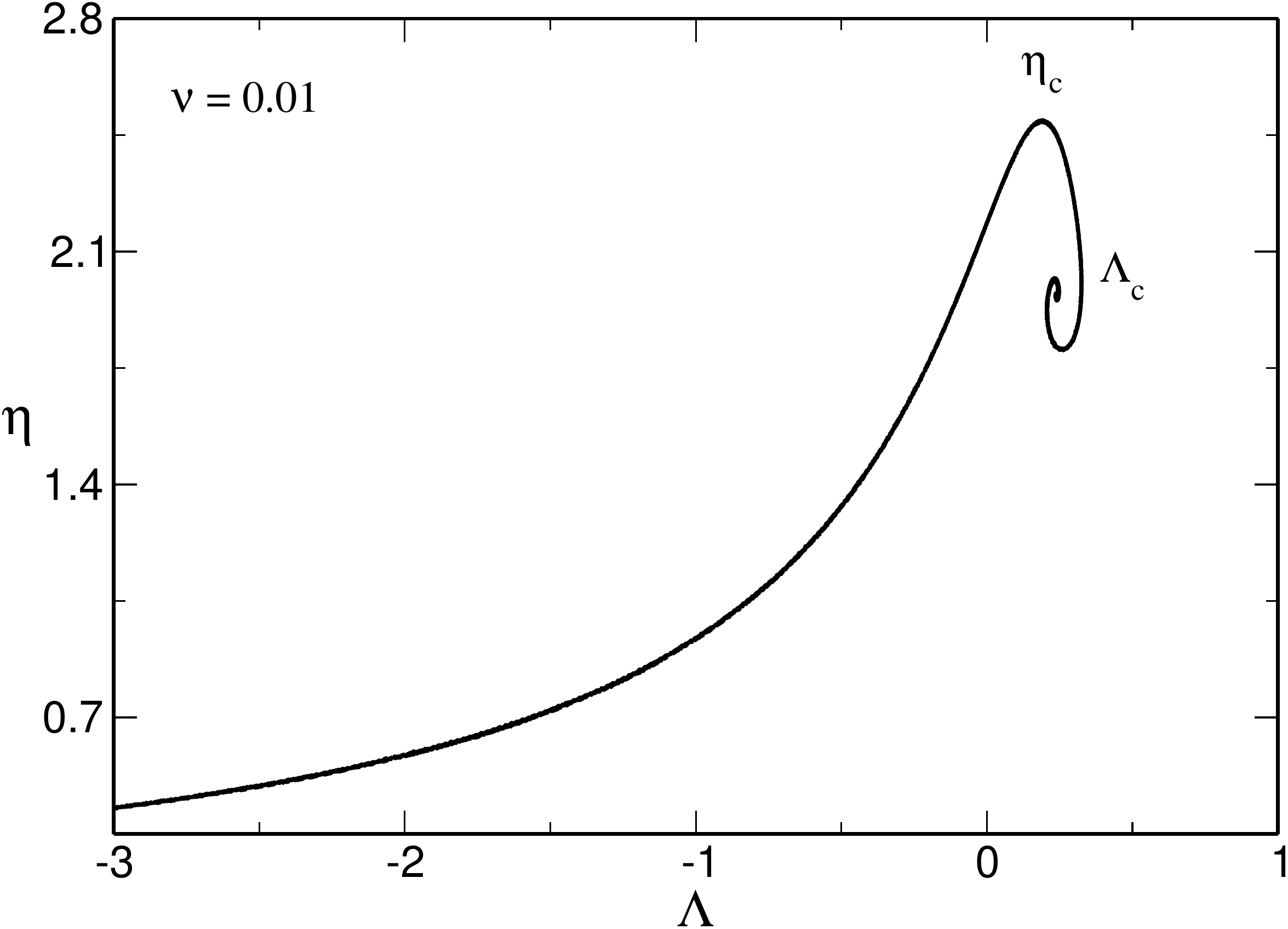}
\caption{Caloric curve of classical self-gravitating systems in
the nonrelativistic limit $\nu\rightarrow 0$ (here $\nu=0.01$). This figures
zooms on the cold spiral.}
\label{Boltzmann_kcal_N0p01dPHH}
\end{center}
\end{figure}

\begin{figure}
\begin{center}
\includegraphics[clip,scale=0.3]{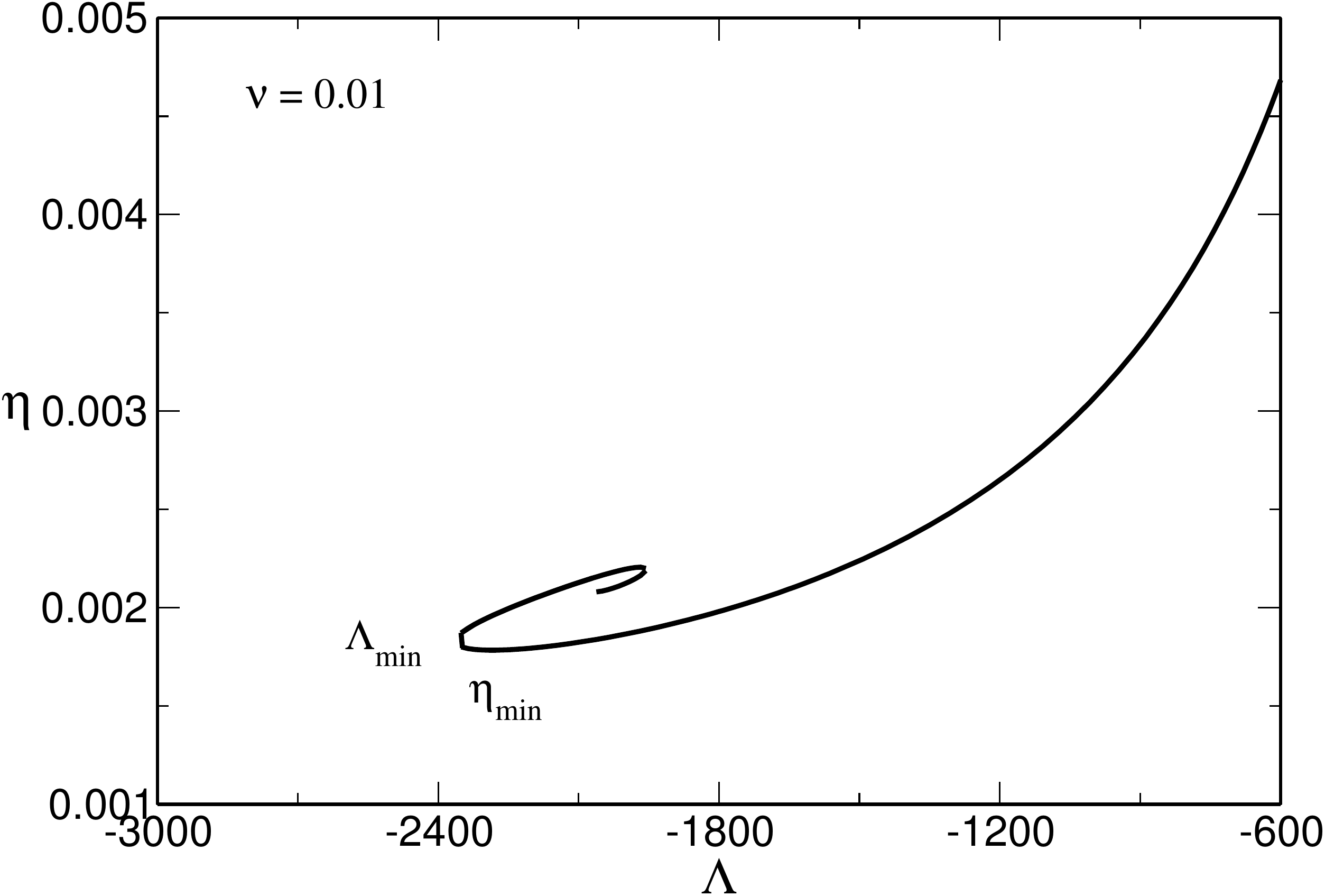}
\caption{Caloric curve of classical self-gravitating systems in
the nonrelativistic limit $\nu\rightarrow 0$ (here $\nu=0.01$). This figures
zooms on the hot spiral.}
\label{Boltzmann_kcal_N0p01sPHH}
\end{center}
\end{figure}

For small but nonzero values of $\nu$ (see Figs.
\ref{Boltzmann_kcal_N0p01dPHH} and \ref{Boltzmann_kcal_N0p01sPHH}), the series
of equilibria forms two
spirals very
distant to each other. When $\nu\rightarrow 0$, the cold spiral tends to a limit
curve  ($\eta_c(\nu)\rightarrow 2.52$, $\Lambda_c(\nu)\rightarrow 0.335$) while
the hot spiral is
rejected to the left at infinity ($\Lambda_{\rm min}(\nu)\sim
-0.24631/\nu^2\rightarrow -\infty$ and $\eta_{\rm min}(\nu)\sim
18.27\,\nu^2\rightarrow 0$) \cite{paper2}. In this
limit, we recover the standard nonrelativistic classical caloric curve of Fig.
\ref{etalambda} plus a spiral at very high energies and very high temperatures
($\Lambda_{\rm min}\rightarrow -\infty$ and $\eta_{\rm min}\rightarrow 0$).

\subsection{The solution of the apparent paradox}

In the two processes described previously, in which $c\rightarrow +\infty$ and
$\hbar\rightarrow 0$, we recover the standard classical nonrelativistic spiral
of Fig. \ref{etalambda}. However,
the manner to obtain it is different depending on the order in which the
limits are taken.

When the nonrelativistic limit  $c\rightarrow +\infty$ is
taken
first, there is no ``hot'' spiral at $\Lambda_{\rm min}$ since the hot spiral is
a general relativity result associated with a form of self-gravitating
radiation. By contrast, there is
always an asymptote at $\Lambda_{\rm max}$ corresponding to  an equilibrium
state at $T=0$ (the ground state of the Fermi gas) because $N<N_{\rm
OV}=+\infty$. In the classical
limit $\hbar\rightarrow 0$, the asymptote at $\Lambda_{\rm max}$ is rejected at
$+\infty$ (while $\Lambda_{*}$ is pushed towards $-\infty$ and
$\eta_*$ towards zero) so the caloric curve has the form of a single spiral.
For
$\hbar$ infinitely small  but finite, we get the classical caloric curve plus
singular branches $\eta\simeq 0$ (horizontal) and $\Lambda_{\rm max}\simeq
+\infty$ (vertical) as described previously. According to the
results of Appendix
\ref{sec_thermo}, we
have the scalings $E_{\min}\propto
-\hbar^{-2}$, $E_{*}\propto
\hbar^{-2}(-\ln \hbar)^{-7/3}$ and  $T_*\propto \hbar^{-2}(-\ln \hbar)^{-1}$
for $\hbar\rightarrow 0$.

When
the classical limit $\hbar\rightarrow 0$ is taken first, there is always a
``hot'' spiral at $\Lambda_{\rm min}$ since the system is relativistic. By
contrast, there is no asymptote at $\Lambda_{\rm max}$, i.e., there is no
equilibrium state at $T=0$ (ground state) because $N>N_{\rm OV}=0$. In the
nonrelativistic limit $c\rightarrow +\infty$, the  ``hot'' spiral is rejected at
infinity so the
caloric curve has the form of a single spiral. For $c$ infinitely
large but finite, we get the nonrelativistic caloric curve plus a spiral at
very high energies $\Lambda_{\rm min}$ and temperatures $\eta_{\rm min}$ as
described
previously. According to the results of \cite{paper2} (see also
footnote 19), we have the scalings $E_{\max}\sim
c^{4}$ and $T_{\rm max}\sim c^{4}$  for $c\rightarrow +\infty$.

The previous considerations lead to the following conclusion.

\begin{figure}
\begin{center}
\includegraphics[clip,scale=0.3]{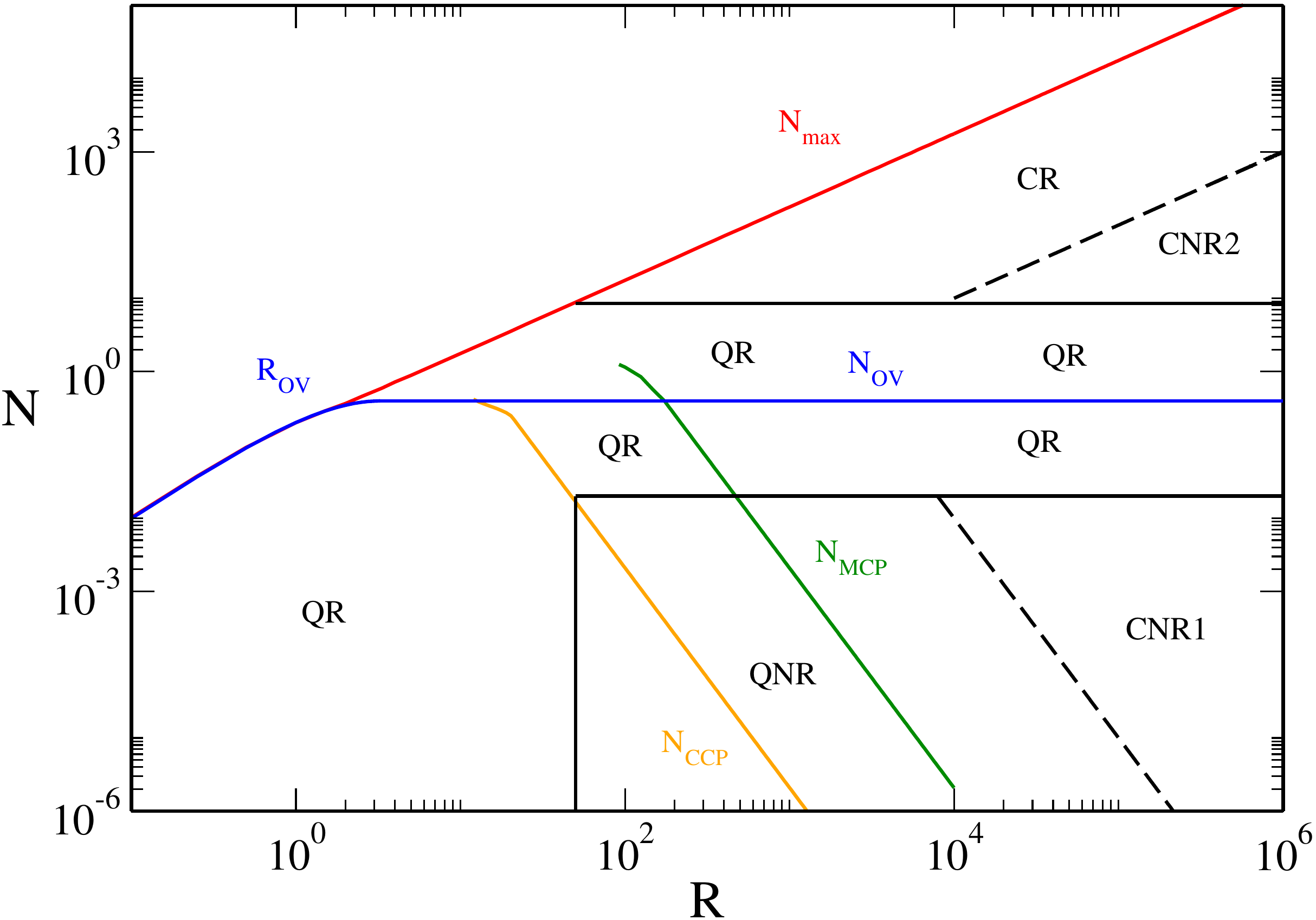}
\caption{Schematic representation of the different regimes of
the self-gravitating Fermi gas. QR: quantum relativistic; QNR: quantum
nonrelativistic; CR: classical
relativistic; CNR: classical nonrelativistic (CNR1 is when
the nonrelativistic limit $c\rightarrow \infty$ limit is taken before the
classical limit $\hbar\rightarrow 0$; CNR2 is when the classical limit
$\hbar\rightarrow 0$ is taken before the nonrelativistic limit $c\rightarrow
\infty$ limit).}
\label{regime}
\end{center}
\end{figure}

The nonrelativistic limit \cite{ijmpb} corresponds to $R\gg R_{\rm OV}$ and
$N\ll N_{\rm
OV}$.  This corresponds to the lower panel QNR of  Fig.
\ref{regime} below $N_{\rm OV}$. In that case, for a fixed radius $R$, the
caloric curve
shows no phase
transition below $N_{\rm
CCP}$, a canonical phase transition above $N_{\rm CCP}$ and a   microcanonical
phase transition (in addition to the canonical phase transition) above $N_{\rm
MCP}$. The classical limit corresponds to $N_{\rm
MCP}\ll N\ll N_{\rm
OV}$. This corresponds to the lower
right panel CNR1 of  Fig. \ref{regime} (far on the right)  below
$N_{\rm OV}$.

The classical limit \cite{roupas,paper2} corresponds to $R\gg R_{\rm OV}$ and
$N\gg N_{\rm
OV}$. This corresponds to the upper panel CR of  Fig.
\ref{regime} above $N_{\rm OV}$. In that case,  for a fixed radius $R$, the
caloric curve shows a double
spiral below $N'_S$,  a
truncated double spiral above $N'_S$,  a loop above $N_S$ and no equilibrium 
states above $N_{\rm max}$. The nonrelativistic limit corresponds to $N_{\rm
OV}\ll N\ll N'_{S}$. This corresponds to the upper right panel
CNR2 of   Fig.
\ref{regime} (far on the right) above $N_{\rm OV}$.

Therefore, the nonrelativistic $+$ classical limit corresponds
to two distinct
regions in the right panel of   Fig.
\ref{regime}, below or
above $N_{\rm OV}$, depending on the order in which the limits are taken.
Note also that quantum and relativistic effects are
{\it both} important only close to $R_{\rm OV}$ or only close to $N_{\rm
OV}$. This corresponds to the region denoted QR in  Fig. \ref{regime}.

\section{Relativistic and quantum corrections}
\label{sec_rqc}

\subsection{Relativistic corrections to the nonrelativistic caloric curves}

We have seen in Sec. \ref{sec_nonrel} that the nonrelativistic caloric curves of
the self-gravitating
Fermi gas correspond to parallel lines of
constant $\mu=(4\sqrt{2}/\pi)(NR^3)^{1/2}$ in the bottom right panel of
Fig. \ref{phase_phaseMUPH}.  On a  line of constant $\mu$, the 
nonrelativistic limit is valid when $R\rightarrow +\infty$ and $N\rightarrow 0$
(physically $R\gg R_{\rm OV}$ and $N\ll N_{\rm OV}$).  For small
values of $R$ and large values of $N$, i.e., at the top of a
$\mu$-line,
relativistic corrections come into play.

\subsubsection{$\mu=10^3$}

\begin{figure}
\begin{center}
\includegraphics[clip,scale=0.3]{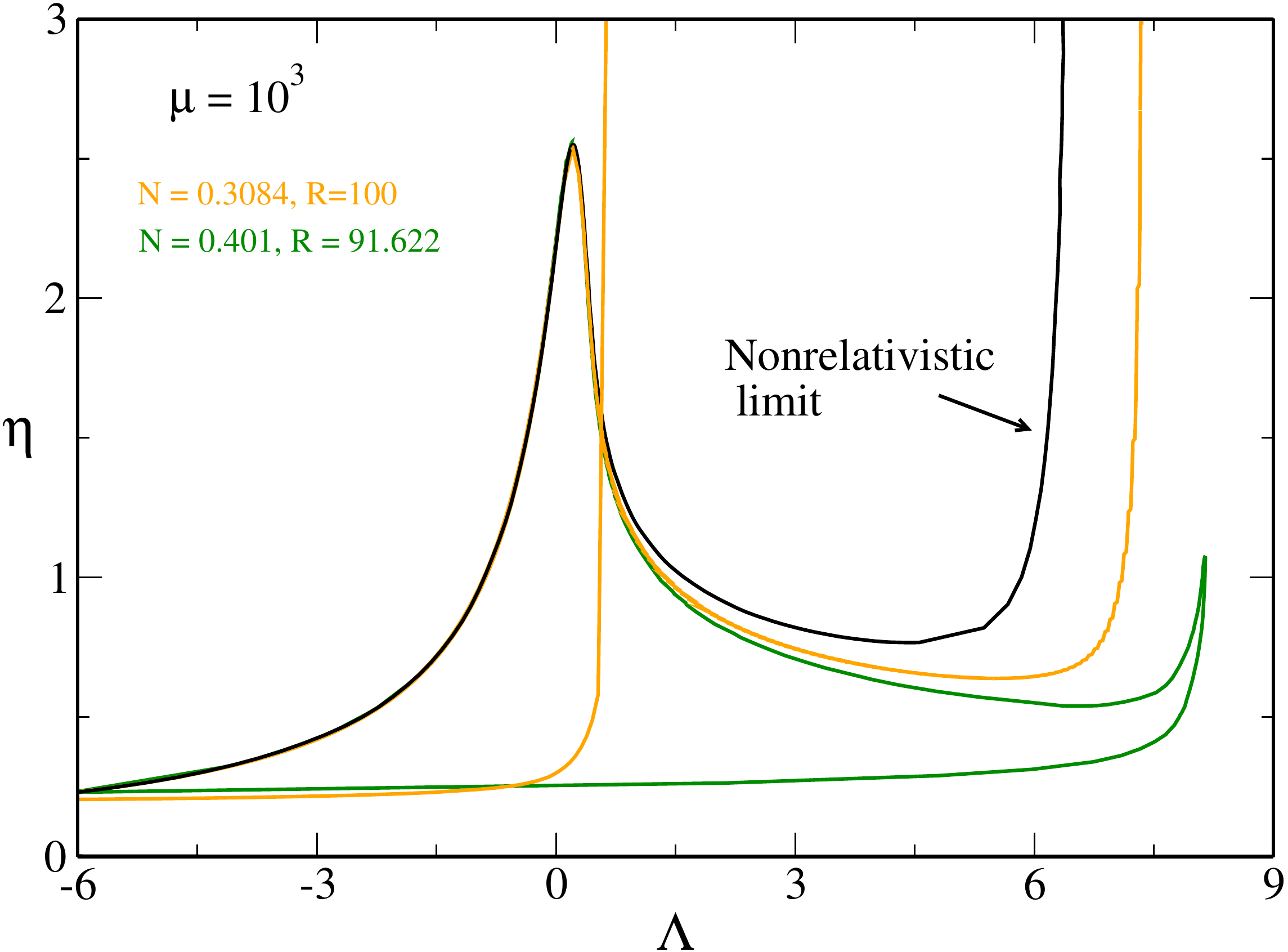}
\caption{Caloric curves for different values of $N$ at
fixed $\mu=(4\sqrt{2}/\pi)(NR^3)^{1/2}= 10^3$. When $N\rightarrow 0$ (black
curve), we recover the nonrelativistic caloric curve with an $N$-shape
structure obtained in Fig. 31 of \cite{ijmpb}. The present figure
illustrates the effect of general relativity on that caloric curve as $N$
increases.}
\label{kcal_mu1000_linked_simplePH}
\end{center}
\end{figure}

\begin{figure}
\begin{center}
\includegraphics[clip,scale=0.3]{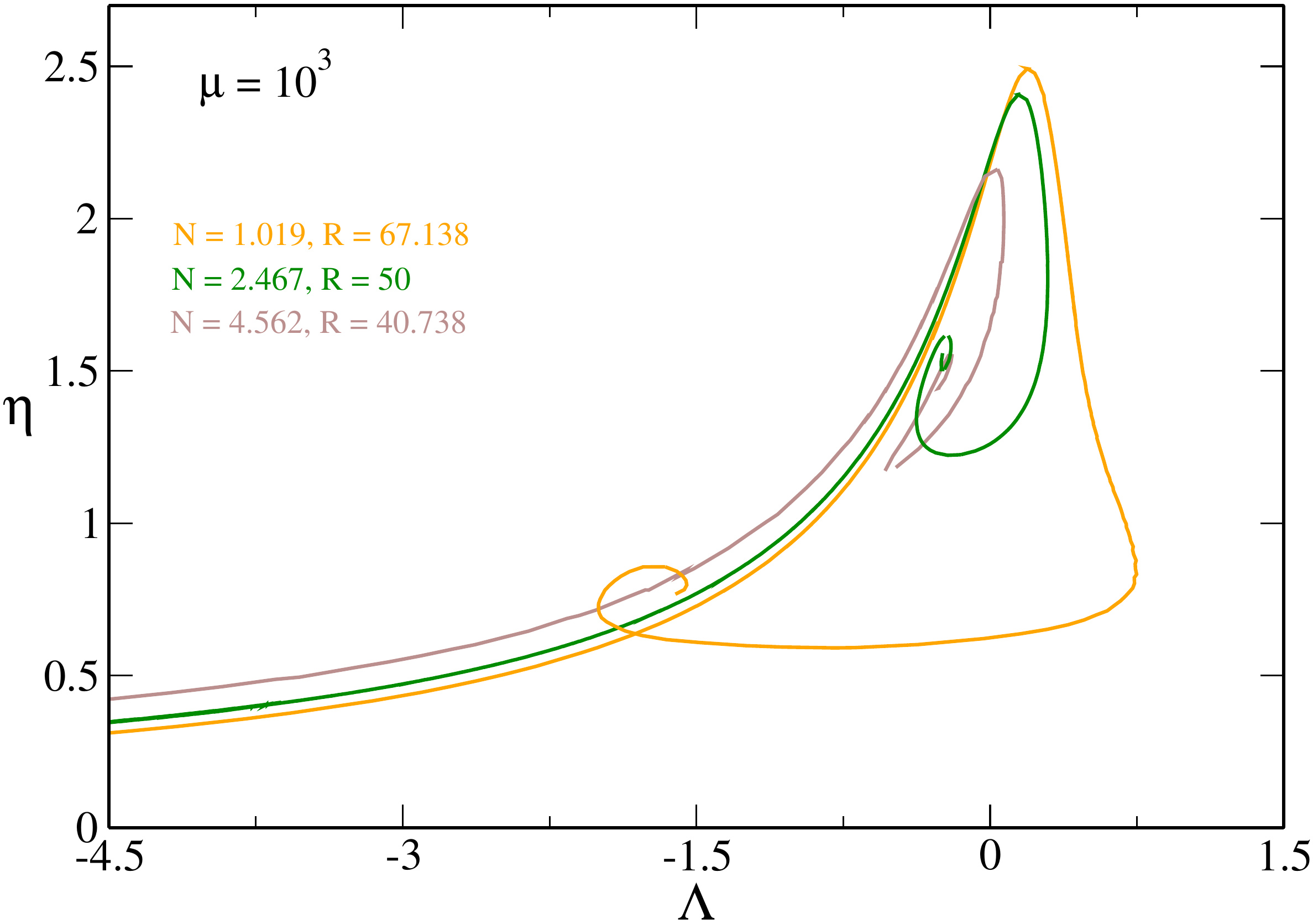}
\caption{Same as Fig. \ref{kcal_mu1000_linked_simplePH} for larger values of
$N$.}
\label{kcal_mu1000_linked_zoomPH}
\end{center}
\end{figure}

Let us first consider the case $\mu=10^3$ corresponding to $\mu_{\rm
CCP}=83 <\mu<\mu_{\rm
MCP}=2670$ (see Figs.
\ref{kcal_mu1000_linked_simplePH} and \ref{kcal_mu1000_linked_zoomPH}). When
$N\rightarrow 0$, we recover the
nonrelativistic caloric curve plotted in Fig. 31 of \cite{ijmpb}. It has a
$N$-shape structure leading to canonical phase transitions. The hot spiral is
rejected at infinity. Let us increase the number of particles
$N$
 at
fixed $\mu$, hence decreasing the box radius $R$ accordingly, in order to see
the relativisitic corrections. The description in the change of the caloric
curves as relativistic effects become more and more important is
qualitatively similar to that given in Sec. \ref{sec_first} for $R=50$ when
$N>N_{\rm CCP}$.
The only
difference is that we work at fixed $\mu$ (with $\mu_{\rm CCP}<\mu<\mu_{\rm
MCP}$) instead of fixed $R$ (with $R_{\rm CCP}<R<R_{\rm
MCP}$). Therefore, in the $(R,N)$ diagram, we follow the
$\mu=10^3$ oblique line  (see Fig. \ref{phase_phaseMUPH}) instead of the $R=50$
vertical line (see Fig. \ref{phase_phase_Newton_corrPH}). As a result, when
$N\rightarrow 0$, we tend towards a
limit curve (the nonrelativistic caloric
curve with $\mu=10^3$ of \cite{ijmpb}) which presents a canonical phase
transition while in the
case $R=50$ studied in Sec. \ref{sec_first}   the canonical phase transition
disappears when $N<N_{\rm
CCP}$.

{\it Remark:} The characteristic particle numbers $N_{\rm X}$ described in
Sec. \ref{sec_chara} now depend on $\mu$ instead of $R$. They can be obtained
by considering
the intersection between the curves $N_{\rm X}(R)$ and the curve
$N=\pi^2\mu^2/(32R^3)$ with fixed $\mu$. In this manner, we obtain $N_{\rm
S}'(\mu)=0.159\sqrt{\mu}$, $N_{\rm
S}(\mu)=0.172\sqrt{\mu}$, and  $N_{\rm max}(\mu)=0.203\sqrt{\mu}$.

\subsubsection{$\mu=10^5$}

\begin{figure}
\begin{center}
\includegraphics[clip,scale=0.3]{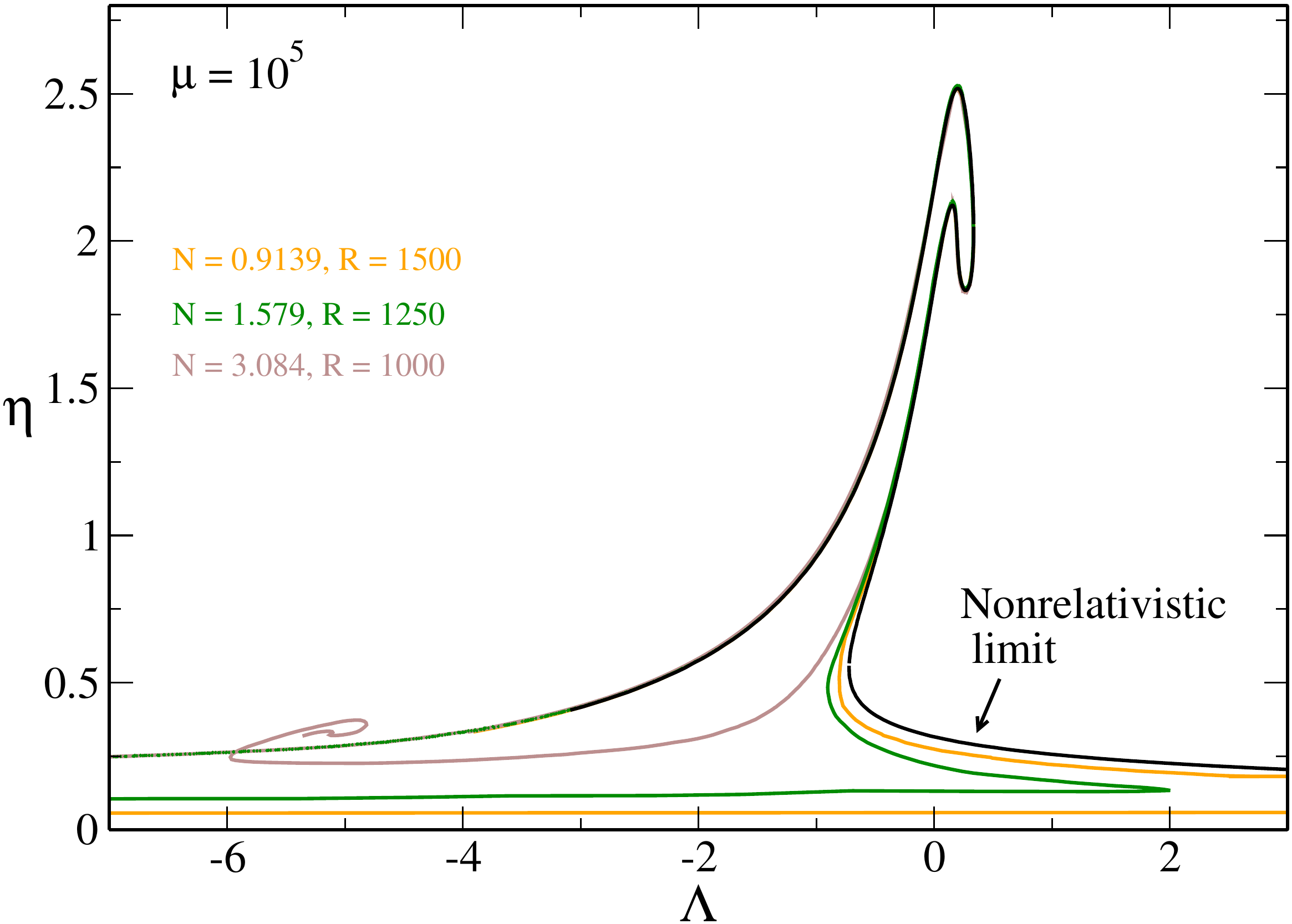}
\caption{Caloric curves for different values of $N$ at
fixed $\mu=(4\sqrt{2}/\pi)(NR^3)^{1/2}= 10^5$. When $N\rightarrow 0$ (black
curve), we recover the nonrelativistic caloric curve with a $Z$-shape
structure obtained in Fig. 21 of \cite{ijmpb}. The present figure
illustrates the effect of general relativity on that caloric curve as $N$
increases.}
\label{kcal_mu100000_quartPH}
\end{center}
\end{figure}

\begin{figure}
\begin{center}
\includegraphics[clip,scale=0.3]{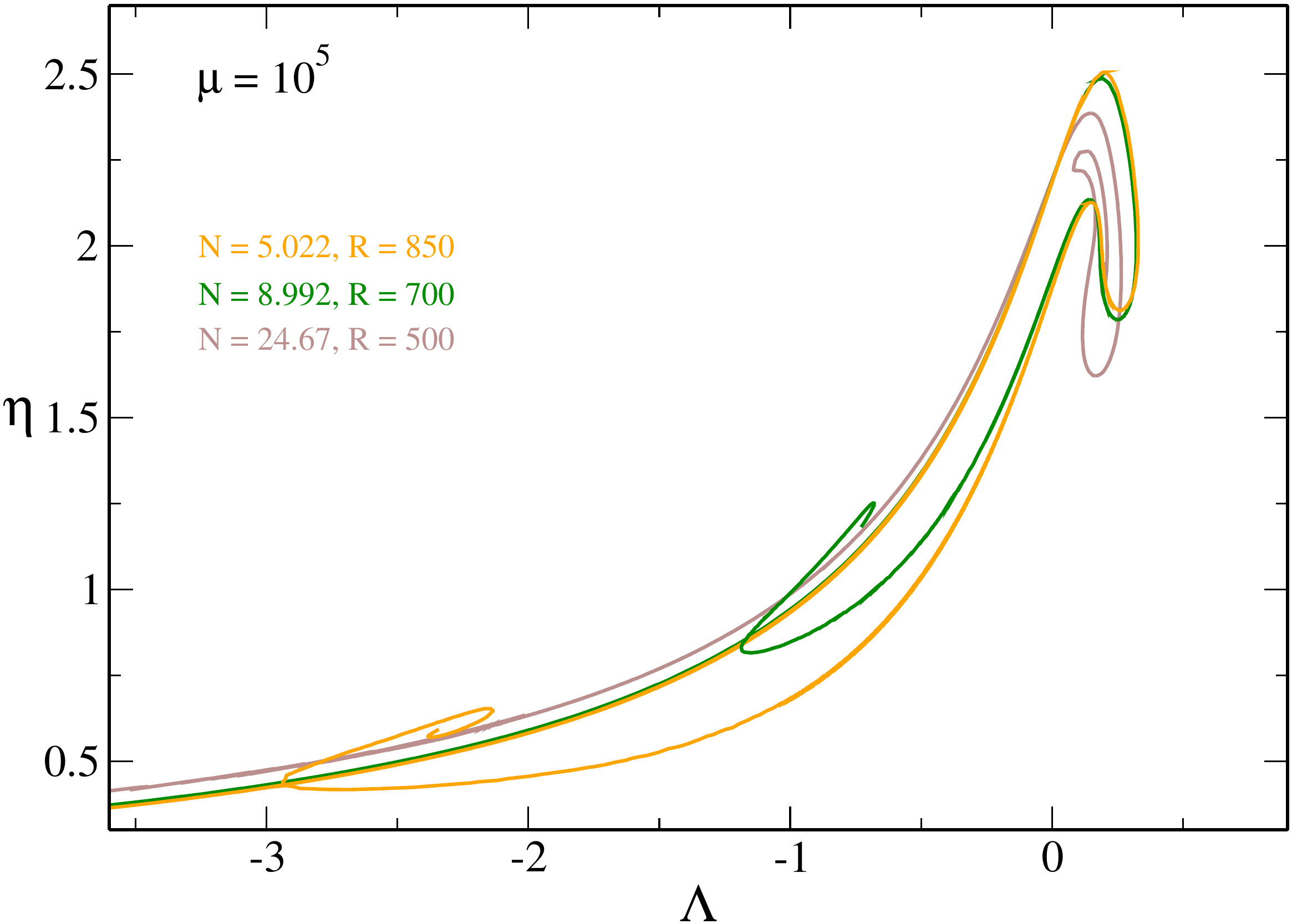}
\caption{Same as Fig. \ref{kcal_mu100000_quartPH} for larger
values of $N$.}
\label{kcal_mu100000_bisPH}
\end{center}
\end{figure}

\begin{figure}
\begin{center}
\includegraphics[clip,scale=0.3]{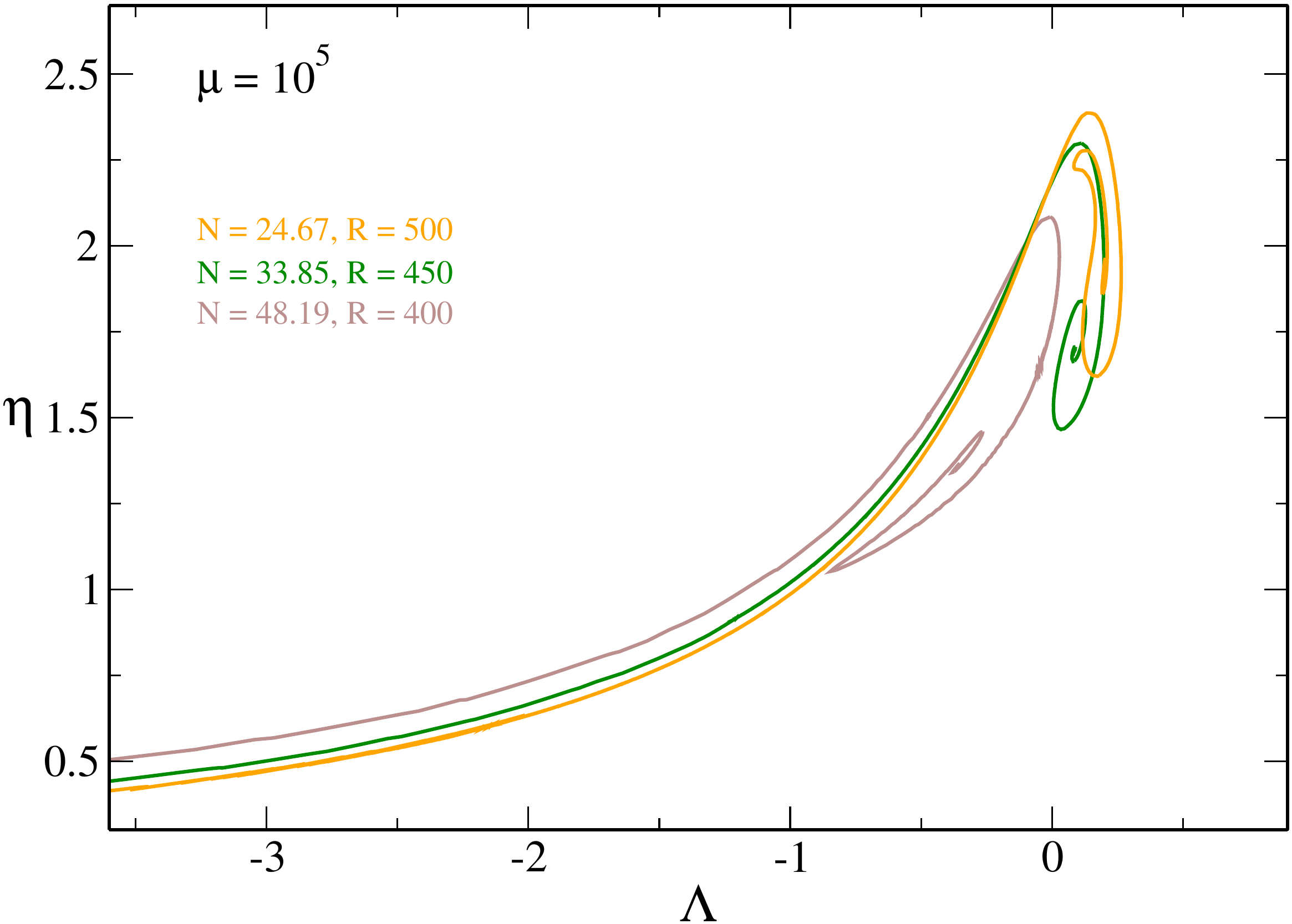}
\caption{Same as Figs. \ref{kcal_mu100000_quartPH} and \ref{kcal_mu100000_bisPH}
for larger
values of $N$.}
\label{kcal_mu100000_terPH}
\end{center}
\end{figure}

Let us now consider the case $\mu=10^5$ corresponding to $\mu>\mu_{\rm
MCP}=2670$  (see Figs.
\ref{kcal_mu100000_quartPH}-\ref{kcal_mu100000_terPH}).
When $N\rightarrow 0$, we recover the
nonrelativistic caloric curve plotted in Fig. 21 of \cite{ijmpb}. It has a
$Z$-shape structure leading to microcanonical phase
transitions.\footnote{The resemblance with a dinosaur's neck in clear on this
figure \cite{ijmpb}.} The
vertical asymptote at $\Lambda_{\rm max}$ is outside of the frame. The
hot spiral is rejected at infinity. Let us
increase the number of particles $N$ at
fixed $\mu$, hence decreasing the box radius $R$ accordingly, in order to see
the relativisitic corrections. The description in the change of the caloric
curves as relativistic effects become more and more important is
qualitatively similar to that given in Sec. \ref{sec_second} for $R=600$ when
$N>N_{\rm MCP}$. The only
difference is that we work at fixed $\mu$ (with $\mu>\mu_{\rm
MCP}$) instead of fixed $R$ (with $R>R_{\rm
MCP}$). Therefore, in the $(R,N)$ diagram, we follow the
$\mu=10^5$ oblique line  (see Fig. \ref{phase_phaseMUPH}) instead of the $R=600$
vertical line (see Fig. \ref{phase_phase_Newton_corrPH}). As a result, when
$N\rightarrow 0$, we tend towards a
limit curve (the nonrelativistic caloric
curve with $\mu=10^5$ of  \cite{ijmpb}) which presents a microcanonical phase
transition while in
the
case $R=600$ studied in Sec. \ref{sec_second}   the microcanonical phase
transition
disappears when $N<N_{\rm
MCP}$.

\subsection{Quantum corrections to the classical caloric curve}

We have seen in Sec. \ref{sec_class} that the classical caloric curves of
the general
relativistic self-gravitating gas correspond to parallel lines of
constant $\nu=N/R$ in the upper right panel of Fig.
\ref{phase_phaseNUPH}.  On a  line of constant $\nu$,
the 
classical limit is valid
when $R\rightarrow +\infty$ and $N\rightarrow +\infty$ (physically $R\gg
R_{\rm
OV}$  and $N\gg N_{\rm OV}$).  For small
values of $R$ and small values of $N$, i.e., on the left of a $\nu$-line,
quantum corrections come into play.

\begin{figure}
\begin{center}
\includegraphics[clip,scale=0.3]{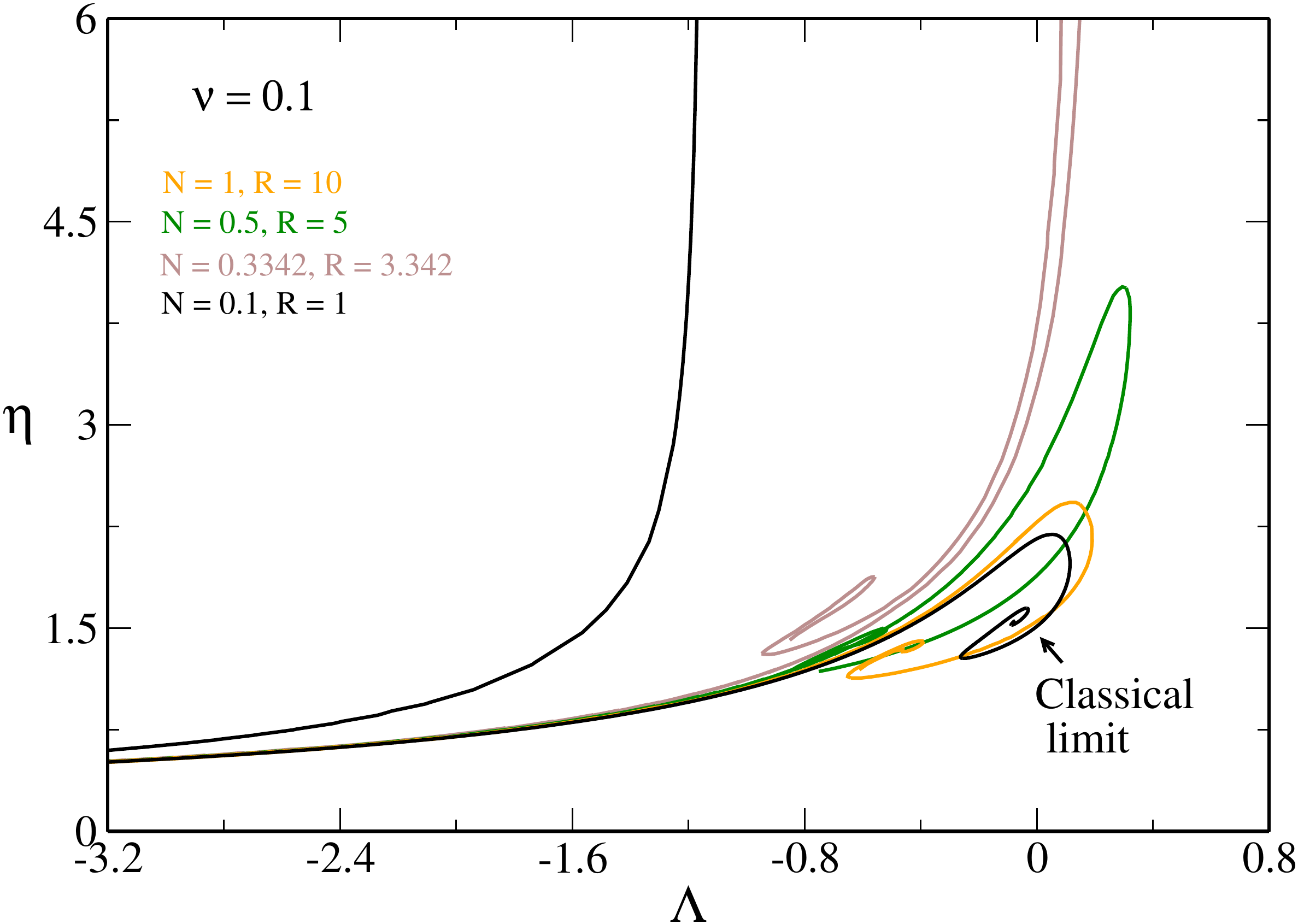}
\caption{Caloric curves for different values of $N$ at
fixed $\nu=N/R=0.1$. When $N\rightarrow +\infty$ (black curve), we recover the
classical caloric curve with a double spiral obtained in Fig. 1
of \cite{paper2}.
The present figure
illustrates the effect of quantum mechanics on that caloric curve as $N$
decreases. This figure is focused on the evolution of the cold spiral.}
\label{kcal_nu01bPH}
\end{center}
\end{figure}

\begin{figure}
\begin{center}
\includegraphics[clip,scale=0.3]{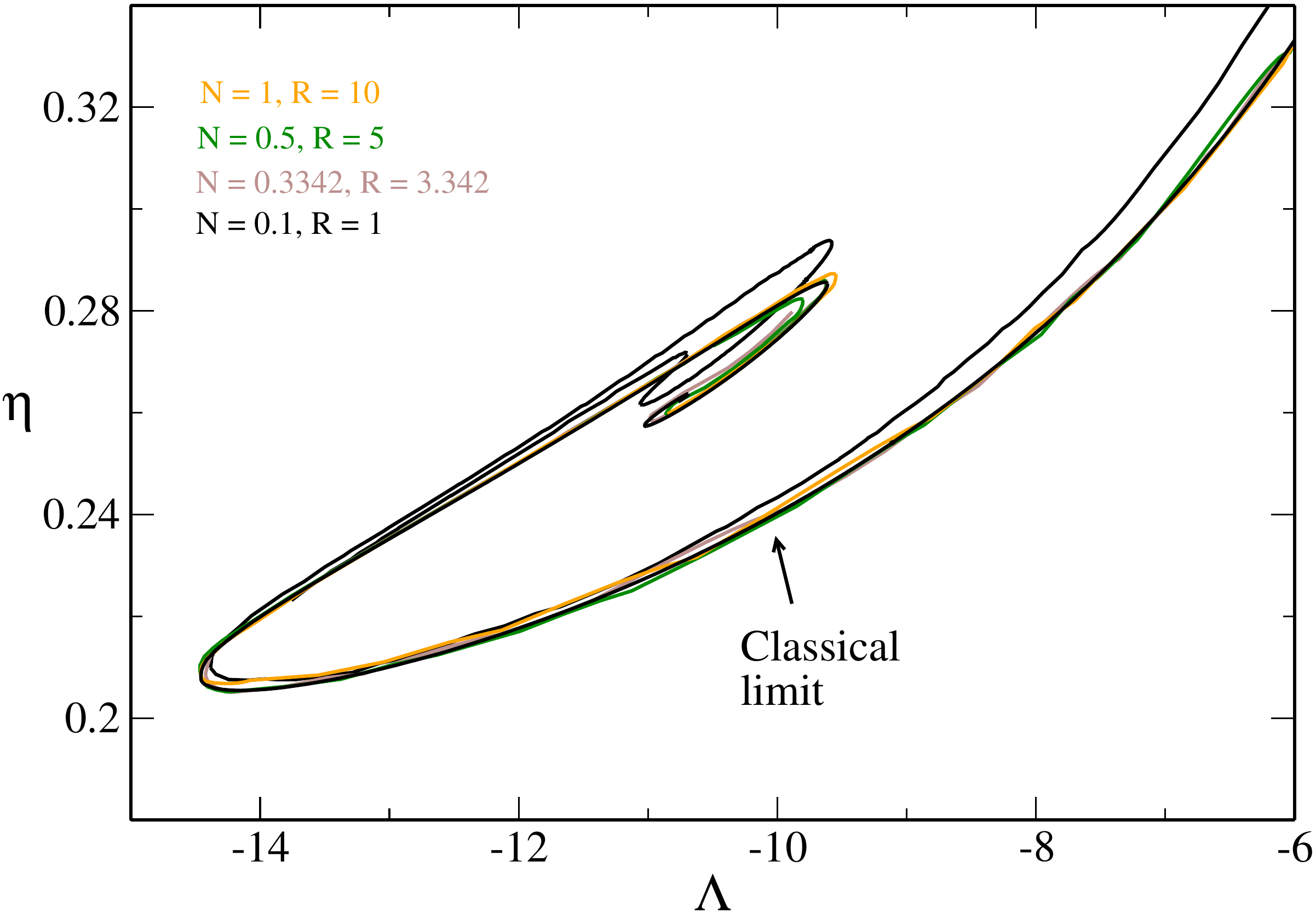}
\caption{Same as Fig. \ref{kcal_nu01bPH}. This figure is focused on the
evolution
of the hot
spiral.}
\label{kcal_nu01aPH}
\end{center}
\end{figure}

Let us consider the case $\nu=0.1$ (see Figs. \ref{kcal_nu01bPH} and
\ref{kcal_nu01aPH}). When $N\rightarrow +\infty$ (black curve) we recover the
classical general
relativistic caloric
curve plotted in Fig. 1 of \cite{paper2}. It has the form of a
double spiral leading to a gravitational collapse for both cold  and hot
systems.\footnote{When
$N\rightarrow +\infty$ the caloric curves of Fig.
\ref{kcal_nu01bPH} and \ref{kcal_nu01aPH} tend
towards a limit curve (the classical general relativistic caloric curve with
$\nu=0.1$ of \cite{paper2}) which presents a double spiral while in the
cases studied in Secs. \ref{sec_first}-\ref{sec_un} the two spirals merge
and
disappear when
$N\rightarrow N_{\rm max}$. This is because in Secs.
\ref{sec_first}-\ref{sec_un} we work at fixed radius
$R$ while in the present case we work at fixed $\nu$ so that the radius $R$
increases as $N$ increases.} Let us decrease the number of particles $N$
 at
fixed $\nu$, hence decreasing the box radius $R$ accordingly, in order to see
the quantum corrections. When $N>N_{\rm OV}=0.39853$ (blue and green curves),
the caloric
curve keeps a similar structure.  When $N_1=0.18131<N<N_{\rm OV}=0.39853$ (red
curve) the caloric curve is made of two branches, each presenting an asymptote
(right)
and a spiral (left). When $N<N_1=0.18131$ (purple curve) the caloric
curve has just one branch presenting an asymptote (right) and a spiral (left).

{\it Remark:} For smaller values of $\nu$,\footnote{This case is
specifically investigated in \cite{rc}.} we have a
richer variety of caloric curves  as $N$ decreases with the appearance of
canonical and microcanonical phase transitions. This can be seen on the phase
diagram of Fig. \ref{phase_phaseNUPH}. The characteristic particle numbers
$N_{\rm X}$ described in
Sec. \ref{sec_chara} now depend on $\nu$ instead of $R$. They can be obtained
by considering
the intersection between the curves $N_{\rm X}(R)$ and the line
$N=\nu R$. In this manner, we obtain $N_{\rm CCP}(\nu)=6.79\, \nu^{3/4}$ and
$N_{\rm
MCP}(\nu)=38.5\, \nu^{3/4}$. There is no microcanonical
phase transition for $\nu>\nu_{\rm MCP}=0.0136$ and there is no canonical
phase transition for $\nu>\nu_{\rm CCP}=0.0353$.

\section{Astrophysical applications}
\label{sec_ap}

\subsection{Nonrelativistic model: collapse towards a fermion
star}
\label{sec_nrm}

In this section, we discuss astrophysical applications of the
nonrelativistic self-gravitating Fermi gas
model \cite{ijmpb}. This model exhibits a phase
transition from a gaseous star to
a compact fermion star. The fermion star can be a white dwarf, a neutron
star or a DM fermion ball.\footnote{In this section we take
$g=2$ in the numerical applications.} We relate this phase
transition to the
onset of red-giant structure and to the supernova phenomenon.

\subsubsection{Canonical ensemble}
\label{sec_nrmce}

Let us consider a system of nonrelativistic self-gravitating fermions in the
canonical ensemble. The canonical phase transition  appears for
$\mu\ge \mu_{\rm CCP}=83$ (canonical critical point) \cite{ijmpb} hence for
\begin{equation}
\label{nrmce1}
R\ge R^{\rm
NR}_{\rm CCP}(N)=0.517 \frac{h^2}{Gm^{8/3}M^{1/3}g^{2/3}}.
\end{equation}
We assume that we are in this situation. In that case, the caloric curve
has the form of Fig. \ref{Xkcal_R50_N0p15_unified2blackPH}.
The natural evolution proceeds along
the series of equilibria towards lower and lower temperatures (see Sec.
\ref{sec_cc}). We
assume that
the system is initially in the metastable gaseous phase. As its
temperature decreases it remains in this phase up to the critical point $T_c$ at
which the metastable gaseous branch disappears.\footnote{We recall that the
collapse takes place at the critical (spinodal) point $T_c$, not
at the transition point $T_t$, because of the tremendously long lifetime of
metastable gaseous states.} This critical temperature is not very sensitive
on
quantum effects (when $\mu\gg 1$) so it can be approximated by its
classical value
$\eta_c=\beta_c GMm/R=2.52$ yielding
\begin{equation}
\label{nrmce2}
k_B T_c=0.397\, \frac{GMm}{R}.
\end{equation} 
At that point, the system collapses and forms  a compact fermion star.
As explained in Appendix \ref{sec_thermoqce}, in the
canonical ensemble, the
fermion star contains almost all the mass ($M_C\simeq M$) and is surrounded by a
tenuous atmosphere. If we
approximate the fermion star by a Fermi gas at $T=0$ (polytrope of
index $n=3/2$) containing all the
mass, its radius is given by (see Appendix \ref{sec_thermoqce}):
\begin{equation}
\label{nrmce3}
R_C= 0.181 \frac{h^2}{Gm^{8/3}M^{1/3}g^{2/3}}.
\end{equation}
The energy
of the gaseous phase at the point of isothermal collapse is not very
sensitive on quantum effects (when $\mu\gg 1$) so it can be approximated by its
classical value
$\Lambda_{\rm gas}=-E_{\rm gas}R/GM^2=0.199$ 
yielding
\begin{equation}
\label{nrmce4}
E_{\rm gas}=-0.199 \frac{GM^2}{R}.
\end{equation}
The energy of the condensed object can be approximated by its value at
$T=0$ yielding (see Appendix \ref{sec_thermoqce}):
\begin{equation}
\label{nrmce5}
E_{\rm min}=-2.36\frac{G^2m^{8/3}M^{7/3}g^{2/3}}{h^2}.
\end{equation}
Finally, the collapse time in the canonical ensemble is of the order of the
dynamical time
\begin{equation}
\label{nrmce6}
t_{D}\sim \frac{R^{3/2}}{(GM)^{1/2}}.
\end{equation}

We note that the collapse in the canonical ensemble corresponds to a pure
implosion since almost all the mass is in the condensate (fermion star), not in
the halo. The
thermodynamical reason for this implosion phenomenon is explained in Appendix
\ref{sec_thermoclass}. It is also corroborated by the following arguments. If we
calculate the density perturbation $\delta\rho$ that triggers the instability at
$T_c$,
we find that it has only one node (see Fig. 10 of \cite{aa}). Therefore, the
instability develops itself in such a way that the density in the core increases
while the
density in the halo decreases. We also find that the velocity profile
$\delta v$ has no node (see Fig. 12 of \cite{aa}) so that it is purely inward.
These two results confirm the implosion phenomenon.

We make below numerical applications to illustrate the preceding results.
They correspond to the nonrelativistic models of neutron stars (made of
neutrons of mass $m_n=0.940\, {\rm GeV/c^2}$) and dark matter
halos (made of fermions of mass $m=17.2\, {\rm keV/c^2}$) respectively studied
by Hertel
and Thirring \cite{ht} and Bilic and Viollier \cite{bvn} as recalled in the
Introduction.

{\it Neutron stars (crude model) \cite{ht}:} We consider a gas of
neutrons of total mass $M=1\, M_{\odot}$.\footnote{The maximum mass and minimum
radius of an ideal neutron star set by general relativity are $M_{\rm
OV}=0.710\,
M_{\odot}$ and $R_{\rm OV}=9.16\, {\rm km}$. Therefore, the value of the mass
chosen by \cite{ht} is larger than the maximum mass. If general relativity were
taken into account (see below) the system would not form a neutron star but
would
collapse towards a black hole (assuming that all the initial mass goes in the
compact object).}
It
contains about $N\sim 10^{57}$ neutrons.
The canonical phase transition appears for $R\ge R^{\rm
NR}_{\rm CCP}(N)=43.1\,
{\rm km}$. For a gaseous star of size $R=100\, {\rm km}$, corresponding to
$\mu=294$,
the collapse
temperature is $T_c=6.39\times 10^{10}\,
{\rm K}$ (the transition temperature is $T_t=7.03\times 10^{10}\,
{\rm K}$ \cite{ht}). The
radius of the neutron star of mass $M_C \simeq 1\, M_{\odot}$ resulting from
the
collapse of the gaseous star is $R_C=15.1\, {\rm
km}$. The energy of the
gaseous star  is $E_{\rm
gas}=-5.25\times 10^{51}\, {\rm erg}$ and the energy of the neutron star
is $E_{\rm min}=-7.49\times 10^{52}\,
{\rm erg}$.  The energy released during the collapse is $\Delta
E=E_{\rm gas}-E_{\rm min}=6.96\times 10^{52}\,
{\rm erg}$. The collapse time is a multiple of $t_D\sim 2.74\times 10^{-3}\,
{\rm s}$ which is very short on an
astrophysical timescale.\footnote{Note
that the prefactor of the collapse time is uncertain and could be of order
$10^3$ or larger. As a result, our estimate of the collapse time is not
inconsistent with
the duration of the supernova
phenomenon which can be as short as a few seconds. For
supernovae, the energy $\Delta E$ may be carried quickly by
neutrinos. The release of gravitational energy in a supernova ($W\sim
GM^2/R\sim Nmc^2$) is comparable with the
energy of fusion processses which kept the star shining during the first
$10^{10}$ years of its life. However, this takes place in a few seconds (or
days) instead of $10^{10}$ years leading to a huge luminosity. This explains why
a
star can become as bright as the whole galaxy. }

{\it Fermion ball \cite{bvn}:} We consider a gas of DM fermions of
total mass $M=10\, M_{\odot}$.\footnote{This is the typical mass of a DM
halo surrounding a baryonic star of mass $\sim 1 \, M_{\odot}$ since the present
fraction of baryons and dark matter are $\Omega_{\rm
b,0}=0.0487273$ and $\Omega_{\rm dm,0}=0.2645$ respectively. Here and in the
following we
consider DM fermions of mass  $m=17.2\, {\rm keV/c^2}$. The maximum mass and
minimum radius set by general
relativity are $M_{\rm OV}=2.12\times 10^9\, M_{\odot}$ and $R_{\rm
OV}=8.86\times 10^{-4}\,
{\rm pc}$. Since $M\ll M_{\rm OV}$, the nonrelativistic model is justified in
that case.}  The canonical
phase transition appears for  $R\ge R^{\rm
NR}_{\rm CCP}(N)=2.78\, {\rm pc}$. 
For a gaseous halo of size $R=41.3\, {\rm pc}$, corresponding to
$\mu=4747$, the collapse
temperature is $T_c=9.18\times 10^{-7}\,
{\rm K}$ (the transition temperature is $T_t=2.83\times 10^{-6}\,
{\rm K}$ \cite{bvn}).  The
radius of the fermion ball of mass $M_C\simeq 10\, M_{\odot}$ resulting from
the
collapse of the gaseous halo is $R_C=0.974\, {\rm
pc}$. The energy of the gaseous halo is $E_{\rm
gas}=-4.12\times 10^{40}\, {\rm erg}$ and the energy of the fermion ball is 
$E_{\rm min}=-3.76\times 10^{42}\,
{\rm erg}$.  The energy released during the collapse
is $\Delta
E=E_{\rm gas}-E_{\rm min}=3.72\times 10^{42}\,
{\rm erg}$. The collapse time is a multiple
of $t_D\sim 1.25\, {\rm Gyrs}$ which is quite long (possibly irrelevant).

{\it Supermassive fermion ball (crude model) \cite{bvn}:} We consider a gas
of DM fermions of total mass $M=10^9\,
M_{\odot}$.\footnote{In that case, general relativity should be taken into
account (see below) since $M\sim M_{\rm max}$.} The canonical
phase transition appears for  $R\ge R^{\rm
NR}_{\rm CCP}(N)=6.00\times 10^{-3}\, {\rm pc}$. 
For a gaseous halo of size $R=1.68\times 10^{-2}\, {\rm pc}$, corresponding to
$\mu=389$, the collapse
temperature is $T_c=2.26\times 10^{5}\,
{\rm K}$ (the transition temperature is $T_t=3.02\times 10^{5}\,
{\rm K}$ \cite{bvn}).  The
radius of the fermion ball of mass $M_C \simeq 10^9\, M_{\odot}$ resulting from
the  collapse of the gaseous halo
is $R_C=2.10\times 10^{-3}\, {\rm
pc}$.  The energy of the gaseous halo is $E_{\rm
gas}=-1.01\times 10^{60}\, {\rm erg}$ and the energy of the fermion ball
is $E_{\rm min}=-1.74\times
10^{61}\,
{\rm erg}$. The energy released during the collapse
is $\Delta
E=E_{\rm gas}-E_{\rm min}=1.64\times 10^{61}\,
{\rm erg}$. The collapse time is a multiple
of $t_D\sim 1.03\, {\rm yrs}$, which is very short on a
cosmological timescale.

\subsubsection{Microcanonical ensemble}
\label{sec_nrmmce}

We now  consider a system of nonrelativistic
self-gravitating fermions in the
microcanonical ensemble. The microcanonical phase transition  appears for
$\mu\ge \mu_{\rm MCP}=2670$ (microcanonical critical point) \cite{ijmpb} hence
for
\begin{equation}
\label{nrmmce1}
R\ge R^{\rm NR}_{\rm MCP}(N)= 5.22 \frac{h^2}{Gm^{8/3}M^{1/3}g^{2/3}}.
\end{equation}
We assume that we are in this situation.\footnote{When $R\le
R^{\rm NR}_{\rm MCP}(N)$ the whole series of equilibria is stable. When 
$R^{\rm NR}_{\rm
CCP}(N)\le R\le
R^{\rm NR}_{\rm
MCP}(N)$ the system evolves, as energy decreases, from the gaseous
states to the core-halo states (with a negative specific heat)
without instability or phase transition.} In that case, the caloric
curve has the form of
Fig. \ref{Xkcal_R600_N0p29_unified_bnewPH}.
The natural evolution proceeds along
the series of equilibria towards lower and lower energies (see
Sec. \ref{sec_cc}). We
assume that
the system is initially in the metastable gaseous phase.  As its energy
decreases it remains in this phase up to the critical point $E_c$ at which the
metastable gaseous branch disappears.\footnote{We recall that the collapse 
takes place at the critical (spinodal) point $E_c$, not
at the transition point $E_t$, because of the tremendously long lifetime of
metastable gaseous states.} This critical energy is not very sensitive on
quantum effects (when $\mu\gg 1$) so it can be approximated by its classical
value $\Lambda_c=-E_cR/GM^2=0.335$ yielding
\begin{equation}
\label{nrmmce2}
E_c=-0.335\frac{GM^2}{R}.
\end{equation}
At that point, the system collapses and forms a compact fermion star.
As explained in Appendix \ref{sec_thermoqmce}, in the
microcanonical ensemble,
the fermion star contains only a fraction of the total mass ($M_C<M$) and is
surrounded by a massive and very hot atmosphere. If we
approximate the fermion star by a Fermi gas at $T=0$ containing a
mass $M_C=\alpha_C M$, its radius is given by (see Appendix
\ref{sec_thermoqmce}):
\begin{equation}
R_C= 0.181 \frac{h^2}{Gm^{8/3}\alpha_C^{1/3}M^{1/3}g^{2/3}}.
\end{equation}
On the other hand, the temperature of the halo in the condensed phase is given
by (see Appendix
\ref{sec_thermoqmce}):
\begin{equation}
\label{tcond}
k_B T_{\rm
cond}=1.57\frac{\alpha_C^{7/3}}{1-\alpha_C}
\frac{ G^2M^ { 4/3 }
m^{11/3}g^{2/3}}{h^2}.
\end{equation}
From the
analytical model developed in \cite{pt} one finds that the fraction
of mass in the core is approximately given by (see Appendix
\ref{sec_thermoqmce}):
\begin{equation}
\label{alpha}
\alpha_C\simeq \frac{7}{4\ln\mu}.
\end{equation}
In many applications, it is sufficient to consider that $\alpha_C\simeq 1/4$
(see footnote 45). 
The temperature
of the gaseous phase at the point of gravothermal catastrophe is not very
sensitive on quantum effects (when $\mu\gg 1$) so it can be approximated by its
classical value
$\eta_{\rm gas}=\beta_{\rm gas}GMm/R=2.03$ yielding
\begin{equation}
\label{tgas}
k_B T_{\rm gas}=0.493 \frac{GMm}{R}.
\end{equation}
The relaxation time depends on the physical process governing the
dynamical evolution of the system so it will not be discussed here.

We note that the collapse in the microcanonical ensemble corresponds to an
implosion of the core and a heating of the halo. In the box
model, the
atmosphere  is held by the walls of the box. Without the box, it would be
expelled at large
distances (see Figs. 38 and 41 of \cite{clm2} for an illustration in the
context of the fermionic King model).
Therefore, in the microcanonical ensemble, the collapse leads to the formation
of a fermion star and to the expulsion of a hot atmosphere. This core-halo
structure  is reminiscent of
the onset of a red-giant before the white dwarf stage. The
implosion of the core and the explosion of the halo is also similar to the
supernova phenomenon leading to the formation of a  neutron
star. These ideas
are further developed and illustrated in \cite{supernova}. The
thermodynamical reason for this implosion-explosion phenomenon is explained in
Appendix \ref{sec_thermoclassmce}. It is also corroborated by the following
arguments. If we
calculate the density perturbation $\delta\rho$ that triggers the instability at
$E_c$,
we find that it has two nodes, corresponding to a core-halo structure (see Fig.
6b of \cite{paddyapj}).
Therefore, the
instability develops itself in such a way that the density increases in the core
and in the
halo while it decreases between them (the intermediate shell is
depopulated). We also find that the velocity profile
$\delta v$ has one node (see Fig. 4.b of \cite{supernova}) so the velocity  is
directed inward
in the
core and outward in the halo. These two results confirm the implosion-explosion
phenomenon similar to the red-giant structure  and to the 
supernova phenomenon previously described.

We make below numerical applications to illustrate the preceding
results.\footnote{In the numerical applications, we have chosen the values of
$M$ and $R$ in
order to have $\mu=10^5$. From Fig. 21 of \cite{ijmpb} we find that
$\eta_{\rm cond}=0.290$. From the relation $\eta_{\rm cond}\sim
7(1-\alpha_C)/(2\lambda\mu^{2/3}\alpha_C^{7/3})$  with $\lambda=0.149736...$,
equivalent to Eq. (\ref{tcond}), we find that $\alpha_{C}=0.220$. This can be
compared to the approximate value 
$\alpha_C\simeq 0.125$ obtained from Eq. (\ref{alpha}). The agreement is
reasonable in view of the numerous approximations and the logarithmic
corrections.}

{\it Neutron stars (crude model) \cite{ht}:} We consider a gas of neutrons of
total
mass $M=1\, M_{\odot}$. 
The microcanonical phase transition appears for $R\ge R^{\rm
NR}_{\rm MCP}(N)=435\,
{\rm km}$. For a gaseous star of size $R=4875\, {\rm
km}$ the
collapse
energy is $E_c=-1.81\times 10^{50}\,
{\rm erg}$ (the transition energy is $E_t=1.51\times
10^{50}\,
{\rm erg}$). The mass of
the neutron star resulting from the collapse of the gaseous star is $M_C=0.220
M_{\odot}$ and its
radius is
$R_C=25.0\, {\rm km}$. The temperature of the gaseous star is $T_{\rm
gas}=1.63\times
10^{9}\, {\rm K}$ and the temperature of
the halo surrounding the neutron star is $T_{\rm cond}=1.14\times 10^{10}\, {\rm
K}$.

{\it Fermion ball \cite{bvn}:} We consider a gas of DM fermions of total mass
$M=10\, M_{\odot}$. The microcanonical phase
transition appears for $R\ge R^{\rm
NR}_{\rm MCP}(N)=28.1\,
{\rm pc}$. For a gaseous halo of size $R=315\, {\rm pc}$ the collapse
energy is $E_c=-9.10\times 10^{39}\,
{\rm erg}$ (the transition energy is $E_t=7.61\times
10^{39}\,
{\rm erg}$). The mass of
the fermion ball resulting from the collapse of the gaseous halo is $M_C=2.20
M_{\odot}$ and its
radius is
$R_C=1.61\, {\rm pc}$. The temperature of the gaseous halo is $T_{\rm
gas}=1.50\times
10^{-7}\, {\rm K}$ and the temperature of
the halo surrounding the fermion ball is $T_{\rm cond}=1.05\times
10^{-6}\, {\rm K}$.
 
{\it Supermassive fermion ball (crude model) \cite{bvn}:} We consider a gas of
DM fermions of total mass $M=10^9\, M_{\odot}$. The microcanonical
phase
transition appears for $R\ge R^{\rm
NR}_{\rm MCP}(N)=6.06\times 10^{-2}\,
{\rm pc}$. For a gaseous halo of size $R=0.679\, {\rm pc}$ the collapse
energy is $E_c=-4.22\times 10^{58}\,
{\rm erg}$ (the transition energy is $E_t=3.53\times
10^{58}\,
{\rm erg}$). The mass of
the fermion ball resulting from the collapse  of the gaseous halo is
$M_C=2.20\times 10^8\,
M_{\odot}$ and its
radius is
$R_C=3.48\times 10^{-3}\, {\rm pc}$. The temperature of the gaseous
halo is $T_{\rm
gas}=6.94\times
10^{3}\, {\rm K}$ and the temperature of
the halo surrounding the fermion ball is $T_{\rm
cond}=4.86\times 10^{4}\, {\rm K}$.

\subsection{Relativistic model with $N<N_{\rm OV}$: Collapse towards a fermion
star}

We now consider the truly general relativistic Fermi gas model. We first assume
that
$N<N_{\rm OV}$ so that the collapse always leads to a fermion star (not a black
hole). The discussion is essentially the same as before. However, we make below
new numerical applications to see the effect of relativistic corrections when
$N\lesssim N_{\rm OV}$. These numerical applications are based on the general
relativistic models of
dark matter halos 
studied by Bilic and Viollier \cite{bvr} as recalled in the
Introduction.\footnote{In terms of the dimensionless variables defined in
Appendix \ref{sec_dq} they take $N=0.38$, which is slightly below the OV
limit $N_{\rm OV}=0.39853$, and $R=100$. The radius of the completely
degenerate fermion ball corresponding to $N=0.38$ is $R_C=4.095=1.22
R_{\rm OV}$ \cite{paper1}.} We consider their adaptation to the case of neutron
stars. We restrict ourselves to the canonical ensemble.

{\it Supermassive fermion ball \cite{bvr}:} For a fermionic
particle of mass
$m=17.2\,
{\rm keV/c^2}$ the OV limits are $N_{\rm
OV}=1.4254\times 10^{71}$, $N_{\rm
OV}m=2.1973\times 10^9\, M_{\odot}$, $M _{\rm
OV}=2.1186\times 10^9\, M_{\odot}$
and $R_{\rm OV}=8.8859\times 10^{-4}\, {\rm pc}$. We
consider a gas of 
$N=0.95350\, N_{\rm OV}$ fermions, corresponding to a rest mass $N
m=2.0951\times 10^9\, M_{\odot}$. According to
Fig. \ref{phase_phase_Newton_corrPH}, the canonical
phase transition appears for $R\ge R_{\rm CCP}(N)=3.93 R_{\rm OV}=3.4818\times
10^{-3}\,
{\rm pc}$ (the nonrelativistic value is $R_{\rm
CCP}^{\rm NR}(N)=4.6894\times
10^{-3}\, {\rm pc}$). For a system of initial size
$R=29.789\, R_{\rm OV}=2.6391\times
10^{-2}\, {\rm pc}$, the
collapse
temperature is $T_c=3.0112\times 10^{5}\,
{\rm K}$ (the transition temperature is
$T_t=0.0043951\, mc^2=8.7725\times 10^5\, {\rm K}$ \cite{bvr}). The collapse of
the gaseous halo leads to a
supermassive fermion ball
containing almost all the particles surrounded by a tenuous atmosphere. Since
the particle number is slightly
below the OV limit, the fermion ball is strongly relativistic. If we
approximate the fermion ball by a Fermi gas at $T=0$ containing all the rest
mass $Nm \sim 2.0951\times 10^9\, M_{\odot}$, we find a radius $R_C=1.220\,
R_{\rm OV}=1.0809\times 10^{-3}\, {\rm pc}$ and
a mass $M_C=0.9577 \, M_{\rm OV}=2.0290\times 10^9\, M_{\odot}$ 
\cite{paper1} (the
nonrelativistic values are $R_C^{\rm NR}=1.6399\times 10^{-3}\, {\rm pc}$ and
$M_C^{\rm NR}= 2.0951\times 10^9\, M_{\odot}$). The energy of the
gaseous halo is
$E_{\rm gas}=-2.8324\times 10^{60}\, {\rm erg}$ and the energy of the fermion
ball is $E_{\rm
min}=(M_C-Nm)c^2=-1.1822\times 10^{62}\,
{\rm erg}$ (the
nonrelativistic value is $E_{\rm min}^{\rm NR}=-9.7925\times 10^{61}\,
{\rm erg}$).  The energy released
during the collapse
is $\Delta
E=E_{\rm gas}-E_{\rm min}=1.1539\times 10^{62}\,
{\rm erg}$ (the
nonrelativistic value is $\Delta
E^{\rm NR}=9.5092\times 10^{61}\,
{\rm erg}$). The collapse time is a multiple
of $t_D\sim 1.3973\, {\rm yrs}$.

{\it Neutron stars:} For the neutrons of mass $m_n=0.940\, {\rm
GeV/c^2}$, the OV limits are $N_{\rm
OV}=8.7448\times 10^{56}$, $N_{\rm
OV}m=0.73636\, M_{\odot}$, $M _{\rm OV}=0.71000\, M_{\odot}$ and $R_{\rm
OV}=9.1614\,
{\rm km}$. We consider a gas of 
$N=0.95350\, N_{\rm OV}$ neutrons, corresponding to a rest mass
$N m=0.70212\, M_{\odot}$. 
The canonical
phase transition appears for $R\ge R_{\rm CCP}(N)=3.93 R_{\rm OV}=36.0\,
{\rm km}$ (the nonrelativistic value is $R_{\rm
CCP}^{\rm NR}(N)=48.5\, {\rm km}$). For a system of
initial size $R=29.789\,
R_{\rm OV}=272.91\, {\rm km}$, the collapse
temperature is $T_c=1.6449\times 10^{10}\,
{\rm K}$ (the transition temperature is $T_t=0.0043951\, mc^2=4.7921\times
10^{10}\, {\rm K}$). The collapse of the gaseous star leads to a neutron star
containing almost all the particles surrounded by a tenuous atmosphere. Since
the particle number is slightly
below the OV limit, the system is strongly relativistic. If we
approximate the neutron star by a Fermi gas at $T=0$ containing all the rest
mass $Nm \sim 0.70212\, M_{\odot}$, we find a radius $R_C=1.220 R_{\rm
OV}=11.177\,
{\rm km}$ and a mass $M_C=0.9577 \, M_{\rm OV}=0.67996\, M_{\odot}$ (the
nonrelativistic values are $R_C^{\rm NR}=16.957\,
{\rm km}$ and $M_C^{\rm NR}=0.70212\, M_{\odot}$). The energy of the gaseous
star is
$E_{\rm gas}=-9.4919\times 10^{50}\, {\rm erg}$ and the energy of the neutron
star is $E_{\rm
min}=(M_C-Nm)c^2=-3.9618\times 10^{52}\,
{\rm erg}$ (the
nonrelativistic value is $E_{\rm min}^{\rm NR}=-3.28167\times 10^{52}\,
{\rm erg}$). The energy released during the
collapse
is $\Delta
E=E_{\rm gas}-E_{\rm min}=3.8669\times 10^{52}\,
{\rm erg}$ (the
nonrelativistic value is $\Delta
E^{\rm NR}=3.1867\times 10^{52}\,
{\rm erg}$). The collapse time is a multiple 
of $t_D\sim 1.4767\times 10^{-2}\, {\rm s}$.

{\it Remark:} For the value of $N$ considered in the previous examples, we find
from Fig. \ref{phase_phase_Newton_corrPH} that the microcanonical
phase transition appears for $R\ge R_{\rm MCP}(N)=52.4\, R_{\rm OV}$ (the
nonrelativistic value is $R_{\rm MCP}^{\rm NR}(N)=53.4\, R_{\rm OV}$). Since
$R$ is below this critical value, the system does not display any phase
transition in the microcanonical ensemble.

\subsection{Relativistic model with  $N_{\rm OV}<N<N_{c}$:
Collapse towards a fermion star followed by a collapse towards a black hole}
\label{sec_crit}

We now assume that $N_{\rm OV}<N<N_{c}$ (where $N_c^{\rm CE}=N_e$ and $N_c^{\rm
MCE}=N_f$) so
that, by cooling, the system undergoes two successive collapses: a
collapse towards
a fermion star followed by a collapse towards a black hole.

We consider a system of relativistic self-gravitating fermions in the canonical
ensemble. We assume that  $N_{\rm OV}<N<N_e(R)$. The caloric curve has the form
of Fig. \ref{kcal_R50_N0p399_unifiedPH}.  We assume
that the system is initially in the gaseous
phase. At its temperature decreases, the system   collapses from the
gaseous phase to the condensed phase at $T_c$ then undergoes a catastrophic
collapse
from the condensed phase to a black hole at $T'_c$. We note that the interval
$(\Delta N)_{\rm CE}=N_e(R)-N_{\rm OV}$ is extremely narrow since $N_{\rm
OV}=0.39853$ and
$N_e=0.40002$ for $R=50$ (we see in  Fig.
\ref{phase_phase_Newton_corrPH} that $N_e(R)$ does not change much with $R$). 
We have $(\Delta N)_{\rm CE}/N_{\rm OV}=3.81\times 10^{-3} \ll 1$ so that
\begin{equation}
\label{mcr1}
N_c^{\rm CE}\simeq N_{\rm OV}.
\end{equation}
The reason why $(\Delta N)_{\rm CE}/N_{\rm OV}\ll 1$ is easy to understand. We
have previously explained that the fermion star contains almost all the
particles ($N_C\sim N$). Therefore, as soon as $N$ is larger than $N_{\rm OV}$
the fermion star becomes unstable ($N_C>N_{\rm OV}$) and collapses towards a
black hole (see Appendix \ref{sec_tdov}).

We consider a system of relativistic self-gravitating fermions in the
microcanonical
ensemble. We assume that  $N_{\rm OV}<N<N_f(R)$. The caloric curve has the form
of Fig. \ref{kcal_R600_N1p3_unifiedPH}.  We assume that the system is initially
in the gaseous
phase. At its energy decreases, the system   collapses from the
gaseous phase to the condensed phase at $E_c$ then undergoes a catastrophic
collapse
from the condensed phase to a black hole at $E'_c$. The interval
$(\Delta N)_{\rm MCE}=N_f(R)-N_{\rm OV}$ is much larger than in the canonical
ensemble since $N_{\rm
OV}=0.39853$ and $N_f\simeq 1.4854$ for $R=600$ (we see in Fig.
\ref{phase_phase_Newton_corrPH} that $N_f(R)$ remains in the range $1-2$). We
have $(\Delta N)_{\rm MCE}/N_{\rm OV}=2.73$ so that
\begin{equation}
\label{mcr2}
N^{\rm MCE}_c\simeq 3.73 \, N_{\rm OV}.
\end{equation}
Again, the reason why $(\Delta N)_{\rm MCE}/N_{\rm OV}\sim 1$ is easy to
understand. We
have previously explained that the fermion star contains only a
fraction of the particles ($N_C\sim N/4$). Therefore, if  $N$ is only
slightly larger
than $N_{\rm OV}$, the fermion ball is stable ($N_C<N_{\rm OV}$). It is only
when $N$ is substantially larger than   $N_{\rm OV}$ (by a factor of $\sim
4$) that the fermion ball
becomes unstable ($N_C>N_{\rm OV}$) and collapses towards a
black hole.

\subsection{Relativistic model with  $N>N_{c}$: Direct collapse
towards
a black hole}

We finally assume that $N>N_{c}$ so
that, by cooling, the system directly collapses towards a black hole, without
forming a fermion star.

We consider a system of relativistic self-gravitating fermions in the canonical
ensemble. We assume that  $N>N_e(R)$. The caloric curve has the form
of Figs. \ref{Xkcal_R50_N0p401_unifiedPH}, \ref{kcal_R50_N0p41_unifiedPH},
\ref{Xkcal_R50_N0p45_unified2PH},
\ref{kcal_R50_N1p5_unifiedPH}, and \ref{kcal_R50_N4_unifiedPH}.  We assume
that the system is initially in the gaseous
phase. As its temperature decreases, the system undergoes a
catastrophic collapse from the gaseous phase to a black hole at $T_c$. 
This situation corresponds to  $N>N_c^{\rm CE}$, where $N_c^{\rm CE}\simeq
N_{\rm OV}$ [see  Eq. (\ref{mcr1})]. 
This result shows that, in the canonical ensemble, there is no condensed
configurations in the general relativisitic Fermi gas at nonzero temperature as
soon as $N$ is slightly larger than $N_{\rm OV}$. The reason is that
almost of all the particles are in the degenerate core, the rest forming a
tenuous isothermal atmosphere.

We consider a system of relativistic self-gravitating fermions in the
microcanonical
ensemble. We assume that  $N>N_f(R)$. The caloric curve has the form
of
Figs. \ref{kcal_R600_N1p5_unifiedPH}, \ref{kcal_R600_N1p65_unifiedPH} and 
\ref{kcal_R600_N5_unified2PH}.  We assume that the system is initially in the
gaseous
phase. At its energy decreases, the system undergoes a
catastrophic collapse from the gaseous phase to a black hole at $E_c$. 
This situation corresponds to  $N>N_c^{\rm MCE}$, where $N_c^{\rm MCE}\simeq
3.73\, N_{\rm OV}$ [see  Eq. (\ref{mcr2})]. This result shows that, in the
microcanonical ensemble, there exist condensed
configurations in the general relativisitic Fermi gas at nonzero temperature
with $N_{\rm OV}<N<3.73 \, N_{\rm OV}$. The reason is that only about $1/4$ of
the particles are in the degenerate core (so that $N_C<N_{\rm OV}$), the rest
forming an isothermal atmosphere.

\subsection{Summary}

In this section, we summarize the possible evolution of a gaseous star when its
temperature (canonical ensemble) or its energy (microcanonical ensemble) is
reduced below a critical value.

\subsubsection{Canonical ensemble}

When $N<N_{\rm OV}$ and $R_{\rm min}(N)<R<R_{\rm CCP}(N)$, there is no collapse.

When $N<N_{\rm OV}$ and $R>R_{\rm CCP}(N)$, the gaseous star collapses
below $T_c$ towards a fermion star (white dwarf, neutron star, DM fermion ball).
The fermion star contains almost all the mass and is surrounded by a very
tenuous atmosphere. Therefore, the collapse corresponds just to an implosion.

When $N_{\rm OV}<N<N^{\rm CE}_{c}\simeq N_{\rm OV}$  and $R>R_{\rm CCP}(N)$ the
gaseous star
first collapses below $T_c$ toward a
fermion star, then the fermion star collapses below $T'_c$ towards a black hole.

When $N_{\rm
OV}<N<N^{\rm CE}_{c}\simeq N_{\rm OV}$  and $R_{\rm min}(N)<R<R_{\rm CCP}(N)$ or
when $N>N^{\rm CE}_{c}\simeq N_{\rm OV}$ the gaseous star directly collapses
towards a black hole.

\subsubsection{Microcanonical ensemble}

When $N<N_{\rm OV}$ and $R_{\rm min}(N)<R<R_{\rm MCP}(N)$, there is no collapse.

When $N<N_{\rm OV}$ and $R>R_{\rm MCP}(N)$, the gaseous star collapses
below $E_c$ towards a fermion star (white dwarf, neutron star, DM fermion ball).
The
fermion star
contains only a fraction of the total mass and is surrounded
by a very hot atmosphere. 
Therefore, the system has a core-halo structure which is similar to the onset
of
red-giants or to the supernova phenomenon. This core-halo structure results
from  an implosion of the core
and an explosion of the halo.

When $N_{\rm OV}<N<N^{\rm MCE}_{c}\simeq 3.73\, N_{\rm OV}$  and $R>R_{\rm
MCP}(N)$ the
gaseous star
first collapses below $E_c$ toward a
fermion star $+$ halo, then the fermion star collapses below $E'_c$
towards a black hole.

When $N_{\rm
OV}<N<N^{\rm MCE}_{c}\simeq 3.73\, N_{\rm OV}$  and $R_{\rm min}(N)<R<R_{\rm
MCP}(N)$ or
when $N>N^{\rm MCE}_{c}\simeq 3.73\, N_{\rm OV}$ the gaseous star directly
collapses
towards a black hole.

\section{Conclusion}
\label{sec_conclusion}

In this paper we have studied the nature of phase transitions in the
general relativistic Fermi gas. This is the most general
situation that we can imagine since both quantum and relativistic effects are
taken into account in a rigorous manner. We have obtained the complete phase
diagram of the system in the $(R,N)$ plane (see Fig.
\ref{phase_phase_Newton_corrPH}).

When $N<N_{\rm OV}$, the results are similar to those obtained in the
nonrelativistic limit (recovered for $N\ll N_{\rm OV}$ and $R\gg
R_{\rm OV}$ with $NR^3$ fixed) \cite{ijmpb}. In that case, there exists an
equilibrium state for any value of the
temperature $T\ge 0$ and any value of the energy $E\ge E_{\rm min}$.
Catastrophic collapse towards a black hole is prevented by quantum mechanics
(Pauli's exclusion principle). Small systems ($R_{\rm
min}<R<R_{\rm CCP}$) do not experience phase transition. Intermediate size
systems ($R_{\rm CCP}<R<R_{\rm MCP}$) experience a canonical phase transition.
Large systems ($R>R_{\rm MCP}$)  experience canonical and microcanonical phase
transitions. A zeroth order phase transition takes place, below a critical
temperature $T_c$ or
below a critical energy $E_c$, from a gaseous phase to a condensed phase. The
gaseous phase corresponds to a radiative star, a molecular cloud or a
primordial DM nebula. The condensed phase corresponds to a compact object
(fermion star) such as a white dwarf, a neutron star or a DM fermion ball. In
the canonical ensemble the fermion star contains almost all the mass (or is
surrounded by a tenuous isothermal atmosphere). The phase transition
corresponds to an implosion. In the microcanonial ensemble the compact object
contains only a fraction of the total mass ($\sim 1/4$ to fix the ideas) and is
surrounded by a hot isothermal
atmosphere that contains the remaining mass. Therefore, the condensed phase
has a core-halo structure. In the box model the atmosphere is held by the walls
of the box. Without the box it would be expelled at large distances. The phase
transition corresponds to an an implosion of the core
and an explosion of the halo.

When $N_{c}<N<N_{\rm max}$ (where $N_c\simeq N_{\rm OV}$ in the canonical
ensemble and $N_c\simeq 3.73\, N_{\rm OV}$ in the microcanonical ensemble), the
results are similar to those obtained in the classical limit (recovered for
$N\gg N_{\rm OV}$ and $R\gg
R_{\rm OV}$ with $N/R$ fixed) \cite{roupas,paper2}. In that case, there is no
condensed phase. Below a critical temperature $T_c$ or below a
critical energy $E_c$, the system undergoes a catastrophic collapse from the
gaseous phase to a black hole (presumably). Indeed, quantum mechanics cannot
prevent this singularity.

When $N_{\rm OV}<N<N_c$, the results are new and more complex
because  the system is both relativistic and quantum. In that
case, the system
can experience two successive collapses: a zeroth order phase transition at
$T_c$ (when $R>R_{\rm
CCP}$) or $E_c$ (when $R>R_{\rm
MCP}$) from the gaseous phase to the condensed phase (representing a white
dwarf, a
neutron star, or a DM fermion ball) followed by a catastrophic collapse at
$T'_c$ or
$E'_c$ from the condensed phase to a black hole (presumably). This behavior
occurs in a very
narrow range of parameters in the canonical ensemble ($N_c\simeq N_{\rm OV}$)
and in a wider range of
parameters ($N_c\simeq 3.73\, N_{\rm OV}$) in the microcanonical ensemble.

The previous results apply to mid and low
values of energy and temperature. At
very high energies and very high temperatures, the system is
ultrarelativistic and behaves like a form of self-gravitating radiation. Above a
maximum energy $E_{\rm max}$ or above a maximum temperature
$T_{\rm max}$, there is no equilibrium state and the system is expected to
collapse towards a
black hole (presumably) \cite{roupas,paper2}.

The astrophysical applications of our results remain limited by the
introduction of an artificial confining box. This is necessary in order to have
isothermal equilibrum states with a finite mass and thus investigate phase
transitions rigorously. A more astrophysically relevant
model with a finite mass is provided by the general relativistic fermionic King
model. Phase transitions in the nonrelativistic fermionic King model
have been studied in detail in \cite{clm1,clm2} and give results that are
qualitatively similar to those obtained with the box model \cite{ijmpb}. We
believe that similar results would be obtained with the  general relativistic
fermionic King model.

We have suggested that the microcanonial phase transitions occurring in the
self-gravitating Fermi gas  may be related to the onset of red-giant
structure
or to the supernova phenomenon. In these spectacular events, the collapse of the
core of the system (resulting ultimately in the formation of a white dwarf or a
neutron star) is accompanied by the explosion and the expulsion of a hot
envelope.\footnote{Newtonian gravity is sufficient to describe white
dwarfs (and the planetary nebula) that follow the red-giant stage while general
relativity is necessary
to describe neutron
stars or black holes formed from supernova explosion.}  Similarly, in the
self-gravitating
Fermi gas, the microcanonical phase
transition from a gaseous state to a condensed state corresponds to the
implosion of the core (leading to a fermion star) and the explosion of 
the halo. These
analogies are further developed in \cite{supernova}. We may wonder if similar
phenomena can occur in the
context of DM as suggested in
\cite{clm2}.
It would also be interesting to develop dynamical models of
gravitational collapse towards a black hole when no equilibrium
state is possible to ensure that the system really forms a black hole. These
topics will be considered in future works.

{\it Acknowledgement:} One of us (PHC) would like to dedicate
this paper to the memory of Donald Lynden-Bell (1935-2018) who was a pioneer in
the statistical mechanics of self-gravitating systems.

\appendix

\section{Thermodynamic limit}
\label{sec_tl}

The thermodynamic limit of nonrelativistic self-gravitating
fermions corresponds to $N\rightarrow +\infty$ in such a way that $\eta=\beta
GMm/R$, $\Lambda=-ER/GM^2$ and $\mu=(gm^4/h^3)\sqrt{512\pi^4G^3NmR^3}$
are $O(1)$. Taking $m\sim h\sim G\sim 1$, this corresponds to $N\rightarrow
+\infty$ with
$R\sim N^{-1/3}$, $T\sim N^{4/3}$,  $E\sim N^{7/3}$, $S\sim N$, and $F\sim
N^{7/3}$. These scalings are given in \cite{htf,ijmpb}. Taking $m\sim h\sim
N/V\sim
1$
(with $V\sim R^3$), this corresponds to $N\rightarrow
+\infty$ with $R\sim N^{1/3}$, $G\sim N^{-2/3}$, $T\sim 1$,  $E\sim N$, $S\sim
N$, and $F\sim
N$.

The thermodynamic limit of relativistic
classical self-gravitating systems corresponds to $N\rightarrow +\infty$ in such
a way that $\eta=\beta
GNm^2/R$, $\Lambda=-ER/GN^2m^2$ (with $E=(M-Nm)c^2$) and $\nu=GNm/Rc^2$
are $O(1)$. Taking $m\sim c\sim G\sim 1$ this corresponds to $N\rightarrow
+\infty$ with
$R\sim N$, $T\sim 1$,  $E\sim N$, $S\sim N$, and $F\sim
N$. To the best of our knowledge, these scalings have not been given previously.
Taking $m\sim c\sim N/V\sim 1$ (with $V\sim R^3$) this corresponds to
$N\rightarrow
+\infty$ with $R\sim N^{1/3}$,  $G\sim N^{-2/3}$, $T\sim 1$,  $E\sim N$, $S\sim
N$,
and $F\sim
N$.

The thermodynamic limit of relativistic self-gravitating
fermions corresponds to $N\rightarrow +\infty$ in such
a way that $\eta=\beta
GNm^2/R$, $\Lambda=-ER/GN^2m^2$ (with $E=(M-Nm)c^2$),
$\mu=(gm^4/h^3)\sqrt{512\pi^4G^3NmR^3}$ and $\nu=GNm/Rc^2$
are $O(1)$. Taking $h\sim c\sim G\sim 1$ this corresponds to $N\rightarrow
+\infty$ with
$R\sim N^{2/3}$, $m\sim N^{-1/3}$, $T\sim N^{-1/3}$,  $E\sim N^{2/3}$, $S\sim
N$, and $F\sim
N^{2/3}$. These scalings were given in  \cite{bvr}. Taking $m\sim c\sim h\sim
N/V\sim 1$ (with $V\sim R^3$) this corresponds to
$N\rightarrow
+\infty$ with $R\sim N^{1/3}$,  $G\sim N^{-2/3}$, $T\sim 1$,  $E\sim N$, $S\sim
N$,
and $F\sim
N$.

The thermodynamic limit of nonrelativistic classical self-gravitating systems is
discussed in Appendix A of \cite{aakin} where several possible scalings are
given.

We note that the different situations considered above can be unified by
considering a
thermodynamic limit of the form $N\rightarrow
+\infty$ with $R\sim N^{1/3}$,  $G\sim N^{-2/3}$, $T\sim 1$,  $E\sim N$, $S\sim
N$, and $F\sim N$, corresponding to $m\sim c\sim h\sim
N/V\sim 1$ (with $V\sim R^3$). This is the standard thermodynamic
limit with a renormalized gravitational constant. To the best of our knowledge
this result has not been highlighted previously.

\section{Dimensionless quantities}
\label{sec_dq}

According to the OV theory \cite{ov}, the maximum mass, the maximum particle
number and the minimum radius of a general relativistic fermion star at $T=0$
are\footnote{Qualitatively, the scaling of the maximum mass $M_{\rm
OV}\sim ({\hbar c}/{G})^{3/2}/{m^2}$
can be obtained from the mass-radius relation $M R^3\sim \hbar^6/(m^8G^3)$
of nonrelativistic fermion stars (see Appendix \ref{sec_bf}) by
determining when the radius $R$ of the configuration becomes comparable to the
Schwarzschild radius $R_S=2GM/c^2$.}
\begin{equation}
\label{dq1}
M_{\rm OV}=0.38426\, \sqrt{\frac{2}{g}}\left (\frac{\hbar c}{G}\right
)^{3/2}\frac{1}{m^2},
\end{equation}
\begin{equation}
\label{dq2}
N_{\rm OV}=0.39853\, \sqrt{\frac{2}{g}}\left (\frac{\hbar c}{G}\right
)^{3/2}\frac{1}{m^3},
\end{equation}
\begin{equation}
\label{dq3}
R_{\rm OV}=8.7360\, \frac{GM_{\rm OV}}{c^2}=3.3569\,
\sqrt{\frac{2}{g}}\left
(\frac{\hbar^3}{Gc}\right )^{1/2}\frac{1}{m^2}.
\end{equation}
 We note that $W_{\rm OV}\sim GM_{\rm OV}^2/R_{\rm
OV}\sim M_{\rm OV}c^2$. We introduce the mass, particle
number and length scales
\begin{equation}
\label{dq4}
M_*=\sqrt{\frac{2}{g}}\frac{M_P^3}{m^2}=\left
(\frac{2\hbar^3c^3}{gm^4G^3}\right )^{1/2},
\end{equation}
\begin{equation}
\label{dq5}
N_*=\frac{M_*}{m}=\sqrt{\frac{2}{g}}\frac{M_P^3}{m^3}=\left
(\frac{2\hbar^3c^3}{gm^6G^3}\right )^{1/2},
\end{equation}
\begin{equation}
\label{dq6}
R_*=\sqrt{\frac{2}{g}}\frac{\hbar M_P}{m^2 c}=
\sqrt{\frac{2}{g}}\frac{M_P^2}{m^2}l_P=
\left
(\frac{2\hbar^3}{gm^4cG}\right )^{1/2},
\end{equation}
where $M_P=(\hbar c/G)^{1/2}$ is the Planck mass and  $l_P=(\hbar G/c^3)^{1/2}$
is the Planck length. We
then define
\begin{equation}
\label{dq7}
r=R_* {\tilde r}, \qquad M=M_* {\tilde M}, \qquad \epsilon=\frac{M_*c^2}{R_*^3}
{\tilde \epsilon},
\end{equation}
\begin{equation}
\label{dq8}
N=N_* {\tilde N}, \qquad n=\frac{N_*}{R_*^3} {\tilde n}, \qquad
\Phi=c^2 {\tilde \Phi},
\end{equation}
\begin{equation}
\label{dq9}
P=\frac{M_*c^2}{R_*^3} {\tilde P}, \qquad
T=\frac{mc^2}{k_B} {\tilde T},\qquad \mu=mc^2 {\tilde \mu},
\end{equation}
\begin{equation}
\label{dq10}
S=N_* k_B {\tilde S},\qquad F=M_* c^2 {\tilde F},
\end{equation}
where the tilde variables are dimensionless. We note that
\begin{equation}
\label{dq10b}
 \frac{M_*c^2}{R_*^3}=\frac{g m^4c^5}{2\hbar^3},\qquad
\frac{N_*}{R_*^3}=\frac{gm^3c^3}{2\hbar^3},
\end{equation}
\begin{equation}
\label{dq11}
N_* R_*^3=\frac{4\hbar^6}{g^2m^9G^3},\qquad \frac{G M_*}{R_* c^2}=1.
\end{equation}
In the main text, in order to simplify the notations, we do not write the tildes
anymore. This amounts to taking $\hbar=c=G=m=g/2=1$ in the dimensional
expressions. In particular, we have
\begin{equation}
\label{dq12}
M_{\rm OV}=0.38426,
\end{equation}
\begin{equation}
\label{dq13}
N_{\rm OV}=0.39853,
\end{equation}
\begin{equation}
\label{dq14}
R_{\rm OV}=3.3569.
\end{equation}

\section{Connexion between the $N_{\alpha}(\Phi_0)$ curves and
the caloric curves $\eta(\Lambda)$}
\label{sec_conn}

In this Appendix, we explain how we obtained the caloric curves of
self-gravitating fermions in general relativity following the method given
by Bilic and Viollier \cite{bvr}.\footnote{This Appendix is technical and can be
skipped for a
first reading. It is nevertheless important to understand where the
critical values of $N$ (such as $N_{1}$, $N_{\rm OV}$, $N_{\rm CCP}$, $N_*$
etc) obtained in the main text come from.}

\begin{figure}
\begin{center}
\includegraphics[clip,scale=0.3]{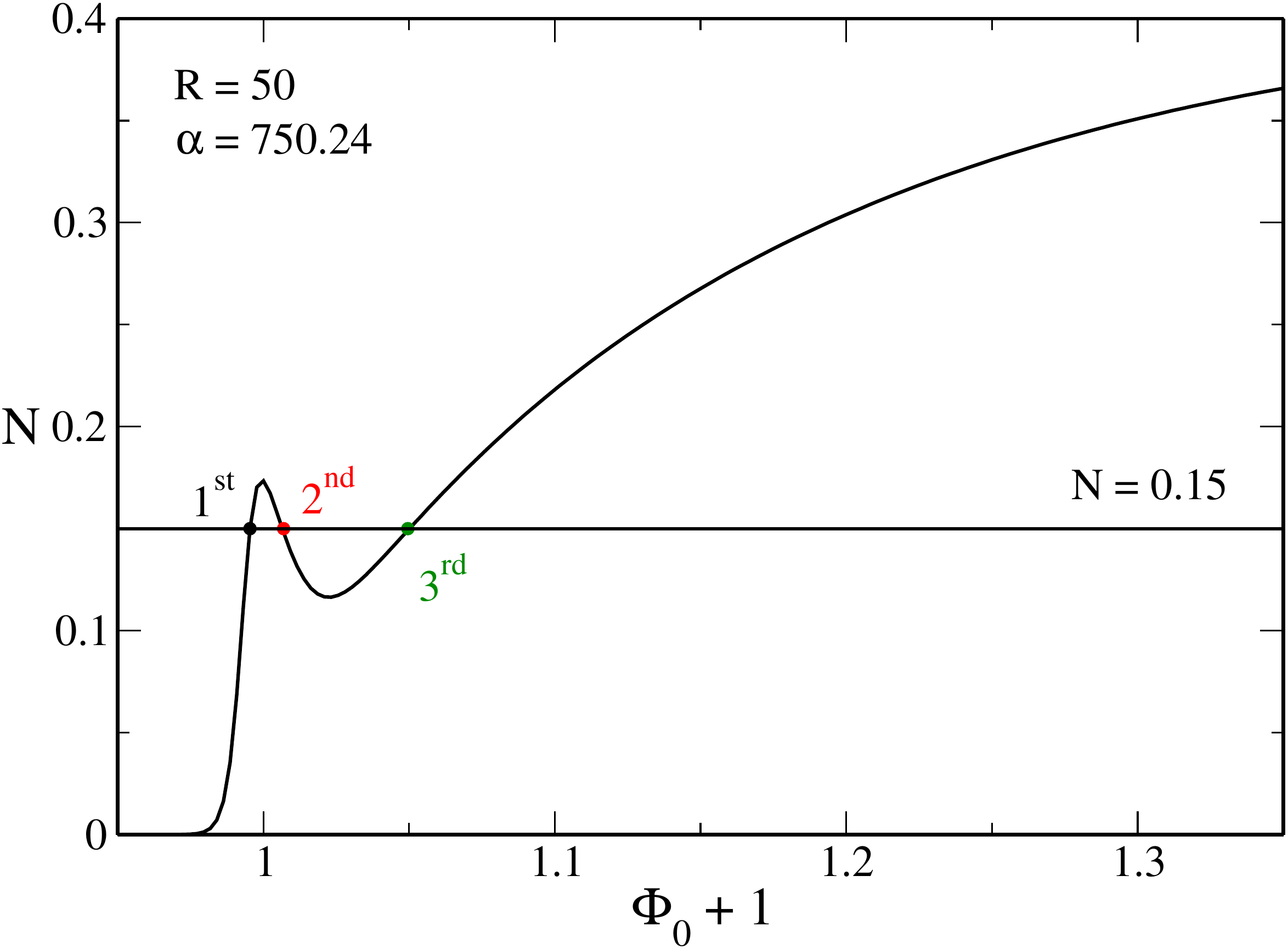}
\caption{Particle number $N$ as a function of the central
potential $\Phi_0$ for $R = 50$ and $\alpha = 750.24$. The intersections
between the curve $N_{\alpha}(\Phi_0)$ and the line level $N=0.15$ determine
three equilibrium states with central potentials $(\Phi_0)_1=0.9954$,
$(\Phi_0)_2=1.0069$ and $(\Phi_0)_3=1.0495$. Their corresponding  energy and
temperature are
$(\Lambda_1,\eta_1)=(-0.0242,2.2725)$,
$(\Lambda_2,\eta_2)=(0.5067,2.2447)$ and
$(\Lambda_3,\eta_3)=(2.4468, 2.2794)$. Each solution is
represented by a
bullet in the caloric curve of Fig. \ref{Xkcal_R50_N0p15_unified2PH}. The first
solution (black) belongs to the gaseous phase, the second solution (red)
belongs to the core-halo phase and the third solution (green) belongs
to the
condensed phase.}
\label{NPhi0_R50_N0p15_bisPH}
\end{center}
\end{figure}

Let us illustrate this procedure with a simple example. In order to
construct the caloric curve $\eta(\Lambda)$ corresponding to
$R=50$ and $N=0.15$ (see Fig. \ref{Xkcal_R50_N0p15_unified2blackPH}), we proceed
as follows. For a given value of $\alpha$ and
$\Phi_0$, we can solve the differential
equations (\ref{b6}) and (\ref{b7}) up to $r=R$ and determine $N$ from Eq.
(\ref{b10}) [we can also determine $\eta$ and
$\Lambda$ with the aid of Eqs. (\ref{b9}), (\ref{b10}), (\ref{b12}) and 
(\ref{b16})]. By varying  the central potential $\Phi_0$ from $-1$ to $+\infty$,
we can obtain
the curve
$N_{\alpha}(\Phi_0)$.  It is usually nonmonotonic and displays
damped oscillations for large values of $\Phi_0$. As example is represented in
Fig.
\ref{NPhi0_R50_N0p15_bisPH} for
$R=50$ and $\alpha=750.24$ (see also Figs. \ref{NPhi0_negative_R50n} and
\ref{NPhi0_positive_R50n} below).

\begin{figure}
\begin{center}
\includegraphics[clip,scale=0.3]{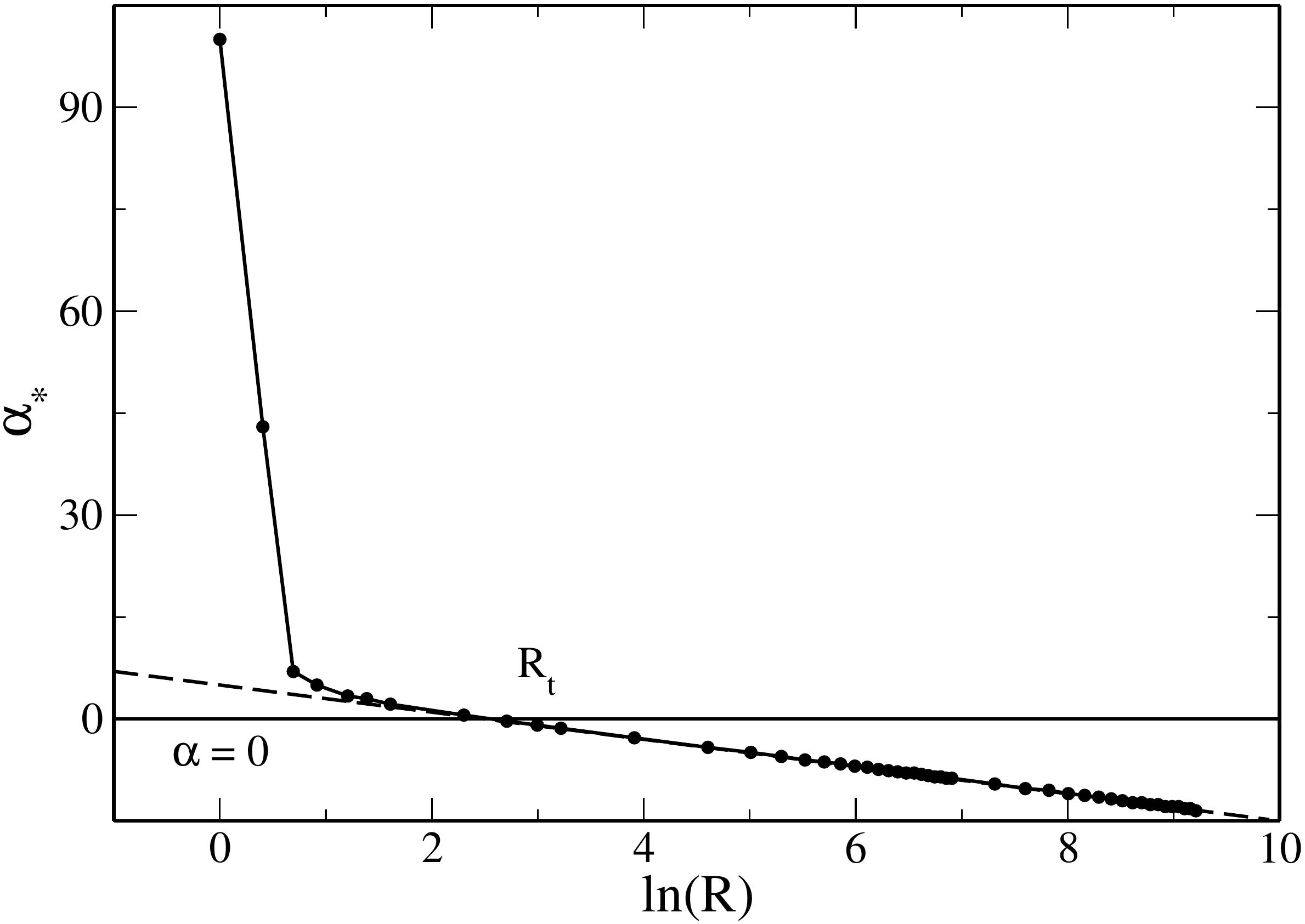}
\caption{Evolution of $\alpha_*(R)$ as a function of $R$. The dashed line
represents the scaling law $\alpha_*(R)=5.0119-2\ln R$ obtained in \cite{paper2}
for
classical systems. In the case of fermions, this law  is asymptotically  valid
for $R\gg 1$. We have
also indicated the radius $R_t=12.255$ at which $\alpha_*(R)$ passes from
negative values to positive values  as the
box radius decreases.}
\label{alpha_NmaxR_bisNEWPH}
\end{center}
\end{figure}

\begin{figure}
\begin{center}
\includegraphics[clip,scale=0.3]{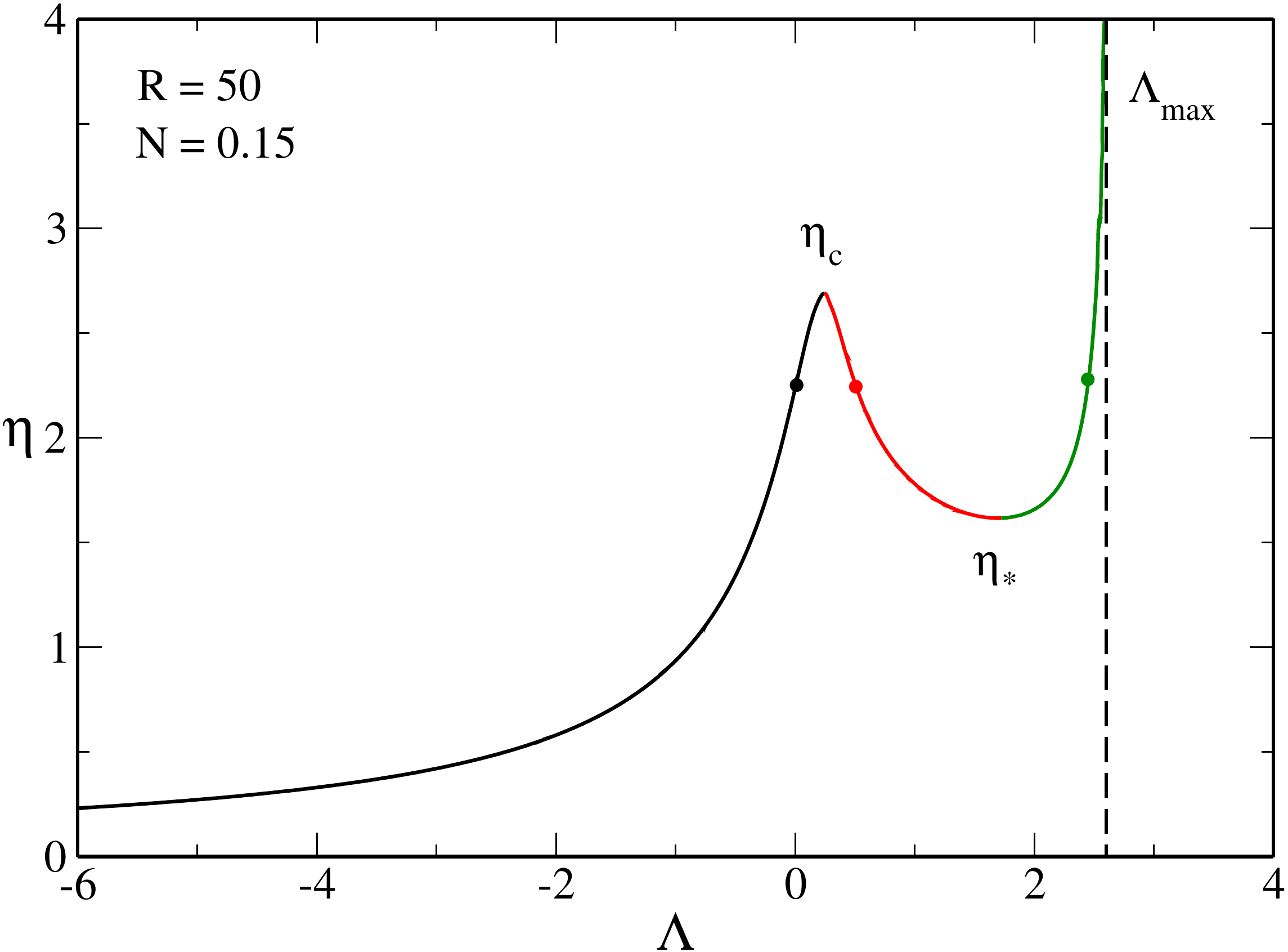}
\caption{Caloric curve for $R = 50$ and $N = 0.15$. By varying $\alpha$ in
Fig. \ref{NPhi0_R50_N0p15_bisPH}, the
first intersections $(\Lambda_1(\alpha),\eta_1(\alpha))$  form the gaseous
branch (black), the second
intersections $(\Lambda_2(\alpha),\eta_2(\alpha))$ form the core-halo branch
(red) and the third intersections $(\Lambda_3(\alpha),\eta_3(\alpha))$ form the
condensed branch (green). It turns out that two branches merge at an extremum of
temperature: $\eta_c$  corresponds to  the merging of the first and second
intersections (for $\alpha=\alpha_M$)  and $\eta_*$ corresponds to the merging
of
the second and third intersections (for $\alpha=\alpha_m$). }
\label{Xkcal_R50_N0p15_unified2PH}
\end{center}
\end{figure}

Let us introduce some notations that will be useful in the following. We call
${\cal N}(\alpha)$ the maximum value of the curve
$N_{\alpha}(\Phi_0)$ and we denote by $\Psi(\alpha)$ the value of the central
potential $\Phi_0$ corresponding to this maximum. By
varying $\alpha$ from
$-\infty$ to $+\infty$, we find that the peaks ${\cal N}(\alpha)$  of the curves
$\lbrace N_{\alpha}(\Phi_0)\rbrace$ reach a maximum $N_{\rm
max}(R)={\cal N}(\alpha_*(R))$ at  $\alpha=\alpha_*(R)$ (see Figs.
\ref{NPhi0_negative_R50n} and
\ref{NPhi0_positive_R50n} below). The
evolution of $\alpha_*$ as a function of $R$ is plotted in
Fig. \ref{alpha_NmaxR_bisNEWPH}.

Let us come back to the curve of Fig. \ref{NPhi0_R50_N0p15_bisPH}. The
intersections $(\Phi_0)_{i\in \lbrace 1,...,n\rbrace}$ between
the curve $N_{\alpha}(\Phi_0)$ and the line level $N=0.15$, and the
corresponding values of
$\Lambda$ and $\eta$ at these intersections, determine $n$ points 
in the caloric curve $\eta(\Lambda)$ of Fig.  \ref{Xkcal_R50_N0p15_unified2PH}. 
In the present exemple, $n=3$. By
varying $\alpha$ these points form $n$ branches in the caloric curve
$\eta(\Lambda)$. These branches have
been represented in color in Fig. \ref{Xkcal_R50_N0p15_unified2PH}. In the
present exemple, they correspond to the gaseous, core-halo and condensed
phases respectively. We have
observed (but not proven mathematically) that when $\alpha>\alpha_*$, the
merging of two intersections between the curves  $\lbrace
N_{\alpha}(\Phi_0)\rbrace$ and the line level $N$, occuring at some
$\alpha_e(N,R)$, corresponds to the merging of two branches in the caloric
curve $\eta(\Lambda)$ occurring  at an extremum of temperature
$\eta_e$.\footnote{By using this result, it is easy to
obtain the curves $\eta_c(N)$, $\eta_*(N)$ and  $\eta'_c(N)$ in the
canonical phase diagram of Fig. \ref{Xphase_eta_R50_def_new2PH}, since they
correspond to $\alpha>\alpha_*$ (see below). Indeed, each extremum of
$N_{\alpha}(\Phi)$
determines an extremum inverse temperature $\eta_e(N)$ for the corresponding
value of $N=N(\alpha)$. Thus, by considering the first three extrema (when
they exist) and running
$\alpha$ from $\alpha_*$ to $+\infty$, we get the full curves $\eta_c(N)$,
$\eta_*(N)$ and $\eta'_c(N)$. Unfortunately, it is
not possible to use a similar method to obtain the curves $\Lambda_c(N)$,
$\Lambda_*(N)$, $\Lambda'_c(N)$ and  $\Lambda''_c(N)$ in the
microcanonical phase diagram of Fig. \ref{Xphase_Lambda_R600_def4aPH}. They
have to be obtained ``by hand'' directly from the caloric curves.} We
have also
observed (but not proven mathematically) that when $\alpha<\alpha_*$, the
merging of two intersections between the curves  $\lbrace
N_{\alpha}(\Phi_0)\rbrace$ and the line level $N$, occuring at some
$\alpha'_e(N,R)$, corresponds to the merging of two branches in the caloric
curve $\eta(\Lambda)$ occurring at an extremum of energy
$\Lambda_e$.\footnote{By using this result, it is easy to
obtain the curve $\Lambda_{\rm min}(N)$ in the
microcanonical phase diagram of Fig. \ref{Xphase_Lambda_R50_def4PH}, since
it
corresponds to $\alpha<\alpha_*$ (see below). Indeed, each extremum of
$N_{\alpha}(\Phi)$ determines
an extremum energy $\Lambda_{e}(N)$ for the corresponding value of
$N=N(\alpha)$. Thus, by considering the first extremum and running $\alpha$ from
$-\infty$ to $\alpha_*$, we get the
full curve $\Lambda_{\rm min}(N)$. Unfortunately, it is
not possible to use a similar method to obtain the curve $\eta_{\rm min}(N)$
in the canonical phase diagram of Fig. \ref{Xphase_eta_R50_def_new2PH}.
It has
to be obtained ``by hand'' directly from the caloric curves.} In
the present example, we have $\alpha>\alpha_*$. As a result the merging of the
first and second intersections in Fig. \ref{NPhi0_R50_N0p15_bisPH}, occurring at
some $\alpha_M$, corresponds to the temperature minimum $\eta_c$ in the
caloric curve of Fig. \ref{Xkcal_R50_N0p15_unified2PH}. Similarly,  the
merging of the
second and third intersections in Fig. \ref{NPhi0_R50_N0p15_bisPH}, occurring  
at some $\alpha_m$, corresponds to the temperature maximum $\eta_*$ in
the caloric curve of Fig. \ref{Xkcal_R50_N0p15_unified2PH} .

We now generalize this procedure to different
values of $N$ and $R$.
The main difference between the case of classical particles
studied in \cite{paper2} and the case of fermions studied in the present paper
is the
following. When quantum mechanics is taken into account, we
find that for $\alpha\rightarrow
+\infty$ the $\lbrace N_{\alpha}(\Phi_0)\rbrace$ curves tend towards an
invariant curve
correponding to the OV-curve $N_{\rm OV}(\Phi_0)$ \cite{paper1}. Close to
this curve, i.e. for $\alpha$ large and $N$ small, the
$\lbrace N_{\alpha}(\Phi_0)\rbrace$
curves can have a complex behavior with several oscillations responsible for the
phase transitions studied in this paper. The nature of these oscillations,
and consequently the nature of phase transitions, depends on the particle
number $N$ and on the radius of the system $R$.  We consider below the cases
treated in the main text.

\subsection{$R=50$}
\label{sec_cit}

In this subsection, we consider a system of size $R=50$ and make the link
between the topological properties of the curves $\lbrace
N_{\alpha}(\Phi_0)\rbrace$ and the caloric curves analysed in Sec.
\ref{sec_first}.

\begin{figure}
\begin{center}
\includegraphics[clip,scale=0.3]{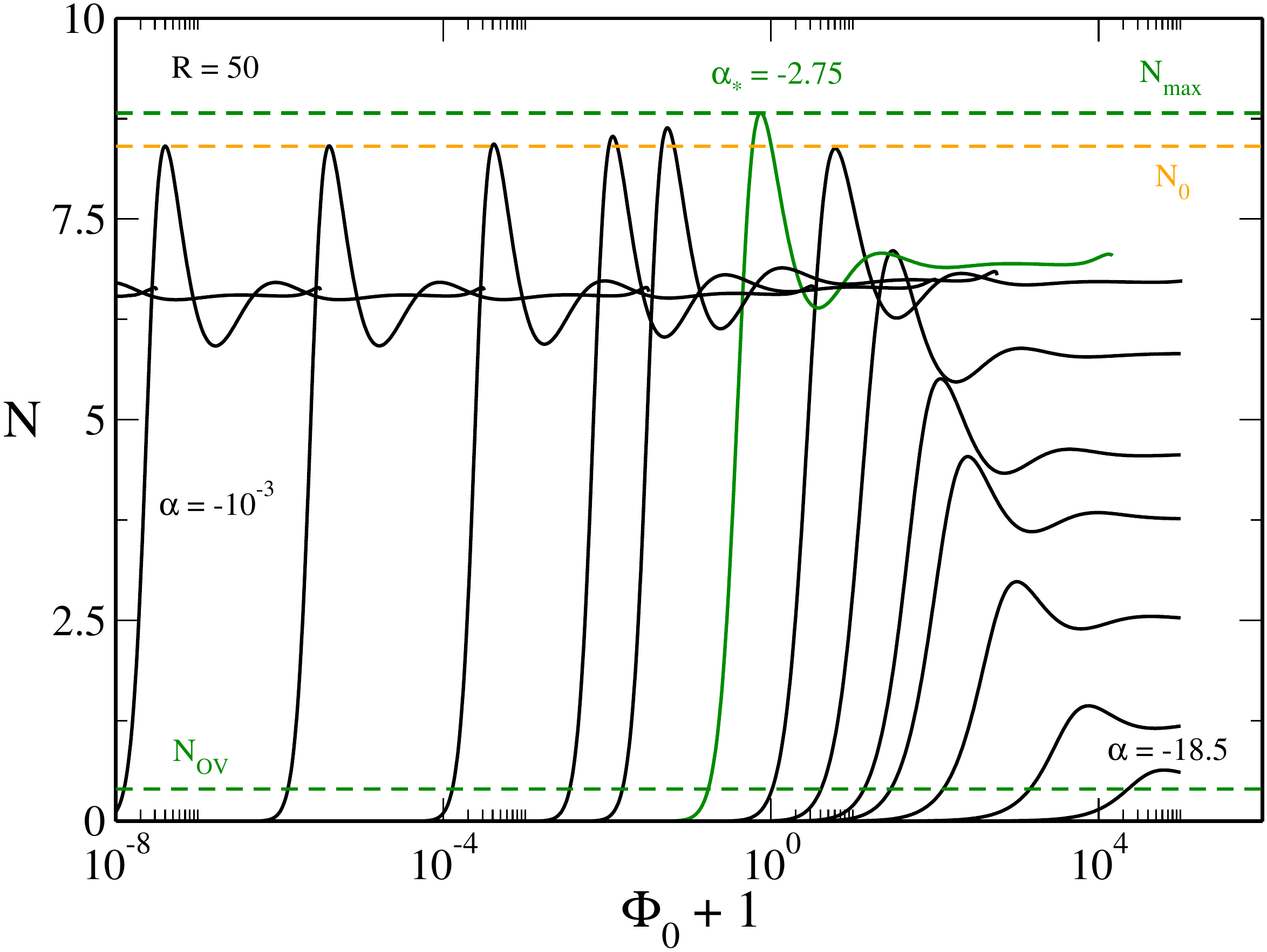}
\caption{Evolution of the curve $N_{\alpha}(\Phi_0)$ for
different values of $\alpha<0$ for $R=50$ (for illustration the curves
go from $\alpha=-18.5$ to $\alpha=-10^{-3}$). We have indicated different
characteristics values of $N$:  $N_0=8.408$ and $N_{\rm max}=8.821$.}
\label{NPhi0_negative_R50n}
\end{center}
\end{figure}

We first consider the case $\alpha<0$ (see Fig. \ref{NPhi0_negative_R50n}).
For $\alpha\rightarrow -\infty$, we find that ${\cal
N}(\alpha)\rightarrow 0$ and 
$\Psi(\alpha)\rightarrow +\infty$.  This corresponds to the ultrarelativistic
regime. As $\alpha$ increases, ${\cal N}(\alpha)$ increases and 
$\Psi(\alpha)$ decreases: the peak of the curve $N_{\alpha}(\Phi_0)$ grows
and moves towards the left.  At $\alpha=\alpha_*=-2.75$, ${\cal N}(\alpha)$
reaches its maximum value $N_{\rm max}=8.821$. For larger values of $\alpha$, 
${\cal
N}(\alpha)$ and $\Psi(\alpha)$ both decrease: the peak of the curve
$N_{\alpha}(\Phi_0)$
decays and moves towards the left. For $\alpha\rightarrow 0^{-}$, ${\cal
N}(\alpha)\rightarrow N_0=8.408$ and $\Psi(\alpha)\rightarrow -1$: the peak of
the curve $N_{\alpha}(\Phi_0)$ is squeezed near
$\Phi_0=-1$.\footnote{Apart from this mathematical property, 
$\alpha=0$ does not play a particular role in the problem. If we had
plotted $N$ as a function of $b_0=1/T_0$ instead of
$\Phi_0$ [see Eq. (\ref{b4})] the specificity of the value $\alpha=0$
would not have arisen.}

\begin{figure}
\begin{center}
\includegraphics[clip,scale=0.3]{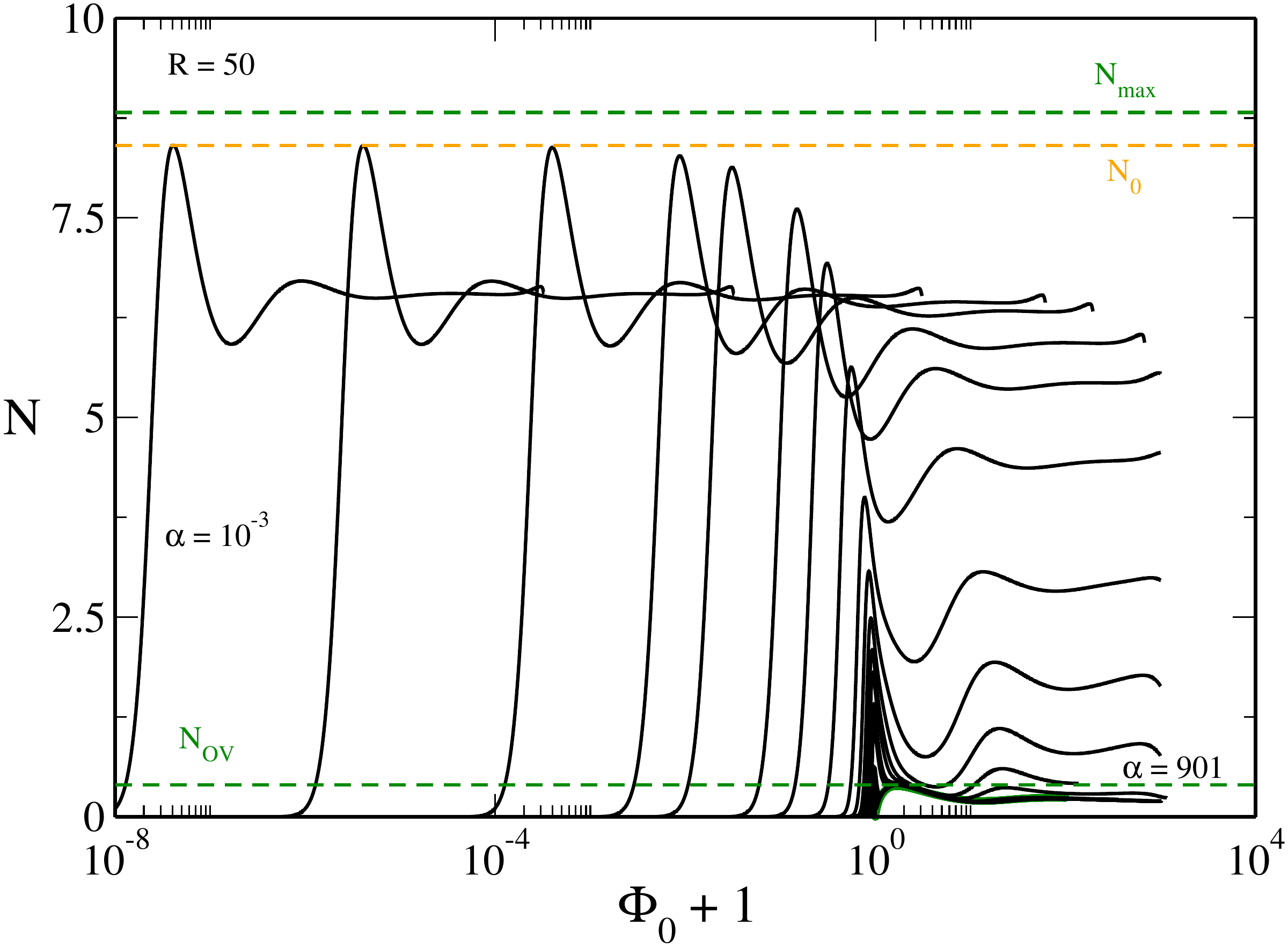}
\caption{Evolution of the curve $N_{\alpha}(\Phi_0)$ for
different values of $\alpha>0$ for $R=50$ (for illustration the curves
go from $\alpha=10^{-3}$ to $\alpha=901$). The OV curve has been plotted in
green. We have indicated different
characteristics values of $N$: $N_{\rm
OV}=0.39853$,  $N_0=8.408$ and $N_{\rm max}=8.821$.}
\label{NPhi0_positive_R50n}
\end{center}
\end{figure}

We now consider the case $\alpha>0$ (see Fig. \ref{NPhi0_positive_R50n}). For
$\alpha>0$, we find
that ${\cal N}(\alpha)$ decreases and $\Psi(\alpha)$ increases: the peak of the
curve $N_{\alpha}(\Phi_0)$ decays and moves towards the
right. For $\alpha\rightarrow +\infty$, the curve $N_{\alpha}(\Phi_0)$ tends
towards the OV-curve $N_{\rm OV}(\Phi_0)$: ${\cal N}(\alpha)$ tends towards 
$N_{\rm OV}=0.39853$ and $\Psi(\alpha)$ tends towards $0.695$.

The curves $\lbrace N_{\alpha}(\Phi_0)\rbrace$ with $\alpha<\alpha_*$ (i.e.
the ones that go up as $\alpha$ increases) are
associated with the hot spiral (radiation) studied in \cite{paper2}.  The hot
spiral corresponds to the ultrarelativistic limit valid for high values of the
energy and of the temperature. This spiral is present for any value of $N$,
except for $N$ close to $N_{\rm max}$ where the caloric curve presents a
different
behavior described in Sec. VI of \cite{paper2} and in Sec.
\ref{sec_mardi} of the present paper. 

\begin{figure}
\begin{center}
\includegraphics[clip,scale=0.3]{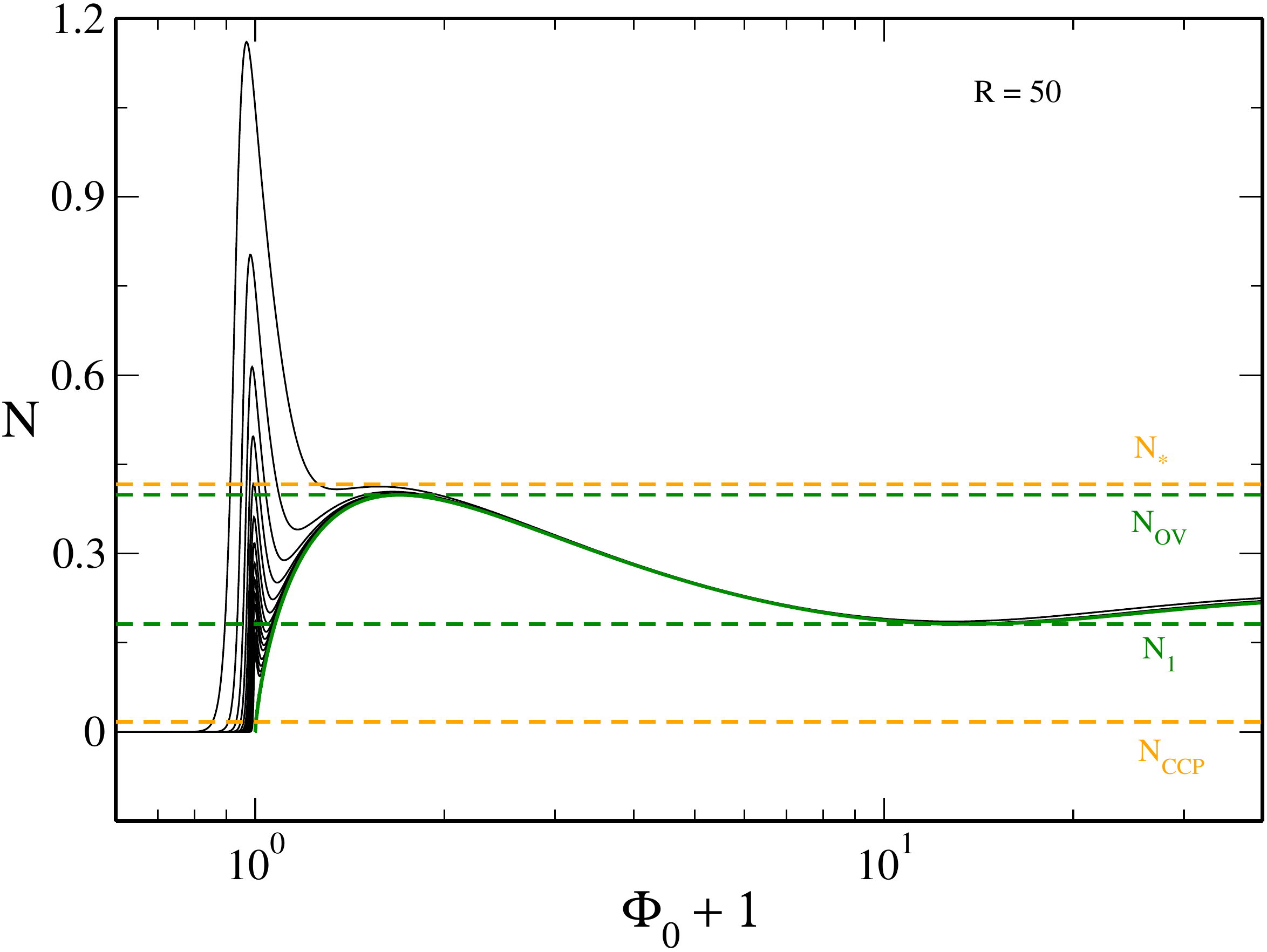}
\caption{Some curves $N_{\alpha}(\Phi_0)$ for $R=50$ together with different
characteristics values of $N$: $N_{\rm CCP}=0.01697$, $N_1=0.18131$, $N_{\rm
OV}=0.39853$, and $N_{*}=0.41637$.}
\label{NPhi0_Nkritik1R50}
\end{center}
\end{figure}

The curves $\lbrace N_{\alpha}(\Phi_0)\rbrace$ with
$\alpha>\alpha_*$ (i.e.
the ones that go down as $\alpha$ increases) are associated with the
part of the caloric curve corresponding to mid and low values of the energy
and of the temperature. Figure \ref{NPhi0_Nkritik1R50} is a zoom of Fig.
\ref{NPhi0_positive_R50n}
close to the OV curve, i.e. for small values of $N$. This is the region of
interest
where canonical phase
transitions appear (they are related to the appearance of turning points of
temperature). We have indicated on this figure 
the different
characteristic
values of $N$ that have been identified in Sec. \ref{sec_first}. They can be
related to the
topological properties of the curves $\lbrace
N_{\alpha}(\Phi_0)\rbrace$ as follows:

(i) $N_{\rm OV}=0.39853$  and $N_1=0.18131$ are
intrinsic properties of the OV curve  $N_{\rm OV}(\Phi_0)$. They correspond
to its
first maximum and to
its  first minimum. 

\begin{figure}
\begin{center}
\includegraphics[clip,scale=0.3]{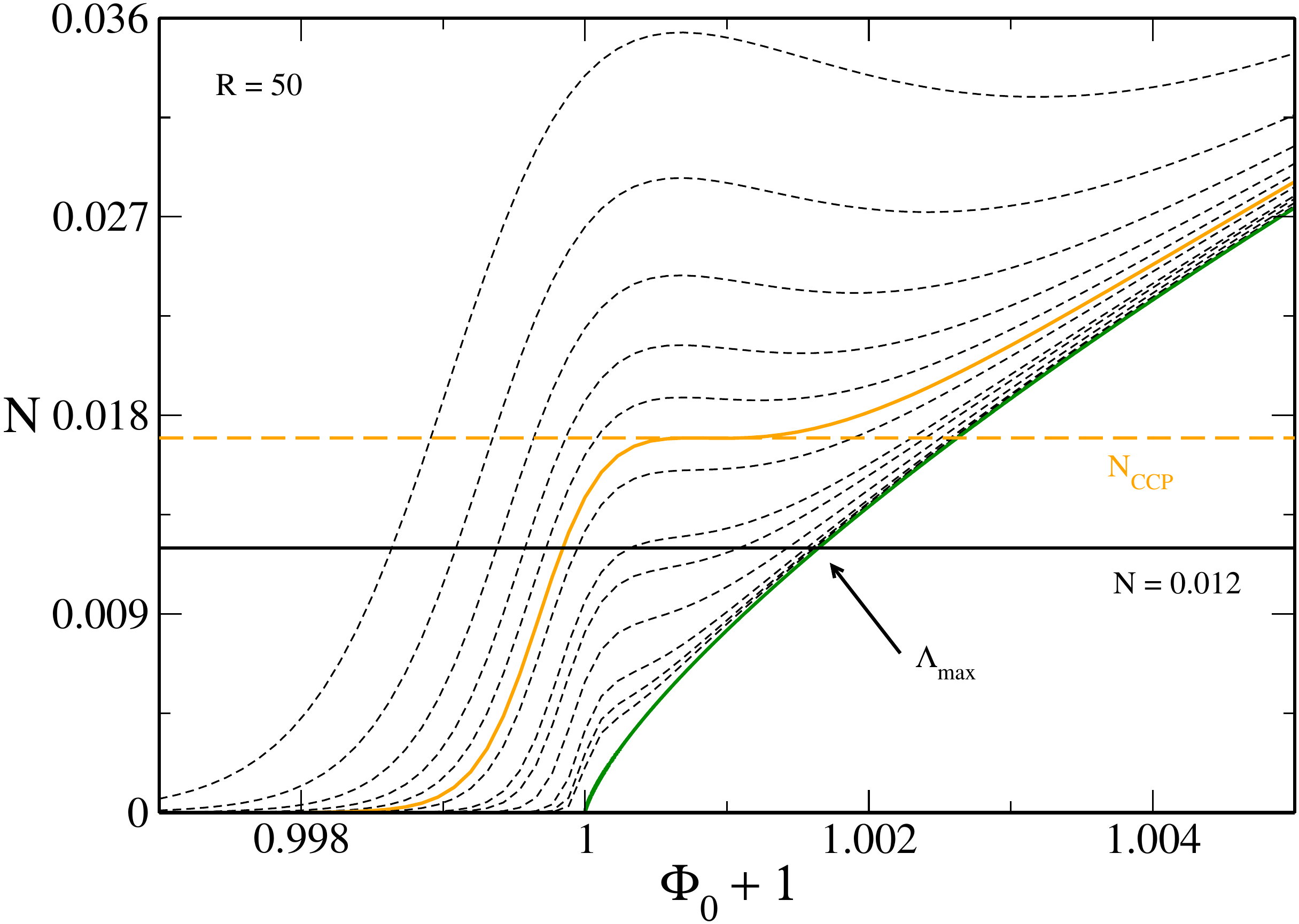}
\caption{Illustration of the intersections in the case $N<N_{\rm CCP}=0.01697$
(specifically $R=50$ and $N=0.012$).}
\label{NPhi0_minor_NCCP}
\end{center}
\end{figure}

(ii) $N_{\rm CCP}=0.01697$ and 
$N_{*}=0.41637$ can be related to the first and second inflexion points of the
curves   $\lbrace N_{\alpha}(\Phi_0)\rbrace$ (see Figs. \ref{NPhi0_minor_NCCP}
and \ref{NPhi0_inter_NOV_Nstar_a}). Indeed, we have seen that an extremum of
temperature in the caloric
curve corresponds to a merging of two intersections in the $N=\lbrace
N_{\alpha}(\Phi_0)\rbrace$ plots. As a
result, the canonical phase transitions appear (at $N=N_{\rm CCP}$) and
disappear
(at $N=N_*$) when the curve  $\lbrace N_{\alpha}(\Phi_0)\rbrace$ presents
an inflexion point. This is how we can precisely determine $N_{\rm CCP}$ and 
$N_{*}$. 

Let us now describe in more detail the different intersections as a function of
$N$.

For $N<N_{\rm CCP}$ (see Fig. \ref{NPhi0_minor_NCCP}), we have just one
intersection between the line level $N$ and the curves $\lbrace
N_{\alpha}(\Phi_0)\rbrace$.
This explains why the caloric curve of Fig. \ref{kcal_R50_N0p012_unified2PH} is
monotonic. We note that the
intersection  between the line level $N$ and the OV-curve
($\alpha\rightarrow +\infty$) corresponds to the
ground state $T=0$ (i.e. $\eta\rightarrow +\infty$). This leads to the
vertical asymptote at $\Lambda=\Lambda_{\rm max}$ in the caloric curve of Fig.
 \ref{kcal_R50_N0p012_unified2PH}. We note that, for $N<N_{\rm CCP}$, 
$\Phi_0\ll 1$
showing that we are in the
Newtonian limit.

\begin{figure}
\begin{center}
\includegraphics[clip,scale=0.3]{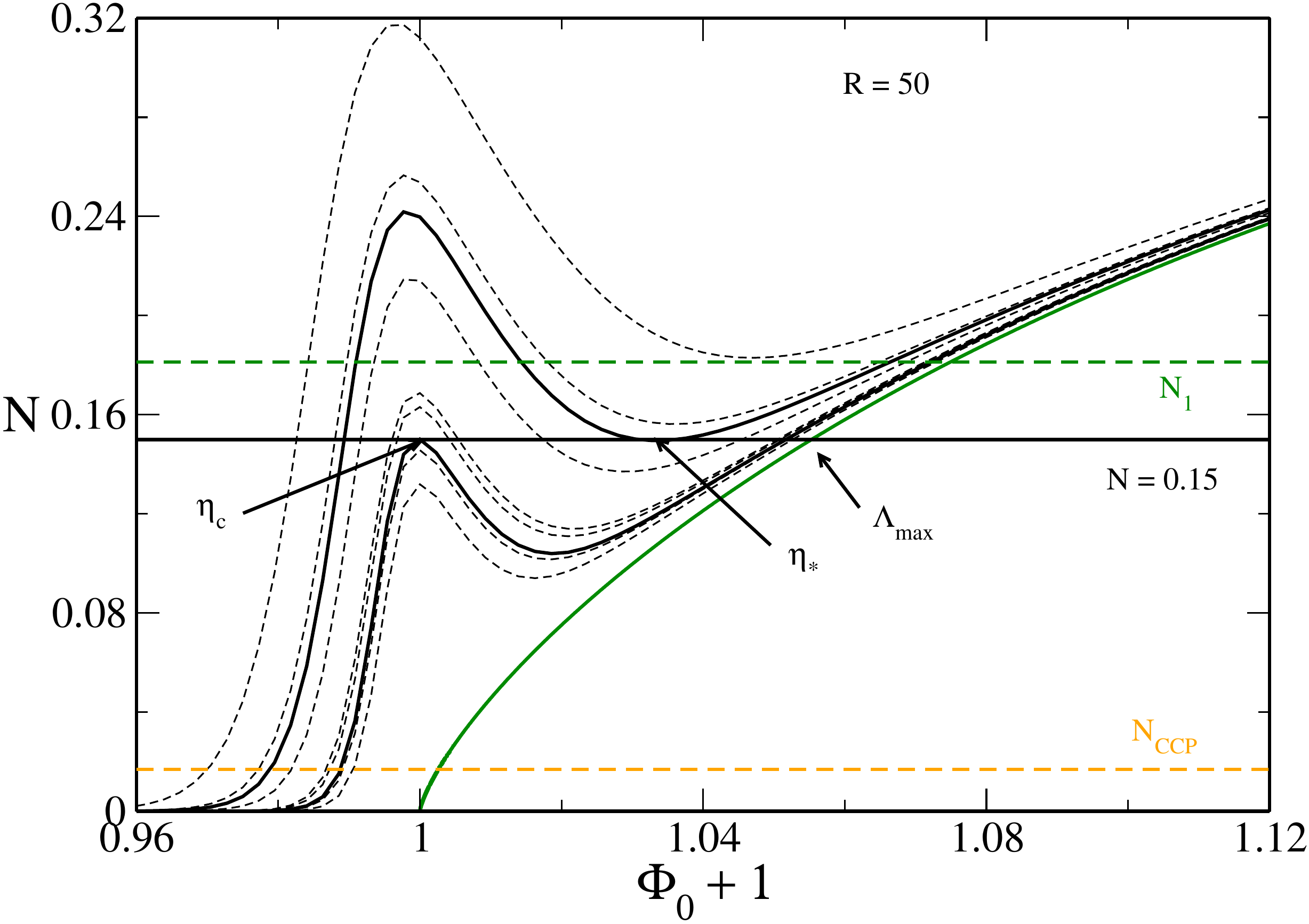}
\caption{Illustration of the intersections in the case $N_{\rm
CCP}=0.01697<N<N_1=0.18131$ 
(specifically $R=50$ and $N=0.15$). }
\label{NPhi0_inter_NCCP_N1}
\end{center}
\end{figure}

For  $N_{\rm CCP}<N<N_1$ (see Fig. \ref{NPhi0_inter_NCCP_N1}), we are above the
first inflexion point so we can have
up to three intersections between the line level $N$ and the curves $\lbrace
N_{\alpha}(\Phi_0)\rbrace$. This determines three branches in the caloric
curve of Fig. \ref{Xkcal_R50_N0p15_unified2blackPH}. This is why it has an
$N$-shape.  The first and second
intersections
merge at $\alpha=\alpha_M$. Correspondingly, the first and second branches in
the caloric curve merge at $\eta_c$, the first turning point of temperature. The
second and third intersections merge at
$\alpha=\alpha_m$. Correspondingly, the second and third branches in
the caloric  curve merge at $\eta_*$, the second turing point of temperature.

\begin{figure}
\begin{center}
\includegraphics[clip,scale=0.3]{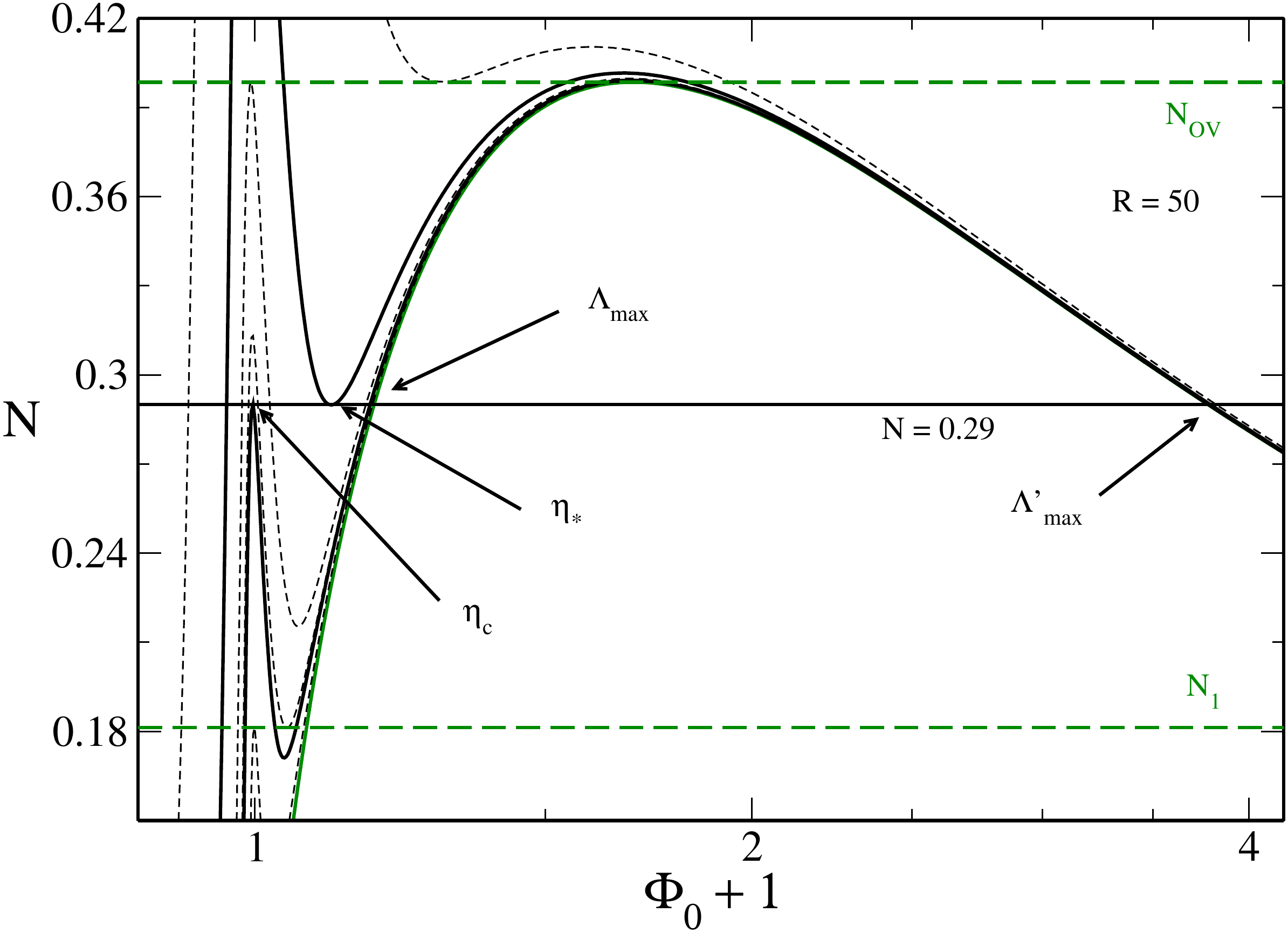}
\caption{Illustration of the intersections in the case  $N_1=0.18131<N<N_{\rm
OV}=0.39853$
(specifically $R=50$ and $N=0.29$).}
\label{NPhi0_inter_N1_NOV}
\end{center}
\end{figure}

For $N_1<N<N_{\rm OV}$ (see Fig. \ref{NPhi0_inter_N1_NOV}), the novelty is that
there is a second intersection between the line level $N$ and  the
OV-curve $N_{\rm OV}(\Phi_0)$. This corresponds to an unstable equilibrium
state at $T=0$ (i.e. $\eta\rightarrow +\infty$). This gives rise  to the second
vertical asymptote at $\Lambda=\Lambda'_{\rm max}$ in the caloric curve of
Fig. \ref{kcal_R50_N0p29_unifiedPH}. There are also secondary intersections
leading to the unstable spiral of Fig. \ref{kcal_R50_N0p29_unifiedPH}
as discussed in the next paragraph.

\begin{figure}
\begin{center}
\includegraphics[clip,scale=0.3]{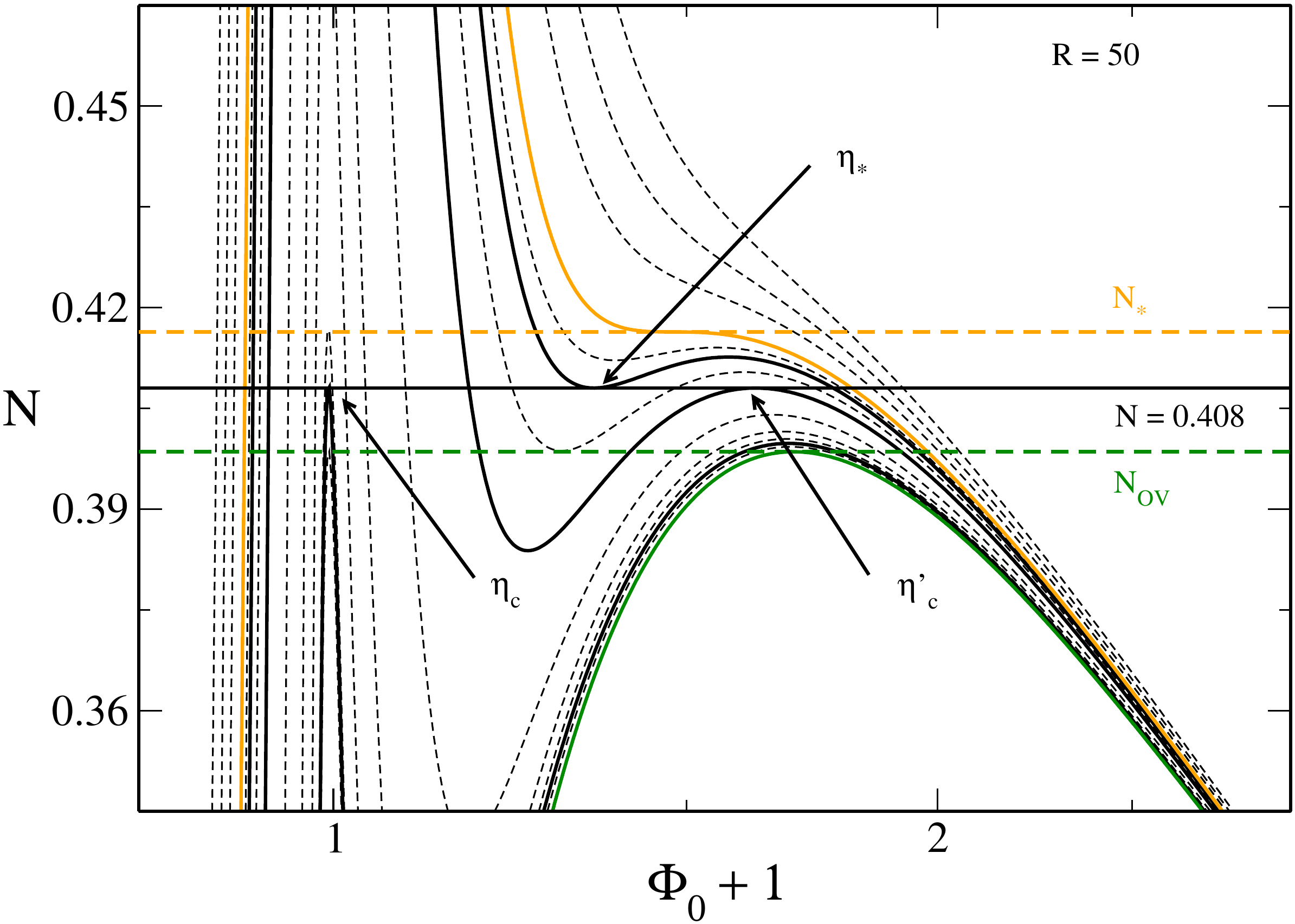}
\caption{Illustration of the intersections in the case $N_{\rm
OV}=0.39853<N<N_*=0.41637$ 
(specifically $R=50$ and $N=0.408$).}
\label{NPhi0_inter_NOV_Nstar_a}
\end{center}
\end{figure}

\begin{figure}
\begin{center}
\includegraphics[clip,scale=0.3]{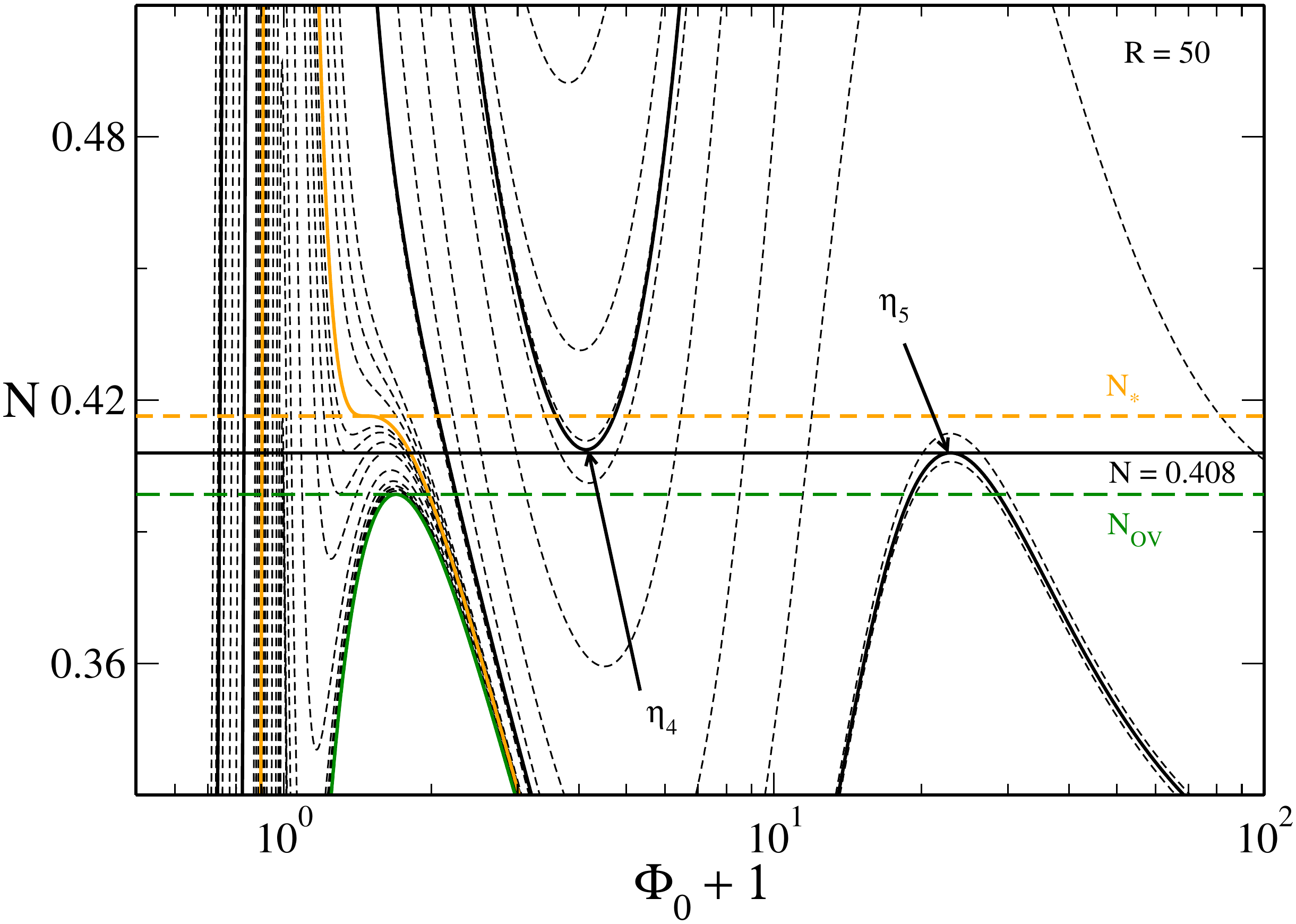}
\caption{Illustration of the intersections in the case $N_{\rm
OV}=0.39853<N<N_*=0.41637$ 
(specifically $R=50$ and $N=0.408$). }
\label{NPhi0_inter_NOV_Nstar_b}
\end{center}
\end{figure}

For $N_{\rm OV}<N<N_*$ (see Figs. \ref{NPhi0_inter_NOV_Nstar_a} and
\ref{NPhi0_inter_NOV_Nstar_b}) we can have up
to four fundamental intersections between the line level $N$ and the
curves $\lbrace
N_{\alpha}(\Phi_0)\rbrace$.  This gives rise to four fundamental branches
in the caloric curve of Fig. \ref{Xkcal_R50_N0p401_unifiedPH}. The first and
second intersections
merge at $\alpha=\alpha_M$. Correspondingly,  the first and second branches in
the caloric curve merge at $\eta_c$, the first turning point of temperature. The
second and third
intersections merge at
$\alpha=\alpha_m$. Correspondingly, the second and third branches in
the caloric  curve merge at $\eta_*$, the second turing point of temperature.
The  third and fourth
intersections merge at
$\alpha=\alpha'_M$. Correspondingly, the third and fourth branches in
the caloric  curve merge at $\eta'_c$, the third turing point of
temperature. Furthermore, there are
additional intersections giving rise to the spiral (that will become
the cold
spiral for larger values of $N$) in the caloric curve of Fig.
\ref{Xkcal_R50_N0p401_unifiedPH}. These intersections are less relevant since
they correspond to
unstable states. Note that there is no intersection with the OV curve so there
is no vertical asymptote corresponding to $\eta\rightarrow +\infty$. 
Finally, we note that $\Phi_0\sim 1$ showing that we are
in the relativistic regime.

\begin{figure}
\begin{center}
\includegraphics[clip,scale=0.3]{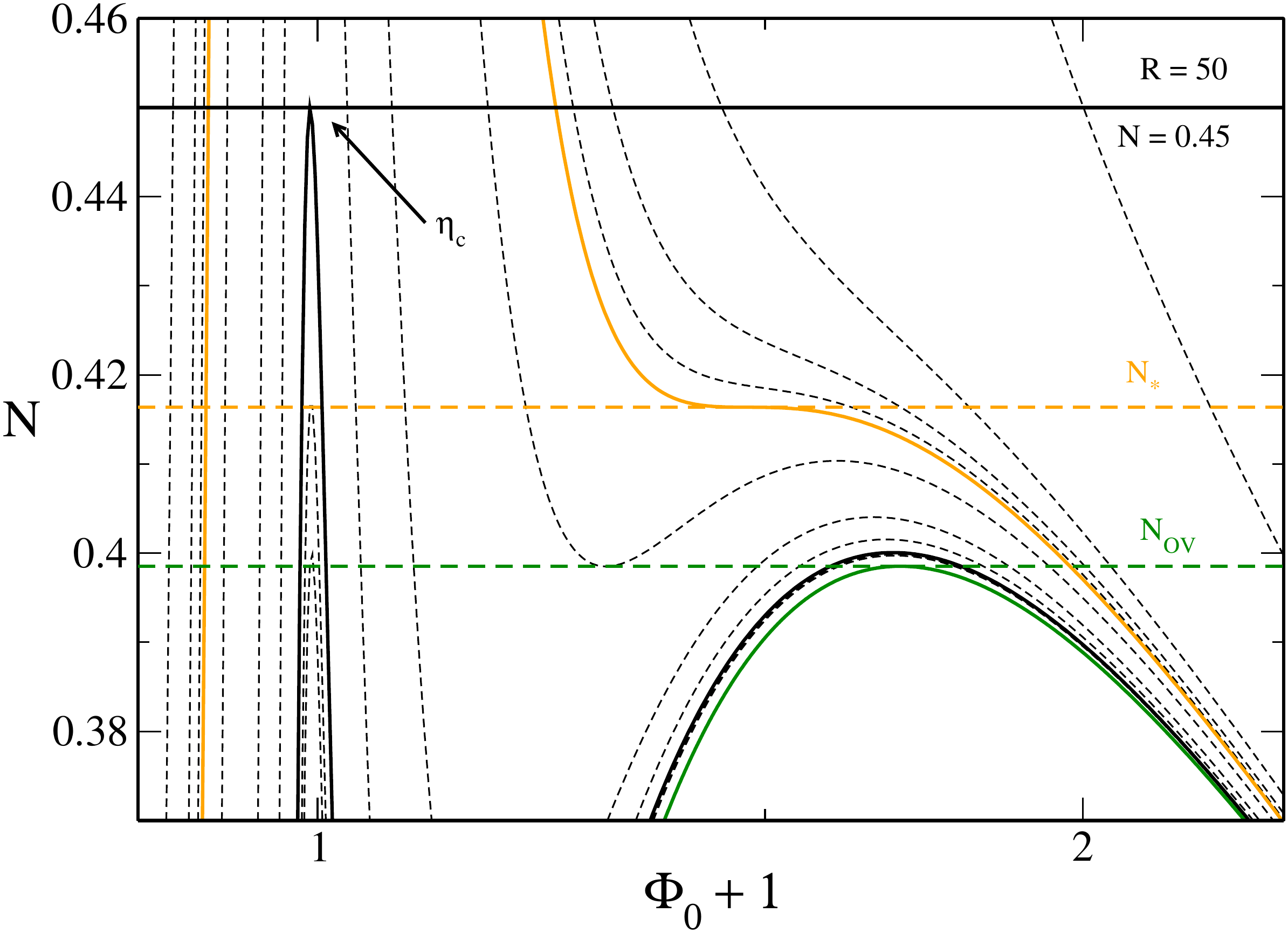}
\caption{Illustration of the intersections in the case $N>N_*=0.41637$ 
(specifically $R=50$ and $N=0.45$).}
\label{NPhi0_major_Nstar}
\end{center}
\end{figure}

For $N>N_*$ (see Fig. \ref{NPhi0_major_Nstar}), we are above the
second inflexion point so   we can have
at most two fundamental intersections between the line level $N$ and the
curves $\lbrace
N_{\alpha}(\Phi_0)\rbrace$.  This determines two fundamental branches in the
caloric
curve of Fig. \ref{Xkcal_R50_N0p45_unified2PH}.  The first and second
intersections
merge at $\alpha=\alpha_M$. Correspondingly, the first and second branches in
the caloric curve merge $\eta_c$, the first turning point of temperature. There
are also secondary intersections giving rise to the spiral (made of unstable
equilibrium
states) as described in the previous paragraph.

For larger values of $N$ the discussion is similar to the one given in
\cite{paper2}.

\subsection{$R=600$}
\label{sec_chin}

We now consider a system of size $R=600$ corresponding to the case analysed
in Sec. \ref{sec_second}. The novelty with respect to the previous situation is
that
microcanonical phase transitions can appear (they are related to
the appearance of turning points of energy).

\begin{figure}
\begin{center}
\includegraphics[clip,scale=0.3]{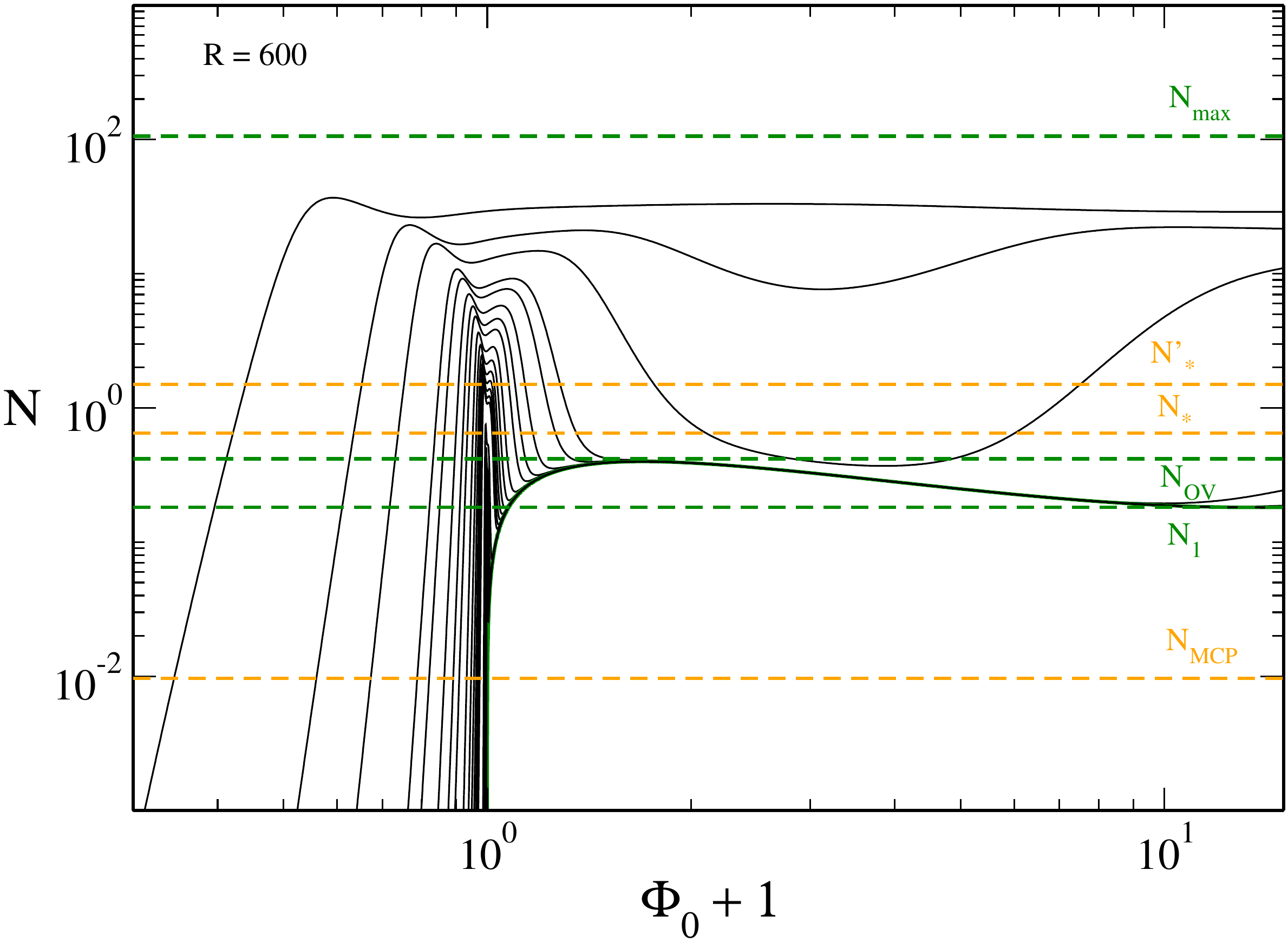}
\caption{Some curves $N_{\alpha}(\Phi_0)$ for $R=600$ together with different
characteristics values of $N$: $N_{\rm CCP} = 9.719 \times 10^{-6}$,
$N_{\rm
MCP}=0.00965$,
$N_1=0.18131$, $N_{\rm OV}=0.39853$, $N_*=0.418$, $N'_*=1.5$ and $N_{\rm
max}=106.057$.}
\label{NPhi0_Nkritik1_R600n}
\end{center}
\end{figure}

Some curves  $N_{\alpha}(\Phi_0)$ are plotted in Fig. \ref{NPhi0_Nkritik1_R600n}
for different values of $\alpha$.  We have indicated on this figure the
different
characteristic values of $N$ that have been identified in Sec.
\ref{sec_second}. As we have
seen in the previous section $N_{\rm max}=106.057$,  $N_{\rm
OV}=0.39853$, $N_1=0.18131$, $N_{\rm CCP}= 9.719 \times 10^{-6}$ and 
$N_{*}=0.418$ can be related to the topological properties of
the curves $\lbrace
N_{\alpha}(\Phi_0)\rbrace$. Unfortunately, $N_{\rm MCP}=0.00965$ and  $N'_*=1.5$
cannot be determined from a simple graphical construction
because there does not seem to be a simple manner to relate a turning point of
energy in the caloric curve $\eta(\Lambda)$ to the topological
properties of the curves $\lbrace
N_{\alpha}(\Phi_0)\rbrace$. Therefore, in Sec. \ref{sec_second}, we
had to determine $N_{\rm
MCP}$ and  $N'_*$ ``by hand'' directly from the study of the caloric curves as a
function
of $N$.

\begin{figure}
\begin{center}
\includegraphics[clip,scale=0.3]{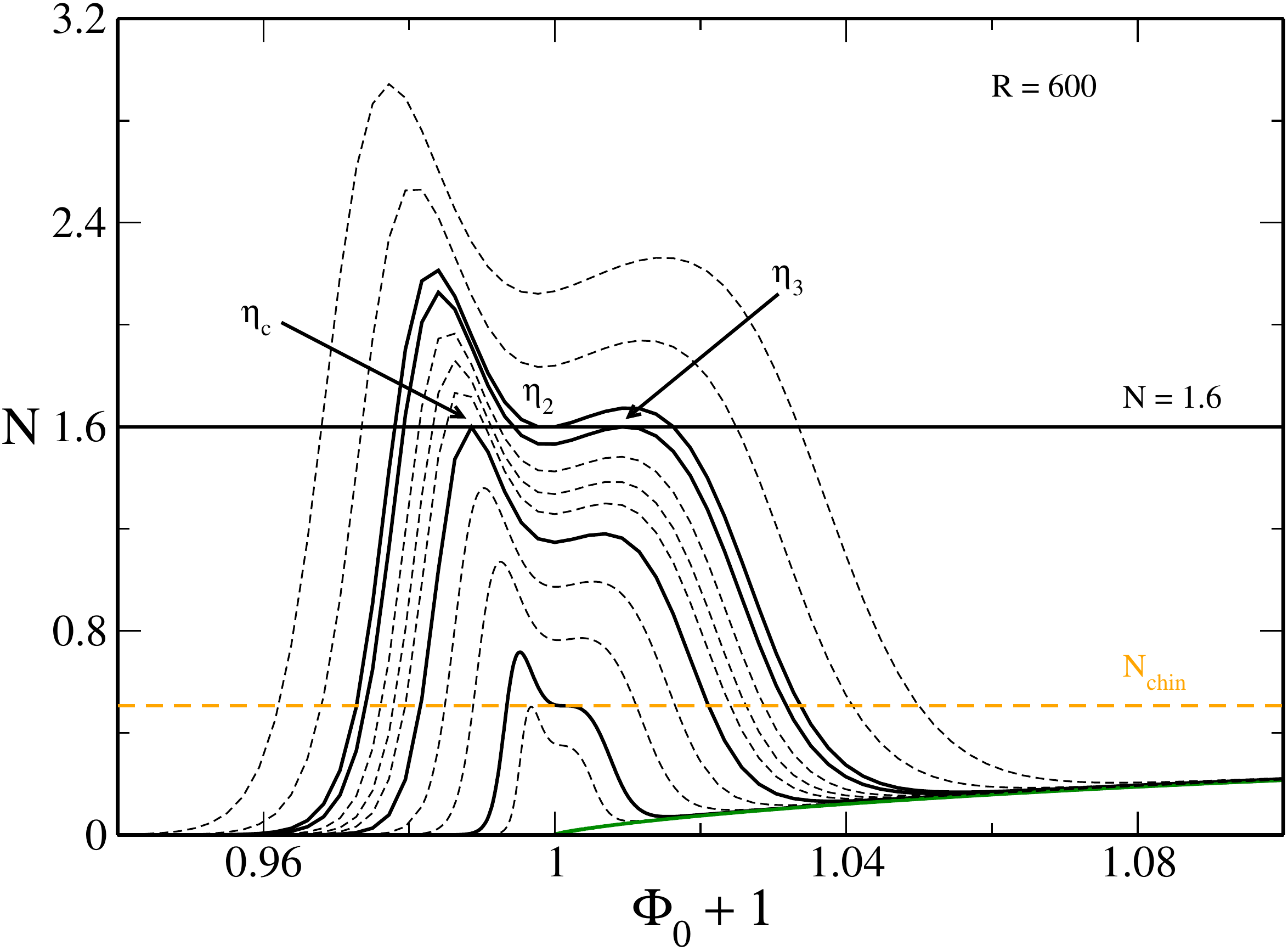}
\caption{Illustration
of the intersections in the case $N>N_{\rm chin}=0.5062$ (specifically $R=600$
and
$N=1.6$).}
\label{pkritik_N1p6_bPH}
\end{center}
\end{figure}

\begin{figure}
\begin{center}
\includegraphics[clip,scale=0.3]{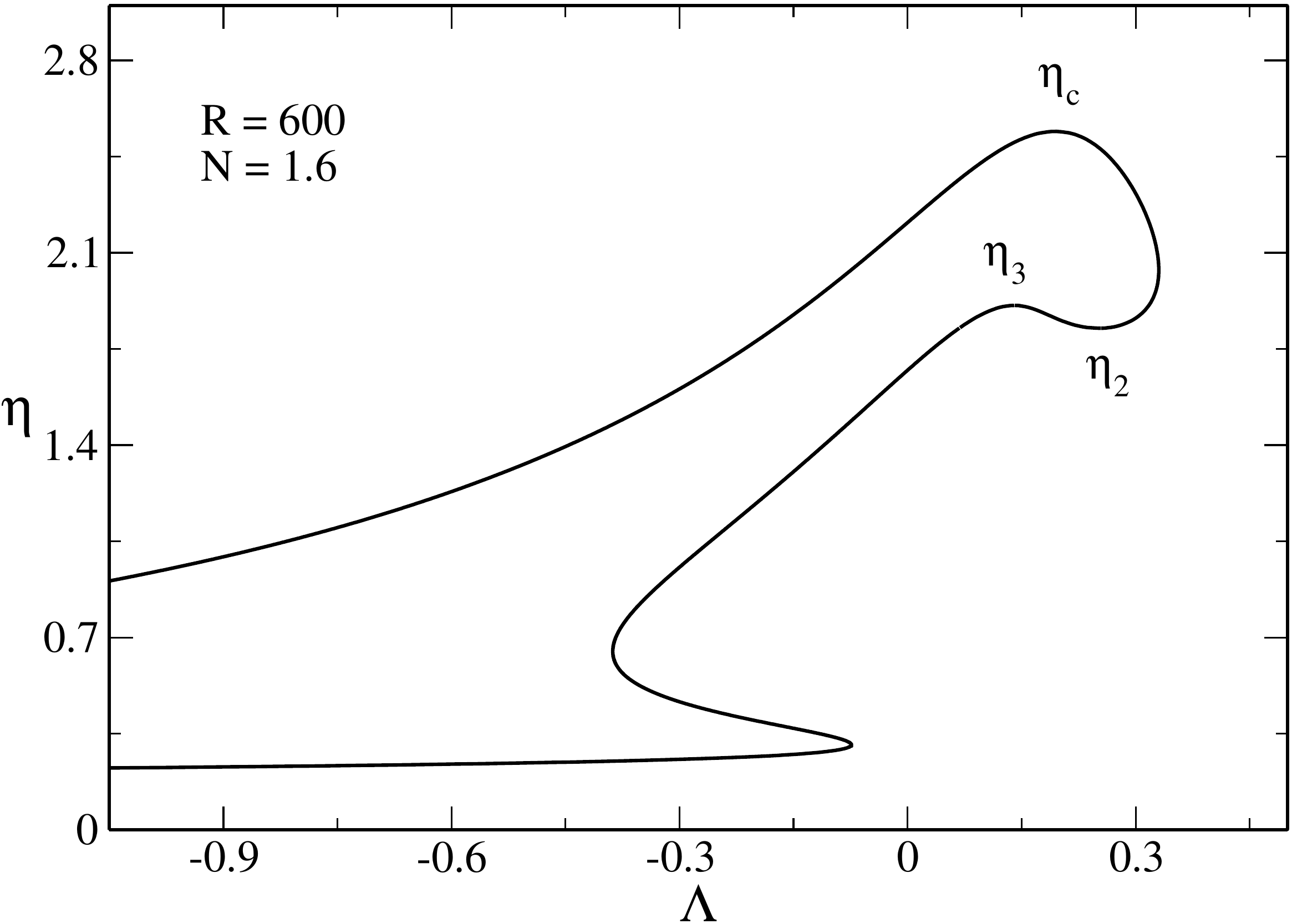}
\caption{Caloric curve for $N>N_{\rm chin}=0.5062$
(specifically $R = 600$ and $N =
1.6$).}
\label{kcal_R600_transition_lambda}
\end{center}
\end{figure}

Apart from the occurence of microcanonical phase transitions, there is another
novelty  with respect to the previous situation ($R=50$). A new inflexion point
appears at $N_{\rm chin}=0.5062$ (see Fig. \ref{pkritik_N1p6_bPH}). For
$N<N_{\rm
chin}$ we have two fundamental intersections in the $N=\lbrace
N_{\alpha}(\Phi_0)\rbrace$  plot leading to two branches of solutions in the
caloric curve $\eta(\Lambda)$  that merge at the temperature minimum $\eta_c$
(see Fig. \ref{Xkcal_R600_N0p29_unified_bnewPH}).\footnote{In this paragraph,
we do not consider the temperature maximum $\eta_*$ that is far away
from the dinosaur's head.} For $N>N_{\rm chin}$ we
have four fundamental intersections  in the $N=\lbrace
N_{\alpha}(\Phi_0)\rbrace$  plot leading to four branches of solutions that
merge at the temperature minimum $\eta_c$, at the temperature maximum $\eta_2$
and at the temperature minimum $\eta_{3}$ repectively (see Fig.
\ref{kcal_R600_transition_lambda}). In that case, the
dinosaur has a ``chin''. This is essentially a curiousity since the
solutions in this part of the caloric curve are unstable.

\subsection{$R=10$}

\begin{figure}
\begin{center}
\includegraphics[clip,scale=0.3]{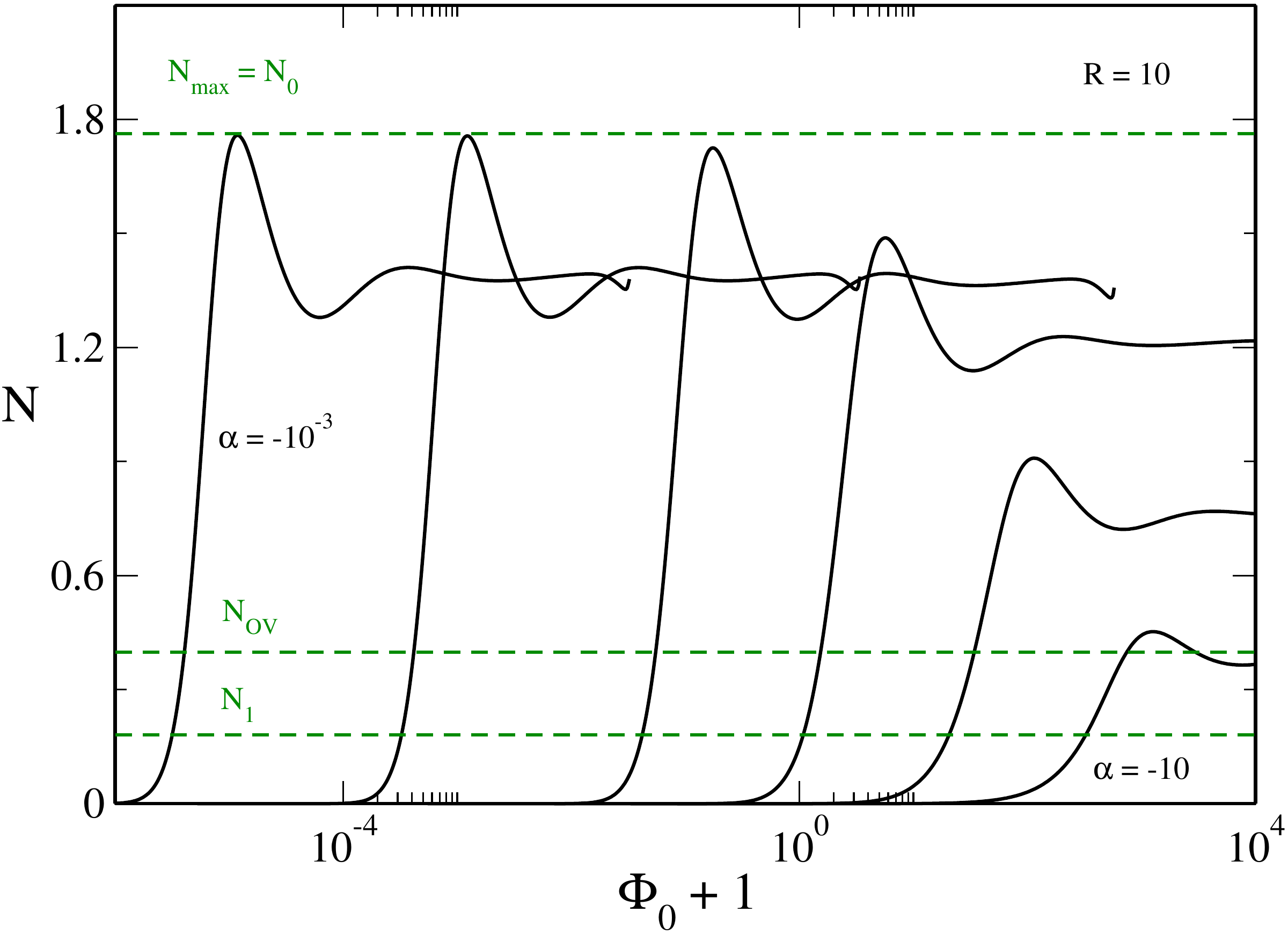}
\caption{Evolution of the curve $N_{\alpha}(\Phi_0)$ for
different values of $\alpha<0$ for $R=10$ (for illustration the curves
go from $\alpha=-10$ to $\alpha=-10^{-3}$).}
\label{NPhi0_R10_a_negn}
\end{center}
\end{figure}

\begin{figure}
\begin{center}
\includegraphics[clip,scale=0.3]{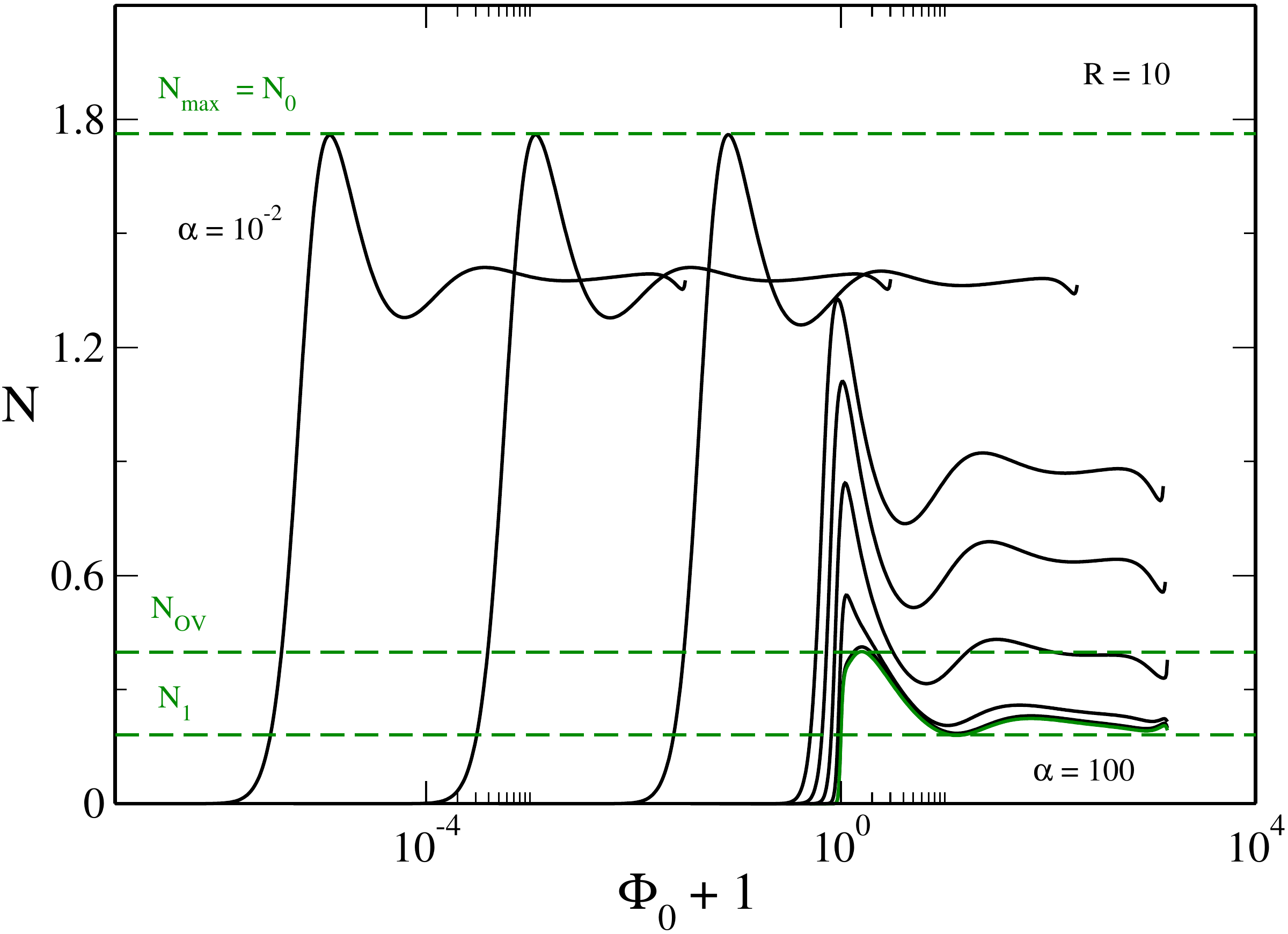}
\caption{Evolution of the curve $N_{\alpha}(\Phi_0)$ for
different values of $\alpha>0$ for $R=10$ (for illustration the curves
go from $\alpha=10^{-2}$ to $\alpha=100$).}
\label{NPhi0_R10_a_posnPH}
\end{center}
\end{figure}

We consider a system of size $R=10$ corresponding to the case analyzed in Sec.
\ref{sec_dix}. Some curves  $N_{\alpha}(\Phi_0)$ are plotted in Figs.
\ref{NPhi0_R10_a_negn} and \ref{NPhi0_R10_a_posnPH} for $\alpha<0$ and
$\alpha>0$ respectively. The difference with the case $R=50$ studied in
Appendix \ref{sec_cit} is that there is no canonical phase transition. This is
manifested by the absence of inflexion points in the family of
curves $\lbrace N_{\alpha}(\Phi_0)\rbrace$. We also note that, for $R=10$,
$N_{\rm max}$ is very close to $N_0$.\footnote{They become equal when
$\alpha_*=0$ corresponding to $R=R_t=12.255$ (see Fig.
\ref{alpha_NmaxR_bisNEWPH}).} However, this is essentially a
mathematical curiosity without physical consequences.

Let us describe the different intersections as a function of $N$. As already
mentioned, the intersections with the curves $\alpha<\alpha_*\simeq 0$
(see Fig. \ref{NPhi0_R10_a_negn}) give rise to the hot
spiral. This case has already been studied in \cite{paper2}. Therefore, we
consider below the intersections with the curves
$\alpha>\alpha_*\simeq 0$ (see Fig. \ref{NPhi0_R10_a_posnPH}).

For $N<N_1$, we have just one intersection between the line level $N$ and the
curves $\lbrace
N_{\alpha}(\Phi_0)\rbrace$. Therefore, the caloric curve is
monotonic and is similar to Fig. \ref{kcal_R50_N0p012_unified2PH}. The
intersection  between the line level $N$ and the OV-curve corresponds to the
ground state $T=0$ (i.e. $\eta\rightarrow +\infty$). This leads to the
vertical asymptote at $\Lambda=\Lambda_{\rm max}$ in the caloric curve.

For $N_1<N<N_{\rm OV}$, there is a second intersection
between the line level $N$ and  the
OV-curve $N_{\rm OV}(\Phi_0)$. This corresponds to an unstable equilibrium
state at $T=0$ (i.e. $\eta\rightarrow +\infty$). This gives rise  to the second
vertical asymptote at $\Lambda=\Lambda'_{\rm max}$ in the caloric curve of
Fig. \ref{kcal_R10_N0p36_unifiedPH}. There are also
secondary intersections
leading to the cold spiral that is apparent on Fig.
\ref{kcal_R10_N0p36_unifiedPH}.

For $N>N_{\rm OV}$, there are two fundamental intersections between the
line level $N$ and the
curves $\lbrace
N_{\alpha}(\Phi_0)\rbrace$.  This gives rise to two fundamental branches
in the caloric curve of Fig. \ref{kcal_R10_N0p4_unifiedPH}. The first and second
intersections
merge at $\alpha=\alpha_M$. Correspondingly,  the first and second branches in
the caloric curve merge at $\eta_c$, the first turning point of temperature.
Furthermore, there are
additional intersections giving rise to the cold spiral in
the caloric curve of Fig. \ref{kcal_R10_N0p4_unifiedPH}. These intersections are
less relevant since
they correspond to
unstable states. Note that there is no intersection with the OV curve so there
is no vertical asymptote corresponding to $\eta\rightarrow +\infty$.

For larger values of $N$ the discussion is similar to the one given  in
\cite{paper2}.

\subsection{$R=1$}

\begin{figure}
\begin{center}
\includegraphics[clip,scale=0.3]{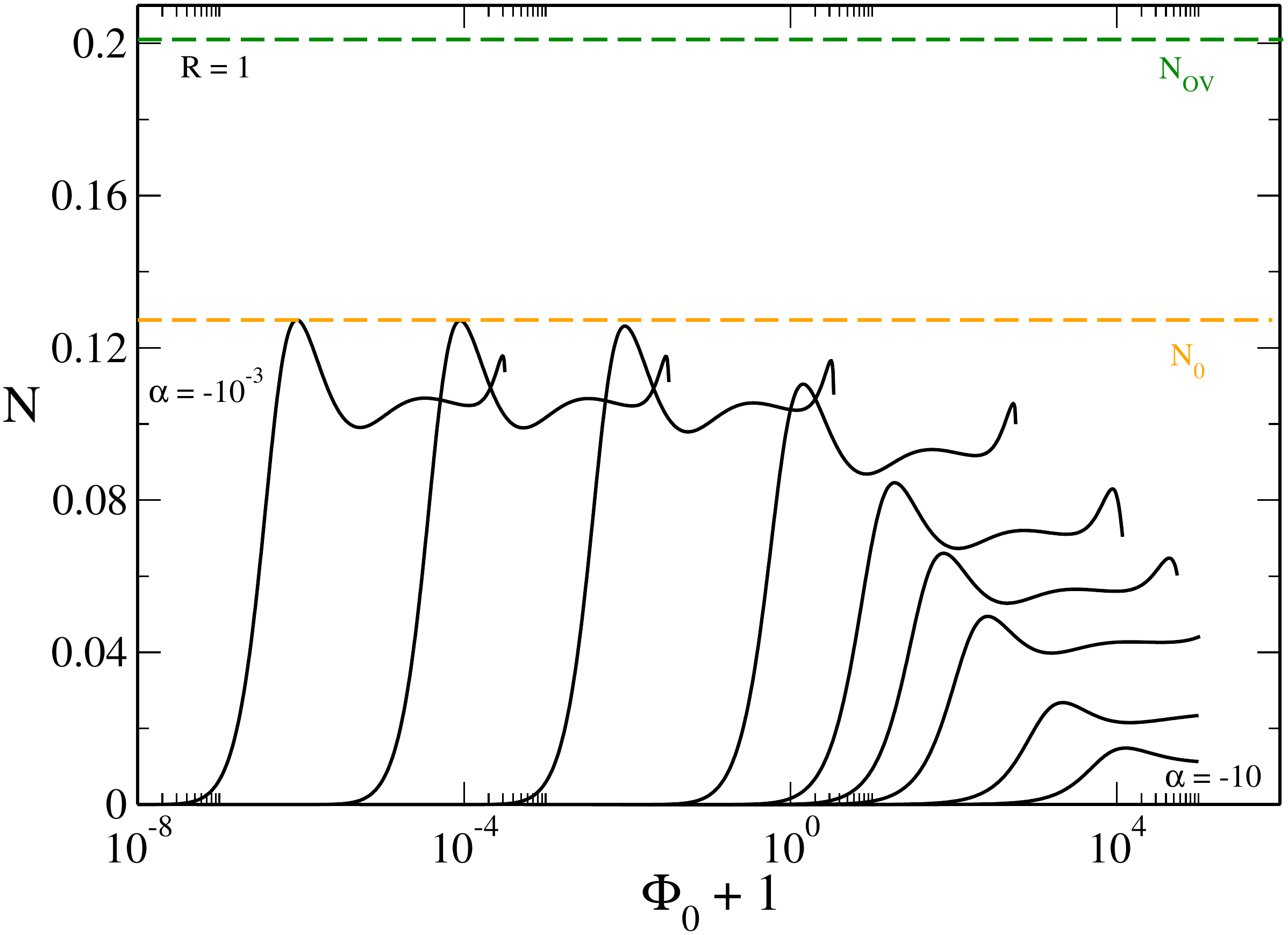}
\caption{Evolution of the curve $N_{\alpha}(\Phi_0)$ for
different values of $\alpha<0$ for $R=1$ (for illustration the curves
go from $\alpha=-10$ to $\alpha=-10^{-3}$).}
\label{NPhi0_power10_R1_negativen}
\end{center}
\end{figure}

\begin{figure}
\begin{center}
\includegraphics[clip,scale=0.3]{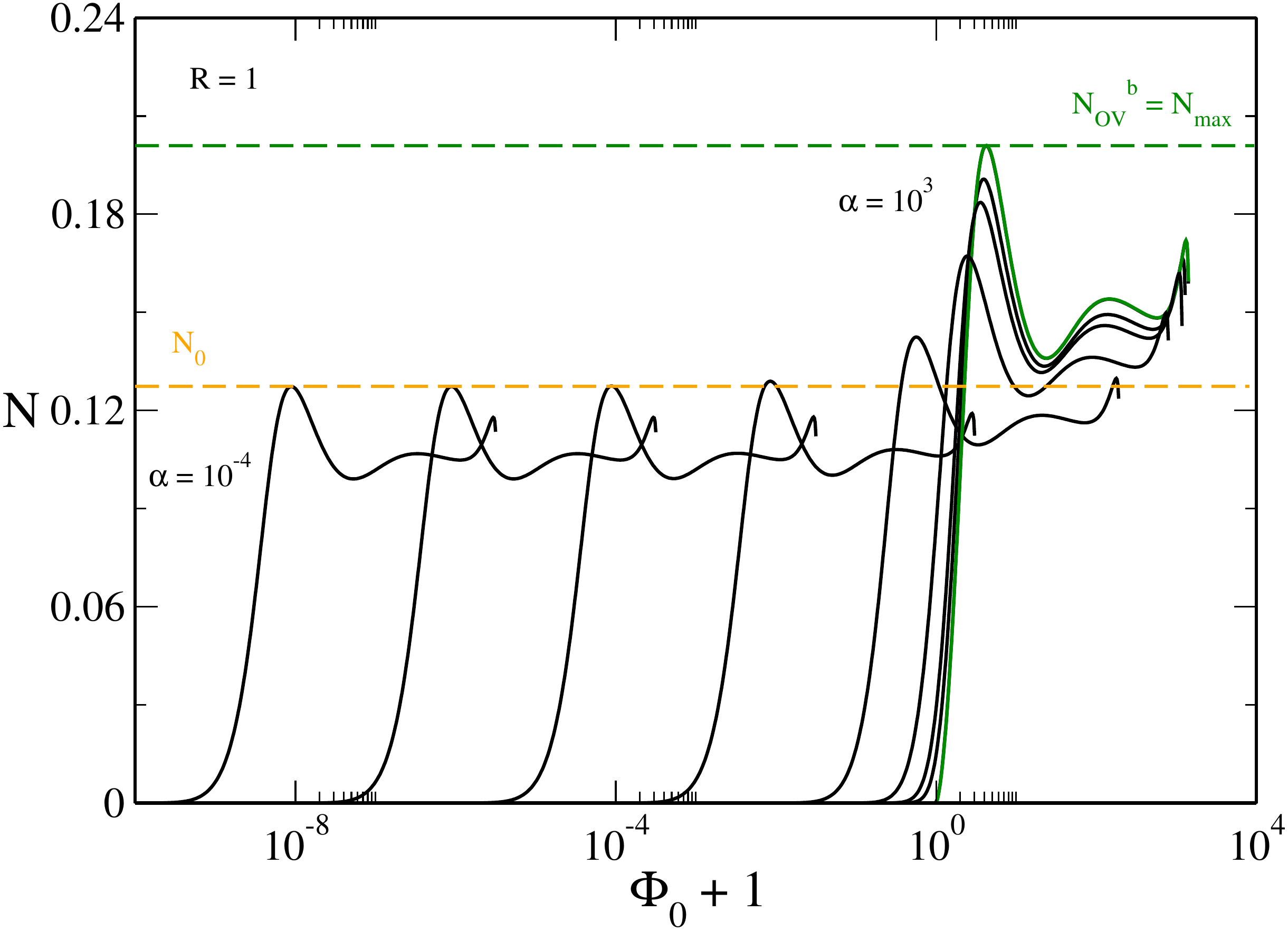}
\caption{Evolution of the curve $N_{\alpha}(\Phi_0)$ for
different values of $\alpha>0$ for $R=1$ (for illustration the curves
go from $\alpha=10^{-4}$ to $\alpha=10^3$).}
\label{NPhi0_power10_R1_positiven}
\end{center}
\end{figure}

We consider a system of size $R=1$ corresponding to the case analyzed in Sec.
\ref{sec_un}. Some curves  $N_{\alpha}(\Phi_0)$ are plotted in Figs.
\ref{NPhi0_power10_R1_negativen} and \ref{NPhi0_power10_R1_positiven} for
$\alpha<0$ and
$\alpha>0$ respectively.\footnote{Since $R=1<R_t=12.255$, implying
$\alpha_*>0$, we note that $N_{\rm max}(\Phi_0)$ is reached after
$N_{0}(\Phi_0)$ (see Figs. \ref{NPhi0_power10_R1_negativen} and
\ref{NPhi0_power10_R1_positiven}) while it was reached before
$N_{0}(\Phi_0)$ in the case $R=50$ (see Figs.
\ref{NPhi0_negative_R50n} and
\ref{NPhi0_positive_R50n}).} The difference with the
previous case is that now
$N_{\rm OV}(\Phi_0)$ is very close to $N_{\rm max}(\Phi_0)$. This is because
$\alpha_*\gg 1$ (see Fig. \ref{alpha_NmaxR_bisNEWPH}). Therefore  $N_{\rm OV}^b$
is very close to $N_{\rm max}$ and  $N_{1}^b$ is very close to
$N'_{\rm S}$.

Let us describe the different intersections as a function of $N$ (note that
most of the curves correspond to $\alpha<\alpha_*$).

\begin{figure}
\begin{center}
\includegraphics[clip,scale=0.3]{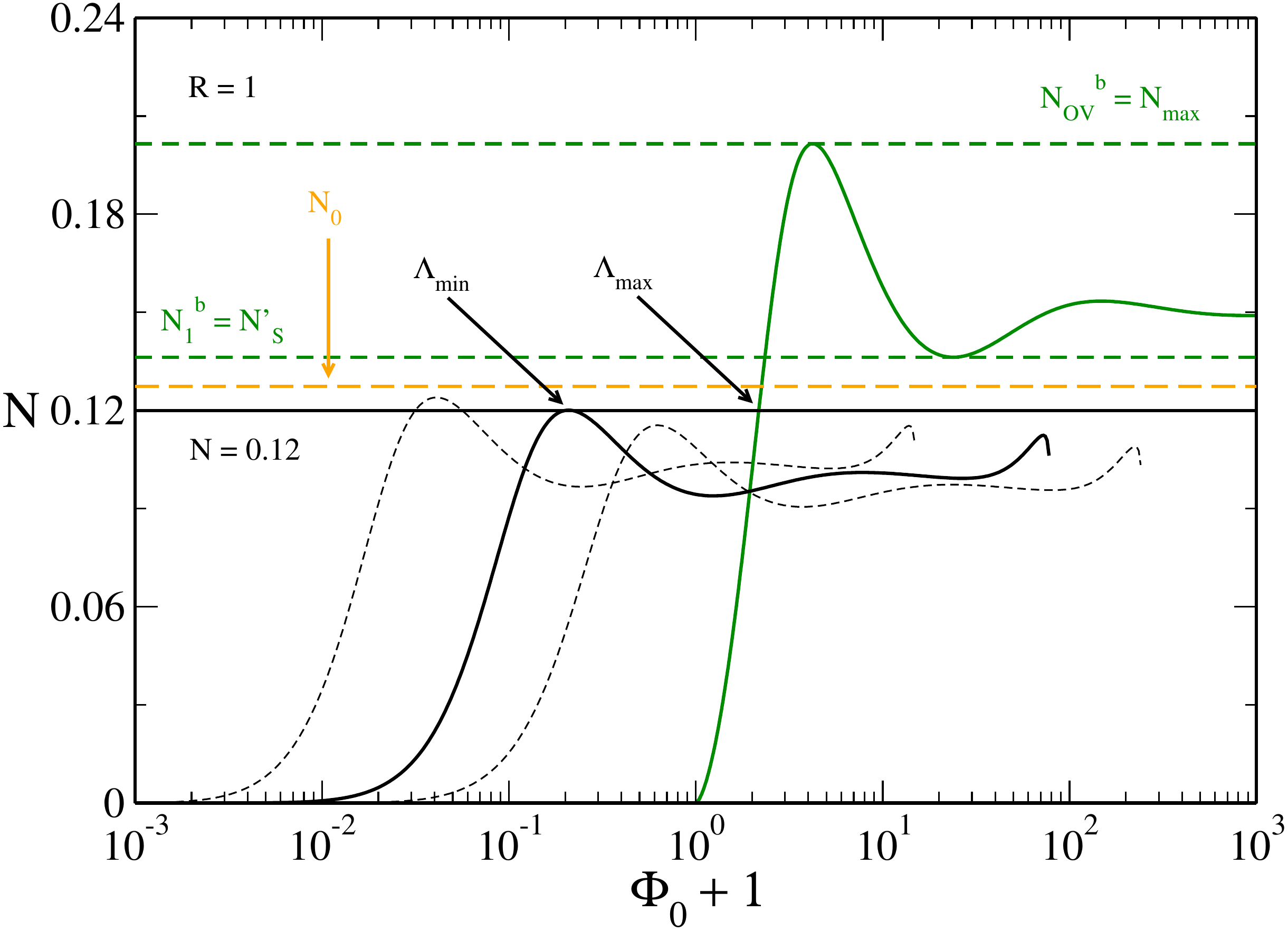}
\caption{Illustration of the intersections in the case $N<N'_{S}\simeq
N_{1}^b$ (specifically $R=1$ and $N=0.12$).}
\label{N012_Phi0_R1ter}
\end{center}
\end{figure}

For  $N<N'_{S}\simeq N_{1}^b$ (see Fig. \ref{N012_Phi0_R1ter}), we have one
intersection between the line level $N$ and the OV-curve which corresponds to
the ground state $T=0$ (i.e. $\eta\rightarrow +\infty$). This leads to the
vertical asymptote at $\Lambda=\Lambda_{\rm max}$ in the caloric curve of
Fig. \ref{kcal_R1_N0p12_intersectionsPH}. In addition, we can have up to an
infinity of
intersections with the curves  $N_{\alpha}(\Phi_0)$ leading to the hot spiral
displayed in the caloric curve of Fig. \ref{kcal_R1_N0p12_intersectionsPH}.

\begin{figure}
\begin{center}
\includegraphics[clip,scale=0.3]{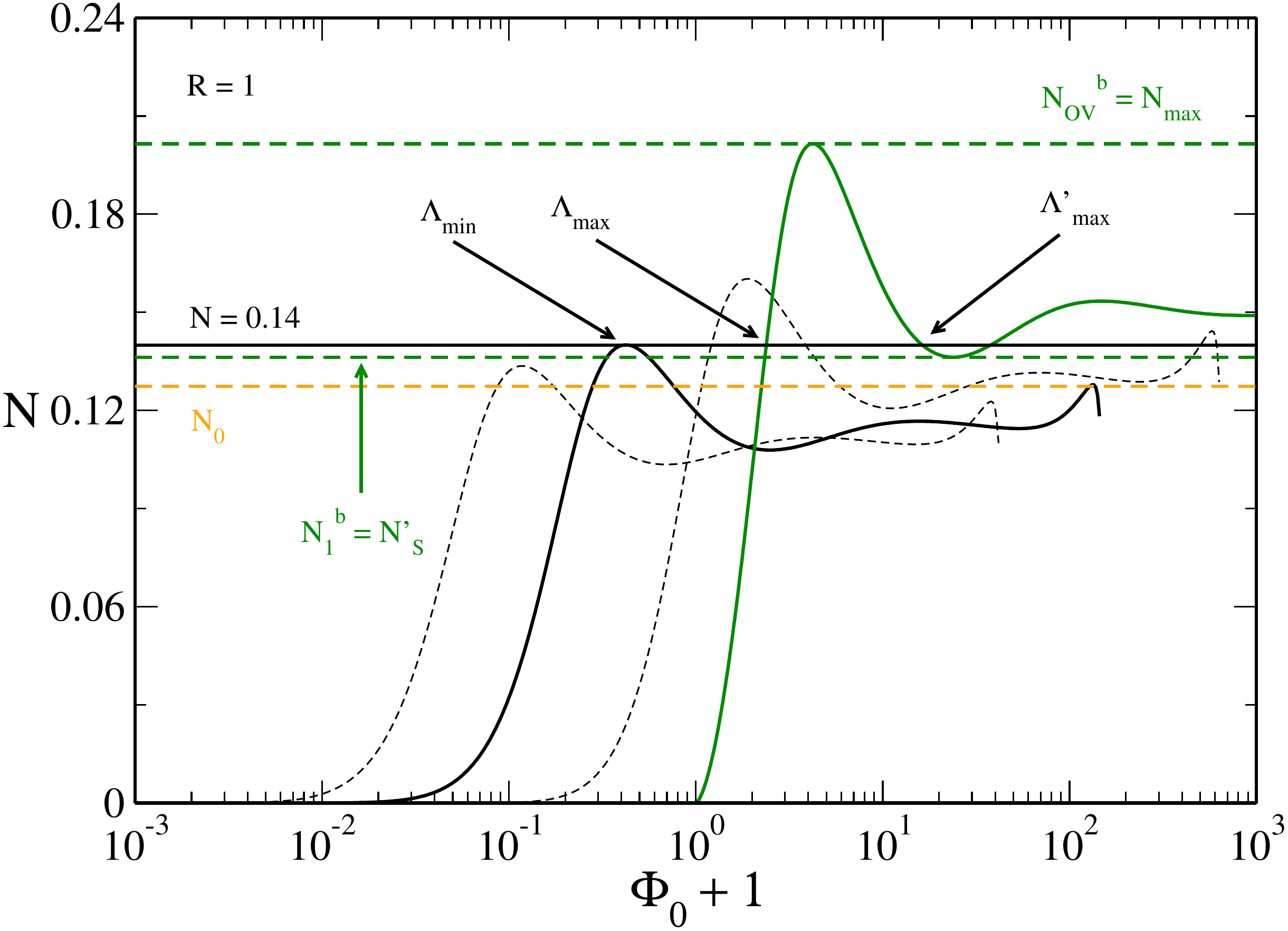}
\caption{Illustration of the intersections in the case $N'_{S}\simeq
N_{1}^b<N<N_{\rm max}\simeq N_{\rm OV}^b$ (specifically $R=1$ and 
$N=0.14$).}
\label{N014_Phi0_R1ter}
\end{center}
\end{figure}

For  $N'_{S}\simeq N_{1}^b<N<N_{\rm max}\simeq N_{\rm OV}^b$ (see
Fig. \ref{N014_Phi0_R1ter}), we have two fundamental intersections between the
line level $N$ and the OV-curve (corresponding to
$\alpha\rightarrow +\infty$) leading to
the asymptotes $\eta=+\infty$ at $\Lambda_{\rm max}$ (stable ground
state) and $\Lambda'_{\rm max}$ (unstable ground state) in the caloric curve of
Fig. \ref{kcal_R1_N0p14_intersectionsPH}. For smaller values of $\alpha$, the
two fundamental
intersections between the line level $N$  and  the curves  $\lbrace
N_{\alpha}(\Phi_0)\rbrace$ lead to two fundamental branches in the caloric
curve of Fig. \ref{kcal_R1_N0p14_intersectionsPH}. These intersections  merge at
$\alpha=\alpha_n$. Since $\alpha_n<\alpha_*$, this is associated with a turning
point of energy at $\Lambda_{\rm
min}$ in the caloric curve of Fig. \ref{kcal_R1_N0p14_intersectionsPH}.
There may also be a third intersection with the OV-curve  and
secondary intersections with the curves $\lbrace
N_{\alpha}(\Phi_0)\rbrace$  forming a third  branch exhibiting an asymptote at
$\Lambda''_{\rm max}$ and a spiral. However, these solutions are not
represented on the caloric curve because they are associated with unstable
states of high order.

\section{Condition to be degenerate}
\label{sec_av}

In the nonrelativistic limit, the system is degenerate
when the thermal pressure $P=\rho k_B T/m$ is small as compared to the quantum
pressure $P=(1/5)(3/4\pi g)^{2/3}h^2\rho^{5/3}/m^{8/3}$
arising from the Pauli exclusion principle. This condition can be written
as
\begin{equation}
\label{av1}
k_B T\ll \frac{1}{5}\left (\frac{3}{4\pi g}\right
)^{2/3}\frac{h^2}{m^{5/3}}\rho^{2/3}
\end{equation}
which is the nonrelativistic Sommerfeld criterion (the
right hand side of Eq. (\ref{av1}) is of the order of the Fermi
temperature $T_F$). 
To get an estimate of the importance of degeneracy, we replace the density
$\rho$ by the average density $\overline{\rho}=3M/4\pi R^3$. In this manner, we
obtain the
condition
\begin{equation}
\label{av2}
\eta\gg 0.917\, \mu^{2/3}.
\end{equation}
As stated above this condition is only valid in an average sense. A system
which does not
satisfy this condition in average may still have a degenerate core and a
nondegenerate halo as in Appendix \ref{sec_thermoq}.

In the ultrarelativistic limit, the system is
degenerate
when the thermal pressure $P=\rho k_B T/m$ is small as compared to the quantum
pressure  $P=(1/4)(3/4\pi g)^{1/3}hc\rho^{4/3}/m^{4/3}$
arising from the Pauli exclusion principle. This can be written as
\begin{equation}
\label{av3}
k_B T\ll \frac{1}{4}\left (\frac{3}{4\pi g}\right
)^{1/3}\frac{hc}{m^{1/3}}\rho^{1/3}
\end{equation}
which is the ultrarelativistic Sommerfeld criterion.
Proceeding as above, we obtain the condition
\begin{equation}
\label{av4}
\eta\gg 1.10\, \left (\frac{M}{M_{\rm OV}}\right)^{2/3},
\end{equation}
where $M_{\rm OV}$ is the OV critical mass given by Eq. (\ref{dq1}).

\section{Thermodynamics of nonrelativistic self-gravitating systems}
\label{sec_thermo}

In this Appendix, we recall and complete important results concerning the 
thermodynamics of nonrelativistic self-gravitating systems (classical particles
and fermions) that are needed in our analysis. We refer to Appendix \ref{sec_bf}
for useful formulae that are used throughout this Appendix.

\subsection{Classical particles}
\label{sec_thermoclass}

We first consider a self-gravitating system of nonrelativistic classical
particles confined within
a spherical box of radius $R$. We show below that there is no statistical
equilibrium state in a strict sense but that long-lived metastable states
can exist under certain
conditions.

\subsubsection{Canonical ensemble}
\label{sec_thermoclassce}

In the canonical ensemble, one can make the free energy
$F=E-TS$ diverge towards $-\infty$ at fixed mass $M$ in the most efficient
manner by approaching all the particles at
the same point (see Appendix B of \cite{sc}). Indeed, let us consider a
homogeneous sphere of radius $a$ containing
all the particles. When
$a\rightarrow 0$ the potential energy
$W=-3GM^2/(5a)$ diverges to $-\infty$. The entropy $S\sim
3Nk_B \ln a$ diverges
to $-\infty$ but it is subdominant. As a result, the free energy $F=E-TS\sim
W\sim -3GM^2/(5a)$
diverges to $-\infty$. Therefore, there is no global minimum of free energy.
In a sense, the most
probable structure in the canonical
ensemble is a Dirac peak containing all the mass.

On the other hand, there exist metastable gaseous states with a temperature
$T>T_c=0.397
GMm/(k_BR)$ \cite{lbw,aa}. They
have a density contrast ${\cal R}<32.1$  \cite{lbw,aa} and they are very
long-lived
\cite{lifetime}. When
$T<T_c$, or when $T>T_c$ and ${\cal R}>32.1$, there are no metastable states
anymore and the system
collapses (isothermal collapse) \cite{aa}. According to the previous
thermodynamical argument, it is expected to form a Dirac peak
containing
all the mass. By solving the Smoluchowski-Poisson equations describing the
canonical evolution of
self-gravitating Brownian particles \cite{sc,post} it is found that the Dirac
peak is
formed
in the postcollapse regime of the dynamics.

\subsubsection{Microcanonical ensemble}
\label{sec_thermoclassmce}

In the microcanonical ensemble, one can make the entropy $S$ diverge towards
$+\infty$ at fixed mass $M$ and
energy $E$ by forming a core-halo structure and
letting the size of the core
go to zero (see Appendix A of \cite{sc}). Indeed, let us consider a
homogeneous core made of $N_C$ particles in a
sphere of radius $R_C$.
Its potential energy  $W_C=-3GM_C^2/(5R_C)$ tends to
$-\infty$ when $R_C\rightarrow 0$. In order to conserve the total energy, the
kinetic energy of the halo $K_h=(3/2)N_h k_B T$ must tend to $+\infty$ like
$K_h\sim -W_C$ meaning that its temperature $T\sim 2GM_C^2/(5R_CN_hk_B)$ tends
to
$+\infty$. As a result, the entropy of the system behaves as
$S\sim -(3/2)k_B(N_h-N_C)\ln R_C$ and tends to $+\infty$ when $R_C\rightarrow
0$. Therefore, there is no global maximum of entropy at fixed energy. We
note that the divergence of the entropy is the most efficient when the core
contains a few particles ($N_C\ll N_h$). Actually, we only need to approach
$2$ particles to each other and  make a tight binary. Its potential energy
$-Gm^2/a$ diverges towards $-\infty$. The released energy serves to heat the
halo made of the $N-2$ other particles. This  produces the most efficient
divergence of entropy (note that
the divergence of entropy is weak --- logarithmic). In a sense, the most
probable structure
in the microcanonical ensemble is a tight binary surrounded by a hot halo. This
can be seen as a Dirac peak of zero mass but infinite potential energy $+$ a hot
halo.

On the other hand, there exist metastable gaseous states with an energy
$E>E_c=-0.335 GM^2/R$
\cite{lbw}. They
have a density contrast ${\cal R}<709$ \cite{antonov,lbw} and they are very
long-lived
\cite{lifetime}. When
$E<E_c$, or when $E>E_c$ and ${\cal R}>709$, there are no metastable states
anymore and the system
collapses (gravothermal catastrophe) \cite{lbw}. It is expected to form a binary
surrounded by a hot halo. Dynamical models describing
the
collisional evolution of globular clusters (fluid equations, orbit-averaged
Fokker-Planck equation...) show that the binary is formed in the postcollapse
regime of the dynamics
\cite{lbe,inagaki,cohn}. The
energy released by the binary can to stop the collapse and
induce a re-expansion of the halo. Then, in
principle, a series of gravothermal oscillations should
follow \cite{oscillations,hr}.

\subsection{Fermions}
\label{sec_thermoq}

We now consider a self-gravitating system of nonrelativistic fermions confined
within a spherical box
of
radius $R$. In that case, there exists a statistical equilibrium state at any
energy $E$ and temperature $T$ \cite{ijmpb}. We consider a situation where we
are close to
the classical limit ($\mu\rightarrow +\infty$ or $\hbar\rightarrow 0$). When
$T>T_c$ and
$E>E_c$ there exist metastable gaseous states that are not affected by quantum
mechanics (see Appendix \ref{sec_thermoclass}). When $T<T_c$ and $E<E_c$ there
are no metastable
states anymore. The system collapses and becomes very dense until quantum
mechanics (Pauli's
exclusion principle) comes into play. Generically, the system forms a core-halo
structure with a completely degenerate fermionic core of mass $M_C$ and
radius $R_C$ surrounded by an (almost classical) isothermal halo of mass
$M_h=M-M_C$ and radius
$R$. We can obtain the value of $M_C$ from a simple analytical model developed
in \cite{pt}. We summarize this model below by using dimensional variables in
order to understand more easily the physical mechanisms at play.

We model the core as a completely degenerate fermion ball. Its
mass-radius relation is
\begin{eqnarray}
\label{pre1}
M_C R_C^3=\chi\frac{h^6}{g^2m^8G^3}.
\end{eqnarray}
Its energy (kinetic $+$ potential) is 
\begin{equation}
\label{pre2}
E_C=-\frac{3}{7\chi^{1/3}}\frac{G^2M_C^{7/3}g^{2/3}m^{8/3}}{h^2}.
\end{equation}
Its entropy is zero: $S_C=0$.

We model the halo by a classical gas at
temperature $T$ with a uniform density.\footnote{We will see that its
temperature is very large in the situations considered. This
justifies the uniform density approximation.} Its kinetic
energy is
\begin{equation}
\label{pre3}
K_h=\frac{3}{2}N_hk_B T.
\end{equation}
Its potential (gravitational) energy, taking into account the presence of the
core,
is
\begin{equation}
\label{pre4}
W_h=-\frac{3GM_CM_h}{2R}-\frac{3GM_h^2}{5R}.
\end{equation}
Its entropy is
\begin{eqnarray}
\label{pre5}
S_{h}=-N_h k_B\ln \left (\frac{M_h}{V}\right )+\frac{3}{2}N_h k_B\ln \left
(\frac{2\pi
k_B T}{m}\right )\nonumber\\
+\frac{5}{2}N_h k_B+N_h k_B\ln\eta_0.
\end{eqnarray}
In the foregoing expressions, we have assumed $R_C\ll R$ which can be checked a
posteriori.

The total mass of the system is $M=M_C+M_h$, its total energy is
$E=E_C+K_h+W_h$, its total entropy is $S=S_h$ and its total free energy is
$F=E-TS$. The mass $M_C$ of the core
is obtained by maximizing the
entropy at fixed mass and energy in the microcanonical ensemble or by
minimizing the free energy at
fixed mass in the canonical ensemble. The extremization problem gives in both
ensembles \cite{pt}:
\begin{eqnarray}
\label{pre6}
-\frac{G^2M_C^{4/3}g^{2/3}m^{8/3}}{\chi^{1/3}h^2}-\frac{3G(M-2M_C)}{2R}
\nonumber\\
+\frac{ 6G(M-M_C)}{5R}
-\frac{k_B T}{m}\ln\left (\frac{M-M_C}{V}\right
)\nonumber\\
+\frac{3}{2}\frac{k_B T}{m}\ln \left (\frac{2\pi k_B T}{m}\right
)+\frac{k_B T}{m}\ln \eta_0=0.
\end{eqnarray}
This equation may have several solutions that have been analyzed in
detail in \cite{pt}. Below, we restrict ourselves to the stable 
condensed state.

\subsubsection{Canonical ensemble}
\label{sec_thermoqce}

In the canonical ensemble, when $h\rightarrow 0$, we expect that the core
contains a
large mass $M_C/M\sim 1$ (see Appendix \ref{sec_thermoclassce}). Guided by this
ansatz, which can be checked a
posteriori, one can see from Eq. (\ref{pre6}) that the core mass is given
by
\begin{equation}
\label{pre7}
1-\frac{M_C}{M}\sim \frac{\eta_0V}{M}\left (\frac{2\pi k_B T}{m}\right
)^{3/2}e^{\frac{3GMm}{2Rk_B T}}
e^{-\frac{G^2M^{4/3}g^{2/3}m^{11/3}}{\chi^{1/3}h^2 k_B T}}.
\end{equation}
When $h\rightarrow 0$, the core mass tends towards $M$ exponentially
rapidly. Therefore the core contains almost all the mass: $M_C\sim M$. Its
radius is given by  
\begin{eqnarray}
\label{pre8}
R_C=\chi^{1/3}\frac{h^2}{g^{2/3}m^{8/3}GM^{1/3}}.
\end{eqnarray}
When $h\rightarrow 0$, it tends to zero as $h^2$. The energy of the core is
\begin{equation}
\label{pre9}
E_C=-\frac{3}{7\chi^{1/3}}\frac{G^2M^{7/3}g^{2/3}m^{8/3}}{h^2}.
\end{equation}
When $h\rightarrow 0$, it tends to $-\infty$ as $-h^{-2}$. The free energy,
which is dominated by the energy of the core, $F\sim E_C$, behaves in a similar
manner. In the
classical limit $h\rightarrow 0$, we recover the Dirac
peak containing all the mass. This structure leads to the divergence of the free
energy in agreement with the arguments of Appendix \ref{sec_thermoclassce}.  We
note that these results are independent of the presence, or not, of the box.

In terms of dimensionless variables \cite{pt}, the preceding
results can be written as
\begin{equation}
\label{end1}
1-\alpha_C\sim
\frac{\sqrt{\pi}}{6}\mu\frac{e^{3\eta/2}}{\eta^{3/2}}e^{-\lambda\eta\mu^{2/3}},
\end{equation}
\begin{equation}
\label{end2}
\frac{R_C}{R}\sim\frac{1}{\lambda\mu^{2/3}},
\end{equation}
\begin{equation}
\label{end3}
\Lambda_C\sim\Lambda_{\rm max}=\frac{3}{7}\lambda\mu^{2/3},
\end{equation}
where $\alpha_C=M_C/M$.
We note that letting $h\rightarrow 0$ (classical limit) in the dimensional
equations is equivalent to letting $\mu\rightarrow +\infty$ in the dimensionless
equations.

{\it Remark:} Using the results of Ref. \cite{pt}, the maximum
temperature of the condensed phase, and the corresponding core mass, are given
by
\begin{equation}
\eta_*\sim \frac{2\ln\mu}{\lambda\mu^{2/3}},\qquad 1-\alpha_*\sim
\frac{3}{8\ln\mu}.
\end{equation}
Coming back to dimensional variables, we get\footnote{Here and in the
following, we give the logarithmic correction in the dominant approximation.
Furthermore, $-\ln h$ should be understood as $(1/3)\ln\mu$ where
$\mu=(gm^4/h^3)\sqrt{512\pi^4G^3MR^3}$ is dimensionless.}
\begin{equation}
k_B T_*\sim \frac{1}{6\chi^{1/3}}\frac{g^{2/3}G^2M^{4/3}m^{11/3}}{h^2 (-\ln h)}.
\end{equation}
On the other hand, the temperature of transition, and the
corresponding core mass, are given by
\begin{equation}
\eta_t\sim \frac{14\ln\mu}{3\lambda\mu^{2/3}},\qquad 1-\alpha_t\sim
\frac{\sqrt{\pi}}{6}\left (\frac{3\lambda}{14}\right
)^{3/2}\frac{1}{\mu^{8/3}(\ln\mu)^{3/2}}.
\end{equation}
Coming back to dimensional variables, we get
\begin{equation}
k_B T_t\sim \frac{1}{14\chi^{1/3}}\frac{g^{2/3}G^2M^{4/3}m^{11/3}}{h^2 (-\ln
h)}.
\end{equation}

\subsubsection{Microcanonical ensemble}
\label{sec_thermoqmce}

In the microcanonical ensemble, we expect that the core contains a small mass
$M_C/M\ll 1$ (see Appendix \ref{sec_thermoclassmce}). Guided by this ansatz,
which can be checked a
posteriori, one can see from Eq. (\ref{pre6})
that the core mass is given by
\begin{eqnarray}
\label{pre10}
\frac{M_C}{M}\sim \frac{7}{12(-\ln h)}.
\end{eqnarray}
When $h\rightarrow 0$, the core mass tends towards $0$ extremely slowly
(logarithmically). Therefore, the core contains a fraction of the total
mass
and this fraction goes to zero as $(-\ln h)^{-1}$ when $h\rightarrow 0$. The 
radius of the core is given
by  
\begin{eqnarray}
\label{pre11}
R_C\sim \left (\frac{12}{7}\right )^{1/3}\chi^{1/3}\frac{h^2(-\ln
h)^{1/3}}{g^{2/3}m^{8/3}GM^{1/3}}.
\end{eqnarray}
When $h\rightarrow 0$, it tends to zero as $h^2 (-\ln h)^{1/3}$. The energy
of the core is
\begin{equation}
\label{pre12}
E_C\sim -\frac{3}{7\chi^{1/3}}\left (\frac{7}{12}\right
)^{7/3}\frac{G^2M^{7/3}g^{2/3}m^{8/3}}{h^2(-\ln
h)^2}.
\end{equation}
When $h\rightarrow 0$, it tends to $-\infty$ as $-h^{-2}(-\ln
h)^{-2}$. Since the energy of the core is very
negative the kinetic energy of the halo must be very positive in order to
conserve the
total energy. It must behave as $K_h\sim -E_C$. Therefore, the temperature of
the halo must be
very large:
\begin{equation}
\label{pre13}
k_B T_{\rm cond}\sim \frac{2}{7\chi^{1/3}}\left
(\frac{7}{12}\right
)^{7/3}\frac{G^2M^{4/3}g^{2/3}m^{11/3}}{h^2(-\ln
h)^2}.
\end{equation}
When $h\rightarrow 0$, it diverges to $+\infty$ as $h^{-2}(-\ln
h)^{-2}$. The entropy behaves as $S\sim -6Nk_B \ln h$. Subtracting the term 
$-3Nk_B \ln h$ that we get even in the absence of gravity (see Appendix
\ref{sec_bf}), we obtain
\begin{equation}
\label{pre14}
\Delta S\sim -3Nk_B \ln h.
\end{equation}
When $h\rightarrow 0$, the entropy diverges to $+\infty$ as $-\ln h$. In the
classical limit $h\rightarrow 0$, we recover the core-halo structure made of a
core having a small mass, a small radius and a huge potential energy (Dirac peak
of zero mass) surrounded by a very hot halo.\footnote{We note
that the collapse at {\it low} energies in the microcanonical ensemble
(gravothermal
catastrophe) produces {\it hot} systems ($T\rightarrow +\infty$) with a
core-halo structure. Actually, although the temperature is uniform throughout
the system, the halo is hot while the core is cold. Indeed, the halo is
nondegenerate (Boltzmannian) because $T\gg T_F$, where $T_F$ is the Fermi
temperature (see Appendix \ref{sec_av}), while the core is
completely degenerate  because $T\ll T_F$.
Fundamentally, this core-halo structure is the consequence of the negative
specific
heat of self-gravitating systems as explained in Ref. \cite{lbw}.} This
core-halo structure
leads to the (logarithmic) divergence of the entropy in agreement with the
results of Appendix
\ref{sec_thermoclassmce}. We
note that 
these results are independent of the presence, or not, of the  box. They are
also independent of the value of the energy $E$ provided that it is not too
extreme.

In terms of dimensionless variables \cite{pt}, the preceding
results can be rewritten as
\begin{equation}
\label{fin1}
\alpha_C\sim \frac{7}{4\ln\mu},
\end{equation}
\begin{equation}
\label{fin2}
\frac{R_C}{R}\sim\frac{1}{\lambda}\left (\frac{4}{7}\right
)^{1/3}\frac{(\ln\mu)^{1/3}}{\mu^{2/3}},
\end{equation}
\begin{equation}
\label{fin3}
\Lambda_C\sim\frac{3}{7}\lambda\left (\frac{7}{4}\right
)^{7/3}\frac{\mu^{2/3}}{(\ln\mu)^{7/3}},
\end{equation}
\begin{equation}
\label{fin4}
\eta_{\rm cond}\sim
\frac{7}{2\lambda}\left
(\frac{4}{7}\right
)^{7/3}\frac{(\ln\mu)^{7/3}}{\mu^{2/3}},
\end{equation}
where $\alpha_C=M_C/M$. More generally (without specifying the value of the core
mass $\alpha_C$), in
the case where the energy of the core tends to
$-\infty$ and the energy of the halo tends to $+\infty$ we have the relation
\begin{equation}
\label{fin5}
\eta_{\rm cond}\sim
\frac{1-\alpha_C}{\alpha_C^{7/3}}\frac{7}{2\lambda\mu^{2/3}}.
\end{equation}
When $\alpha_C$ is given by Eq. (\ref{fin1}) obtained from Eq. (\ref{pre6}),
we recover Eq. (\ref{fin2}). Finally, we note that letting $h\rightarrow 0$
(classical limit) in the dimensional
equations is equivalent to letting $\mu\rightarrow +\infty$ in the dimensionless
equations.

{\it Remark:} Using the results of Ref. \cite{pt}, the maximum energy of the
condensed phase, and the corresponding core mass, are given
by
\begin{equation}
\Lambda_*\sim -\frac{9\lambda}{28}\frac{\mu^{2/3}}{(\ln\mu)^{7/3}},\qquad
\alpha_*\sim
\frac{1}{\ln\mu}.
\end{equation}
Coming back to dimensional variables, we get
\begin{equation}
E_*\sim \frac{1}{28(3\chi)^{1/3}}\frac{g^{2/3}G^2M^{7/3}m^{8/3}}{h^2
(-\ln h)^{7/3}}.
\end{equation}
The energy of transition $\Lambda_t$, and the
corresponding core mass $\alpha_t$, have the same scalings.

\section{Useful formulae}
\label{sec_bf}

In this Appendix, we regroup basic formulae that are
useful in our study.

\subsection{Energy and entropy}

The energy of a nonrelativistic self-gravitating system can be written as
\begin{equation}
\label{bf1}
E=\int f\frac{v^2}{2}\, d{\bf r}d{\bf v}+\frac{1}{2}\int\rho\Phi\,
d{\bf r}=K+W,
\end{equation}
where $K$ is the kinetic energy and $W$ the potential (gravitational)
energy \cite{bt}.

The Fermi-Dirac entropy functional is given by
\begin{equation}
\label{bf2}
S=-k_B\frac{\eta_0}{m}\int\Biggl \lbrace \frac{f}{\eta_0} \ln
\frac{f}{\eta_0}+\left (1-\frac{f}{\eta_0}\right )\ln \left
(1-\frac{f}{\eta_0}\right )\Biggr\rbrace\, d{\bf r}d{\bf v},
\end{equation}
where $\eta_0=g{m^4}/{h^3}$ is the maximum allowed value of the distribution
function $f({\bf r},{\bf v})$ fixed by Pauli's exclusion principle \cite{ptd}.
In the classical (nondegenerate) limit $f\ll \eta_0$, it reduces to the
Boltzmann entropy functional
\begin{equation}
\label{bf3}
S=-k_B\int\left ( \frac{f}{m} \ln
\frac{f}{\eta_0}-\frac{f}{m}\right )\, d{\bf r}d{\bf v}.
\end{equation}

The distribution function that maximizes the Boltzmann
entropy at fixed density $\rho$ and  energy
$E$ is the Boltzmann distribution 
\begin{equation}
\label{bf4}
f({\bf r},{\bf v})=\left (\frac{m}{2\pi k_B T}\right )^{3/2}\rho({\bf r})\,
e^{-\frac{mv^2}{2k_B T}},
\end{equation}
where $T$  is the temperature \cite{paddy,aa}. Using Eq. (\ref{bf4}),
the kinetic energy and the entropy of a nonrelativistic classical isothermal
self-gravitating system are
\begin{equation}
\label{bf5}
K=\frac{3}{2}Nk_B T
\end{equation}
and
\begin{eqnarray}
\label{bf6}
S=-k_B\int\frac{\rho}{m}\ln\rho\, d{\bf r}+\frac{3}{2}Nk_B\ln\left (\frac{2\pi
k_B T}{m}\right )\nonumber\\
+\frac{5}{2}Nk_B+Nk_B\ln\eta_0.
\end{eqnarray}

{\it Remark:} We note that the Boltzmann entropy defined by Eq.
(\ref{bf3}) diverges like $S\sim Nk_B\ln\eta_0\sim -3Nk_B\ln\hbar\rightarrow
+\infty$ when $\hbar\rightarrow 0$. This divergence is present in the famous
Sackur-Tetrode formula for the entropy of a perfect gas (without self-gravity).
In order to see the absence of statistical equilibrium states for classical
self-gravitating systems, marked by the divergence of the entropy when
$\hbar\rightarrow 0$, we first have to subtract the term $-3Nk_B \ln h$ from
the total entropy (see Appendix \ref{sec_thermoqmce}).

\subsection{Homogeneous sphere}

The potential (gravitational) energy of a spatially homogeneous sphere of mass
$M$ and
radius $R$ is \cite{bt}:
\begin{equation}
\label{bf7}
W=-\frac{3GM^2}{5R}.
\end{equation}
Using Eqs. (\ref{bf1}), (\ref{bf5}) and (\ref{bf7}), the total energy of a
nonrelativistic classical isothermal self-gravitating system with a
unifom density is
\begin{equation}
\label{bf8}
E=\frac{3}{2}Nk_B T-\frac{3GM^2}{5R}.
\end{equation}
Using Eq. (\ref{bf6}) its entropy is
\begin{eqnarray}
\label{bf9}
S_{B}=-Nk_B\ln \left (\frac{M}{V}\right )+\frac{3}{2}Nk_B\ln \left (\frac{2\pi
k_B T}{m}\right )\nonumber\\
+\frac{5}{2}Nk_B+Nk_B\ln\eta_0,
\end{eqnarray}
where $V=(4/3)\pi R^3$ is the volume of the system.

\subsection{Completely degenerate nonrelativistic self-gravitating Fermi gas}

The mass-radius relation of a completely degenerate fermion star
($T=0$) in the nonrelativistic limit is 
\begin{equation}
\label{bf10}
M_CR_C^3=\chi\frac{h^6}{g^2m^8G^3},\quad
R_C=\chi^{1/3}\frac{h^2}{g^{2/3}m^{8/3}GM_C^{1/3}}
\end{equation}
with
\begin{eqnarray}
\label{bf11}
\chi=\frac{1}{8}\left (\frac{3}{4\pi}\right
)^2\frac{\omega_{3/2}}{16\pi^2}=5.97241\times 10^{-3},
\end{eqnarray}
where $\omega_{3/2}=132.3843$ \cite{chandrabook}.

Its energy (kinetic $+$ potential) is 
\cite{chandrabook}:
\begin{equation}
\label{bf12}
E_C=-\frac{3GM_C^2}{7R_C}.
\end{equation}
Combined with the mass-radius relation (\ref{bf10}), we get
\begin{equation}
\label{bf13}
E_C=-\frac{3}{7\chi^{1/3}}\frac{G^2M_C^{7/3}g^{2/3}m^{8/3}}{h^2}.
\end{equation}
This is the energy of the ground state.

{\it Remark:} In terms of dimensionless variables \cite{pt}, the mass-radius
relation can be written as
\begin{eqnarray}
\label{bf14}
\frac{R_C}{R}=\frac{1}{\lambda\alpha_C^{1/3}\mu^{2/3}},
\end{eqnarray}
where $\alpha_C=M_C/M$ and 
\begin{equation}
\label{bf15}
\lambda=\frac{1}{(512\pi^4\chi)^{1/3}}=0.149736...
\end{equation}
Similarly, the energy-mass relation can be written as
\begin{eqnarray}
\label{bf16}
\Lambda_C=\frac{3}{7}\lambda\alpha_C^{7/3}\mu^{2/3}.
\end{eqnarray}
In writing these expressions, we have implicitly assumed that the fermion
star of mass $M_C$ and radius $R_C$ constitutes the core of a larger system of
mass $M$ and radius $R$ as in Appendix \ref{sec_thermo}.

\subsection{Ground state of a self-gravitating Fermi gas in a box}

In terms of dimensionless variables \cite{pt}, the minimum energy (ground
state) of a  nonrelativistic self-gravitating  Fermi gas enclosed within a box
is given by (see Eq. (\ref{bf16}) with $\alpha_C=1$):
\begin{equation}
\label{bf17}
\Lambda_{\rm max}=\frac{3}{7}\lambda\mu^{2/3}.
\end{equation}
This expression is valid for a self-confined fermion star such that
$R_C<R$ (i.e., the density of the fermion star vanishes before reaching the
box). Using Eq. 
(\ref{bf14}) with $\alpha_C=1$, we find that Eq. (\ref{bf17}) is valid for
$\mu>\lambda^{-3/2}=17.26$. When $\mu<17.26$, the fermion star at $T=0$ (ground
state) is
box-confined ($R_C>R$) and its energy $\Lambda_{\rm max}^b(\mu)$ is represented
in
Fig. 2 of \cite{ptd}.

Introducing the normalized variables of Appendix \ref{sec_dq} and using Eq.
(\ref{nl1}), we  find from Eq. (\ref{bf17}) that the minimum energy of a 
nonrelativistic self-gravitating  Fermi gas is given by 
\begin{equation}
\label{bf20}
\frac{\Lambda_{\rm max}}{R}=\frac{3}{7}\lambda \left
(\frac{4\sqrt{2}}{\pi}\right )^{2/3}N^{1/3}=0.0950\, N^{1/3}.
\end{equation}
This expression is valid for $N>91.9/R^3$ so that the fermion star is
self-confined ($R_C<R$). This
equation can be used to locate the vertical asymptote $\Lambda_{\rm max}$ in the
caloric curves of this paper. However, it is only valid in the nonrelativistic
regime $N\ll N_{\rm OV}$.  In the relativistic regime, the  minimum energy
$\Lambda_{\rm max}$ of the self-gravitating  Fermi gas, as well as the energy
$\Lambda'_{\rm max}$ of the unstable fermion star at $T=0$, are represented
in Fig. 14 of \cite{paper1}. At the point $N=N_{1}$ where the
second asymptote (corresponding to the unstable fermion star at $T=0$) appears,
we find that
\begin{equation}
\label{bf22}
\frac{\Lambda'_{\rm max}}{R}=-0.53617,\qquad \frac{\Lambda_{\rm
max}}{R}=0.0570.
\end{equation}
At the point $N=N_{\rm OV}$ where the two asymptotes meet each other, we find
that
\begin{equation}
\label{bf21}
\frac{\Lambda'_{\rm max}}{R}=\frac{\Lambda_{\rm max}}{R}=0.08985.
\end{equation}

\section{Temperature-dependent OV maximum
particle number}
\label{sec_tdov}

For $R=50$ and  $N\rightarrow
N_{\rm
OV}^+$, we find from
Fig. \ref{Xphase_eta_R50_defZOOM_new2PH} that
\begin{eqnarray}
\label{td1}
\eta'_c(N)\sim 0.104 \, (N-N_{\rm OV})^{-1/2}.
\end{eqnarray}
For a given normalized temperature
$\eta>\eta_c\simeq 2.52$, the system collapses towards a black hole when
$\eta'_c(N)<\eta$, i.e., when
$N\ge N_{\rm
OV}(\eta)$ with
\begin{eqnarray}
\label{td2}
N_{\rm
OV}(\eta)=N_{\rm
OV}+0.0108/\eta^{2}.
\end{eqnarray}
This can be seen as a temperature-dependent OV maximum
particle number. We note that $ N_{\rm
OV}(\eta)$ is very close to $N_{\rm OV}$ since $N_{\rm OV}(\eta_c)=N_c^{\rm
CE}=1.00427\, N_{\rm
OV}$ (see Sec. \ref{sec_crit}). 
The relation (\ref{td1}) remains valid, with a different prefactor, for other
values of $R>R_{\rm CCP}$. On the other hand, for $R_{\rm OV}<R<R_{\rm
CCP}$, we have a similar relation for $\eta_c(N)$ close to $N_{\rm OV}$ (see
Fig. \ref{Xphase2_R10_newPH} for $R=10$):
\begin{eqnarray}
\label{td3}
\eta_c(N)\sim 0.516 \, (N-N_{\rm OV})^{-1/2}.
\end{eqnarray}
Using the same argument as before, this yields
\begin{eqnarray}
\label{td4}
N_{\rm
OV}(\eta)=N_{\rm
OV}+0.266/\eta^{2}.
\end{eqnarray}
More generally, writing Eqs. (\ref{td1}) and (\ref{td3}) under the form
\begin{eqnarray}
\label{td5}
\eta_c^{(')}(N)\sim a(R) \, (N-N_{\rm OV})^{-1/2},
\end{eqnarray}
we get
\begin{eqnarray}
\label{td6}
N_{\rm
OV}(\eta)=N_{\rm
OV}+a(R)^2/\eta^{2}.
\end{eqnarray}
If we substitute $\eta=\beta GNm^2/R$ into Eq. (\ref{td6}) and replace $N$ by
$N_{\rm OV}$ at leading order, we obtain
\begin{eqnarray}
\label{td7}
N_{\rm
OV}(T_{\infty})=N_{\rm
OV}+a(R)^2\frac{R^2 (k_B T_{\infty})^2}{G^2N_{\rm OV}^2m^4}.
\end{eqnarray}
We make the guess that the product $a(R)R$ in Eq. (\ref{td7}) is independent of
$R$. This can be checked on the two values that we have computed since the
products $0.104\times 50=5.2$ and $0.516\times 10=5.16$ are almost the same. As
a result, we guess that $a(R)(R/R_{\rm OV})\simeq 5.2/3.3569\simeq 1.5$, i.e.,
$a(R)\simeq 1.5(R_{\rm OV}/R)$. Substituting this relation into Eq. (\ref{td7}),
we finally obtain (using the results of
Appendix \ref{sec_dq}):
\begin{eqnarray}
\label{td5b}
N_{\rm
OV}(T_{\infty})\simeq N_{\rm
OV}+160 \left (\frac{k_B T_{\infty}}{mc^2}\right )^2.
\end{eqnarray}
This relation is expected to be valid for $k_B T_{\infty}\ll mc^2$. It gives
the first order correction to the OV maximum number due to thermal effects.

\end{document}